\documentclass[prd,aps,showpacs,preprintnumbers,amsmath,amssymb,nofootinbib,eqsecnum, preprintnumbers]{revtex4}  
\usepackage{graphicx}
\usepackage{epsf}
\usepackage{amsmath}
\usepackage{epstopdf}
\usepackage{bm}
\usepackage{color}
\usepackage{tabularx}
\usepackage{enumitem}
\usepackage{float}
\usepackage{array,booktabs}
\usepackage{footnote}
\usepackage{threeparttable}
\usepackage{graphicx} 
\usepackage{etoolbox}
\usepackage{hyperref}
\usepackage{amssymb,epsf}
\usepackage{latexsym}
\usepackage{epstopdf}
\usepackage{epsfig}

\begin{document}

\title{Criticality and extended phase space thermodynamics of AdS black holes in higher curvature massive gravity}
\author{Seyed Hossein Hendi$^{1,2}$\footnote{email address: hendi@shirazu.ac.ir},
Ali Dehghani$^1$\footnote{email address: ali.dehghani.phys@gmail.com}} \affiliation{$^1$
Physics Department and Biruni Observatory, College of Sciences,
Shiraz University, Shiraz 71454, Iran \\
$^2$ Research Institute for Astrophysics and Astronomy of Maragha
(RIAAM), P.O. Box 55134-441, Maragha, Iran}

\begin{abstract}
Considering de Rham-Gabadadze-Tolley theory of massive gravity coupled with (ghost free) higher curvature terms arisen from the Lovelock Lagrangian, we obtain charged AdS black hole solutions in diverse dimensions. We compute thermodynamic quantities in the extended phase space by considering the variations of the negative cosmological constant, Lovelock coefficients ($\alpha_{i}$) and massive couplings ($c_{i}$), and prove that such variations is necessary for satisfying the extended first law of thermodynamics as well as associated Smarr formula. In addition, by performing a comprehensive thermal stability analysis for the topological black hole solutions, we show in what regions thermally stable phases exist. Calculations show the results are radically different from those in Einstein gravity. Furthermore, we investigate $P-V$ criticality of massive charged AdS black holes in higher dimensions, including the effect of higher curvature terms and massive parameter, and find that the critical behavior and phase transition can happen for non-compact black holes as well as spherically symmetric ones. The phase structure and critical behavior of topological AdS black holes are drastically restricted by the geometry of event horizon. In this regard, the universal ratio, i.e. $\frac{{{P_c}{v_c}}}{{{T_c}}}$, is a function of the event horizon topology. It is shown the phase structure of AdS black holes with non-compact (hyperbolic) horizon could give birth to three critical points corresponds to a reverse van der Waals behavior for phase transition which is accompanied with two distinct van der Waals phase transitions. For black holes with spherical horizon, the van der Waals, reentrant and analogue of solid/liquid/gas phase transitions are observed.

\end{abstract}

\pacs{04.40.Nr, 04.20.Jb, 04.70.Bw, 04.70.Dy} \maketitle

\section{Introduction}

Einstein's General Relativity (GR, also known as Einstein gravity) has been astonishingly regarded as the most successful description of gravitation and a well supported by numerous experiments since was proposed \cite{PoundRebka1960,GenzelEckart1996,LIGO2016} (see also these reviews \cite{Will2006,review2015GR}). Theoretically inconsistency appears when GR is supposed to be reconciled with the laws of quantum physics for producing the theory of quantum gravity. From the experimental side, Einstein gravity has problem
 with the accelerated expansion of the universe in the large scale structure since it needs an unknown source of energy (the so called dark energy) captured by the cosmological constant \cite{Riess1998,Perlmutter1999,Huterer1999,PerlmutterPRL1999}. In this regard, various attempts have been made to find an alternative such that it modifies Einstein gravity in the large scales (IR limit). Massive gravity is one of the alternatives that modifies Einstein gravity by giving the graviton a mass, and provides a possible explanation for the accelerated expansion of the universe without requirement of dark energy component \cite{D'Amico 2011,Akrami2013,Akrami2015}. Assuming that gravitons are dispersed in vacuum like massive particles, gravitational waves' observation of the coalescence for a pair of stellar-mass black holes (GW170104) has bounded the graviton mass to ${m_g} \le 7.7 \times {10^{ - 23}}eV/{c^2}$ \cite{LIGO2017}. On the other hand, depending on the exact model of massive gravity, the graviton mass is typically bounded to be a few times the Hubble parameter today, i.e., ${m_g} \le \times {10^{ - 30}-10^{ - 33}}eV/{c^2}$, in which for graviton mass region $ m_{g}\ll 10^{ - 33}eV/{c^2}$ , its observable effects would be undetectable \cite{deRhamREVIEW2014} (for more details on different mass bounds see \cite{deRham2017}). Massive gravitons, if they exist, are yet to be found; but, according to the recent data of LIGO, such an assumption is experimentally logical and therefore deserves to be explored theoretically \cite{D'Amico 2011,Akrami2013,Akrami2015,LIGO2017,deRhamREVIEW2014,deRham2017}. \\

Depending on what features of GR is accepted unchanged, various theories of gravity have been created. Modification of GR is characterized by a deformation parameter such as Lovelock coefficients $\alpha_{i}$'s in Lovelock gravity (which determines the strength of higher curvature terms) and graviton mass parameter in massive gravity models. Based on the nature of deformation parameter, the original theory (GR) can be recovered by taking some limits (e.g. the zero limit of graviton mass parameter must recover GR and its associated outcomes). In order to have a generalized well-defined theory, we should take care of ghosts. Although the first linear version ofthe massive theory (i.e., the Fierz-Pauli model \cite{FierzPauli1939}) is ghost-free, it does not lead to linearized GR as the graviton mass goes to zero which is known as the vDVZ discontinuity \cite{vDVZ1970a,vDVZ1970b}. Vainshtein discovered that such discontinuity appears as a consequence of working with the linearized theory of GR \cite{Vainshtein1972}, and by employing the Stueckelberg trick it can be found that all of degrees of freedom introduced by the graviton mass do not decouple in the zero limit of graviton mass \cite{Hinterbichler2012Review}. On the other hand, Boulware and Deser showed some specific nonlinear models of massive gravity suffer from ghost instabilities however they could restore continuity with GR \cite{BDghost1972}. Eventually, the de Rham-Gabadadze-Tolley (dRGT) theory of fully nonlinear massive gravity resolved the ghost problem in four dimensions by adding higher order graviton self-interactions with appropriately tuned coefficients \cite{deRham2010Gabadadze, dRGT, deRham2012PLB,HassanRosen2012PRL}. The higher dimensional extension of the massive (bi)gravity has been discussed in \cite{TQDo2016massive,TQDo2016bigravity}, which confirms the absence of ghost fields using the Cayley- Hamilton theorem. Interestingly, in massive gravity framework, spherically symmetric black hole solutions were found in \cite{Koyama2011(PRL),Nieuwenhuizen2011(PRD)} and in the limit of vanishing graviton mass they go smoothly to the Schwarzschild and Reissner-Nordstr\"{o}m black holes. Furthermore, asymptotically flat black hole solutions were found in \cite{Gruzinov2011Mirbabayi}, but the curvature diverges near the horizon of these solutions. In this regard, black hole solutions with non-singular horizon were introduced in \cite{deRham2011BHs} with the identification of the unitary gauge to the coordinate system in which black hole has no horizon (for more details see \cite{deRham2011BHs}). The other interesting solutions related to the cosmology, gravitational waves and (time dependent) black holes were found in \cite{D'Amico 2011,HassanRosen2011,Chamseddine2011Volkov,Gumrukcuoglu2011Lin,Mohseni2011,deRham2011Pirtskhalava,Arraut2015,Rosen2017BHs} which will not discussed in this paper. Of interesting case for us is Vegh's black hole solution \cite{Vegh2013} in which the general covariance preserves in the "$t$" and "$r$" coordinates, but, is broken in the other spatial dimensions. In \cite{Cai2015}, this solution was generalized to the topological black holes in higher dimensions. Inspired by the interesting features of these solutions, black hole solutions of massive gravity coupled to the higher curvature terms, dilaton and nonlinear electromagnetic fields were constructed and studied in details \cite{Hendi2016Annalen-massive,Hendi2016J.X.Mo,Hendi2016CQG,Hendi2015BI-MassiveGravity,Hendi2016JHEP,Hendi2018Fortschr,Hendi2018Momennia,HendiMomennia2018}. \\

From the string theory point of view and also brane world cosmology perspective, Lovelock gravity \cite{Lovelock1971,Lovelock1972}, as a natural deformation of GR in higher dimensions \cite{Deruelle1990}, has essential role. The most important motivation to study such a theory is related to superstring theory models which lead to ghost-free nontrivial gravitational interactions in higher dimensions \cite{Zwiebach1985}. The low-energy limit of type II string theory and $E_{8}\times E_{8}$ heterotic superstring give rise to effective models of gravity in higher dimensions which contains higher powers of the Riemann curvature (e.g., ${R^2}\,,{R^3}\,,{R^{\mu \nu }}{R_{\mu \nu }},\,{R^{\mu \nu \gamma \delta }}{R_{\mu \nu \gamma \delta }},\,...$) in addition to the usual Einstein and cosmological constant terms \cite{ScherkSchwarz1974,WittenStromingerHarowitzCandelas1985,Gross1985(PRL)HeteroticString,GrossWitten1986}. It is notable that the ghost-free combinations of these terms are proportional to the Euler invariant \cite{Zwiebach1985,Zumino1986} which is exactly the same as the Lovelock Lagrangian \cite{Lovelock1971,Lovelock1972}. The Lagrangian of the Lovelock gravity is given by a sum of dimensionally extended Euler densities as
\begin{equation}
{\cal L} = \sum\limits_{k = 0}^{\left[ {(d - 1)/2} \right]} {{\alpha _k}{{\cal L}_k}} \,,\,\,\,\,{{\cal L}_k} = \frac{1}{{{2^k}}}\delta _{{\rho _1}\,{\sigma _1}\,...\,{\rho _k}\,{\sigma _k}}^{{\mu _1}\,{\nu _1}\,...\,{\mu _k}\,{\nu _k}}{R_{{\mu _1}\,{\nu _1}}}^{{\rho _1}\,{\sigma _1}}...\,{R_{{\mu _k}\,{\nu _k}}}^{{\rho _k}\,{\sigma _k}},
\end{equation}
in which $\delta _{{\rho _1}\,{\sigma _1}\,...\,{\rho _k}\,{\sigma _k}}^{{\mu _1}\,{\nu _1}\,...\,{\mu _k}\,{\nu _k}}$ and ${R_{{\mu _k}\,{\nu _k}}}^{{\rho _k}\,{\sigma _k}}$ are the
generalized totally antisymmetric Kronecker delta and the Riemann tensor respectively. In a 4-dimensional spacetime, the Lovelock theory of gravity reduces to GR, and in spacetime dimensions with $d\geqslant5$ the Gauss-Bonnet term appears, and for $d\geqslant7$ the third order Lovelock term has contribution besides the Einstein and Gauss-Bonnet terms. Lovelock gravity is also ghost free with second order field equations and admits black hole solutions and the associated thermodynamics as expected \cite{BoulwareDeser1985StringModels,Wheeler1986GBsolution,Wiltshire1986GB,Wiltshire1988GB,Myers1988Simon,Banados1994,Cai1999Lovelock,Cai2002AdSGaussBonnet,Cai2004dSGaussBonnet,Cai2004LovelockGravity,Neupane2002Cho,Odintsov2001Nojiri,Odintsov2002Nojiri,Dehghani2005Shamirzaie,Dehghani2006Mann,Dehghani2008Alinejadi,Asnafi2011,LovelockNEDs2015,LovelockRainbow2017}.\\

Of interesting case for theoretical physicists is the thermodynamic properties of black holes in comparison with ordinary systems in the nature. In fact, black hole mechanics obeys the same laws as the laws of thermodynamics \cite{BardeenCarterHawking1973}, and many investigations have confirmed this statement for more complicated black hole spacetimes in modified gravities. In addition, a wealth of results in the context of black hole thermodynamics have been presented which show black holes in Einstein gravity can imitate some thermodynamic properties of ordinary systems such as the van der Waals phase transition which represents a liquid/gas (first order) phase transition \cite{vanderWaals1873}, the reentrant phase transition in multicomponent fluid systems \cite{Hudson1904,Narayanan1994} and the triple point in solid/liquid/gas phase transition. The phase space of Schwarzschild-AdS black holes admits the so-called Hawking-Page phase transition \cite{HawkingPage1983} which is interpreted as a confinement/deconfinement transition in the dual boundary gauge theory (SYM plasma) \cite{Witten1998b}. Remarkably, Reissner-Nordstr\"{o}m-AdS (RN-AdS) and Kerr-Newman-AdS black holes possess a first order phase transition which closely resembles the well-known van der Waals phase transition in fluids \cite{EmparanChamblin1999a, EmparanChamblin1999b,Caldarelli2000KerrNewmanAdS}. Interestingly, Born-Infeld-AdS black holes as a nonlinearly electromagnetic generalized version of RN-AdS ones display a phase structure relating to the mass ($M$) and the charge ($Q$) of the black holes similar to the solid-liquid-gas phase diagram \cite{Fernando(2006)BI-AdS}. These considerations were done in the presence of the cosmological constant as a fixed parameter and recently are referred in literature as non-extended phase space. In fact, as stated in \cite{KubiznakMann2012}, these mathematical analogies are confusing since some black hole intensive (extensive) quantities have to be identified with a irrelevant extensive (intensive) quantities in the fluid system, for example, identification between the fluid temperature and the charge of the black hole is puzzling. \\

In thermodynamic systems, some quantities are thermodynamic variables and the others are fixed parameters which cannot vary. Only experiment can determine that a quantity (parameter) can vary or held fixed. From the theoretical perspective, one can assume the variation of a fixed parameter of a theory and then see its consequences. In this regard, the later mismatch between extensive and intensive quantities of the black hole and fluid systems can be solved if one treats the cosmological constant ($\Lambda$) as thermodynamic variable, i.e., pressure \cite{KubiznakMann2012}. This idea (which first established in \cite{Kastor2009CQG} and then developed in \cite{Dolan2011CQG1,Dolan2011CQG2,Dolan2011PRD,CveticKubiznak2011Gibbons-PRD,DolanKastorMann2013}) leads to extension of the phase space thermodynamics and the exact analogy between quantities of  black hole and liquid-gas systems at the critical point. For example, a transition occurs in $P-T$ plane for the both of RN-AdS (small/large) black hole and liquid-gas systems. In addition, the variation of $\Lambda$ in the first law of black hole thermodynamics solves the inconsistency between the Smarr formula and the tradition form of first law since in the presence of a fixed cosmological constant the scaling argument \cite{Kastor2009CQG} is no longer valid. This motivates consideration of the first law of black hole thermodynamics with varying $\Lambda$ which is referred as the extended phase space thermodynamics in community. Regarding the extended phase space thermodynamics, reentrant phase transition has been observed for Born-Infeld-AdS and singly spinning Kerr-AdS black holes in the context of Einstein gravity \cite{Mann2012Gunasekaran, Mann2012Altamirano}. For a black hole system, it is interpreted as large/small/large black hole phase transition. Moreover, the analogue of solid/liquid/gas phase transition were found for doubly spinning Kerr-AdS black holes which is interpreted as small/intermediate/large black hole transition \cite{Mann2014TriplePoint,MannSherkatghanad}. \\

The objective of this paper is to construct the higher curvature massive gravity in order to study the effects of higher order curvatures on the black hole solutions of massive gravity and investigate the associated criticality and thermodynamics in the extended phase space. Indeed, some thermodynamic features of black holes, e.g. universality ratio, may depend on the specific choice of the gravitational theory. Therefore it is so important to understand the effect of modified gravities. We select the Lovelock gravity up to third order (referred as TOL gravity) as the higher curvature framework for our investigations. When the Lovelock massive theory of gravity (LM gravity) is constructed, in principle, the parameters $\alpha$ (Lovelock coefficient) and $m$ (graviton mass) are considered as deformations of GR, and by taking the limits $m\rightarrow 0$ and $\alpha\rightarrow 0$, GR is naturally recovered. According to scaling argument, any dimensionful parameter in a given theory has a thermodynamic interpretation and as a result Smarr formula and the first law of black hole thermodynamics must be modified. According to this fact, thermodynamically, more interesting phenomena can take place in a more complicated theory of gravity such as Lovelock and massive gravities which have a finite number of dimensionful parameters. One can observe that modified gravities such as massive and Lovelock theories exhibit a rich black hole phase space structure with respect to the those counterparts in Einstein gravity. The existence of higher order curvatures based on the third order Lovelock (TOL) gravity can lead to critical behavior and phase transition for AdS black holes with hyperbolic horizon topology \cite{PV2014Lovelock, PV2014LovelockBI-Mo, PV2015LovelockBI-Belhaj, Frassino2014} in contrast to Einstein gravity which only spherically symmetric AdS black holes admits phase transitions. Remarkably, hyperbolic vacuum black holes in Lovelock gravity expose non-standard critical exponents at a special isolated critical point which are different from those of van der Waals ones \cite{DolanMann2014Lovelock}. Until writing this paper, a wealth of evidence has been indicating that all the black hole solutions in Einstein gravity in the presence of any matter field have the same critical exponents as the van der Waals fluid \cite{KubiznakMann2012,Mann2012Gunasekaran,HendiVahidinia2013,BI-AdSBH-PV2014}. Interestingly, a "$\lambda$-line" phase transition occurs for a class of AdS-hairy black holes with hyperbolic horizon in Lovelock gravity where a real scalar field is conformally coupled to gravity \cite{HennigarMann2017PRL}. In addition, for charged black branes, the inclusion of higher curvature gravities based on a generalized quasi-topological class could lead to phase transition and critical behavior with the standard critical exponents \cite{PV-BlackBranes-2017Hennigar}. These indications reveal the rich phase space structure of Lovelock gravity's black holes. On the other hand, in the massive gravity framework, phase transition and critical behavior could take place for all kinds of topological black holes \cite{Hendi2017Mann-PRD}. In this regard, the van der Waals and reentrant phase transitions were found for AdS black holes \cite{PVmassive2015PRD, Reentrant-dRGTmassive-2017}, and in the presence of Born-Infeld (BI) nonlinear electromagnetic fields, the triple point emerges and the corresponding large/intermediate/small transition could take place \cite{Triple-BI-massive-2017}. \\

Taking these considerations seriously, in this paper, we mainly focus on the critical behavior and phase transitions of AdS black hole solutions in the Lovelock massive (LM) gravity. By constructing this model, besides its novel phase structure, we could be able to figure out what characteristic features of Lovelock and massive gravities persist or ruin. Thus, we have organized this paper as follows: First, in Sec. \ref{general formalism} , we give a brief review of thermodynamics in extended phase space, stability analysis and phase transition for AdS black holes in the context of Einstein gravity. After, in Sec. \ref{Lovelock-massive}, we construct the  LM gravity by introducing the action and associated filed equations and then present a new class of  charged-AdS black hole solutions in arbitrary dimensions. By computing thermodynamic quantities, we prove the traditional first law of black hole thermodynamics is satisfied. After that, in Sec. \ref{phase transition-massive Lovelock}, we perform a thermal stability analysis for the obtained black hole solutions in the canonical ensemble. Furthermore, we reconsider the first law of thermodynamics in the extended phase space and then study $P-V$ criticality and phase transition(s) of black holes to complete our discussion. Finally, In Sec. \ref{conclusion}, we finish our paper with some concluding remarks.\\


\section{General Formalism: Thermodynamics, stability and phase transition for AdS black holes \label{general formalism}}

In this section, we will develop the basic framework that we need to study the critical behavior and thermodynamic properties of AdS black holes in the next sections. Some useful issues will be briefly reviewed such as gravitational partition function, black hole thermodynamics, local thermodynamic stability, phase transition and critical behavior of black holes. Throughout this paper, we use the geometric units, ${G_{N}}=\hbar =c={k_{\rm{B}}}=1$. In these units,
$[\rm{Energy}]=[\rm{Mass}]=[\rm{Length}]^{-1}=[\rm{Time}]^{-1}$, and therefore there is only one dimensionful unit. Moreover, our convention of metric signature is $(-,+,+,+,...)$.

\subsection{Basic set up: partition function and action \label{general formalism}}

According to an old idea of unification, it is believed that all known forces (strong, weak, electromagnetisms and gravitation) in the nature might be unified in the so called "theory of everything". For many years, physicists have been looking for a consistent theory which all forces in the nature to be eventually described using path integral formalism of quantum field theory (QFT), like QCD and electroweak theories. Respecting such approach, we expect generating functional of quantum theory of gravity could be defined by an Euclidean path integral over a dynamical metric (tensor field), ${g_{\mu \nu }}$, as follows
\begin{equation} \label{generating functional}
{\cal Z} = \int {{\cal D}[g,\phi]} \,{e^{ - {{\cal I}_G}[g,\phi]}} \simeq {e^{ - {{\cal I}_G}}}\,(\rm{on-shell}),
\end{equation}
where $\phi$ is considered as matter fields and ${\cal I}_G$ represents the on-shell gravitational action which is obtained by substituting the classical solutions of $g$. The generating functional $\cal Z$ contains a complete summary of the theory which its dominant contribution originates from classical solution of the action by applying the stationary phase method (also known as steepest descent method or saddle-point approximation). Since the Euclidean formalism is obtained by applying the Wick rotation (${t_E} = it$) on the Lorentzian version, the Euclidean spacetime would be periodic in time. Following the method proposed by Matsubara \cite{Matsubara1955}, one can use the mapping

\begin{equation} \label{filed/statistical identification}
\frac{{it}}{\hbar } = \frac{{{t_E}}}{\hbar } \leftrightarrow \beta  = \frac{1}{{{k_B}T}},
\end{equation}
to calculate the partition function of a thermodynamic system applying the techniques of calculus in QFT, and vice versa (for more specific details see \cite{Kaputsa2006-book,Zee2010-book,Zinn-Justin1996-book}). As a result, considering the substitution (\ref{filed/statistical identification}), it is natural to regard $\cal Z$ as the statistical mechanical partition function of a gravitational system such as black hole. Comparing the eq. (\ref{generating functional}) with the free energy ($F =  - {k_B}T\ln {\cal Z}$), we find our main relation as
\begin{equation}
{{\cal I}_G} = \beta F.
\end{equation}
It should be noted that $F$ can be identified with Helmholtz or Gibbs free energy depending on the ensemble and thermodynamic variables of the system. Therefore, thermodynamic quantities associated to the gravitational system can be directly extracted by using of the standard methods in statistical mechanics \cite{Huang2009-book,Natsuume2015-book}.

For systems in the nature, one can define an appropriate action which its dynamical equation of motion is determined by the variational principle. The total action (${\cal I}_G$) for a gravitational system, here black holes, consists of three terms as
\begin{equation} \label{total action}
{{\cal I}_G} = {{\cal I}_b} + {{\cal I}_s} + {{\cal I}_{ct}},
\end{equation}
where ${\cal I}_b$,  ${\cal I}_s$ and ${\cal I}_{ct}$ are, respectively, called the bulk action, the surface term (boundary action), and the counterterm action. The surface term is needed to have well-defined variational principle and remove the derivative terms of ${g_{\mu \nu }}$ normal to the boundary, and the counterterm actually is an alternative surface term that comes from renormalization method in QFT to eliminate UV divergences and only can regulate asymptotically AdS spacetimes \cite{Kraus1999Balasubramanian,Kraus1999Larsen}. For asymptotically flat spacetimes, one can use the subtraction method to cancel divergences \cite{BrownYork1993,BrownMann1994}. A finite number of surface terms and counterterms are always needed to have a set of well-defined field equations and a finite total action.

Since we intend to study critical behavior of black holes and understand the theory dependence behind this phenomenon, we devote the rest of this paper to explore in these objects. Of interesting case is $d$-dimensional topological AdS black holes with the following metric ansatz
\begin{equation} \label{line element}
d{s^2} =  - \psi (r)d{t^2} + \frac{{d{r^2}}}{{\psi (r)}} + {r^2}{h_{ij}}d{x_i}d{x_j}\,\,\,\,\,\,(i,j = 1,2,3,...,d-2),
\end{equation}
where the line element ${h_{ij}}d{x_i}d{x_j}$ is the metric of $(d-2)$-dimensional (unit) hypersurface with the constant curvature $d_{1}d_{2}k$ and volume $\omega_{d_{2}}$ with the following forms
\begin{equation}
{h_{ij}}d{x_i}d{x_j} =\left\{
\begin{array}{cc}
dx_1^2 + \sum\limits_{i = 2}^{d_{2}} {\prod\limits_{j = 1}^{i - 1} {{{\sin }^2}{x_j}dx_i^2} }  & k=1 \\
dx_1^2 + {\sinh ^2}{x_1}dx_2^2 + {\sinh ^2}{x_1}\sum\limits_{i = 3}^{d_{2}} {\prod\limits_{j = 2}^{i - 1} {{{\sin }^2}{x_j}dx_i^2} } & k=-1 \\
\sum\limits_{i = 1}^{d_{2}} {dx_i^2} & k=0%
\end{array}%
\right.,
\end{equation}
in which $d_{i}=d-i$ (in what follows we will use this notation). The different values of the topological factor (i.e., $k=-1, 0, +1$) determine the topology of event horizon and could be positive (spherical, ${S^{n}}$), zero (Planar, $R^{n}$), or negative (hyperbolic, $H^{n}$). The details of metric function $\psi (r)$ depends on the theory that we pick out. In this paper, we always consider our line element ansatz as above.

Now, we focus on the charged (static) AdS black holes in Einstein's GR, briefly. The bulk action for the Einstein gravity on the $d$-dimensional background manifold ${\cal M}$ in the presence of negative cosmological constant and (Maxwell) electromagnetic filed is
\begin{equation} \label{buck action - Einstein}
{{\cal I}_b} =  - \frac{1}{{16\pi }}\int_{\cal M} {{d^d}x\sqrt { - g} (R - 2\Lambda  - {\cal F})},
\end{equation}
where $g$ is the determinant of metric tensor ${g_{\mu \nu }}$, $\Lambda  = -(d - 1)(d - 2)/2{\ell ^2}$ with the AdS radius $\ell$, and  ${\cal F} = {F^{\mu \nu }}{F_{\mu \nu }}$ is the Maxwell invariant in which ${F_{\mu \nu }} = {\partial _\mu }{A_\nu } - {\partial _\nu }{A_\mu }$ is the electromagnetic field tensor (Faraday tensor) with the gauge potential ${A_\mu }$. The Einstein bulk action has to be accompanied by boundary action(s) and counterterm(s). The Gibbons-Hawking boundary action and the counterterm for regulating divergences of the Einstein bulk action have been introduced in  \cite{GibbonsHawking1977} and \cite{Kraus1999Larsen}, respectively. In this paper, we leave out the details of these terms and refer the readers to the above references where the relevant details can be found.

Gravitational field equations are obtained by varying eq. (\ref{buck action - Einstein}) with respect to the metric tensor ${g_{\mu \nu }}$ as
\begin{equation} \label{gravitational field equation - Einstein Maxwell}
{G_{\mu \nu }} + \Lambda {g_{\mu \nu }} =  - \frac{1}{2}{g_{\mu \nu }}{\cal F} +2 {F_{\mu \lambda }}{F_\nu }^{\,\lambda },
\end{equation}
where ${G_{\mu \nu }} = {R_{\mu \nu }} - \frac{1}{2}{g_{\mu \nu }}R$ is the Einstein tensor.

Considering the line element ansatz (\ref{line element}), the so called RN-AdS black hole solutions of gravitational field equations (\ref{gravitational field equation - Einstein Maxwell}) are given by
\begin{equation} \label{metric function RN-AdS}
\psi {(r)_{Einstein}} = k - \frac{{2\Lambda {r^2}}}{{{d_1}\,{d_2}}} - \frac{{{m_0}}}{{{r^{{d_3}}}}} + \frac{{2{q^2}}}{{{d_2}\,{d_3}\,{r^{2{d_3}}}}},
\end{equation}
where $q$ and $m_{0}$ are integration constants related to the electric charge and finite mass of the black holes, respectively. Calculating the Kretschmann scalar, one can find an essential singularity. In fact, the Kretschmann scalar diverges only at $r =0$ which is covered with an event horizon and thus one can interpret it as a black hole. The larger root of $\psi (r_{+})=0$ with positive slope determines the event horizon of black holes. In Einstein gravity, numerical calculations show the metric function may have two real positive roots (Reissner-Nordestr\"{o}m black holes), one extreme root (extreme black holes) or it may be positive definite (naked singularity). These results still hold for higher curvature gravities such as the Lovelock gravity \cite{Dehghani2005Shamirzaie,Dehghani2006Mann,Dehghani2008Alinejadi,LovelockNEDs2015}, but situation is different for the massive gravity which the metric function may have more than three roots \cite{Hendi2015BI-MassiveGravity,Hendi2016Annalen-massive,Hendi2016JHEP} as shown in section \ref{blackhole}.


\subsection{Black hole thermodynamics \label{thermodynamics}}

Since the four laws of black hole mechanics were formulated in the pioneering paper \cite{BardeenCarterHawking1973}, a number of evidence has been indicating that there exists a sort of analogy between black hole mechanics and the standard thermodynamics in ordinary systems. Motivated by this analogy, one can regard charged (static) black holes as thermodynamic systems with the following first law
\begin{equation}
dM = TdS+ \Phi dQ,
\end{equation}
in which a formal equivalence between the physical temperature $T$ and the surface gravity ($\kappa$) were established in \cite{Zeldovich1972,Hawking1975} by studying quantum fields and particle creation near the event horizon for different observers in the black hole background. The notion of black hole, like ordinary thermodynamical systems, should be assigned an entropy and a temperature. Entropy of the black holes can be calculated by
\begin{equation} \label{entropy-area law}
{S_{{\rm{Einstein}}}} = \frac{1}{4}\int {{d^{d - 2}}x\sqrt {\tilde g} }  = \frac{A}{4},
\end{equation}
where $\tilde g$ is the determinant of the induced metric ${\tilde{g}_{\mu \nu }}$ on $(d-2) $-dimensional boundary and $A$ denotes its area. Entropy formula (\ref{entropy-area law}) satisfies the so-called area law ($S = A/4$) in Einstein gravity and this may be modified for higher derivative gravities (see Refs. \cite{Dehghani2005Shamirzaie,Dehghani2006Mann,Dehghani2008Alinejadi} for the Lovelock gravity). In addition, respecting the first law of thermodynamics, for all gravitational theories, one can always obtain the entropy using the following relation
\begin{equation}
S = \int_{}^{{r_ + }} {\frac{1}{T}\left( {\frac{{\partial M}}{{\partial {r_ + }}}} \right)} \,d{r_ + },
\end{equation}
which needs the functional forms of mass ($M$) and temperature ($T$). Besides, one can use the definition of the surface gravity with the Killing vector $\chi$ as
\begin{equation} \label{surface gravity}
\kappa  = \sqrt { - \frac{1}{2}({\nabla _\mu }{\chi _\nu })({\nabla ^\mu }{\chi ^\nu })},
\end{equation}
to calculate the Hawking temperature as
\begin{equation} \label{temperature definition}
T = \frac{\kappa }{{2\pi }} = {\left. {\frac{1}{{4\pi }}\frac{{\partial \psi (r)}}{{\partial r}}} \right|_{r = {r_ + }}}.
\end{equation}
It is worth mentioning that $\chi  = {\partial _t}$ is the temporal Killing vector for static spacetimes. In order to define the electric potential, it is necessary to select a reference. Naturally, electric potential can be measured at the horizon with respect to infinity as a reference, i.e.,
 \begin{equation} \label{potential definition}
\Phi  = {\left. {{A_\mu }{\chi ^\mu }} \right|_{r \to \infty }} - {\left. {{A_\mu }{\chi ^\mu }} \right|_{r \to {r_ + }}}.
\end{equation}
Moreover, the electric charge is an extensive quantity corresponds to the potential as an intensive quantity. Using the Gauss' law as
\begin{equation} \label{charge definition}
Q = \frac{{{\omega _{{d_2}}}}}{{4\pi }}\int {{d^n}x{F^{\mu \nu }}d{A_{\mu \nu }}},
\end{equation}
and calculating the flux of the electromagnetic field at infinity, the elctric charge is obtained.

Finally, the mass $M$ of the static black hole can be written down as measured by a faraway observer using Ashtekar-Magnon-Das (AMD) formula as \cite{Ashtekar1984AMD,Ashtekar2000AMDmass,Emparan2008Reall(review)}
\begin{equation} \label{mass definition}
M = \frac{{{d_2}\,{\omega _{{d_2}}}}}{{16\pi }}{m_0},
\end{equation}
in which $m_{0}$ is a positive integration constant (see eq. \ref{metric function RN-AdS}) and easily obtained from $\psi (r_{+})=0$ (its details depends on the parameters of the gravitational theory). This formula still holds for higher curvature gravities like Lovelock theory \cite{Dehghani2005Shamirzaie,Dehghani2008Alinejadi,LovelockNEDs2015} and can be calculated using the behavior of the metric at large $r$ (for asymptotically flat black holes) or counterterm method (for asymptotically AdS black holes). In the following table, we summarize the analogy between the standard thermodynamics and black holes mechanics. 
\begin{center} 
	\begin{tabular}{|c|c|} 
		\hline \hline
		Standard thermodynamics	& Black hole variables\\ 
		\hline \hline
		Internal energy	& $M$ \,\,\,\ (Mass) \\ 
		\hline
		Temperature & $T=\kappa/2\pi$  \,\,\,\ (Surface gravity) \\
		\hline 
		Entropy & $S=A/4$  \,\,\,\ (Horizon area)\\
		\hline
	\end{tabular} 
\end{center}

Obviously, in traditional treatment of the first law of black hole thermodynamics, the work term "$PdV$" is missed. Recent developments \cite{Kastor2009CQG, Dolan2011CQG1, Dolan2011CQG2, Dolan2011PRD, CveticKubiznak2011Gibbons-PRD, KubiznakMann2012} show that one can extend the thermodynamic phase space and insert the volume-pressure term in the first law by use of proposing a definition of thermodynamic volume as follows
\begin{equation} \label{thermodynamic volume}
V = {\left( {\frac{{\partial H}}{{\partial P}}} \right)_{{X_i}}} = {\left( {\frac{{\partial M}}{{\partial P}}} \right)_{{X_i}}},
\end{equation}
in which the quantity $H$ is enthalpy of the black hole system and in our case $X_{i}$'s are the extensive quantities $Q$ and $S$. Regarding this, as implicitly pointed out in eq. (\ref{thermodynamic volume}), the mass of black hole has to be interpreted as the enthalpy.

Remarkably, a generalized first law of black hole thermodynamics could be obtained by treating the negative cosmological constant as a thermodynamic variable. As a result, one has to interpret the quantity $P =  - \Lambda /8\pi $ as the (positive definite) thermodynamic pressure of the black hole system. It has to be noted that the gravitational background necessarily will be an asymptotically anti de Sitter (AdS) spacetime to have a positive definite pressure. In conclusion, the first law of black hole thermodynamics in the extended phase space may be written as
\begin{equation} \label{extended first law}
dM = TdS + \Phi dQ + VdP.
\end{equation}
Now, it can be tested that the thermodynamic quantities of RN-AdS black holes satisfy the first law. Thermodynamic quantities for RN-AdS black holes read
\begin{equation} \label{temperature RN-AdS}
T = \frac{{{d_2}{d_3}kr_ + ^2 - 2\Lambda r_ + ^4 - 2{q^2}r_ + ^{ - 2{d_4}}}}{{4\pi {d_2}\,{r_ + }^3}},
\end{equation}
\begin{equation} \label{entropy - Einstein gravity}
S = \frac{{{\omega _{{d_2}}}}}{4}r_ + ^{{d_2}},
\end{equation}
\begin{equation} \label{charge - RN-AdS}
Q = \frac{{{\omega _{{d_2}}}}}{{4\pi }}q,
\end{equation}
\begin{equation} \label{potintial - RN-AdS}
\Phi  = \frac{q}{{{d_3}\,r_ + ^{{d_3}}}},
\end{equation}
and
\begin{equation}
M = \frac{{{d_2}\,{\omega _{{d_2}}}}}{{16\pi }}\left( {kr_ + ^{{d_3}} + \frac{{2{q^2}}}{{{d_2}\,{d_3}r_ + ^{{d_3}}}} - \frac{{2\Lambda r_ + ^{{d_1}}}}{{{d_1}\,{d_2}}}} \right).
\end{equation}

 We use the first law of thermodynamics to define temperature and electric potential as the intensive parameters conjugate to the entropy and electric charge. Computing $T = {\left( {\partial M/\partial S} \right)_{{X_i}}}$ and $\Phi = {\left( {\partial M/\partial Q} \right)_{{X_i}}}$, one can find the same quantities as (\ref{temperature RN-AdS}) and (\ref{potintial - RN-AdS}). Therefore, we deduce all intensive and corresponding extensive parameters satisfy the extended first law of thermodynamics as eq. (\ref{extended first law}).


\subsection{Thermal stability of black holes \label{stabilty}}

Here, we are going to explain the basic treatment for studying thermal stability of black holes. Analyzing the local thermal stability of a black hole can be performed in the both of canonical and grand canonical ensembles. It can be done by studying the behavior of entropy $S$ near a sufficiently small neighborhood of a given point in the phase space of possible extensive variable $X_{i}$. Entropy is a smooth function of the extensive variables which in our case these variables are $M$ and $Q$. Local thermal stability states that entropy must be a smooth concave function of the extensive variables ($S=S(M,Q)$). This is equivalent to have a negative definite value for the determinant of Hessian matrix of the entropy $S$, i.e. $\textbf{H}_{{X_i},{X_j}}^S \equiv \left[ {{\partial ^2}S/\partial {X_i}\partial {X_j}} \right]$ \cite{Gubser1999Cvetic}. Also, since one could invert $S=S(M,Q)$ picture to $M=M(S,Q)$ one, the stability criterion $\textbf{H}_{{X_i},{X_j}}^S \le 0$ may be expressed, differently, as the determinant of Hessian matrix of $M$ being positive definite \cite{Gubser2001Mitra}, i.e.
\begin{equation} \label{Hessian-canonical}
\left| {\textbf{H}_{{X_i},{X_j}}^M} \right| \equiv \left| {\frac{{{\partial ^2}M}}{{\partial {X_i}\partial {X_j}}}} \right| \ge 0.
\end{equation}
In what follows, we will use this later criterion for analyzing thermal stability of AdS black holes. 

In \textit{the canonical ensemble}, the total charges are held fixed and, consequently, the Hessian matrix has one component as $\textbf{H}_{S,S}^M \equiv \left[ {{\partial ^2}M/\partial {S^2}} \right]$. As we have shown below,
\begin{equation} \label{Hessian-heat capacity}
\textbf{H}_{S,S}^M = {\left( {\frac{{{\partial ^2}M}}{{\partial {S^2}}}} \right)_Q} = {\left( {\frac{{\partial T}}{{\partial S}}} \right)_Q} = \frac{T}{{{C_Q}}}\,\,\,\, \Rightarrow \,\,\,{C_Q} = T{\left( {\textbf{H}_{S,S}^M} \right)^{ - 1}}\geqslant 0,
\end{equation}
the Hessian matrix will be a function of the heat capacity $C_{Q}$. It is possible that we confront with a situation that the both of $T$ and $H_{S,S}^M$ being negative definite, and therefore, get a positive value for $C_{Q}$. In order to avoid this problem one has to always take care of positivity of the temperature and heat capacity, simultaneously. In conclusion, positivity of the heat capacity ensures the local thermal stability in the phase space of allowed physical black hole quantities with $T, M\geqslant 0$. 
 
In \textit{the grand canonical ensemble}, the chemical potentials are held fixed; therefore in our case, the Hessian matrix has the following explicit form
\begin{equation} \label{Hessian-grand canonical}
\textbf{H}_{{X_i},{X_j}}^M \equiv \left( {\begin{array}{*{20}{c}}
	{\frac{{{\partial ^2}M}}{{\partial {S^2}}}}&{\frac{{{\partial ^2}M}}{{\partial S\,\partial Q}}}\\
	{\frac{{{\partial ^2}M}}{{\partial Q\,\partial S}}}&{\frac{{{\partial ^2}M}}{{\partial {Q^2}}}}
	\end{array}} \right).
\end{equation}

\begin{figure}[!htbp]
	$%
	\begin{array}{ccc}
	\epsfxsize=5.5cm
	 \epsffile{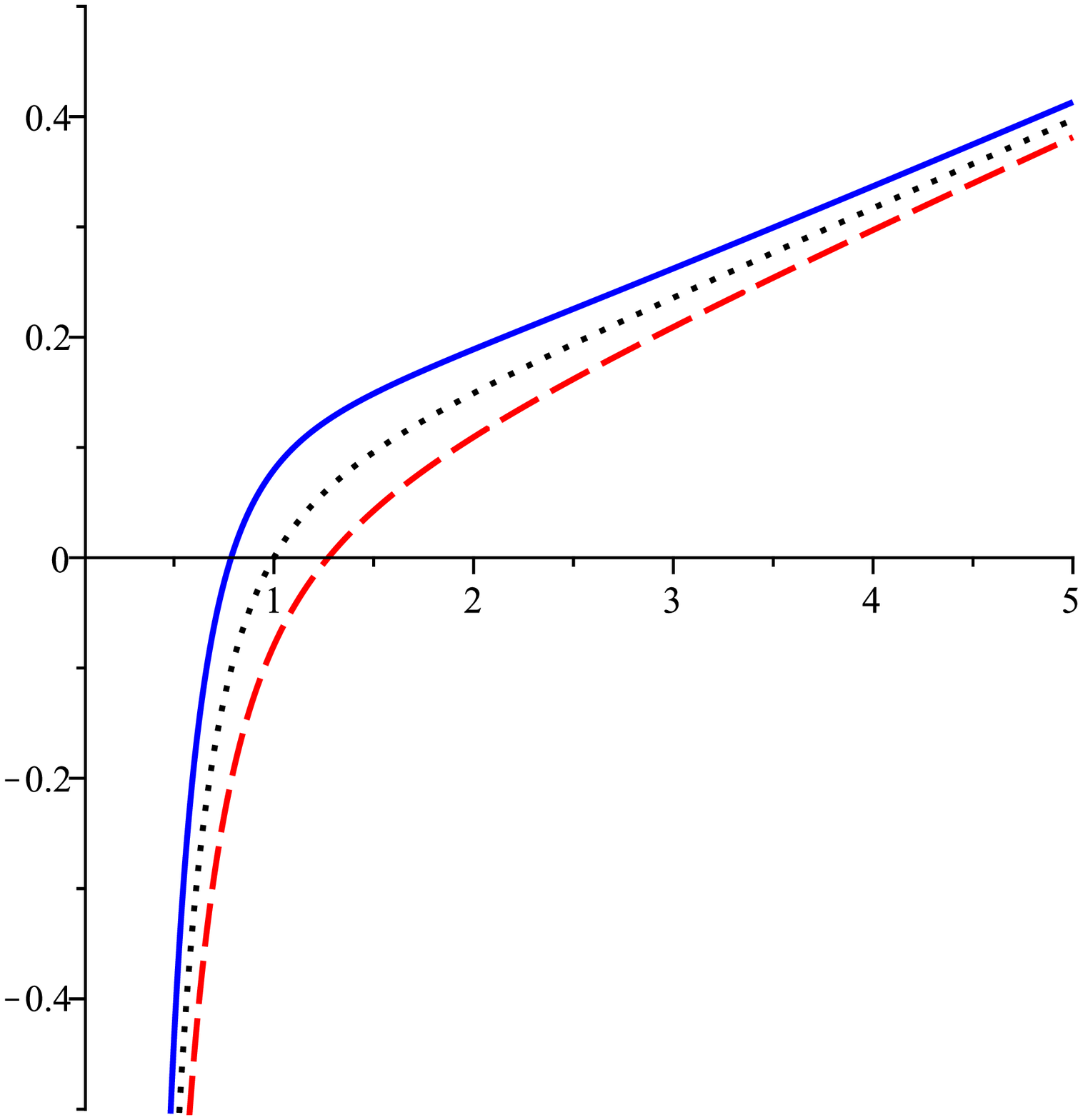}
	\epsfxsize=5.5cm
	\epsffile{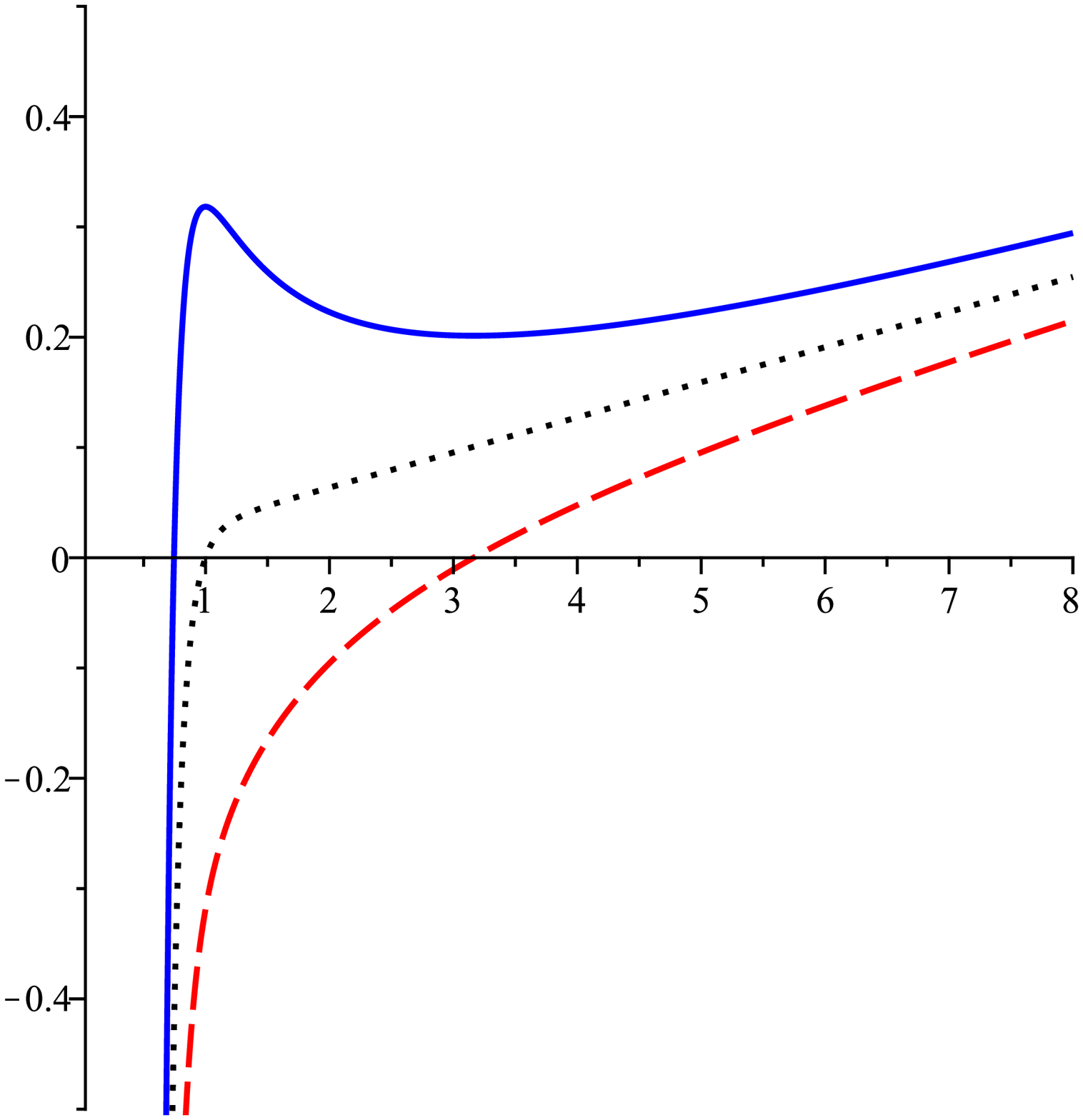}
	\epsfxsize=5.5cm
	 \epsffile{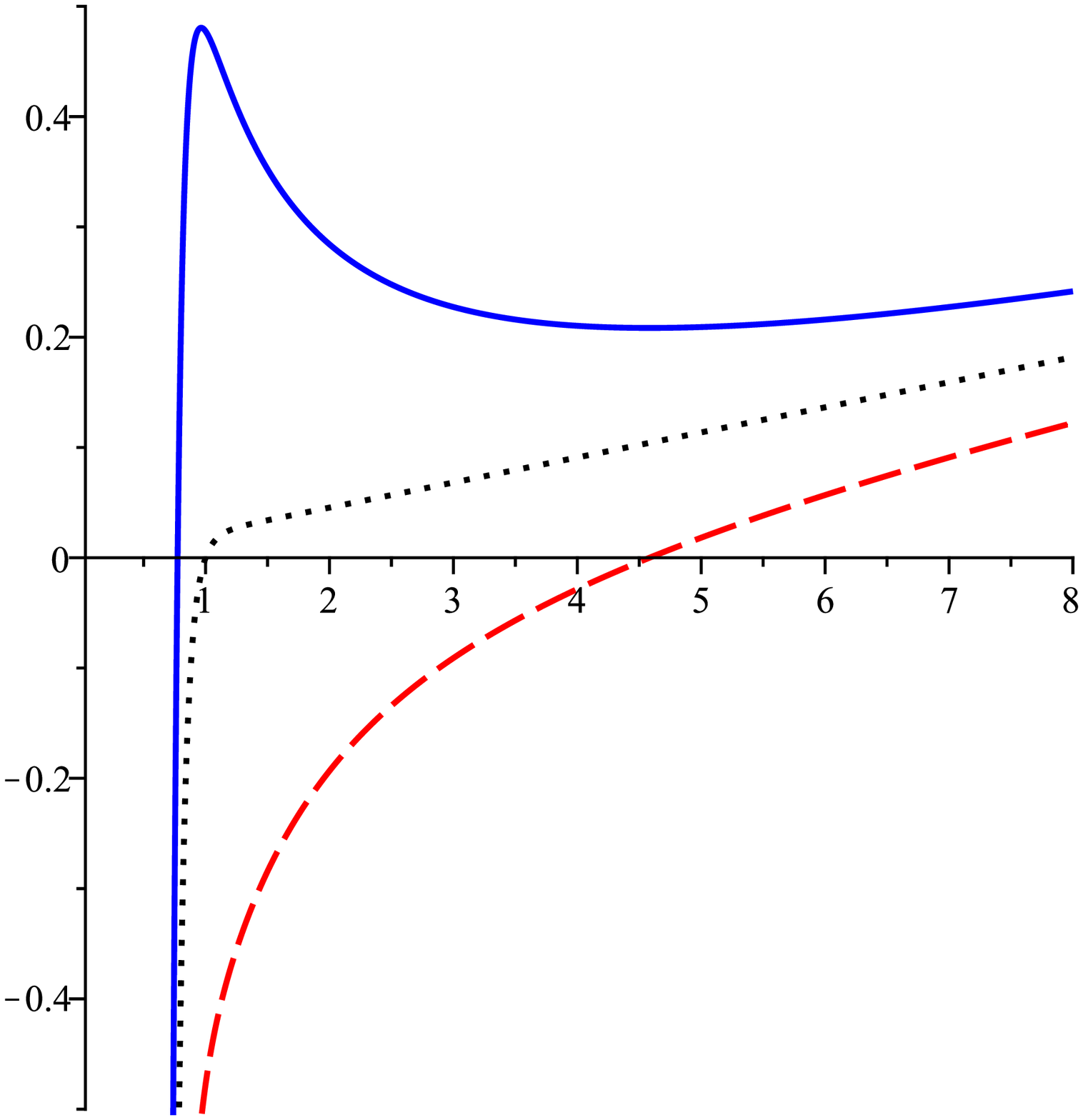}
	&  & 
	\end{array}
	$%
	\caption{ $T$ versus ${r_ + }$ for $q=1$, $\Lambda = - 1$, $k=+1$ (solid lines), $k=0$ (dotted lines) and $k=-1$ (dashed lines). $d=4$ (\textbf{left panel}), $d=7$ (\textbf{middle panel}) and $d=9$ (\textbf{right panel}).}
	\label{Temp-GR}

\end{figure}

Now, we perform a thermal stability analysis for RN-AdS black holes. To determine thermally stable regions for black holes, first, one has to find in what regions the associated temperature is positive. That could depend on the topology of event horizons. In Fig. \ref{Temp-GR}, typical behaviors of temperature are depicted for different horizon topologies ($k=1,0,-1$) in four and higher dimensions. Clearly, there always is a bound point for radius of event horizon ($r_{b}$) in which the temperature of black hole is positive for $r_{+}>r_{b}$. The topological type of event horizon ($k=1,0,-1$) can change the value of $r_{b}$. The more value for $k$ results into the lower value for $r_{b}$. In addition, in the region $r_{+}>r_{b}$ the mass of the black holes ($M$) is always positive.

In the canonical ensemble, the positivity of heat capacity ensures the local stability in regions where black hole temperature is positive. The heat capacity of RN-AdS black holes can be calculated as
\begin{equation} \label{heat capacity RN-AdS}
{C_Q} = \frac{{{d_2}\left( {2{q^2}r_ + ^{{d_2}} - r_ + ^{3d - 8}({d_2}{d_3}k - 2\Lambda r_ + ^2)} \right)}}{{4r_ + ^{2{d_3}}({d_2}{d_3}k + 2\Lambda r_ + ^2) - 16{d_{5/2}}{q^2}}}.
\end{equation}
 Using the heat capacity, one can obtain some information about phase transition. For example, divergence point of heat capacity is a sign of possible phase transition. Since temperature and heat capacity are both increasing functions of $r_{+}$, thus the large black holes are thermally stable. Here, we examine the heat capacity for RN-AdS black holes with different topological factor ($k$) in more details. This will complete our discussion of local thermal stability.\\
 
\begin{figure}[!htbp]
	$%
	\begin{array}{ccc}
	\epsfxsize=5.5cm
	 \epsffile{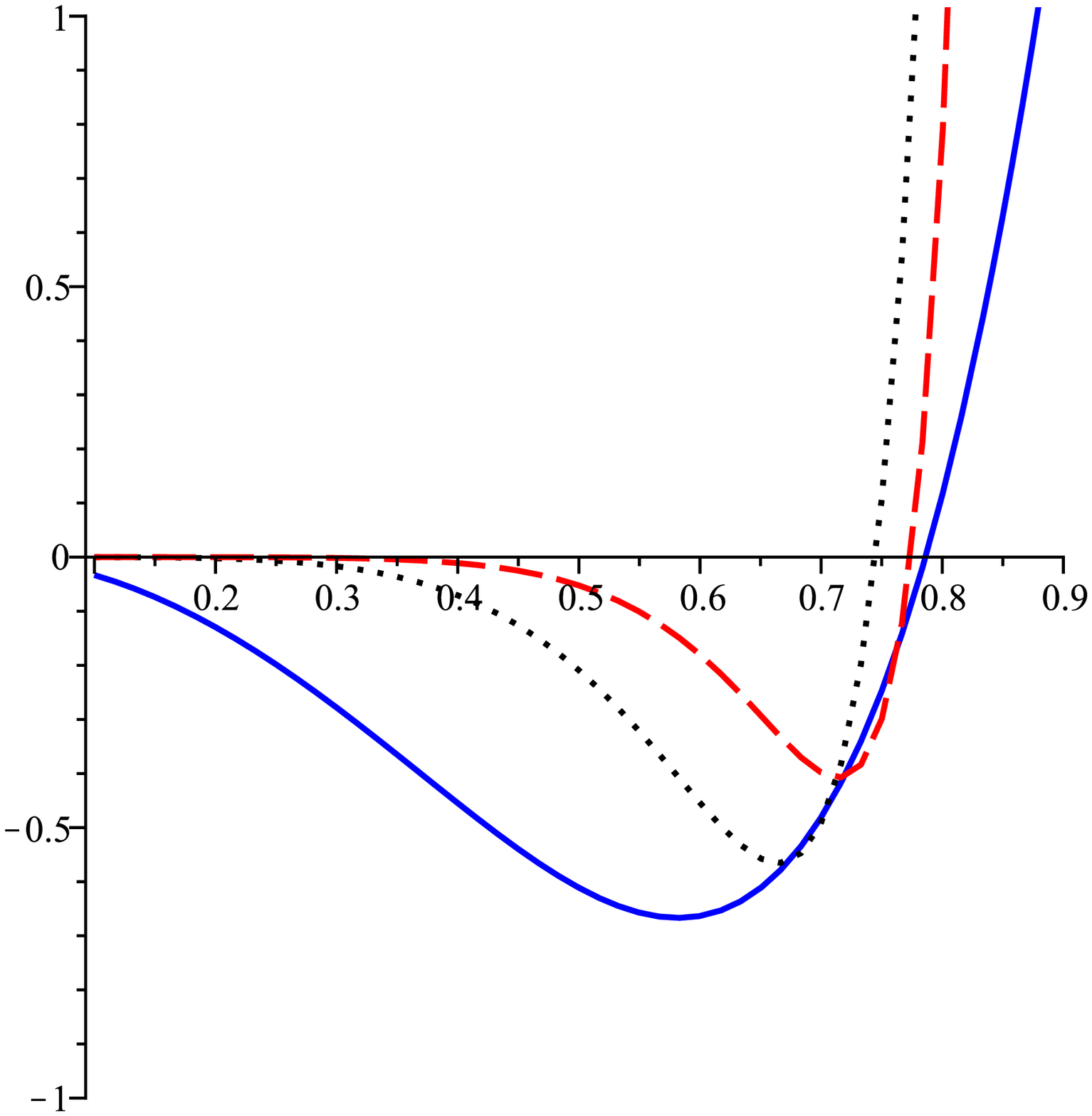}
	\epsfxsize=5.5cm 
	\epsffile{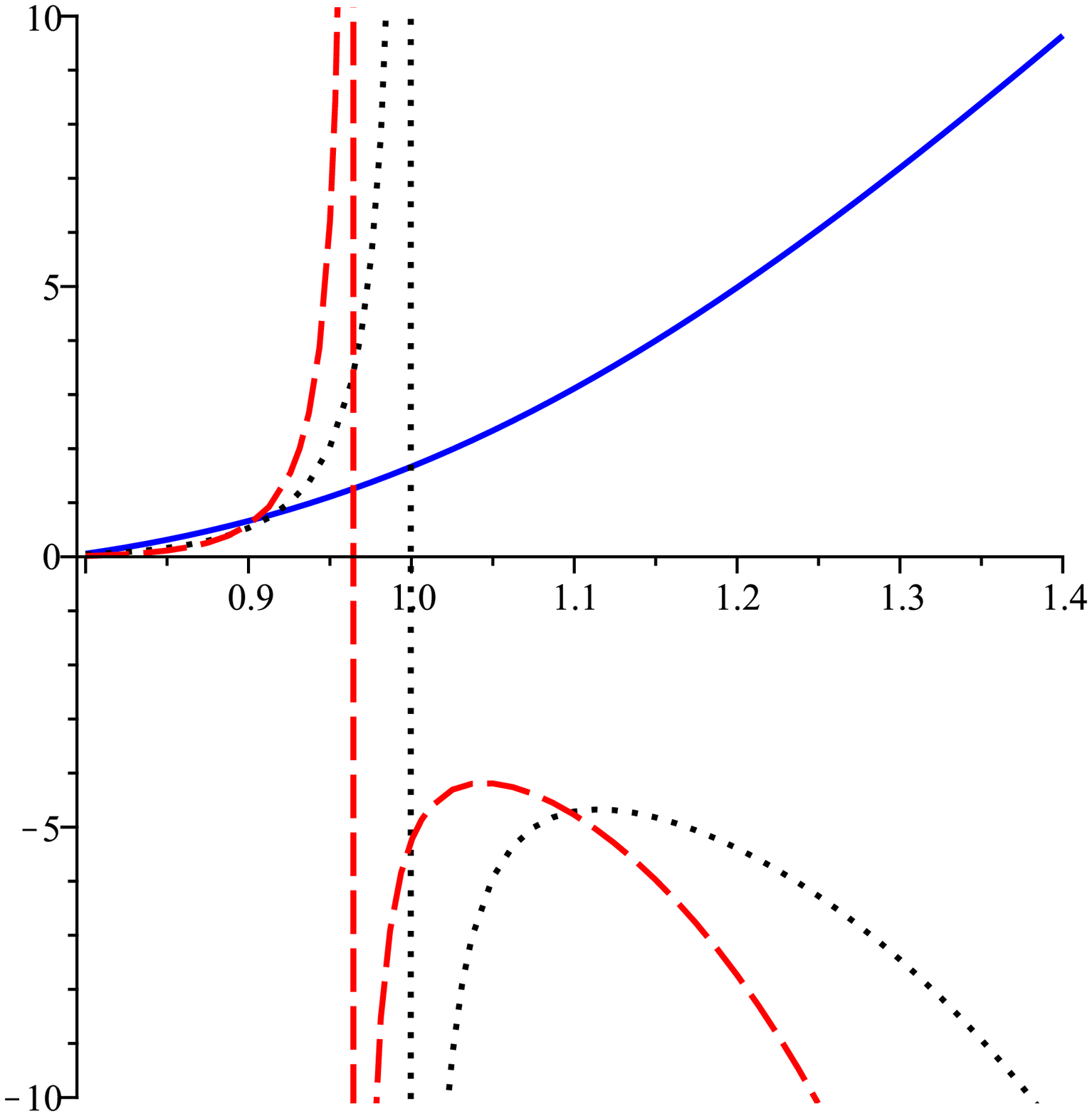}
	\epsfxsize=5.5cm 
	\epsffile{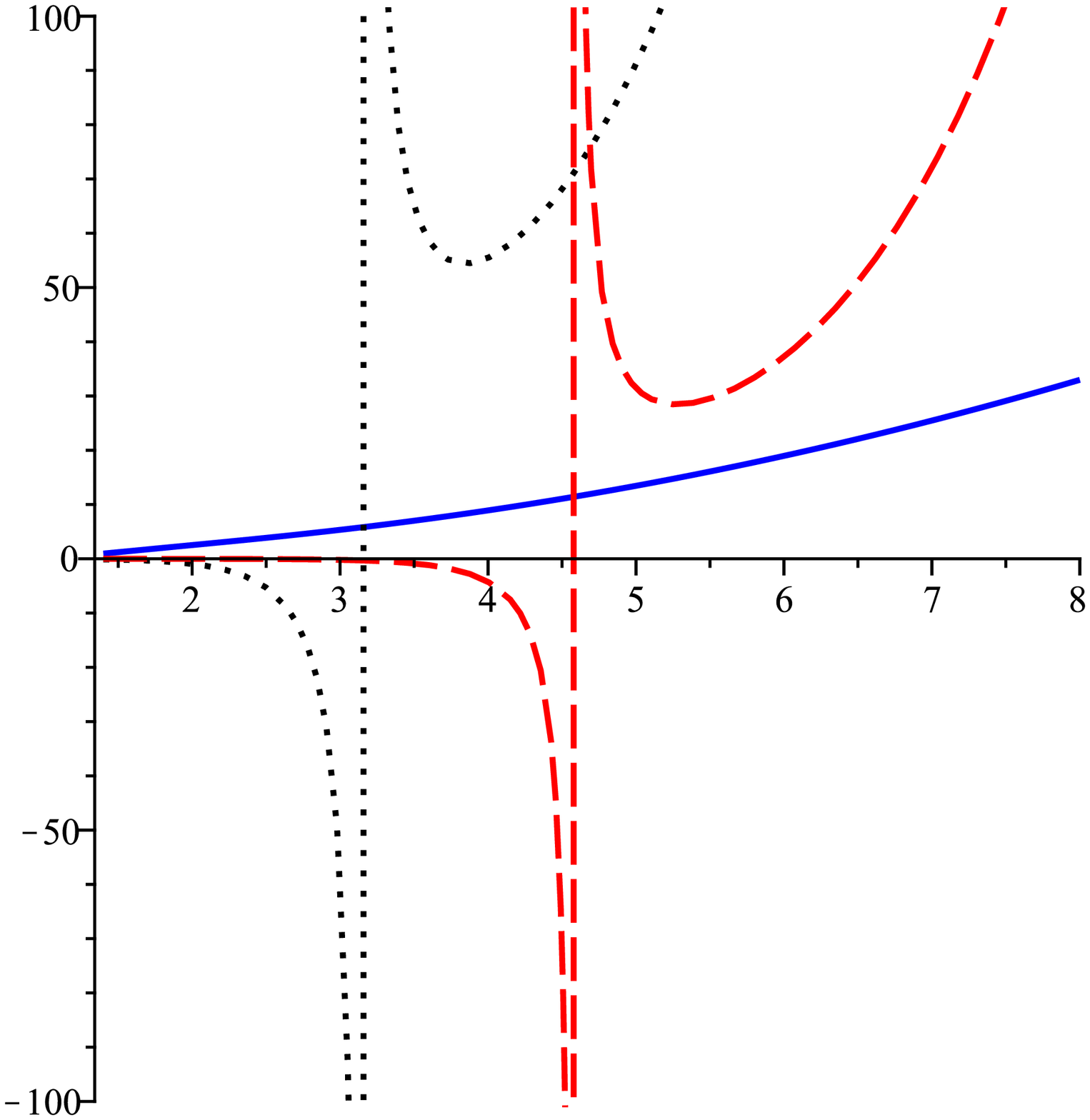}
	&  & 
	\end{array}
	$%
	\caption{ $C_{Q}$ versus ${r_ + }$ for $k=1$, $q=1$, $\Lambda = - 1$, $d=4$ (solid lines), $d=7$ (dotted lines) and $d=9$ (dashed lines).
			\textbf{Different scales:} \textit{left panel} ($0<r_{+}<0.9$), \textit{middle panel} ($0.8<r_{+}<1.4$) and \textit{right panel}
          	($1.4<r_{+}<8$).}
	\label{HC-GR-k=1}
\end{figure}

\textbf{Spherical black holes ($k=1$):}
In this case, the heat capacity has only one positive root ($r_{b}$) in which $C_{Q}$ is negative definite for regions $r_{+}<r_{b}$. For regions $r_{+}>r_{b}$, depending on the values of $q$ and spacetime dimensions ($d$), three possibilities may happen: i) $C_{Q}$ is an increasing function of the $r_{+}$, thus, in regions $r_{+}>r_{b}$, the heat capacity will be positive and black holes are thermally stable. ii) $C_{Q}$ may have two divergent points for the RN-AdS black holes, where between divergent points the heat capacity is negative definite (unstable black hole region), thus a thermal phase transition can happen. iii) $C_{Q}$ has one divergence point which is positive around such divergency. Such single divergence point, with positive $C_{Q}$ around it, may indicate critical behavior of the system. In Fig. \ref{HC-GR-k=1}, the possibilities of items i) and ii) are depicted. We refer to the first and second divergent points as $r_{m}$ and $r_{u}$ respectively ($r_{b}<r_{m}<r_{u}$). In regions $r_{b}<r_{+}<r_{m}$ and $r_{+}>r_{u}$, RN-AdS black holes are thermally stable (a phase transition could occur between these two thermally stable regions) and for the other regions are unstable.\\

\begin{figure}[!htbp]
	$%
	\begin{array}{ccc}
	\epsfxsize=5.5cm 
	\epsffile{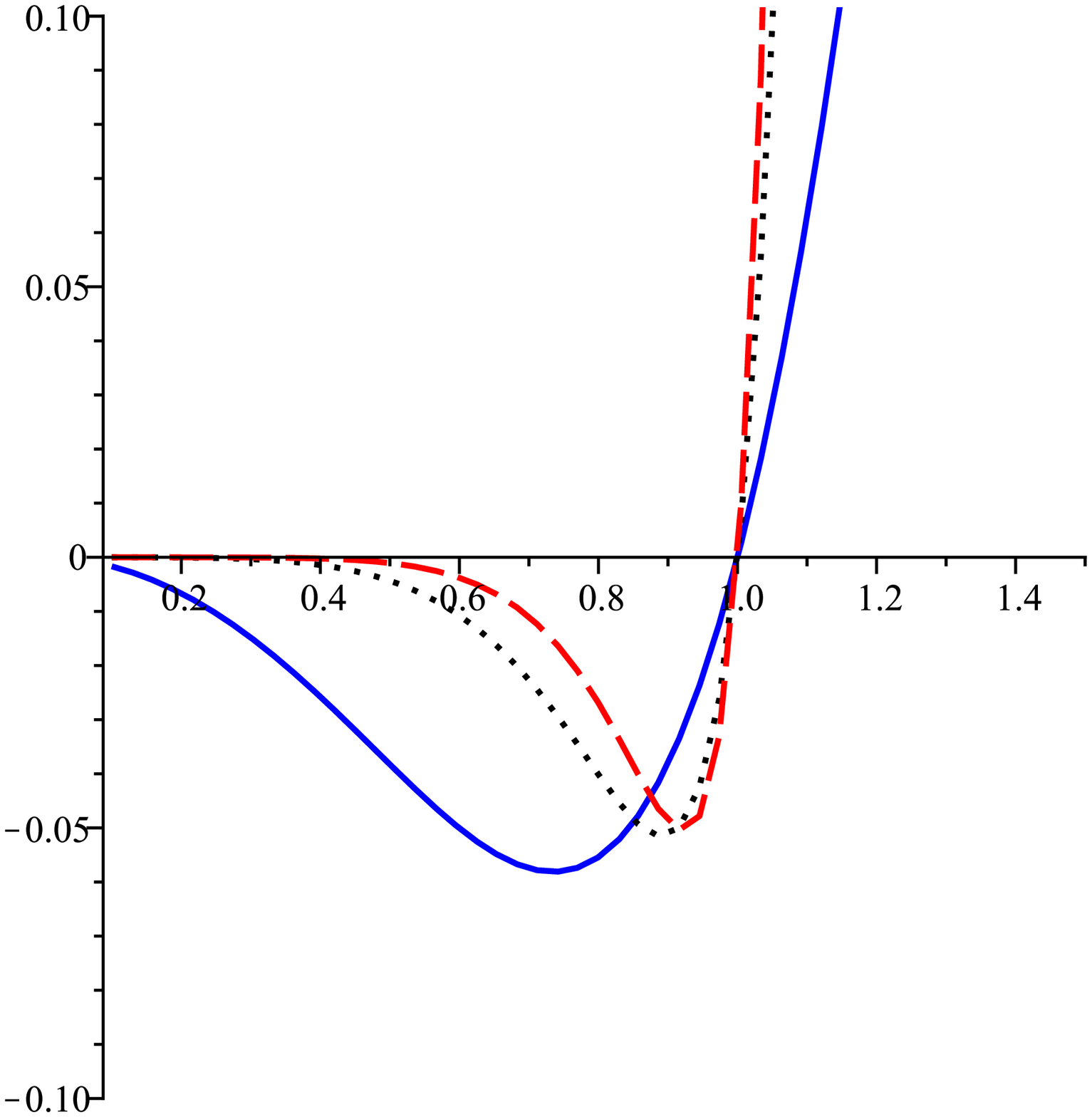}
	\epsfxsize=5.5cm 
	\epsffile{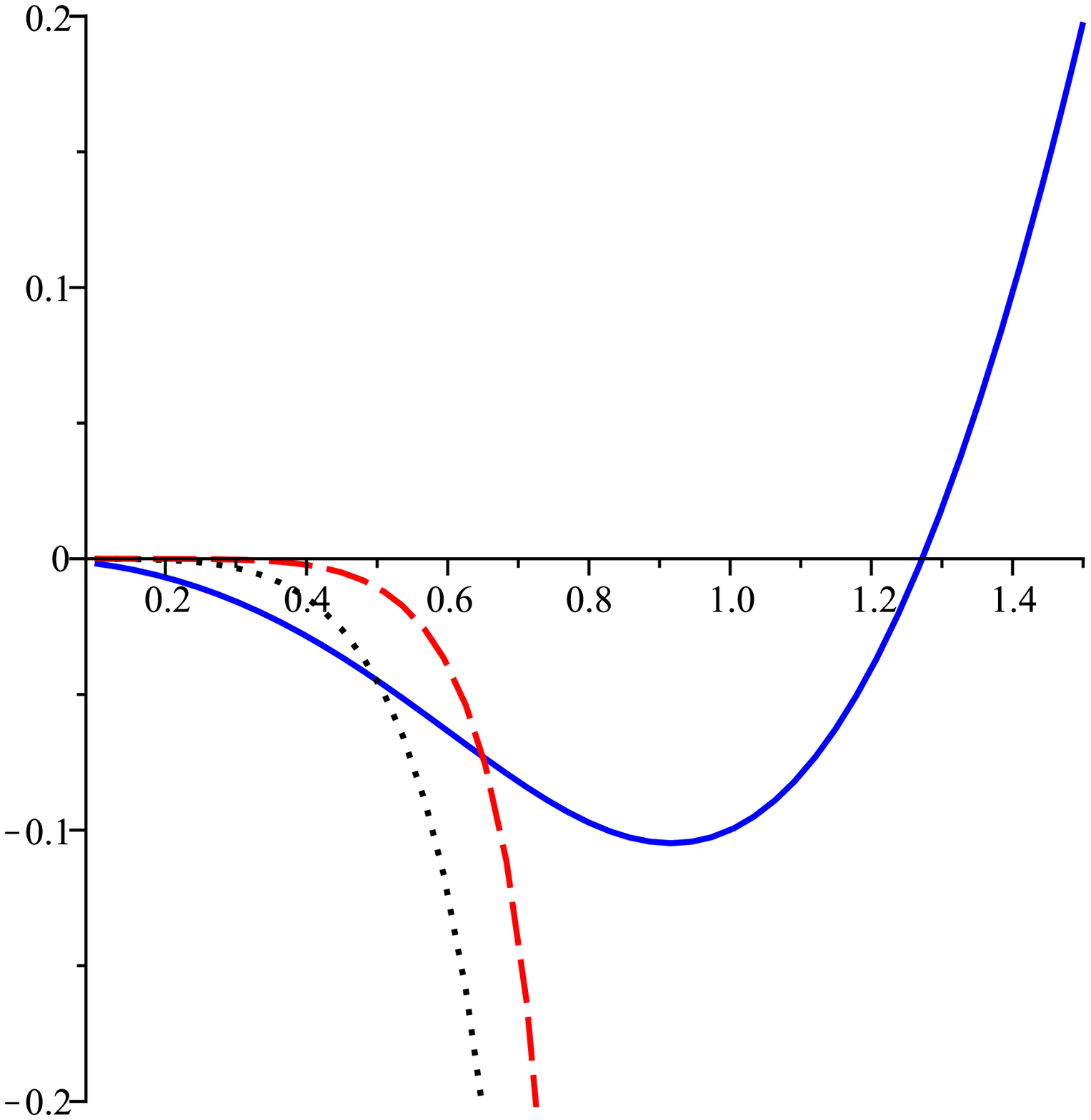}
	\epsfxsize=5.5cm 
	\epsffile{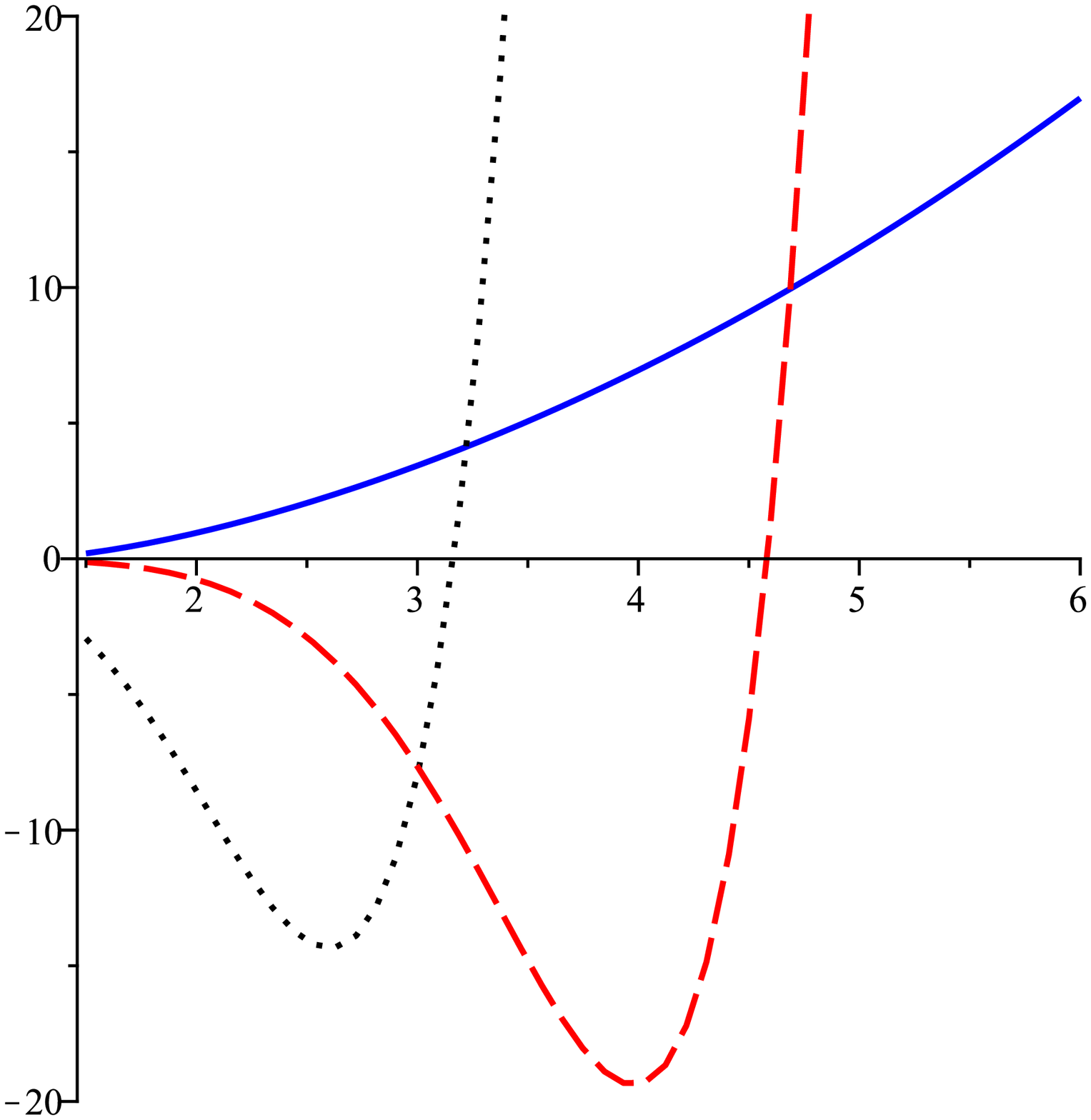}
	&  & 
	\end{array}
	$%
	\caption{ $C_{Q}$ versus ${r_ + }$ for $q=1$, $\Lambda = - 1$, $d=4$ (solid lines), $d=7$ (dotted lines) and $d=9$ (dashed lines).
	\textbf{Left panel} for $k=0$. \textbf{Middle panel} ($0<r_{+}<1.5$) and \textbf{right panel} ($1.4<r_{+}<6$) for $k=-1$ with different scales.}
	\label{HC-GR-k=0-1}
\end{figure}

\textbf{Ricci flat black holes ($k=0$):}
In this case, the heat capacity has only one positive root , $r_{b}$, in which $C_{Q}$ is negative definite for regions $r_{+}<r_{b}$ and positive for $r_{+}>r_{b}$. According to eq. \ref{heat capacity RN-AdS}, the heat capacity does not diverge for finite values of $r_{+}$ since the denominator of the heat capacity cannot have any root (see the left panel of Fig. \ref{HC-GR-k=0-1}). As a result, Ricci flat black holes are thermally stable for regions $r_{+}>r_{b}$ and no phase transition takes place.\\

\textbf{Hyperbolic black holes ($k=-1$):}
In this case, as well as Ricci flat black holes ($k=0$), there is only one root ($r_{b}$) for the heat capacity. Again, hyperbolic black holes are thermally stable in regions $r_{+}>r_{b}$ since the heat capacity is always positive and no phase transition takes place (see the middle and right panels of Fig. \ref{HC-GR-k=0-1}).

\subsection{Phase transition for RN-AdS black holes \label{PV for RN-AdS}}

In this section, we review the essence of critical behavior and phase transition in AdS black holes. In recent years, the interesting analogy between liquid-gas and small/large black hole phase transitions have attracted the attention of many authors. In this regard, the exact analogy between liquid-gas system (van der Waals fluid) and charged AdS black hole was first completed by \cite{KubiznakMann2012} in the context of Einstein-Maxwell gravity. In fact, RN-AdS black holes exhibit first-order phase transition with the same critical exponents as the van der Waals system. Here, we generalize the results in \cite{KubiznakMann2012} for higher dimensions and various event horizon topologies and show $P-V$ criticality only exists for spherically symmetric black holes. 

Now, we show the extension of thermodynamic phase space, by introducing negative cosmological constant as thermodynamic pressure (i.e. $P =  - \Lambda /8\pi $), indicates the phase transition for charged AdS black holes. As stated, the thermodynamic volume is the conjugate quantity of the thermodynamic pressure and can be straightforwardly obtained as $V = {\left( {\partial M/\partial P} \right)_{{X_i}}}$. In this regard, using eq. (\ref{temperature RN-AdS}), equation of state reads as
\begin{equation} \label{pressure - RN-AdS}
P = \frac{{{d_2}T}}{{4{r_ + }}} - \frac{{{d_2}{d_3}k}}{{16\pi r_ + ^2}} + \frac{{{q^2}}}{{8\pi r_ + ^{2{d_2}}}},
\end{equation}
for RN-AdS black holes. Following the method introduced in \cite{KubiznakMann2012}, one can translate the geometric equation of state eq. (\ref{pressure - RN-AdS}) to a physical version by performing dimensional analysis and substituting the following quantities
\begin{equation} \label{state equation-RN-AdS}
P \Rightarrow {P_{phys}} = \frac{{\hbar c}}{{\ell _{\rm{P}}^{{d_2}}}}P,\,\,\,T \Rightarrow {T_{phys}} = \frac{{\hbar c}}{{{k_{\rm{B}}}}}T,
\end{equation}
in which $P_{phys}$ and $T_{phys}$ denote the physical pressure and temperature, respectively. As a result, physical equation of state is given as 
\begin{equation} \label{state equation - physical}
{P_{phys}} = \frac{{{d_2}}}{{4{r_ + }\ell _{\rm{P}}^{{d_2}}}}{k_{\rm{B}}}{T_{phys}} + ... .
\end{equation}
Equation (\ref{state equation-RN-AdS}) should be compared with the following equation of state of van der Waals fluid
\begin{equation} \label{van der Waals}
\left( {P + \frac{a}{{{v^2}}}} \right)\left( {v - b} \right) = {k_{\rm{B}}}T,
\end{equation}
in which $a$ (molecular interaction forces) and $b$ (molecular size) are positive definite quantities, and $v$ is the specific volume of  van der Waals fulid. To do so, rewriting van der Waals equation in terms of $P$ and then expanding for $v \gg b$ lead to
\begin{equation} \label{van der Waals expansion}
P = \left( {\frac{{{k_{\rm{B}}}T}}{v} - \frac{a}{{{v^2}}}} \right) + \frac{{{k_{\rm{B}}}T}}{{{v^2}}}b + \frac{{{k_{\rm{B}}}T}}{{{v^3}}}{b^2} + O(\frac{{{b^3}}}{{{v^4}}}).
\end{equation}
Comparing eq. (\ref{state equation - physical}) with the above expansion implies that the horizon radius (not the thermodynamic volume $V$) is associated with the van der Waals fluid specific volume, i.e.
\begin{equation}
v = \frac{{4{r_ + }\ell _{\rm{P}}^{{d_2}}}}{{{d_2}}}.
\end{equation}
It should be emphasized, in this analogy, the same physical quantities are compared with each other. To sum up, we summarize the analogy between van der Waals fluid and RN-AdS black hole in the following table. 

\begin{center}
	\begin{tabular}{|c|c|}  
		\hline \hline
		van der Waals fluid	& RN-AdS black hole \\ 
		\hline \hline
		Temperature	& $T$ \\ 
		\hline
		Pressure & $P =  - \Lambda /8\pi $ \\
		\hline 
		Volume & ${r_ + } = {(3\,V/4\pi )^{1/3}}$ \\
		\hline
	\end{tabular}
\end{center}

The critical point occurs at the spike like divergence of specific heat at constant pressure (i.e., an inflection point in the $P-V$ diagram) and can be found by considering the following equations, simultaneously
\begin{eqnarray} \label{critical point equation}
{\left( {\frac{{\partial P}}{{\partial v}}} \right)_T} = 0\,\,\, &\Longleftrightarrow& \,\,\,{\left( {\frac{{\partial P}}{{\partial {r_ + }}}} \right)_T} = 0, \nonumber\\
{\left( {\frac{{{\partial ^2}P}}{{\partial {v^2}}}} \right)_T} = 0\,\,\, &\Longleftrightarrow& \,\,\,\left(\frac{{{\partial ^2}P}}{{\partial {r_ +^2}}} \right)_T= 0.
\end{eqnarray}
Regarding eq. (\ref{critical point equation}) with the equation of state, (\ref{pressure - RN-AdS}), one find that the critical quantities are obtained as
\begin{equation} \label{critical specific volume}
{v _c} = \frac{4}{{{d_2}}}{\left( {\frac{{2{q^2}(2d - 5)}}{{{d_3}k}}} \right)^{\frac{1}{{2{d_3}}}}}, \quad
{P_c} = \frac{{d_3^2k}}{{d_2^2\pi v_c^2}}, \quad
{T_c} = \frac{{4d_3^2k}}{{(2d - 5){d_2}\pi {v_c}}}.
\end{equation}
The thermodynamic quantities ${P_c}$, ${T_c}$ and ${v_c}$ must be positive definite. Evidently, for uncharged AdS BHs ($Q=0$) or non-spherical horizon solutions ($k=0,-1$), there are no phase transition and critical behavior. In fact, the electric charge $Q$ and topological factor $k$ play crucial rules to have critical behavior for AdS black holes in the Einstein gravity. In section \ref{phase transition-massive Lovelock}, we will show this is not the case in Lovelock or massive gravities and phase transition can happen even for uncharged black holes or black holes with Ricci flat/hyperbolic horizons. 

The thermodynamic quantities at the critical point satisfy the following universal ratio
\begin{equation} \label{universal ratio-RN-AdS}
\frac{{{P_c}{v_c}}}{{{T_c}}} = \frac{{2d - 5}}{{4{d_2}}}, 
\end{equation}
and of course, is only valid for spherical black holes ($k=1$). Amazingly, the universal ratio for RN-AdS black holes coincides with the van der Waals fluid (the universality is equal to $3/8$ in 4-dimensions) and is only dependent of spacetime dimensions.

\begin{figure}[!htbp]
	$%
	\begin{array}{ccc}
	\epsfxsize=5.5cm \epsffile{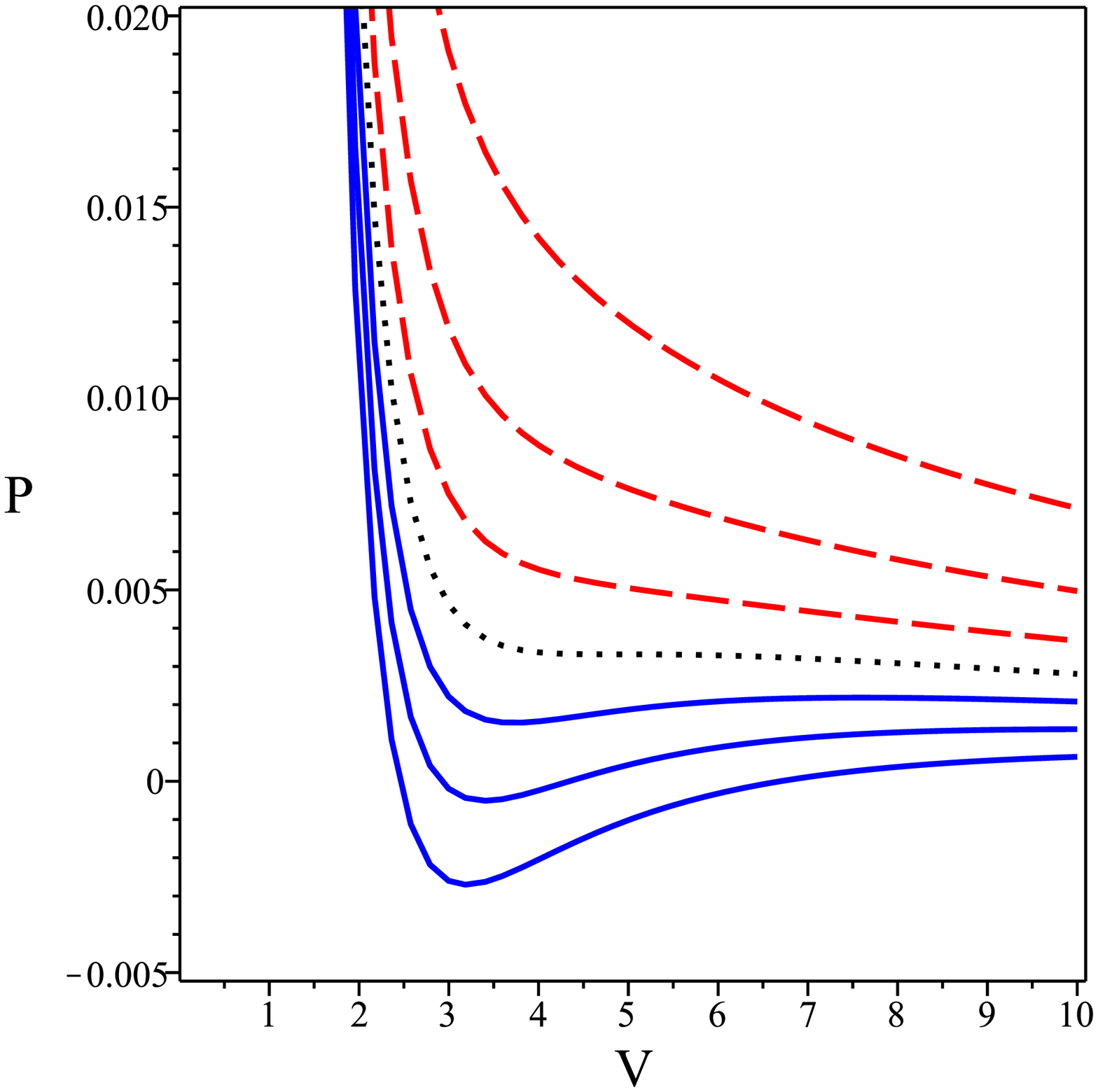}
	\epsfxsize=5.5cm \epsffile{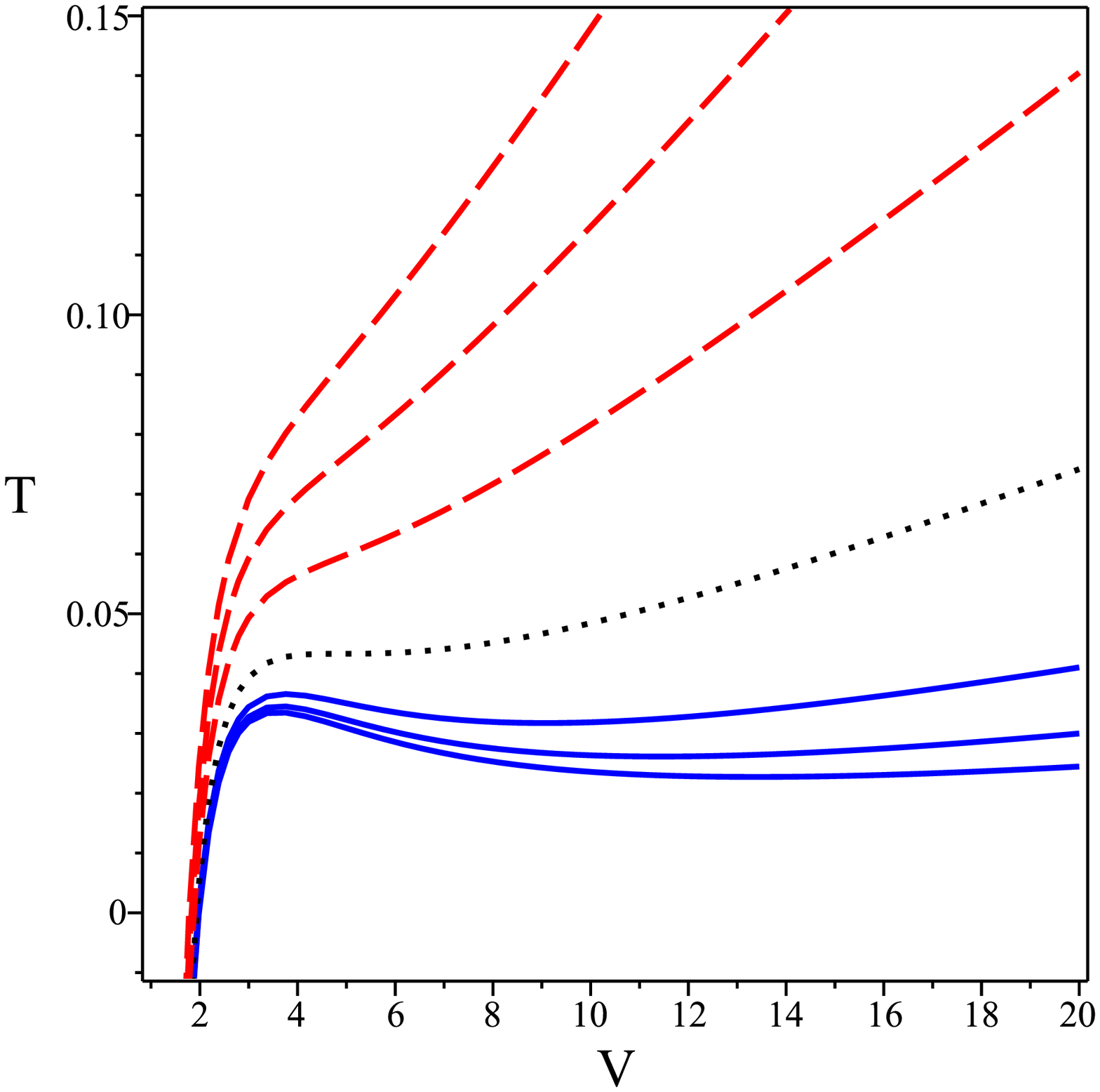}
	\epsfxsize=5.5cm \epsffile{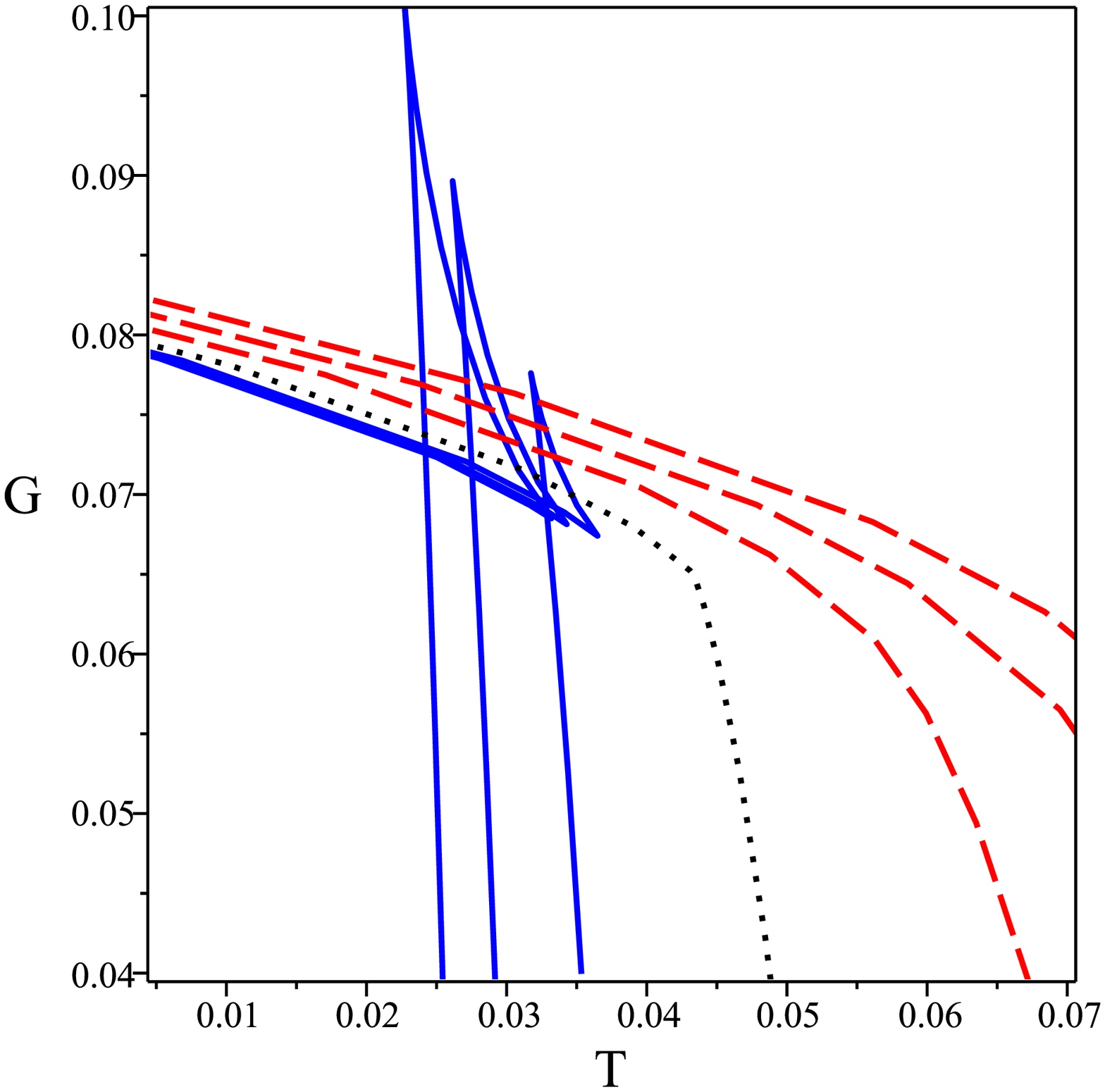}
	&  & 
	\end{array}
	$%
	\caption{\textbf{RN-AdS black hole:} $P-V$ (left), $T-V$
		(middle) and $G-T$ (right) diagrams for $k=1$, $d=4$ and $q=1$. \newline
		\textbf{Left panel:} $T<T_{c}$ (continuous lines), $T=T_{c}$ (dotted line) and $T>T_{c}$ (dashed lines).\newline
		\textbf{Middle and right panels:} $P<P_{c}$ (continuous lines), $P=P_{c}$ (dotted lines) and $P>P_{c}$ (dashed lines).}
	\label{PV-RN-AdS}
\end{figure}

Now, by finding and analyzing the free energy of RN-AdS black hole system, we complete our discussion on the criticality. In the fixed charged ensemble (canonical ensemble), the on-shell action is identified with the Gibbs free energy (since $\Lambda$ is a thermodynamic variable in the theory). Since $Q$ is held fixed, one has to add a surface term (for electromagnetic field) to fix charge on the boundary. As a result, the total action \ref{total action} has to be accompanied with the following boundary term
\begin{equation} \label{surface term, canonical ensemble }
{{\cal I}_{s_{2}}} = - \frac{1}{{4\pi }}\int_{\partial {\cal M}} {{d^{d - 1}}x\sqrt {-h} {n_\mu }{F^{\mu \nu }}{A_\nu }}.
\end{equation}
The Gibbs free energy can be obtained using the Legendre transformation or calculating the on-shell action as follows
\begin{equation} \label{Gibbs free energy - RN-AdS}
G = M - TS = \frac{{{\omega _{{d_2}}}}}{{16\pi }}\left( {kr_ + ^{{d_3}} - \frac{{16\pi P\,r_ + ^{{d_1}}}}{{{d_1}{d_2}}} + \frac{{(4d - 10){q^2}}}{{{d_2}{d_3}r_ + ^{{d_3}}}}} \right).
\end{equation}
The qualitative behavior of Gibbs free energy as a function of temperature is depicted in the right panel of Fig. \ref{PV-RN-AdS}. Obviously, the swallow-tail behavior demonstrates the first order phase transition (exactly the same as van der Waals fluid). For the sake of completeness, in the left and middle panels of Fig. \ref{PV-RN-AdS}, $P-V$ and $T-V$ diagrams are plotted. Evidently, the RN-AdS equation of state (\ref{pressure - RN-AdS}) mimic the behavior of the van der Waals fluid for any fixed $Q$.

\section{Lovelock Massive (LM) gravity with Maxwell field \label{Lovelock-massive}}

\subsection{Action and field equations \label{basic equations}}
In this section, we start our investigation of black holes in the LM gravity framework. Again, the total action consists of three terms, ${{\cal I}_G} = {{\cal I}_b} + {{\cal I}_s} + {{\cal I}_{ct}}$. The Lovelock boundary terms (${\cal I}_s$) to have a well-defined gravitational action and the counterterms for static solutions with Ricci flat and curved horizons (${\cal I}_{ct}$) to regulate divergences of the Lovelock buck action are constructed before and the relevant details can be found in \cite{Myers1987SurfaceTerm,Dehghani2006Mann,Mahdizadeh2015LovelockCounterterm}. We add generic mass terms for the gravitons in the Lovelock theory of gravity by supplementing higher order interaction terms of massive gravitons (which first proposed in \cite{dRGT}). Considering this theory on the $d$-dimensional background manifold ${\cal M}$ with the dynamical metric ${g_{\mu \nu }}$, the bulk action for LM gravity with a $U(1)$ gauge field (Maxwell field) may be written as

\begin{equation}\label{bulk action - massive Lovelock gravity}
{{\cal I}_b} =  - \frac{1}{{16\pi }}\int_{\cal M} {{d^d}x\sqrt { - g} \left[ {{\alpha _0}{{\cal L}_0} + {\alpha _1}{{\cal L}_1} + {\alpha _2}{{\cal L}_2} + {\alpha _3}{{\cal L}_3} - {\cal F} + {m^2}\sum\limits_{i \ge 1} {{c_i}{{\cal U}_i}(g,f)} } \right]},
\end{equation}
where ${\alpha _0}{{\cal L}_0} =  - 2\Lambda$ and ${\alpha _1}{{\cal L}_1} = R$ are, respectively, the negative cosmological constant and the Ricci scalar, and ${\cal L}_2$ and ${\cal L}_3$ are called the Gauss-Bonnet (GB) and the third order Lovelock (TOL) Lagrangians in literature which have the following forms

\begin{equation} \label{GB and TOL Lagrangians}
\begin{array}{l}
{{\cal L}_2} \equiv {{\cal L}_{{\rm{GB}}}} = {R^{\mu \nu \gamma \delta }}{R_{\mu \nu \gamma \delta }} - 4{R^{\mu \nu }}{R_{\mu \nu }} + {R^2},\\
{{\cal L}_3} \equiv {{\cal L}_{{\rm{TOL}}}} = 2{R^{\mu \nu \sigma \kappa }}{R_{\sigma \kappa \rho \tau }}R_{\,\,\,\,\,\,\,\,\mu \nu }^{\rho \tau } + 8R_{\,\,\,\,\,\,\,\,\,\sigma \rho }^{\mu \nu }R_{\,\,\,\,\,\,\,\,\,\nu \tau }^{\sigma \kappa }R_{\,\,\,\,\,\,\,\,\mu \kappa }^{\rho \tau } + 24{R^{\mu \nu \sigma \kappa }}{R_{\sigma \kappa \nu \rho }}R_{\,\,\,\,\,\mu }^\rho \\
\,\,\,\,\,\,\,\,\,\,\,\,\,\,\,\,\,\,\,\,\,\,\,\,\, + 3R{R^{\mu \nu \sigma \kappa }}{R_{\sigma \kappa \mu \nu }} + 24{R^{\mu \nu \sigma \kappa }}{R_{\sigma \mu }}{R_{\kappa \nu }} + 16{R^{\mu \nu }}{R_{\nu \sigma }}R_{\,\,\,\,\,\mu }^\sigma  - 12R{R^{\mu \nu }}{R_{\mu \nu }} + {R^3}.
\end{array}
\end{equation}

The Lovelock coefficients $\alpha _{2}$ and $\alpha _{3}$, which are positive definite constants \cite{BoulwareDeser1985StringModels}, indicate the strength of the second and third order curvature terms. In the action (\ref{bulk action - massive Lovelock gravity}), the last term is called massive interaction terms in which $m$ is the graviton mass parameter, and $f$ is the auxiliary reference metric which is a fixed rank-2 symmetric tensor. $c_{i}$'s are constants and ${\cal U}_i$'s (interaction terms) are symmetric polynomials of the eigenvalues of $d \times d$ matrix ${\cal K}_{\,\,\,\,\,\nu }^\mu  = \sqrt {{g^{\mu \alpha }}{f_{\alpha \nu }}}$ with the following explicit forms \cite{deRham2010Gabadadze,dRGT,Vegh2013}

\begin{equation}\label{U}
\begin{array}{l}
{{\cal U}_1} = [{\cal K}],\\
{{\cal U}_2} = {[{\cal K}]^2} - [{{\cal K}^2}],\\
{{\cal U}_3} = {[{\cal K}]^3} - 3\,[{\cal K}]\,[{{\cal K}^2}] + 2\,[{{\cal K}^3}],\\
{{\cal U}_4} = {[{\cal K}]^4} - 6\,{[{\cal K}]^2}\,[{{\cal K}^2}] + 8\,[{\cal K}]\,[{{\cal K}^3}] + 3\,{[{{\cal K}^2}]^2} - 6\,[{{\cal K}^4}],\\
{{\cal U}_5} = {[{\cal K}]^5} - 10\,{[{\cal K}]^3}\,[{{\cal K}^2}] + 20{[{\cal K}]^2}\,[{{\cal K}^3}] - 20[{{\cal K}^2}]\,[{{\cal K}^3}]\\
\,\,\,\,\,\,\,\,\,\,\, + 15\,[{\cal K}]{[{{\cal K}^2}]^2} - 30\,[{\cal K}]\,[{{\cal K}^4}] + 24\,[{{\cal K}^5}]\\
\,\,  \vdots 
\end{array}
\end{equation}
 where the square root in ${\cal K}$ stands for matrix square root, i.e. ${\cal K}_{\,\,\,\,\,\nu }^\mu  = (\sqrt {\cal K} )_{\,\,\,\,\lambda }^\mu (\sqrt {\cal K} )_{\,\,\,\,\nu }^\lambda $, and the rectangular bracket denotes the trace $[{\cal K}] = {\cal K}_{\,\,\,\,\mu }^\mu $. We restrict our study to ${\cal U}_i$ up to the fourth interaction term (${\cal U}_4$), while considering the higher order terms is straightforward. It should be noted that the massive coupling coefficients $c_{i}$'s required to be negative if $m^{2}>0$, but, the theory is stable even for $m^{2}<0$ if the squared mass obeys the corresponding Breitenlohner-Freedman bounds \cite{Breitenlohner1982,BreitenlohnerFreedman1982,Cai2015}. This is permissible in AdS spacetimes, and according to AdS/CFT correspondence, a mass term in a (gravitational) bulk theory corresponds to adding an operator with a different scaling dimension in the boundary theory. In this paper, we relax the restriction on the massive couplings  ($c_{i}$'s) and at the end one can restrict them to satisfy some certain limitations. 
 
 Using the variational principle, the electromagnetic and gravitational field equations of LM-Maxwell gravity can be obtained as
 
 \begin{equation}\label{gravitational field equation}
{G_{\mu \nu }} + \Lambda {g_{\mu \nu }} + G_{\mu \nu }^{{\rm{(GB)}}} + G_{\mu \nu }^{({\rm{TOL}})} + {m^2}{{\cal X}_{\mu \nu }} =  - \frac{1}{2}{g_{\mu \nu }}{\cal F} + 2{F_{\mu \lambda }}{F_\nu }^{\,\lambda },
 \end{equation}
\begin{equation} \label{electromagnetic field equation}
{\nabla _\mu }{F^{\mu \nu }} = 0,
\end{equation}
in which ${G_{\mu \nu }} = {R_{\mu \nu }} - \frac{1}{2}{g_{\mu \nu }}R$ is the Einstein tensor and $G_{\mu \nu }^{{\rm{(GB)}}}$ and $G_{\mu \nu }^{({\rm{TOL}})}$ are, respectively, the second (GB) and third order Lovelock (TOL) tensors given as
 \begin{equation} \label{GB tensor}
G_{\mu \nu }^{({\rm{GB}})} = 2(R{R_{\mu \nu }} - {R_{\mu \sigma \kappa \tau }}{R^{\kappa \tau \sigma }}_\nu  - 2{R_{\mu \rho \nu \sigma }}{R^{\rho \sigma }} - 2{R_{\mu \sigma }}{R^\sigma }_\nu ) - \frac{1}{2}{g_{\mu \nu }}{{\cal L}_{{\rm{GB}}}},
\end{equation}
\begin{eqnarray} \label{TOL tensor}
G_{\mu \nu }^{(\rm{TOL})} &=&-3[4R^{\tau \rho \sigma \kappa }R_{\sigma \kappa
	\lambda \rho }R_{\phantom{\lambda }{\nu \tau \mu}}^{\lambda }-8R_{%
	\phantom{\tau \rho}{\lambda \sigma}}^{\tau \rho }R_{\phantom{\sigma
		\kappa}{\tau \mu}}^{\sigma \kappa }R_{\phantom{\lambda }{\nu \rho \kappa}%
}^{\lambda }+2R_{\nu }^{\phantom{\nu}{\tau \sigma \kappa}}R_{\sigma \kappa
	\lambda \rho }R_{\phantom{\lambda \rho}{\tau \mu}}^{\lambda \rho }  \nonumber
\label{TOL} \\
&&-R^{\tau \rho \sigma \kappa }R_{\sigma \kappa \tau \rho }R_{\nu \mu }+8R_{%
	\phantom{\tau}{\nu \sigma \rho}}^{\tau }R_{\phantom{\sigma \kappa}{\tau \mu}%
}^{\sigma \kappa }R_{\phantom{\rho}\kappa }^{\rho }+8R_{\phantom
	{\sigma}{\nu \tau \kappa}}^{\sigma }R_{\phantom {\tau \rho}{\sigma \mu}%
}^{\tau \rho }R_{\phantom{\kappa}{\rho}}^{\kappa }  \nonumber \\
&&+4R_{\nu }^{\phantom{\nu}{\tau \sigma \kappa}}R_{\sigma \kappa \mu \rho
}R_{\phantom{\rho}{\tau}}^{\rho }-4R_{\nu }^{\phantom{\nu}{\tau \sigma
		\kappa }}R_{\sigma \kappa \tau \rho }R_{\phantom{\rho}{\mu}}^{\rho
}+4R^{\tau \rho \sigma \kappa }R_{\sigma \kappa \tau \mu }R_{\nu \rho
}+2RR_{\nu }^{\phantom{\nu}{\kappa \tau \rho}}R_{\tau \rho \kappa \mu }
\nonumber \\
&&+8R_{\phantom{\tau}{\nu \mu \rho }}^{\tau }R_{\phantom{\rho}{\sigma}%
}^{\rho }R_{\phantom{\sigma}{\tau}}^{\sigma }-8R_{\phantom{\sigma}{\nu \tau
		\rho }}^{\sigma }R_{\phantom{\tau}{\sigma}}^{\tau }R_{\mu }^{\rho }-8R_{%
	\phantom{\tau }{\sigma \mu}}^{\tau \rho }R_{\phantom{\sigma}{\tau }}^{\sigma
}R_{\nu \rho }-4RR_{\phantom{\tau}{\nu \mu \rho }}^{\tau }R_{\phantom{\rho}%
	\tau }^{\rho }  \nonumber \\
&&+4R^{\tau \rho }R_{\rho \tau }R_{\nu \mu }-8R_{\phantom{\tau}{\nu}}^{\tau
}R_{\tau \rho }R_{\phantom{\rho}{\mu}}^{\rho }+4RR_{\nu \rho }R_{%
	\phantom{\rho}{\mu }}^{\rho }-R^{2}R_{\nu \mu }]-\frac{1}{2}g_{\mu \nu }%
\mathcal{L}_{\rm{TOL}}.
\end{eqnarray}
In addition, $\cal X_{\mu\nu}$ is

\begin{equation}
\begin{array}{l}
{{\cal X}_{\mu \nu }} =  - \frac{{{c_1}}}{2}({{\cal U}_1}{g_{\mu \nu }} - {{\cal K}_{\mu \nu }}) - \frac{{{c_2}}}{2}({{\cal U}_2}{g_{\mu \nu }} - 2{{\cal U}_1}{{\cal K}_{\mu \nu }} + 2{\cal K}_{\mu \nu }^2) - \frac{{{c_3}}}{2}({{\cal U}_3}{g_{\mu \nu }} - 3{{\cal U}_2}{{\cal K}_{\mu \nu }} + 6{{\cal U}_1}{\cal K}_{\mu \nu }^2 - 6{\cal K}_{\mu \nu }^3)\\
\,\,\,\,\,\,\,\,\,\,\,\,\,\, - \frac{{{c_4}}}{2}({{\cal U}_4}{g_{\mu \nu }} - 4{{\cal U}_3}{{\cal K}_{\mu \nu }} + 12{{\cal U}_2}{\cal K}_{\mu \nu }^2 - 24{{\cal U}_1}{\cal K}_{\mu \nu }^3 + 24{\cal K}_{\mu \nu }^4) + ... .
\end{array}
\end{equation}

It is helpful to consider the gravitational field equations (\ref{gravitational field equation}) in the weak-field limit. The metric of spacetime is slightly curved, i.e., ${g_{\mu \nu }} = {\eta _{\mu \nu }} + {h_{\mu \nu }}$, with the property $|{h_{\mu \nu }}|\ll 1$ for the metric perturbation (${h_{\mu \nu }}$). In the weak-field limit for the geometric tensors one can find ${G_{\mu \nu }} \propto O(h)\,,\,\,\,\,G_{\mu \nu }^{{\rm{GB}}} \propto O({h^2})\,,\,\,\,\,G_{\mu \nu }^{{\rm{TOL}}} \propto O({h^3})$, and, regardless of cosmological term and energy-momentum tensor, to first order in the metric perturbation ${h_{\mu \nu }}$, the Fierz-Pauli field equations \cite{FierzPauli1939} are recoverd in higher dimensions.

In the next section, we will obtain the charged AdS black hole solutions of the fully nonlinear gravitational field equations (\ref{gravitational field equation}) and study their geometric properties.


\subsection{LM charged-AdS black holes
\label{blackhole}}

Here, we intend to obtain static black hole solutions of LM gravity. In order to achieve the topological charged AdS black holes, again, we consider the following $d$-dimensional line element ansatz
\begin{equation}
d{s^2} =  - \psi (r)d{t^2} + \frac{{d{r^2}}}{{\psi (r)}} + {r^2}{h_{ij}}d{x_i}d{x_j}\,\,\,\,\,\,(i,j = 1,2,3,...,d-2).
\end{equation}
We make use of the appropriate ansatz for the reference metric $f_{\mu\nu}$ with the following form \cite{Vegh2013,Cai2015,Hendi2016JHEP}
\begin{equation} \label{reference metric}
{f_{\mu \nu }} = diag\left( {0,0,c_0^2{h_{ij}}} \right),
\end{equation}
where $c_{0}$ is a positive constant. It is important to note that this choice of reference metric $f_{\mu \nu }$, first, cannot produce any infinite value for the bulk action eq.(\ref{bulk action - massive Lovelock gravity}) (since the bulk action only contains non-negative powers of $f_{\mu \nu }$), second, does not preserve general covariance in the transverse spatial coordinates $x_{1}, x_{2}, ...$ (since $f_{\mu \nu }$ only depends on the spatial components $h_{ij}$ of the spacetime metric). Regarding eqs. (\ref{U}) and (\ref{reference metric}), the interaction terms ${\cal U}_i$'s can be calculated as
\begin{equation}
{{\cal U}_1} = \frac{{{d_2}\,{c_0}}}{r}\,,\,\,\,\,{{\cal U}_2} = \frac{{{d_2}\,{d_3}\,c_0^2}}{{{r^2}}},\,\,\,\,{{\cal U}_3} = \frac{{{d_2}\,{d_3}\,{d_4}\,c_0^3}}{{{r^3}}},\,\,\,\,{{\cal U}_4} = \frac{{{d_2}\,{d_3}\,{d_4}\,{d_5}\,c_0^4}}{{{r^4}}},\,\,\,\,{{\cal U}_5} = \frac{{{d_2}\,{d_3}\,{d_4}\,{d_5}\,{d_6}\,c_0^5}}{{{r^5}}},\,\,...
\end{equation}

Using the electromagnetic field equation (\ref{electromagnetic field equation}), the gauge potential is obtained as follows
\begin{equation}  \label{Amu}
{A_\mu } = \frac{{\ q}}{{d_{3}r^{d_{3}}}}\delta _\mu ^0,
\end{equation}
where $q$ is an integration constant related to the electric charge. As a result, the non-zero components of the Faraday tensor are
\begin{equation}
{F_{tr}}=-{F_{rt}}=\frac{q}{{{r^{d_{2}}}}}.  \label{Ftr}
\end{equation}
Now, by use of the gravitational field equations (\ref{gravitational field equation}), we are going to obtain the AdS black hole solutions. It is well known that solutions of TOL gravity with different Lovelock coefficients $\alpha _2$ and $\alpha _3$ are mathematically too long; therefore not appropriate to be studied. These complete solutions have been proposed in \cite{Dehghani2005Shamirzaie}. The black hole solutions can still be found if $\alpha _2$ and $\alpha _3$ were dependent to each other. In order to have practical black hole solutions, we consider the following special case for Lovelock coefficients
\begin{equation} \label{coefficient condition}
{\alpha _2} = \frac{\alpha }{{{d_3}\,{d_4}}}\,\,,\,\,\,\,\,{\alpha _3} = \frac{{{\alpha ^2}}}{{3\,{d_3}\,{d_4}\,{d_5}\,{d_6}}},
\end{equation}
which is a standard simple choice \cite{Dehghani2005Shamirzaie,Dehghani2006Mann,Dehghani2008Alinejadi,LovelockNEDs2015,LovelockRainbow2017,Asnafi2011}. Considering eq. (\ref{coefficient condition}), one may find one real and two complex (conjugate) solutions for the metric function $\psi(r)$. The real valued solution for the metric function $\psi(r)$ is obtained as
\begin{equation} \label{metric function}
\psi (r) = k + \frac{{{r^2}}}{\alpha }\left( {1 - {{\left[ {1 + \frac{{3\alpha {m_0}}}{{{r^{{d_1}}}}} + \frac{{6\alpha \Lambda }}{{{d_1}\,{d_2}}} - \frac{{6\alpha {q^2}}}{{{d_2}\,{d_3}\,{r^{2{d_2}}}}} - 3\alpha {m^2}{\cal A}(r)} \right]}^{\frac{1}{3}}}} \right),
\end{equation}
with
\begin{equation} \label{A-metric}
{\cal A}(r) = \frac{{{c_0}\,{c_1}}}{{{d_2}\,r}} + \frac{{c_0^2{c_2}}}{{{r^2}}} + \frac{{{d_3}\,c_0^3{c_3}}}{{{r^3}}} + \frac{{{d_3}\,{d_4}\,c_0^4{c_4}}}{{{r^4}}} + O\left( {\frac{{{c_5}}}{{{r^5}}}} \right),
\end{equation}
in which satisfies all components of the field equations (\ref{gravitational field equation}) simultaneously. It is notable that the parameter $m_{0}$ is a positive integration constant related to the finite mass. The obtained solutions reduce to the Lovelock-Maxwell black hole solutions as $m\rightarrow0$ \cite{Dehghani2005Shamirzaie}. In addition, for $\alpha\rightarrow0$, static black hole solutions of (Einstein) massive gravity \cite{Cai2015} can be recovered as follows
\begin{equation}
\psi {(r)_{massive}} = k - \frac{{2\Lambda {r^2}}}{{{d_1}\,{d_2}}} - \frac{{{m_0}}}{{{r^{{d_3}}}}} + \frac{{2{q^2}}}{{{d_2}\,{d_3}\,{r^{2{d_3}}}}} + {m^2}{r^2}{{\cal A}(r)}.
\end{equation}

\begin{figure}[!hbbp]
	$%
	\begin{array}{cc}
	\epsfxsize=7cm \epsffile{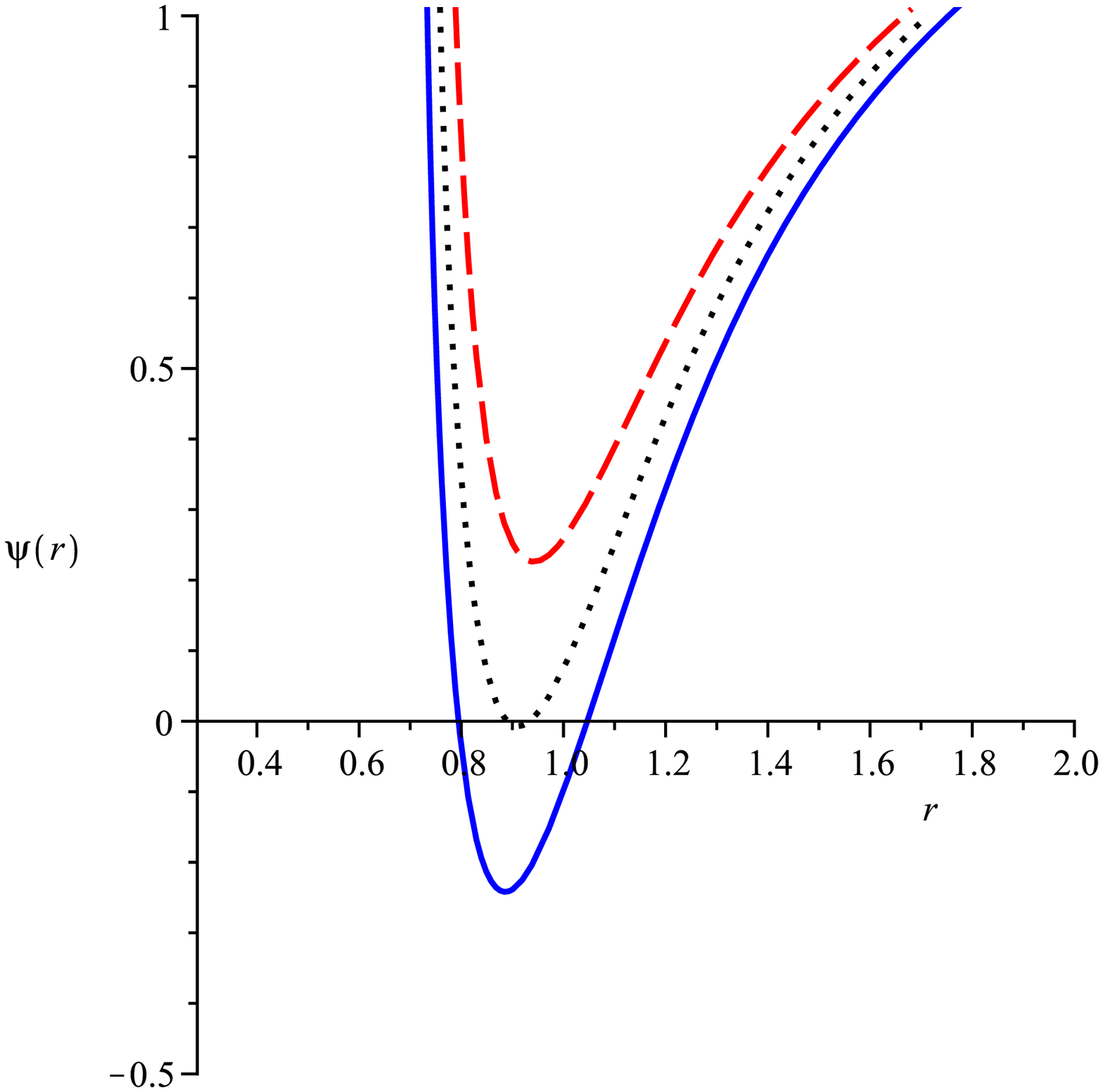} \epsfxsize=7cm %
	\epsffile{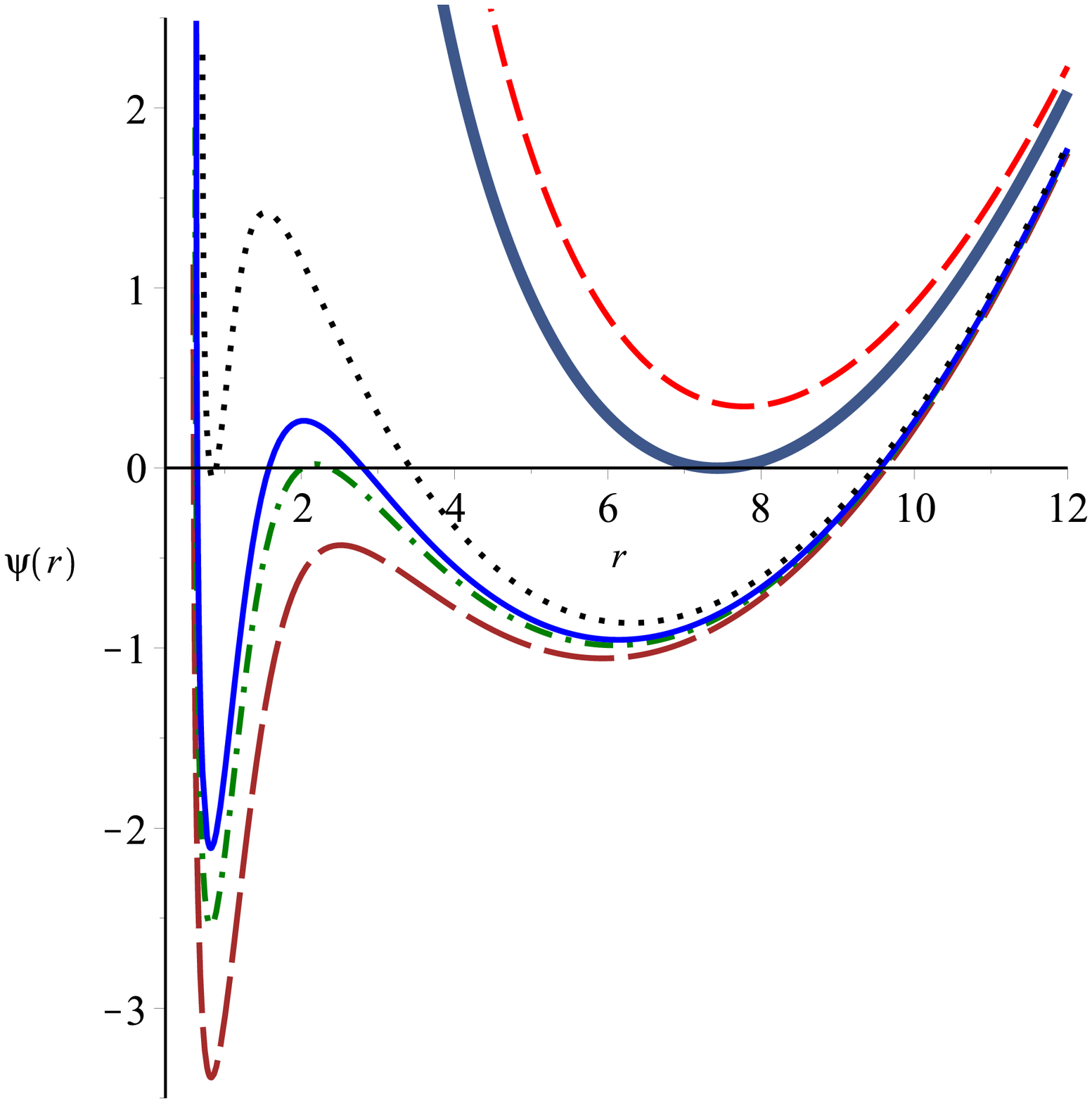}\epsfxsize=7cm %
	& 
	\end{array}
	$%
	\caption{ $\psi(r)$ versus ${r }$ for $k=1$, $d=7$, $\Lambda=-1$, $q=2.4$, $\alpha=0.2$. \textbf{Left panel:} for Lovelock gravity with $m_{0}=2$ (solid line), $m_{0}=1.75$ (dotted line) and $m_{0}=1.5$ (dashed line). 
	\textbf{Right panel:} for Lovelock massive (LM) gravity with $m_{0}=1.2$, $m=0.8$, $c_{0}=1$, $c_{1}=-5$, $c_{2}=-3$, $c_{3}=4$, $c_{4}=-2$ (long-dashed line), $c_{4}=-1.65$ (dash-dotted line), $c_{4}=-1.5$ (thin-solid line), $c_{4}=-1.02$ (dotted line), $c_{4}=4.4$  (bold-solid line) and , $c_{4}=7$ (dashed line).}
	\label{horizon}
\end{figure}

The existence of the possible curvature singularity could be explored by use of calculating the Kretschmann scalar which is
\begin{equation} \label{Kretschmann}
{R^{\alpha \beta \gamma \delta }}{R_{\alpha \beta \gamma \delta }} = {\left( {\frac{{{\partial ^2}\psi (r)}}{{\partial {r^2}}}} \right)^2} + 2{d_2}{\left( {\frac{1}{r}\frac{{\partial \psi (r)}}{{\partial r}}} \right)^2} + 2{d_2}{d_3}{\left( {\frac{{\psi (r) - k}}{{{r^2}}}} \right)^2}.
\end{equation}
Taking into account the metric function $\psi(r)$, one may find ${R^{\alpha \beta \gamma \delta }}{R_{\alpha \beta \gamma \delta }} \propto {r^{ - 4{d_2}}}$ near the origin ($r\rightarrow0$). In addition, the Kretschmann scalar is finite for $r>0$ and diverges at $r=0$, and in conclusion, we regard the origin as an essential singularity of the curvature. This physical singularity can be covered by an event horizon. The roots of the metric function $\psi (r) = {g^{rr}} = 0$ specify the number of horizons. Surprisingly, numerical calculation shows the metric function $\psi(r)$ could have more than two roots for all horizon topologies in contrast to the usual solutions of Lovelock and Einstein gravities (reported in refs. \cite{Katsuragawa2015,Hendi2016Annalen-massive,Hendi2015BI-MassiveGravity} as well). Evidently, this is due to the massive interaction terms. In Fig. \ref{horizon} we have depicted diverse cases for the possibility of the existence of the multi-horizon solutions in massive gravity. In the left panel of Fig. \ref{horizon}, we have displayed a typical example for the behavior of metric function $\psi(r)$ in Lovelock gravity (with zero mass for gravitons). As shown in Fig. \ref{horizon}, in LM gravity (right panel), the metric function may have i) four (ordinary) roots, ii) one extreme and two  roots, iii) one extreme root, iv) two roots, or v) without any root (which all the roots are real and positive). Hence presented solutions may be interpreted as the black holes with (three) four horizons, extreme black holes, black holes with one non-extreme horizon, or naked singularity. Hereafter, we assume that $r_{+}$ is the event horizon radius of the black hole solutions (\ref{metric function}) and can be numerically computed by finding the largest real positive root of $\psi(r)=0$.

In order to investigate the asymptotic behavior of spacetime, we consider the metric function $\psi(r)$ at large $r$, i.e.
\begin{equation}
{\left. {\psi (r)} \right|_{asymp.}} = k + \frac{{{r^2}}}{\alpha }\left( {1 - {{\left[ {1 + \frac{{6\alpha \Lambda }}{{{d_1}\,{d_2}}}} \right]}^{\frac{1}{3}}}} \right).
\end{equation}
We find that the obtained solutions are asymptotically ${\rm{Ad}}{{\rm{S}}_d}$ (${\Lambda _{eff}} < 0$) with $SO(2,d-1)$ invariance, dS (${\Lambda _{eff}} > 0$, $k=1$) with $SO(1,d)$ invariance or flat (${\Lambda _{eff}} = \Lambda  = 0$, $k=1$), if we replace $\Lambda$ with the effective cosmological constant ${\Lambda _{eff}} = \frac{{\Lambda [{d_1}{d_2} - 2\alpha \Lambda ]}}{{{d_1}{d_2}}}$.

We are more interested in studying AdS black holes since these types of black objects admit dual interpretation and also possess certain phase transition(s) in the extended phase space. Hereafter, we assume only AdS black holes.


\subsection{Thermodynamics of LM charged-AdS black holes \label{thermodynamics-massive Lovelock}}

Here, we examine traditional form of the first law of black hole thermodynamics in which the cosmological constant is considered as a fixed in the theory. We calculate the conserved charges and thermodynamic quantities associated with the charged AdS black hole solutions (\ref{metric function}) in LM gravity. First, using the definition of surface gravity, eq. (\ref{surface gravity}), the Hawking temperature can be obtained as
\begin{equation} \label{temperature - massive AdS BH}
T = \frac{{{d_2}k\left( {{d_7}{\alpha ^2} + 3{d_5}k\alpha r_ + ^2 + 3{d_3}r_ + ^4} \right) - 6\Lambda r_ + ^6 - 6{q^2}r_ + ^{ - 2{d_5}} + 3{d_2}{m^2}r_ + ^6{{\cal B}_ + }}}{{12\pi {d_2}\,{r_ + }{{\left( {k\alpha  + r_ + ^2} \right)}^2}}},
\end{equation}
where
\begin{equation}
{{\cal B}_ + } = {c_0}{c_1}r_ + ^{ - 1} + {d_3}c_0^2{c_2}r_ + ^{ - 2} + {d_3}{d_4}c_0^3{c_3}r_ + ^{ - 3} + {d_3}{d_4}{d_5}c_0^4{c_4}r_ + ^{ - 4} + O({c_5},r_ + ^{ - 5}).
\end{equation}

Entropy is the conjugate (extensive) quantity of the temperature. Since the area law of the black hole entropy is generally not valid for higher curvature gravities, one should use another approach to calculate it. For asymptotically flat spacetimes, one can easily apply the Wald method \cite{Wald1993} as
\begin{equation} \label{Lovelock entropy Formula}
{S_{{\rm{Lovelock}}}} = \frac{1}{4}\sum\limits_{k = 1}^{[{d_1}/2]} {k{\alpha _k}} \int {{d^{d - 2}}x\sqrt {\tilde g} {{\tilde {\cal L}}_{k - 1}}},
\end{equation}
where $\tilde g$ is the determinant of the induced metric ${\tilde{g}_{\mu \nu }}$ on $(d-2) $-dimensional boundary, ${\tilde {\cal L}}_{k - 1}$ is the $ k$-th order of Lovelock Lagrangian constructed from ${\tilde{g}_{\mu \nu }}$ and ${[{d_1}/2]}$ denotes the integer part of ${{d_1}/2}$. Moreover, as shown in \cite{Mahdizadeh2015LovelockCounterterm}, one can still use the Wald formula to calculate the entropy for asymptotically AdS spacetimes. Using this method, in TOL gravity the entropy may be written as
\begin{equation} \label{TOL entropy}
S = \frac{1}{4}\int {{d^{d - 2}}x\sqrt {\tilde g} \left[ {1 + 2{\alpha _2}\tilde R + 3{\alpha _3}\left( {{{\tilde R}^{\mu \nu \gamma \delta }}{{\tilde R}_{\mu \nu \gamma \delta }} - 4{{\tilde R}^{\mu \nu }}{{\tilde R}_{\mu \nu }} + {{\tilde R}^2}} \right)} \right]}
\end{equation}
where ${{{\tilde{R}}_{\mu \nu \gamma \delta }}}$, ${{{\tilde{R}}_{\mu \nu }}}$ and ${\tilde{R}}$ are, respectively, the Riemann tensor, the Ricci tensor and the Ricci scalar of the induced metric ${\tilde{g}_{\mu \nu }}$ on $(d-2)$-dimensional boundary. The modified entropy of TOL gravity is obtained as
\begin{equation} \label{entropy - massive AdS BH}
S = \frac{{{\omega_{{d_2}}}}}{4}r_ + ^{{d_2}}\left( {1 + \frac{{2{d_2}k\alpha }}{{{d_4}r_ + ^2}} + \frac{{{d_2}{k^2}{\alpha ^2}}}{{{d_6}r_ + ^4}}} \right).
\end{equation}

The electric charge and its conjugate potential can be found by use of definitions (\ref{potential definition}) and (\ref{charge definition}), yielding
\begin{equation} \label{potential - massive AdS BH}
\Phi  = \frac{q}{{{d_3}\,r_ + ^{{d_3}}}},
\end{equation}
and
\begin{equation} \label{charge - massive AdS BH}
Q = \frac{{{\omega _{{d_2}}}}}{{4\pi }}q.
\end{equation}

Using $\psi(r_{+})=0$ and eq. (\ref{mass definition}), the finite mass of $d$-dimensional AdS black holes with different horizon topologies can be derived as
\begin{equation} \label{mass of massive AdS black holes}
M(S,Q) = \frac{{{d_2}{\omega _{{d_2}}}}}{{16\pi }}\left( {\frac{1}{3}{k^3}{\alpha ^2}r_ + ^{{d_7}} + {k^2}\alpha r_ + ^{{d_5}} + kr_ + ^{{d_3}} + \frac{{32{\pi ^2}{Q^2}}}{{{d_2}{d_3}r_ + ^{{d_3}}}} - \frac{{2\Lambda r_ + ^{{d_1}}}}{{{d_1}\,{d_2}}} + {m^2}r_ + ^{{d_1}}{{\cal A}_ + }} \right),
\end{equation}
where $r_{+}=r_{+}(S)$ and
\begin{equation}
{{\cal A}_ + } \equiv {\cal A}({r_ + }) = \frac{{{c_0}\,{c_1}}}{{{d_2}\,{r_ + }}} + \frac{{c_0^2{c_2}}}{{r_ + ^2}} + \frac{{{d_3}\,c_0^3{c_3}}}{{r_ + ^3}} + \frac{{{d_3}\,{d_4}\,c_0^4{c_4}}}{{r_ + ^4}} + O\left( {\frac{{{c_5}}}{{r_ + ^5}}} \right),
\end{equation}
as introduced in eq. (\ref{A-metric}).

Now, it can be straightforwardly checked that by computing the following intensive quantities
\begin{eqnarray} \label{chain derivative}
&&T=\left( \frac{\partial M}{\partial S}\right) _{Q}=\frac{\left( \frac{%
		\partial M}{\partial r_{+}}\right) _{Q}}{\left( \frac{\partial S}{\partial
		r_{+}}\right) _{Q}}, \\
&&\Phi =\left( \frac{\partial M}{\partial Q}\right) _{S}=\frac{\left( \frac{%
		\partial M}{\partial q}\right) _{r_{+}}}{\left( \frac{\partial Q}{\partial q}%
	\right) _{r_{+}}},
\end{eqnarray}
we obtain the same quantities as those found before for $T$ (\ref{temperature - massive AdS BH}) and $\Phi$ (\ref{potential - massive AdS BH}). In conclusion, we have a prescription at hand for the first law of black hole thermodynamic which takes the form
\begin{equation} \label{first law}
dM = TdS + \Phi dQ.
\end{equation}

This prescription for first law is not consistent with Smarr formula based on scaling argument. In sec. \ref{extended thermodynamics-massive}, we reconsider the first law of thermodynamics in the extended phase space and resolve this inconsistency.


\subsection{Thermal stability of LM charged-AdS black holes \label{stability}}
 
\begin{figure}[!htbp]
	$%
	\begin{array}{ccc}
	\epsfxsize=5.5cm \epsffile{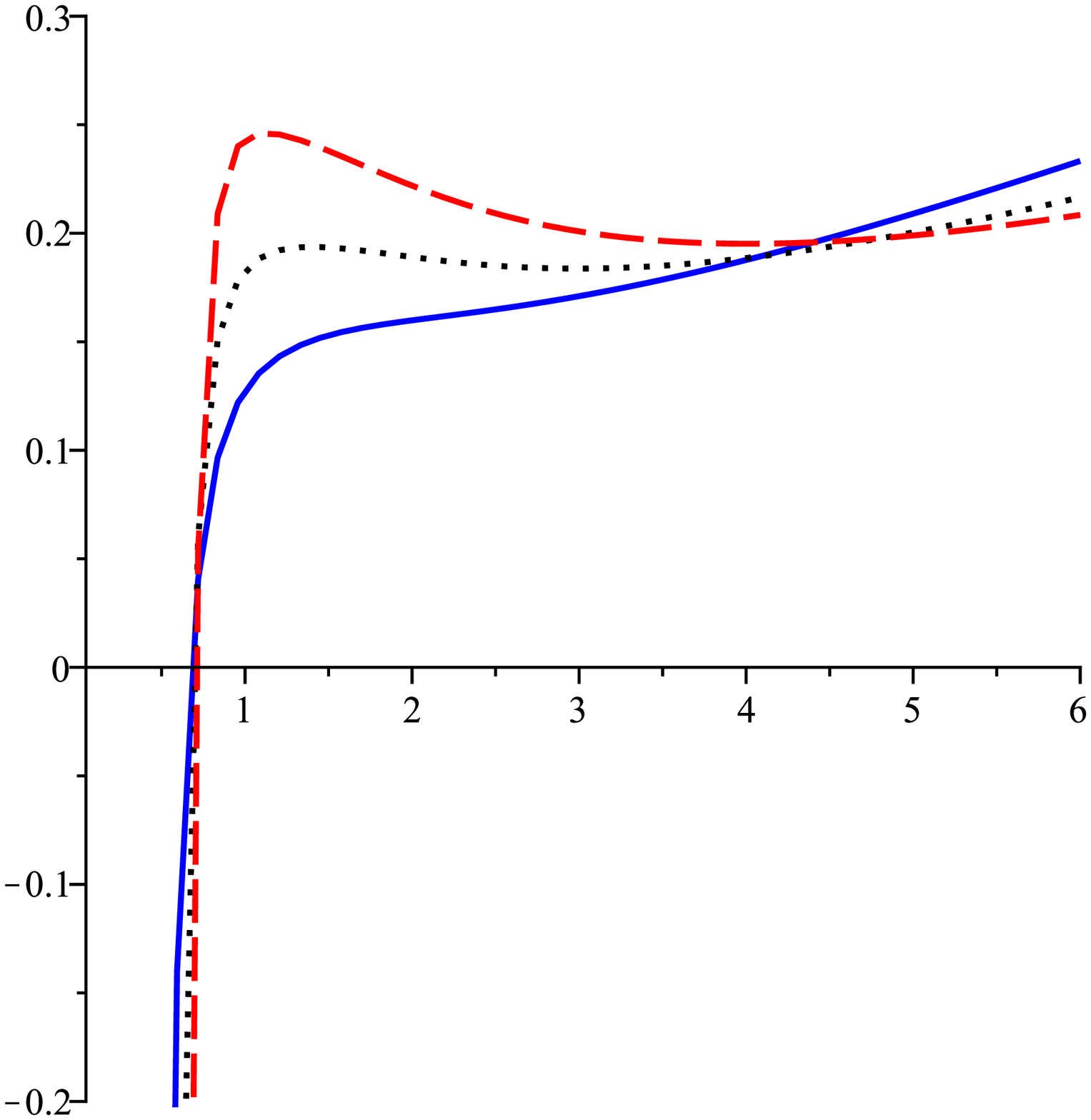}
	\epsfxsize=5.5cm \epsffile{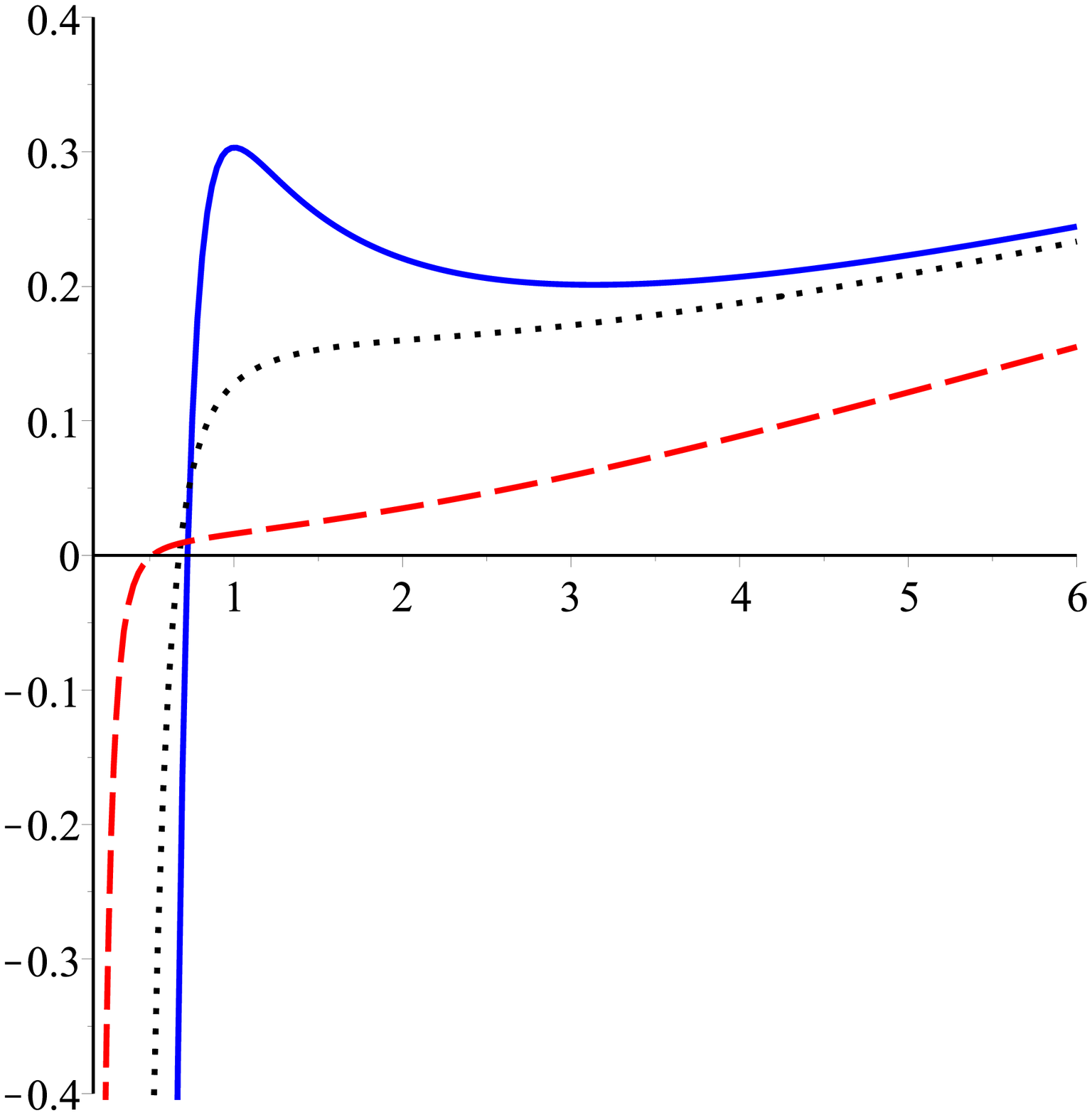}
	\epsfxsize=5.5cm \epsffile{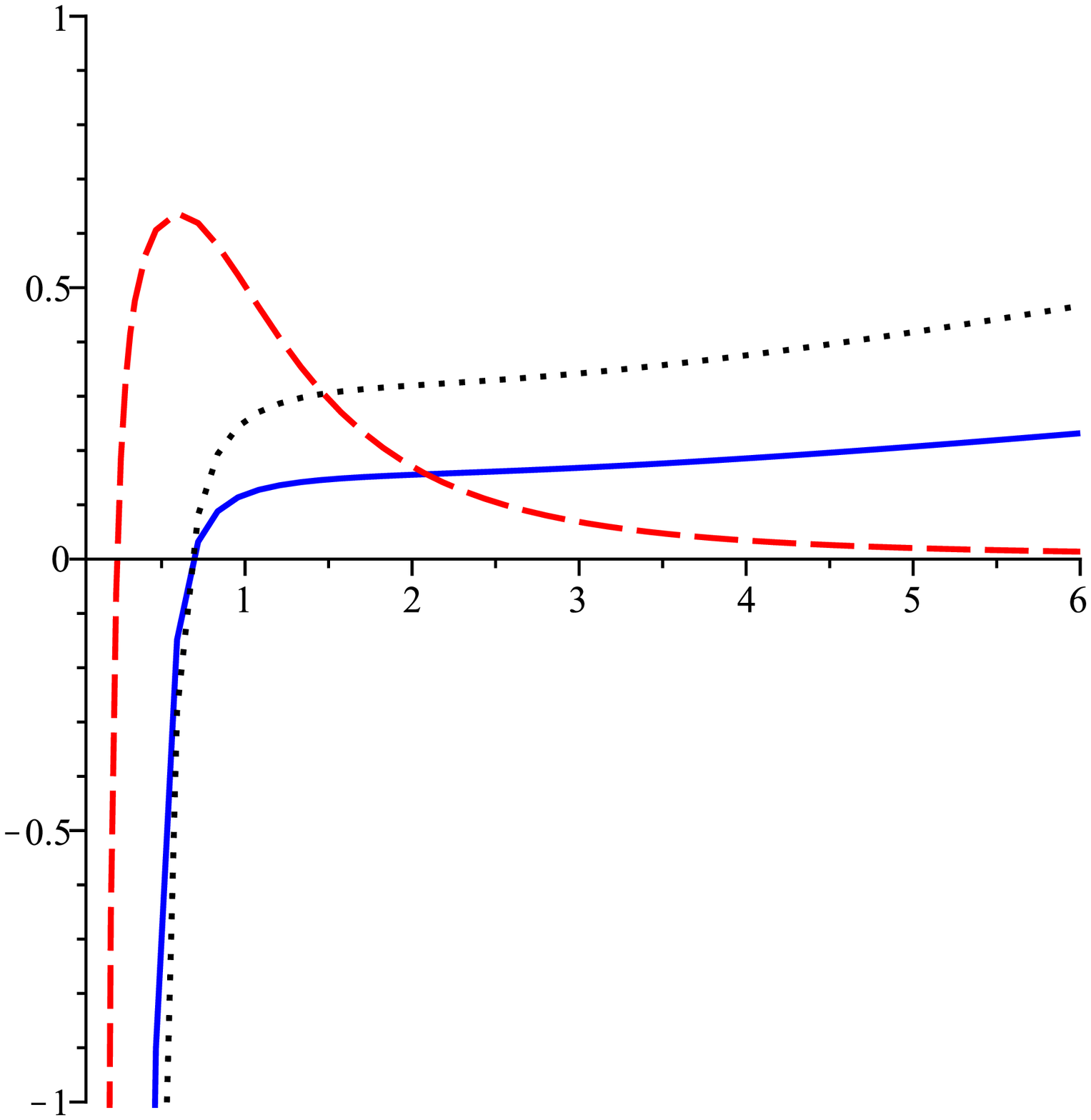}
	&  & 
	\end{array}
	$%
	\caption{ $T$ versus ${r_ + }$ for $k=1$, $q=1$, $\Lambda = - 1$, $m=0.1$, $c_{1}=c_{2}=c_{3}=c_{4}= 1$.
		\textbf{Left panel:} for $\alpha=1$, $c_{0}=1$ and $d=7$ (solid line), $d=8$ (dotted line) and $d=9$ (dashed line).
		\textbf{Middle panel:} for $d=7$, $c_{0}=1$ and $\alpha = 0.1$ ({solid line}), $\alpha = 1$ ({dotted line}) and $\alpha = 10$ ({dashed line}).
		\textbf{Right panel:} for $d=7$, $\alpha=1$ and $c_{0}=0.1$ ({solid line}), $c_{0}=1$ ({dotted line}) and $c_{0}=10$ ({dashed line}).}
	\label{Temp-k=1}
\end{figure}

\begin{figure}[!htbp]
	$%
	\begin{array}{ccc}
	\epsfxsize=5.5cm \epsffile{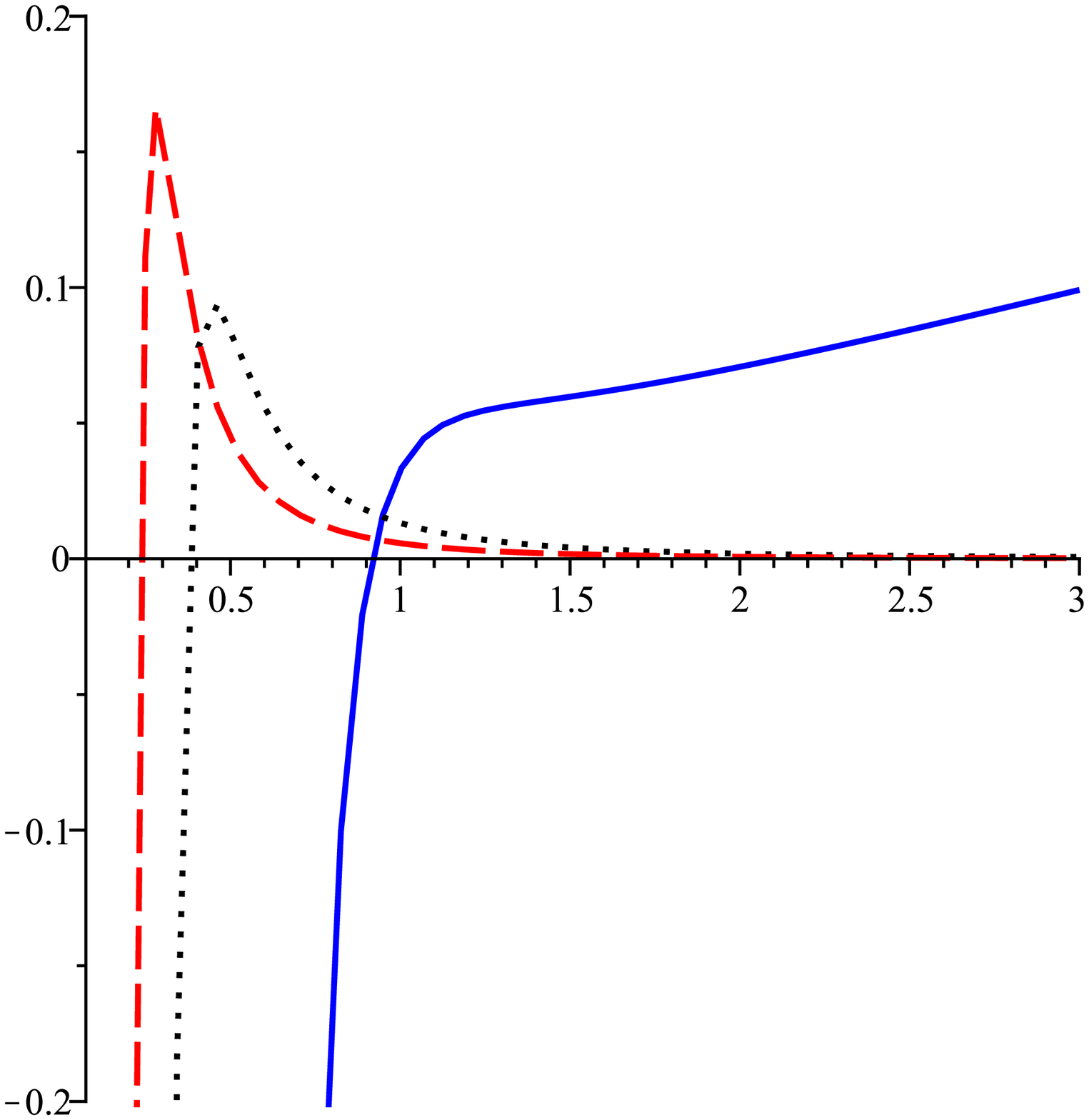}
	\epsfxsize=5.5cm \epsffile{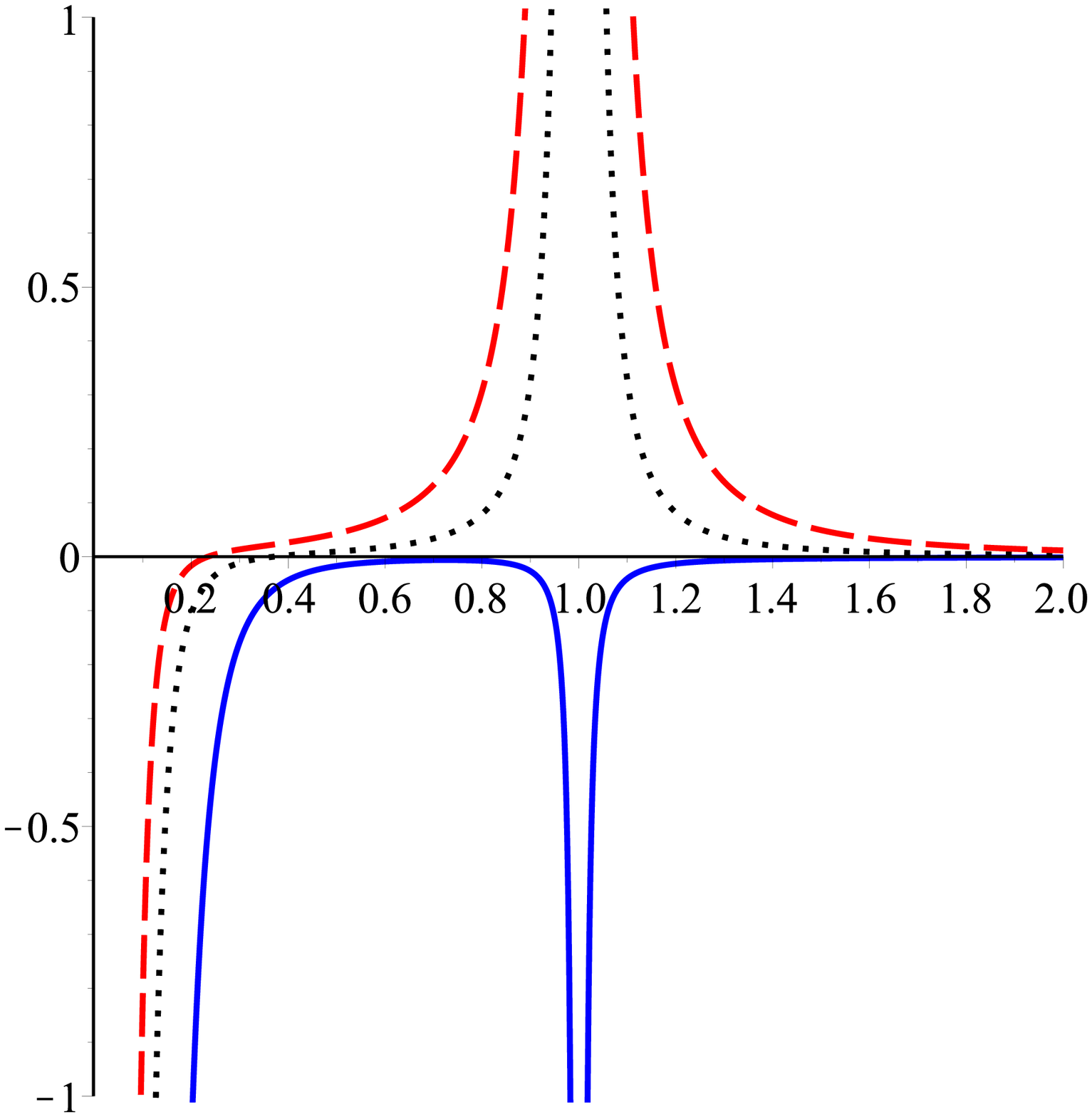}
	\epsfxsize=5.5cm \epsffile{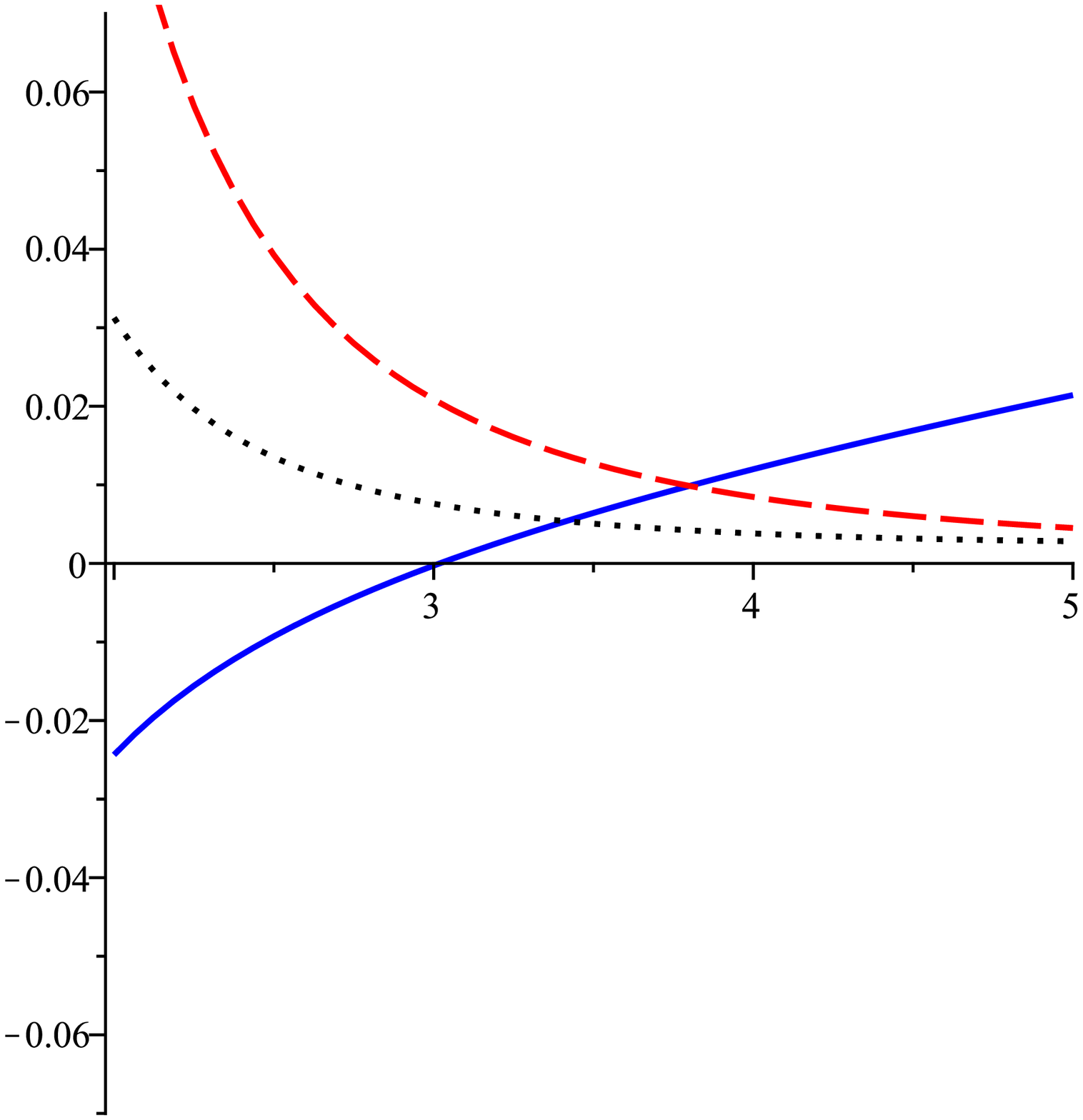}
	&  & 
	\end{array}
	$%
	\caption{ $T$ versus ${r_ + }$ for $q=1$, $\Lambda = - 1$, $m=0.1$, $\alpha=1$, $c_{1}=c_{2}=c_{3}=c_{4}= 1$, $c_{0}=1$ ({solid lines}), $c_{0}=5$ ({dotted lines}) and $c_{0}=10$ ({dashed lines}).
		\textbf{Left panel} for $k=0$. \textbf{Middle panel} ($0<r_{+}<2$) and \textbf{right panel} ($2<r_{+}<5$) for $k=-1$ with different scales.}
	\label{Temp-k=0-1}
\end{figure}

 In this section, we perform a thermal stability analysis for massive charged AdS black holes in the canonical ensemble, so the conserved (electric) charge $Q$ will be regarded as a fixed parameter. The local thermodynamic stability requires that ${C_Q} = T{\left( {H_{s,s}^M} \right)^{ - 1}} \ge 0$ which is equivalent to the internal energy, $M(S,Q)$, be a concave function of its extensive quantities. In this regard, we calculate the heat capacity as
\begin{equation}
{C_Q} = T{\left( {\frac{{\partial S}}{{\partial T}}} \right)_Q} = \frac{D}{F},
\end{equation}
where
\begin{equation}
D = 3{d_2}\left[ {{d_2}k({d_3} + {d_5}k\alpha r_ + ^{ - 2} + {d_7}{\alpha ^2}r_ + ^{ - 4}/3)r_ + ^{2{d_3}} + {d_2}{m^2}{\cal B}r_ + ^{2d - 5} - 2{q^2} - 2r_ + ^{2{d_2}}\Lambda } \right]{(\alpha  + r_ + ^2)^3}r_ + ^{{d_6}},
\end{equation}
\begin{equation}
\begin{array}{l}
F = 4\Big[12{d_2}k\alpha {m^2}{c_0}{c_1}r_ + ^{2d - 5} + 3{d_2}{d_3}(3k\alpha  - r_ + ^2)(k + {m^2}c_0^2{c_2})r_ + ^{2{d_3}} + 6{d_2}{d_3}{d_4}{m^2}c_0^3{c_3}(k\alpha  - r_ + ^2)r_ + ^{2d - 7}\\
\,\,\,\,\,\,\,\,\,\, + 3{d_2}{d_5}(k\alpha  - 3r_ + ^2)({k^2}\alpha  + {d_3}{d_4}{m^2}c_0^4{c_4})r_ + ^{2{d_4}} - {d_2}{d_7}k{\alpha ^2}(k\alpha  + 5r_ + ^2)r_ + ^{2{d_5}} - 6\Lambda (5k\alpha  + r_ + ^2)r_ + ^{2{d_2}}\\
\,\,\,\,\,\,\,\,\,\, + 6{q^2}((2d - 9)k\alpha  + (2d - 5)r_ + ^2)\Big].
\end{array}
\end{equation}
As mentioned before, we have restricted our study to ${\cal U}_i$ up to the fourth interaction term (${\cal U}_4$). To study local thermodynamic stability of black holes, first, we explore the physical temperature for topological black holes. Numerical calculations for thermal analysis show that for a black hole with definite mass $M$, there always exists a lower value for the radius of event horizon, i.e. $r_{b}$, in which the black hole temperature is always positive for $r_{+}>r_{b}$. According to eq. (\ref{temperature - massive AdS BH}), it can be seen that temperature always has a root regardless of horizon geometry. In Fig. \ref{Temp-k=1}, the typical behavior of temperature is depicted for spherical black holes with various values for spacetime dimensions ($d$), Lovelock coefficient ($\alpha$) and massive parameter $c_{0}$. As seen, the value of $r_{b}$ depends on many parameters in the theory. For instance, it is an increasing function of  spacetime dimensions ($d$) and a decreasing function of $\alpha$ and $c_{0}$. In addition, we plot Fig. \ref{Temp-k=0-1} to investigate the qualitative behavior of temperature for Ricci flat and hyperbolic black holes. For Ricci flat black holes, the behavior of temperature is qualitatively similar to spherical black holes (see left panel of Fig. \ref{Temp-k=0-1}). The result is radically different for hyperbolic black holes. In this case, interestingly, an infinity is observed at the divergent point ${r_{ i}} = \sqrt \alpha$. Again, the black hole temperature is always positive for $r_{+}>r_{b}$, in which $r_{b}$ is a lower value for the radius of event horizon (see middle and right panels of Fig. \ref{Temp-k=0-1}). If $r_{i}>r_{b}$, an infinite (positive) temperature is observed for hyperbolic black holes with horizon radius of $r_{+}={r_{i}} = \sqrt \alpha$.

It should emphasize that the temperature behaves as $T\propto r_{+}$ for large values of the event horizon $r_{+}$ (see eq. \ref{temperature - massive AdS BH}). Therefore, temperature diagrams (Figs. \ref{Temp-k=1} and \ref{Temp-k=0-1}) diverge at $r_{+}\rightarrow \infty$ for massive AdS black holes with diverse horizon topologies.\\

\begin{figure}[!htbp]
	$%
	\begin{array}{ccc}
	\epsfxsize=5.5cm \epsffile{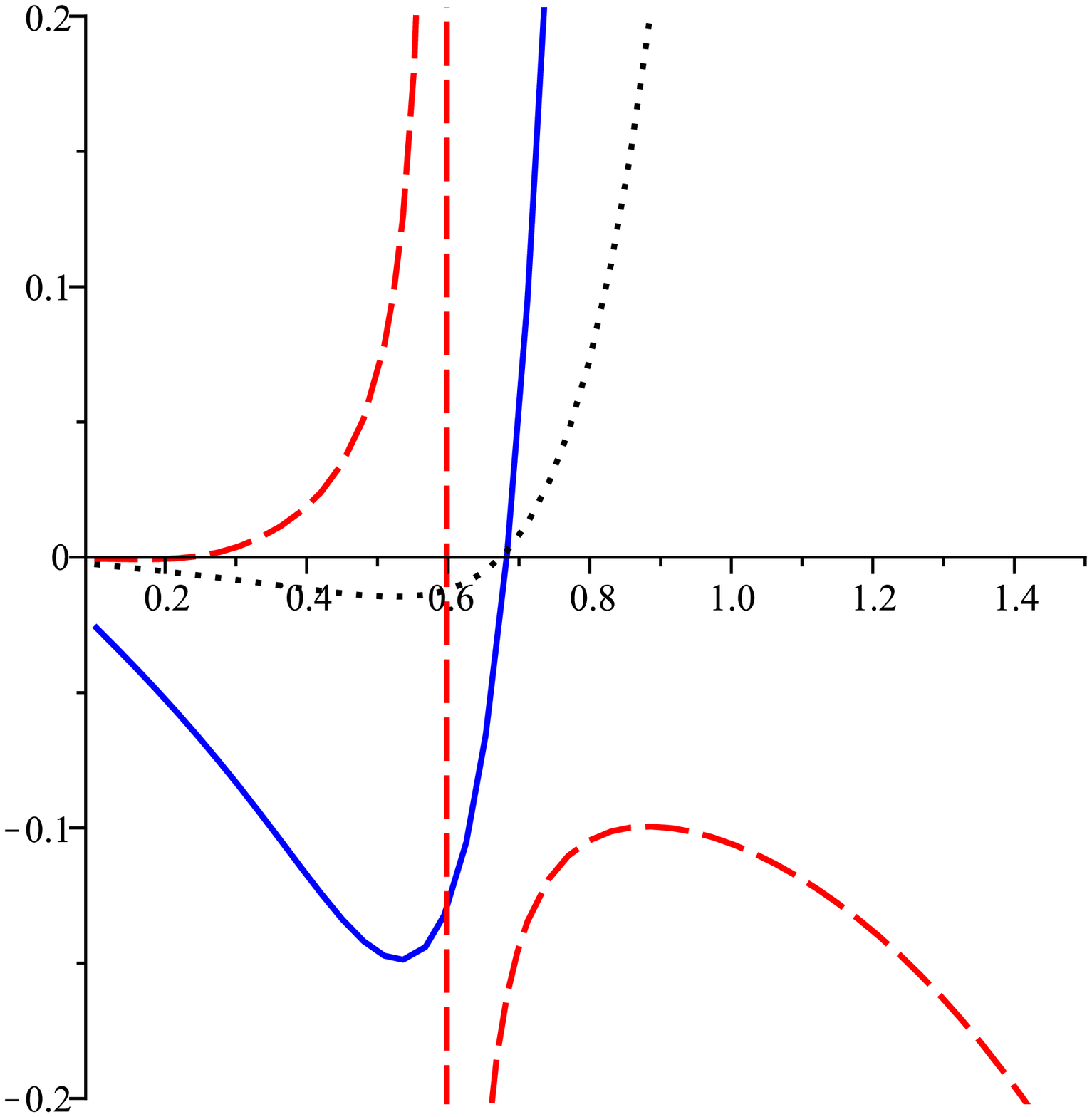} \epsfxsize=5.5cm %
	\epsffile{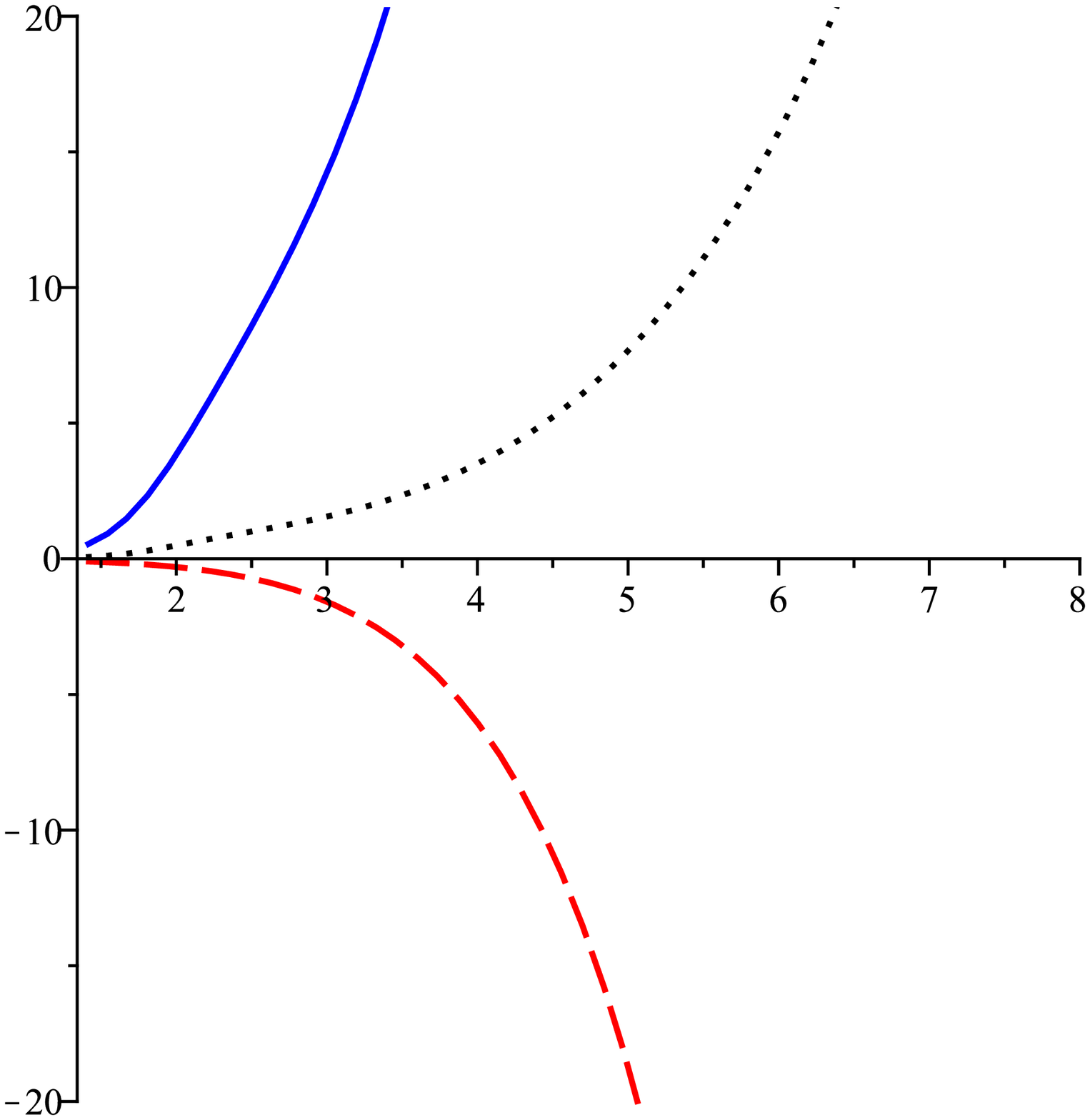}\epsfxsize=5.5cm %
	\epsffile{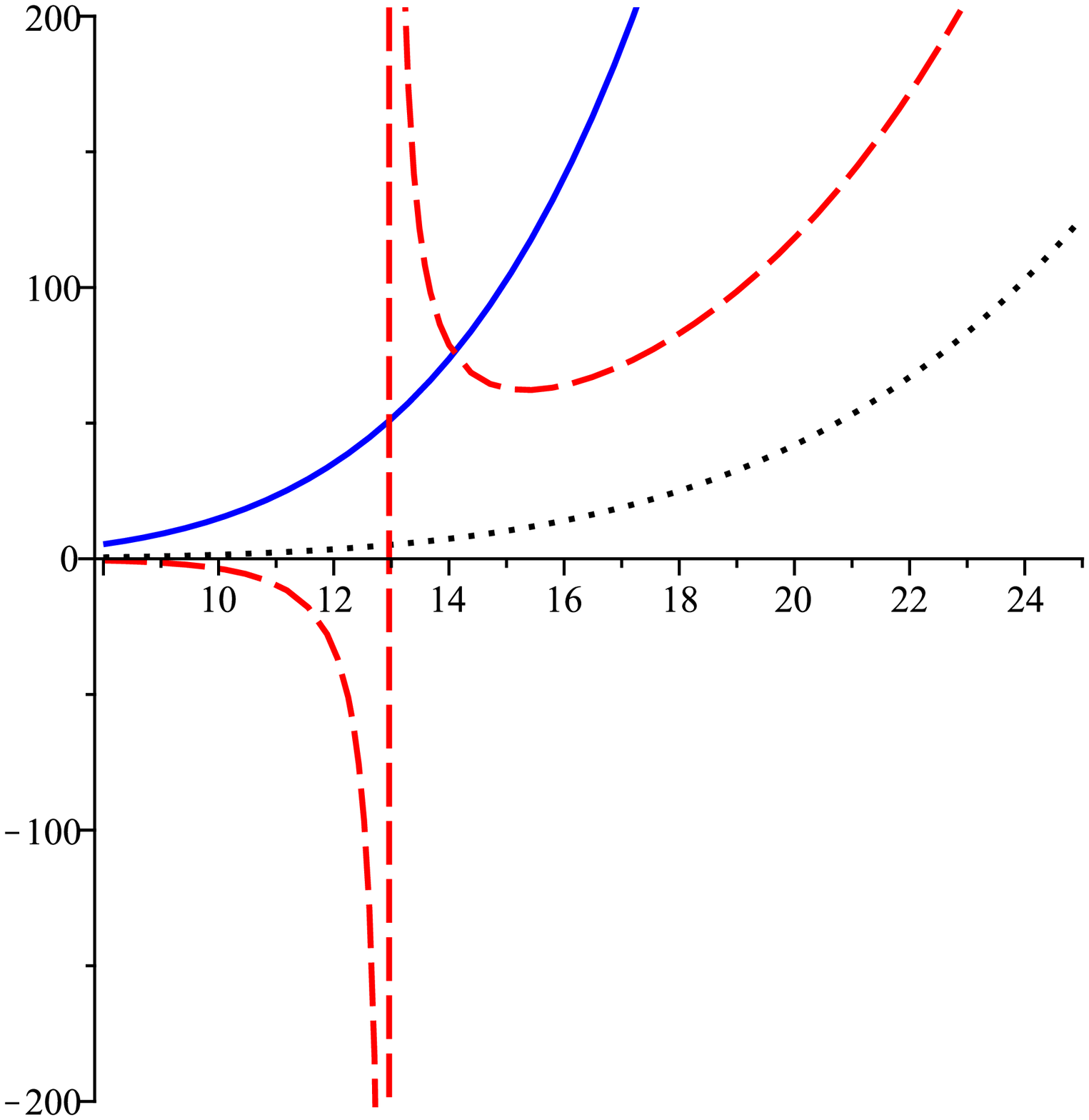} &  &
	\end{array}
	$%
	\caption{ $C_{Q}$ versus ${r_ + }$ for $k=1$, $d=7$, $q=1$, $\alpha = 1$, $\Lambda = - 1$, $m=0.1$,  $c_{1}=c_{2}=c_{3}=c_{4}= 1$ and $c_{0}=0.1$ ({solid line}), $c_{0}=1$ ({dotted line}) and $c_{0}=10$ ({dashed line}).
		\textbf{Different scales:} \textit{left panel} ($0<r_{+}<1.5$), \textit{middle panel} ($1.4<r_{+}<8$) and \textit{right panel}
		($8<r_{+}<25$).}
	\label{HC-massive}
\end{figure}

\begin{figure}[!htbp]
	$%
	\begin{array}{ccc}
	\epsfxsize=5.5cm \epsffile{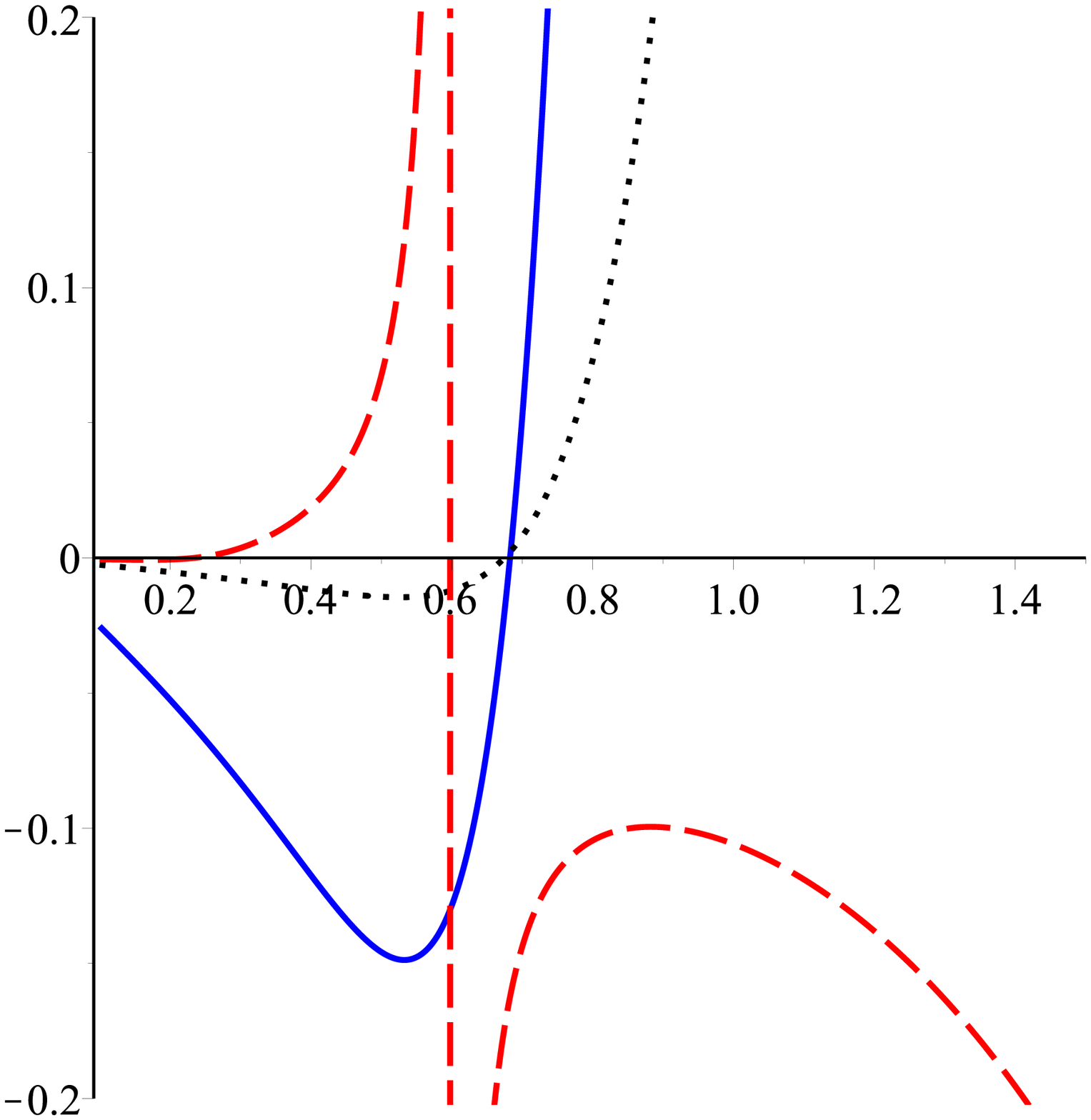} \epsfxsize=5.5cm %
	\epsffile{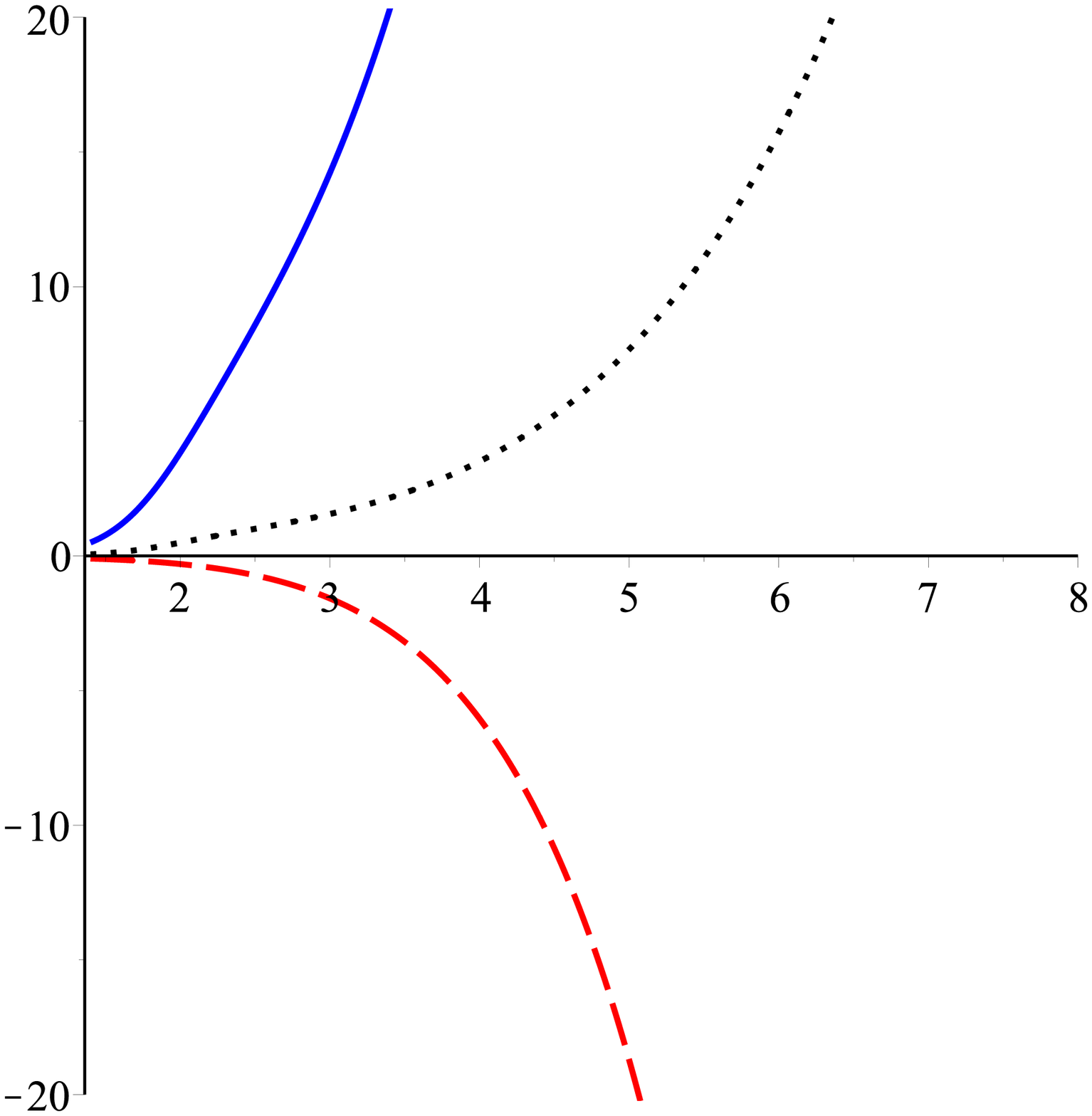}\epsfxsize=5.5cm %
	\epsffile{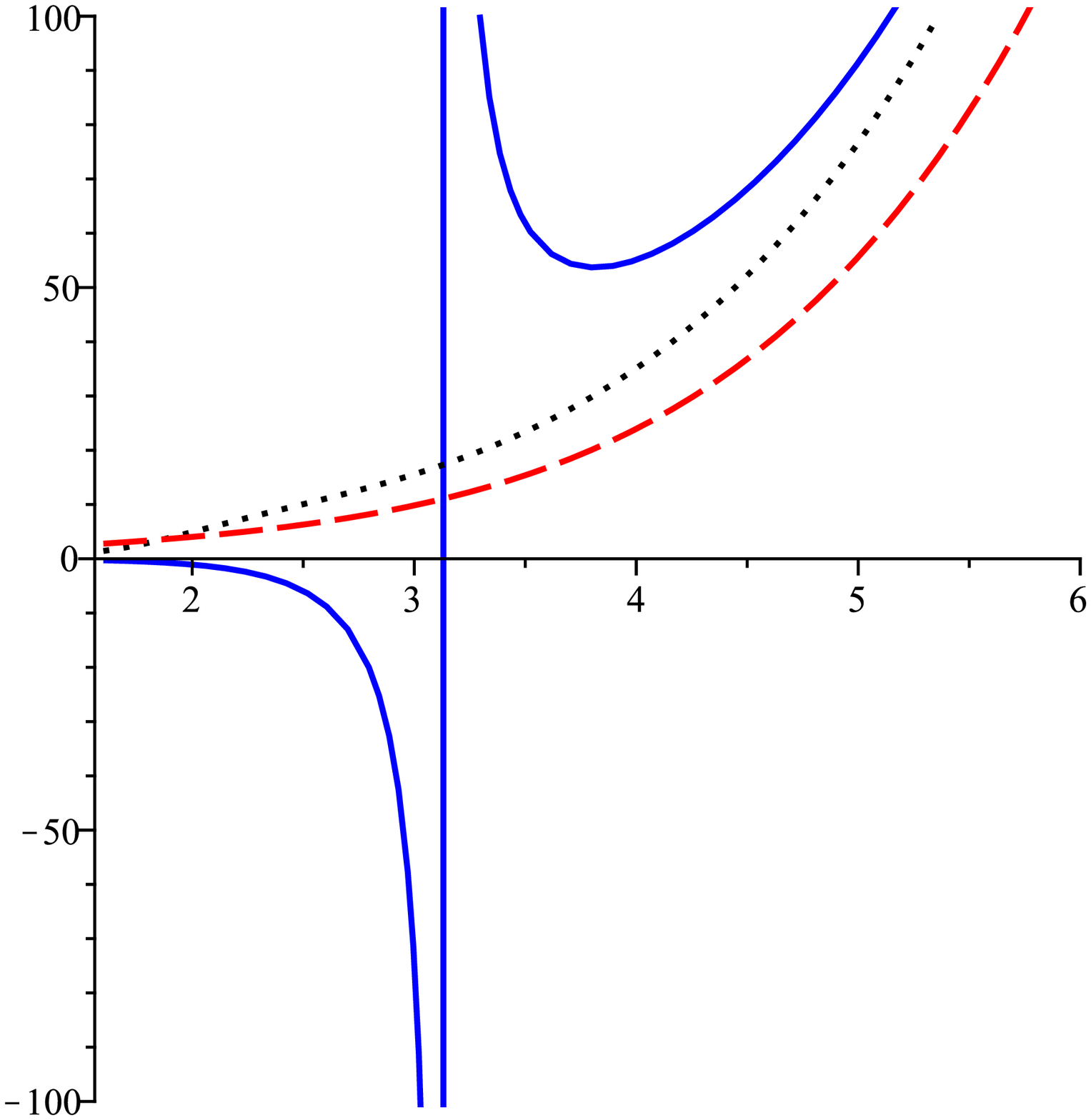} &  &
	\end{array}
	$%
	\caption{ $C_{Q}$ versus ${r_ + }$ for $k=1$, $d=7$, $q=1$, $\Lambda = - 1$, $m=0.1$, $c_{0}=c_{1}=c_{2}=c_{3}=c_{4}= 1$ and $\alpha = 0.1$ ({solid line}), $\alpha = 1$ ({dotted line}) and $\alpha = 10$ ({dashed line}).
		\textbf{Different scales:} \textit{left panel} ($0<r_{+}<0.9$), \textit{middle panel} ($0.8<r_{+}<1.6$) and \textit{right panel}
		($1.5<r_{+}<6$).}
	\label{HC-alpha}
\end{figure}

\begin{figure}[!htbp]
	$%
	\begin{array}{ccc}
	\epsfxsize=5.5cm \epsffile{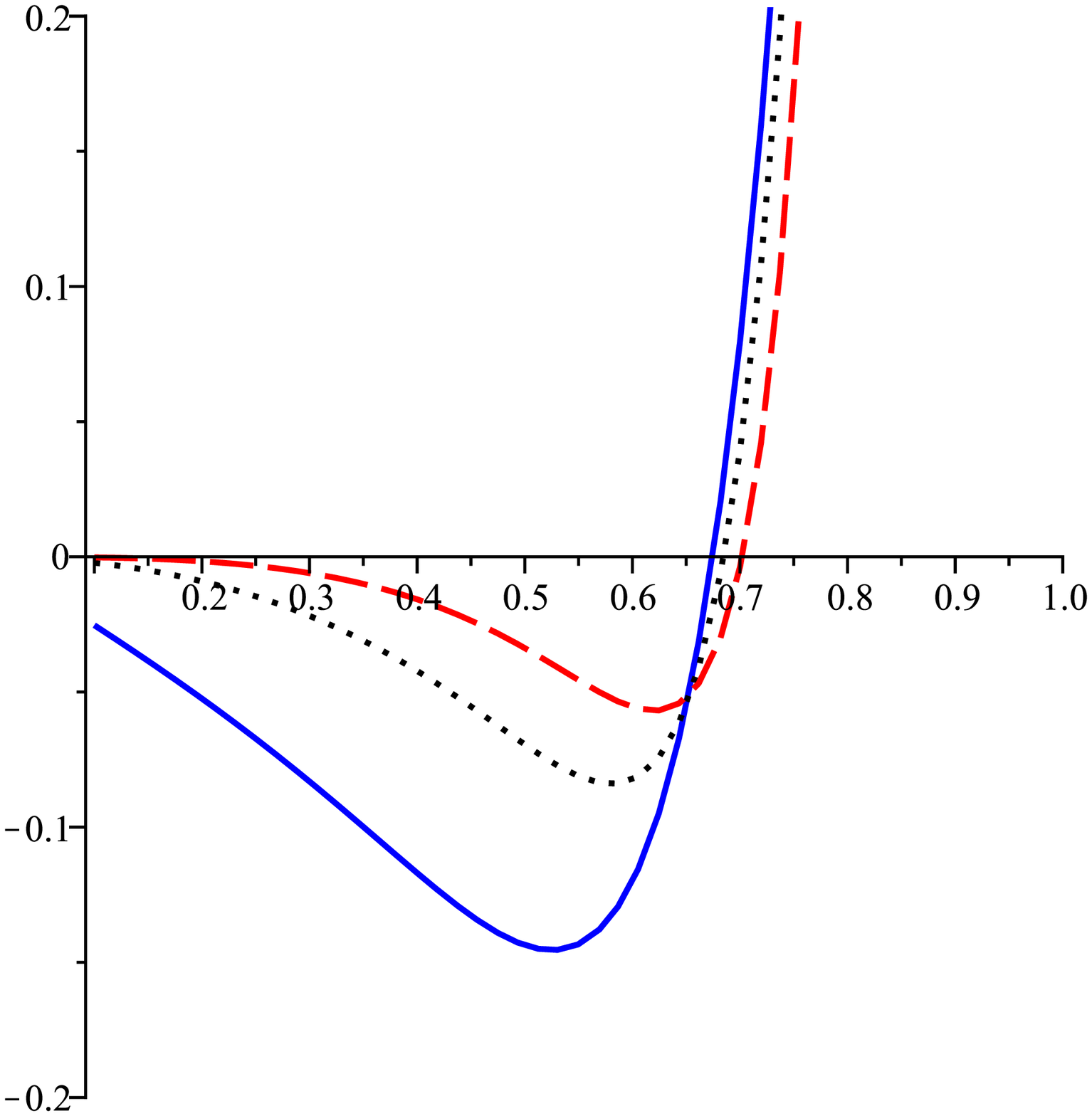} \epsfxsize=5.5cm %
	\epsffile{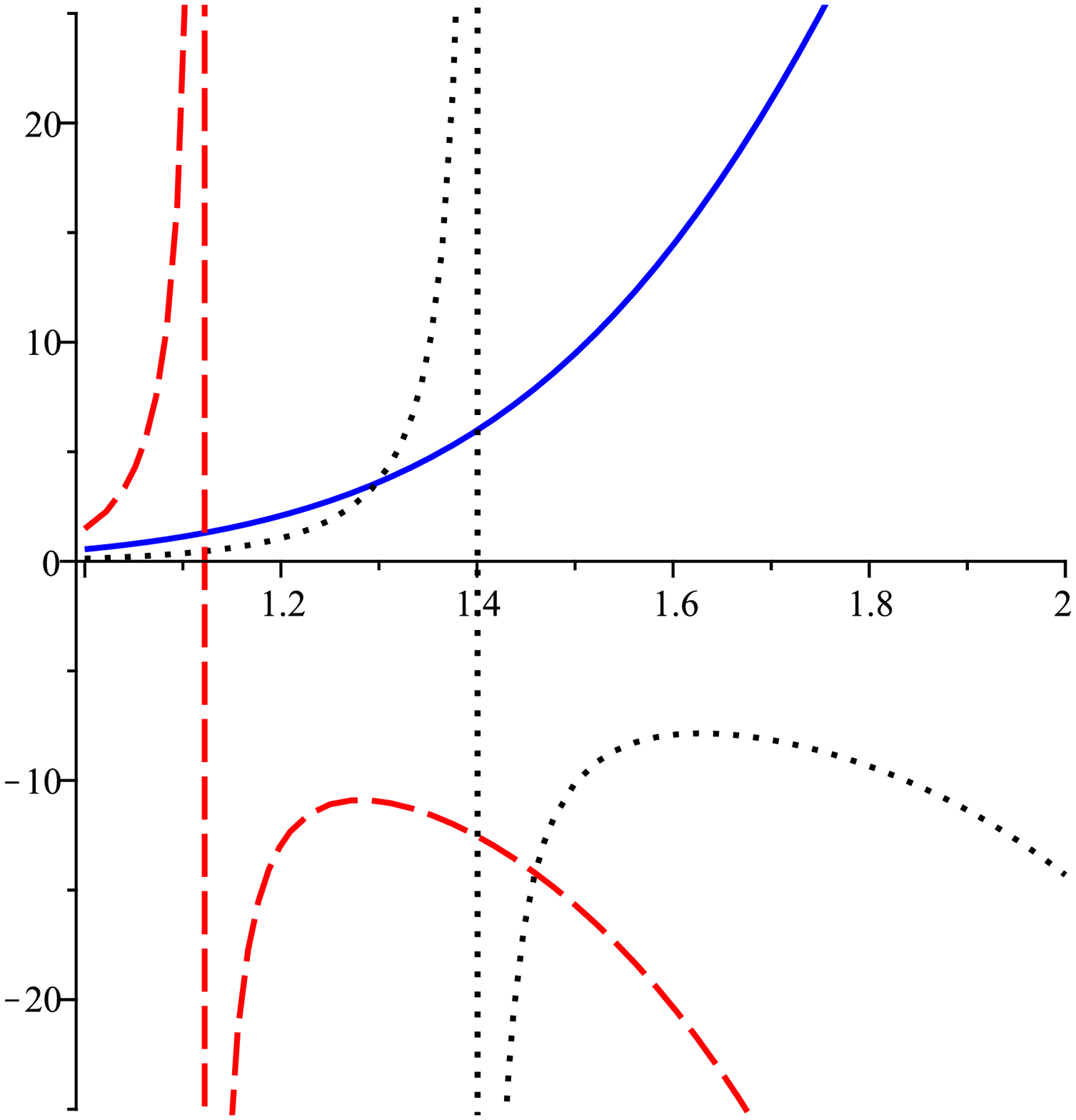}\epsfxsize=5.5cm \epsffile{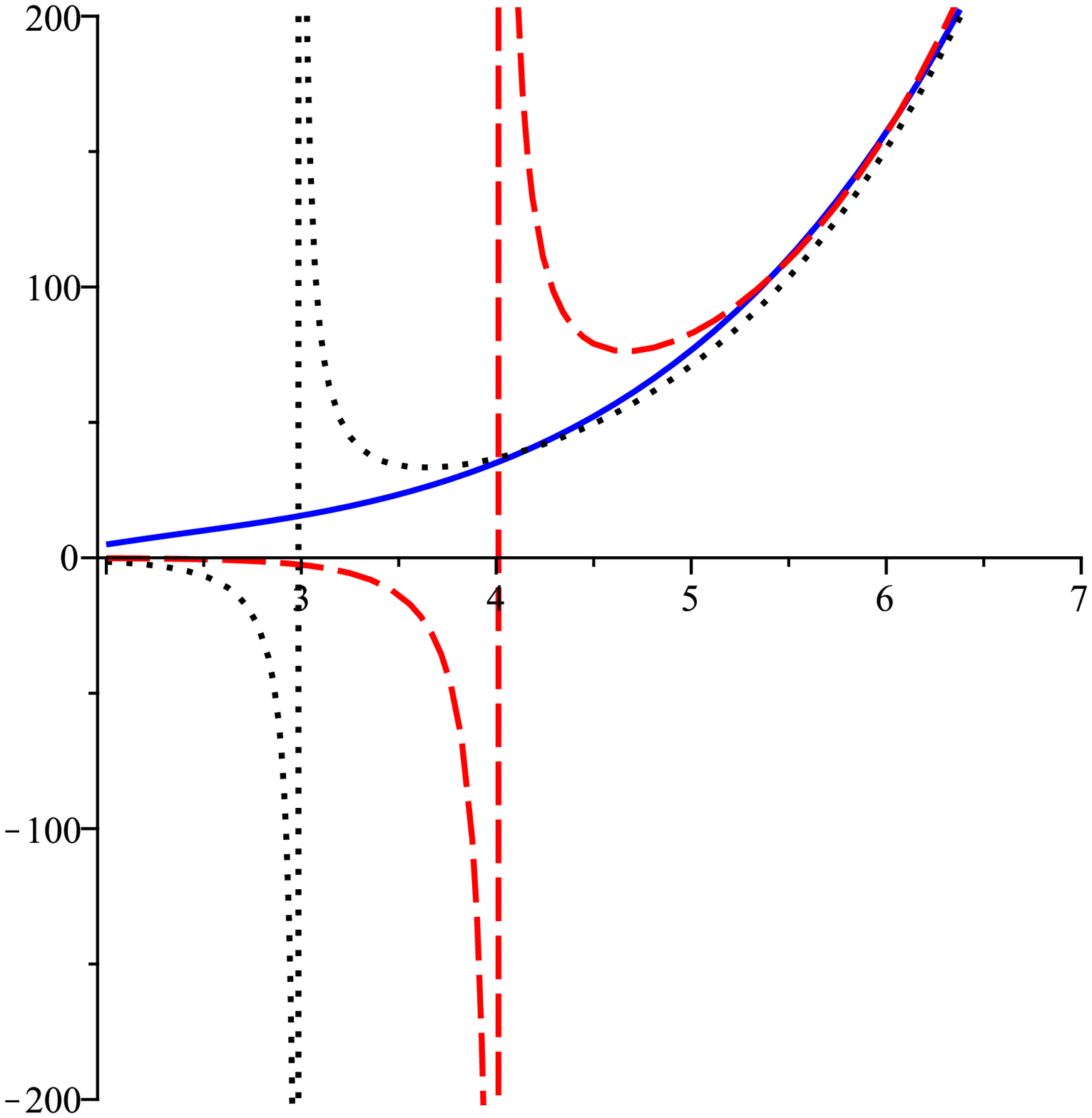}
	&  &
	\end{array}
	$%
	\caption{ $C_{Q}$ versus ${r_ + }$ for $k=1$, $q=1$, $\alpha = 1$, $\Lambda = - 1$, $m=0.1$, $c_{0}=c_{1}=c_{2}=c_{3}=c_{4}= 1$ and $d=7$ ({solid line}), $d=8$ ({dotted line}) and $d=9$ ({dashed line}).
		\textbf{Different scales:} \textit{left panel} ($0<r_{+}<1$), \textit{middle panel} ($1<r_{+}<2$) and \textit{right panel} ($2<r_{+}<7$).}
	\label{HC-dimensions}
\end{figure}

\textbf{Spherical black holes ($k=1$):}
In this case, there exists a lower value for event horizon radius, $r_{b}$, in which black holes are unstable for regions $r_{+}<r_{b}$. In Figs. \ref{HC-massive}-\ref{HC-dimensions}, we have displayed various cases for behavior of $C_{Q}$ with respect to $r_{+}$. As seen, the heat capacity always has only one root ($r_{b}$) in which it is negative definite for regions $r_{+}<r_{b}$. For regions $r_{+}>r_{b}$, so as before in Einstein gravity, there are two possibilities. Depending on the values of $q$, $\alpha$, $m$ (graviton mass), massive gravity couplings ($c_{1}$, $c_{2}$, $c_{3}$, $c_{4}$) and spacetime dimensions ($d$), the heat capacity i) may be an increasing function of the $r_{+}$, or ii) may have two divergent points. Massive AdS black holes would be thermally stable if $C_{Q}$ were an increasing function without any divergence or root in regions $r_{+}>r_{b}$. For the second possibility, again, we refer to the first and second divergent points as $r_{m}$ and $r_{u}$ respectively ($r_{b}<r_{m}<r_{u}$). As seen in Figs. \ref{HC-massive}-\ref{HC-dimensions}, for regions $r_{b}<r_{+}<r_{m}$ and $r_{+}>r_{u}$, massive AdS black holes are thermally stable and for regions $r_{m}<r_{+}<r_{u}$ are unstable.

For the sake of completeness, Figs. \ref{HC-massive}-\ref{HC-dimensions} are plotted for different cases to investigate the effects of Lovelock coefficient and massive graviton parameter ($m$) on thermal stability of the solutions in higher dimensions. Since the quantities $m$ and $c_{0}$ are always coupled, thus we would rather vary $c_{0}$ instead of $m$. The results are qualitatively similar.We found that thermally stable regions are drastically affected by both of the deformation parameters, i.e., $\alpha$ and $m$.\\



\begin{figure}[!hbbp]
	$%
	\begin{array}{cccc}
	\epsfxsize=4.5cm \epsffile{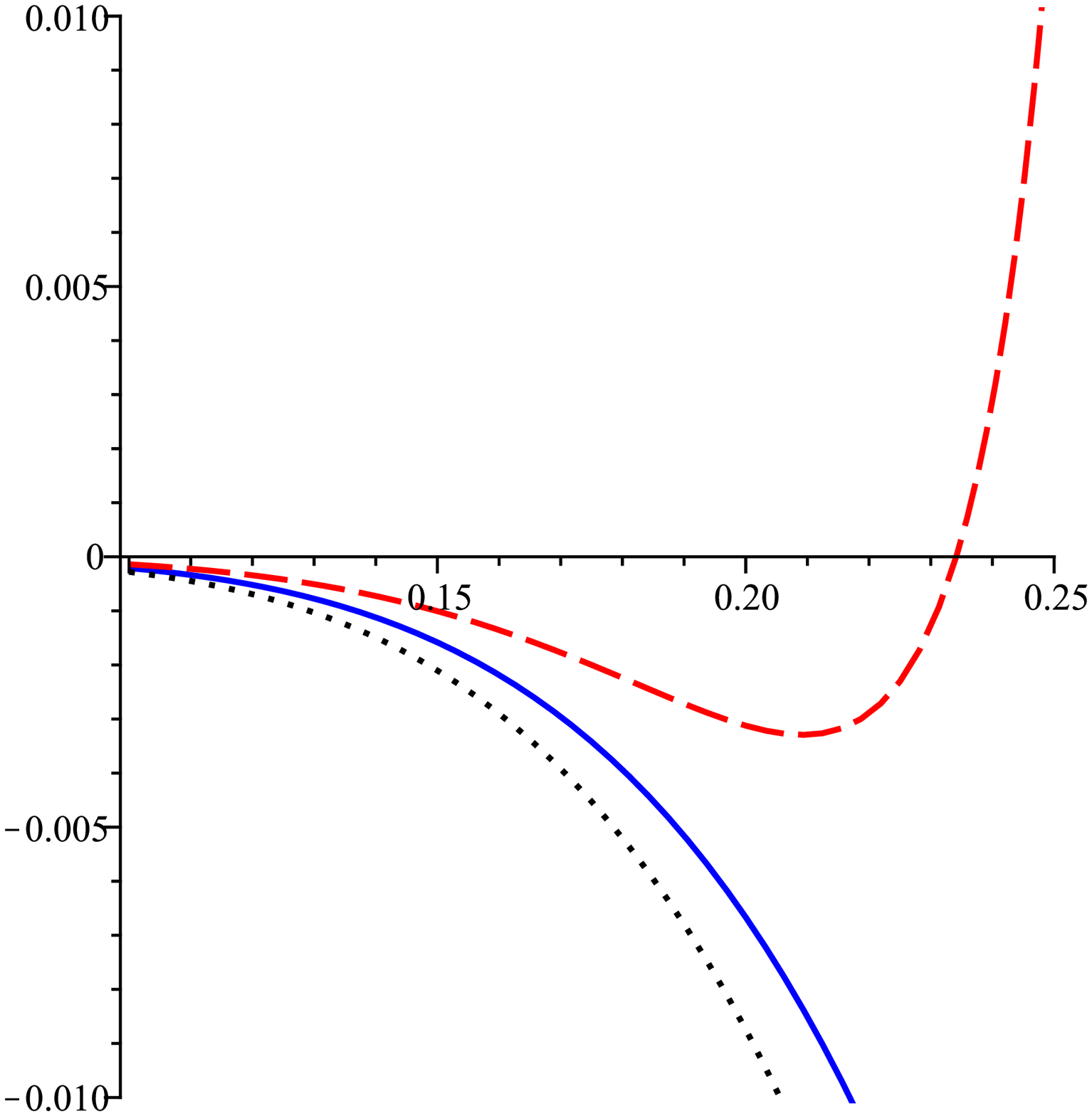}
	\epsfxsize=4.5cm \epsffile{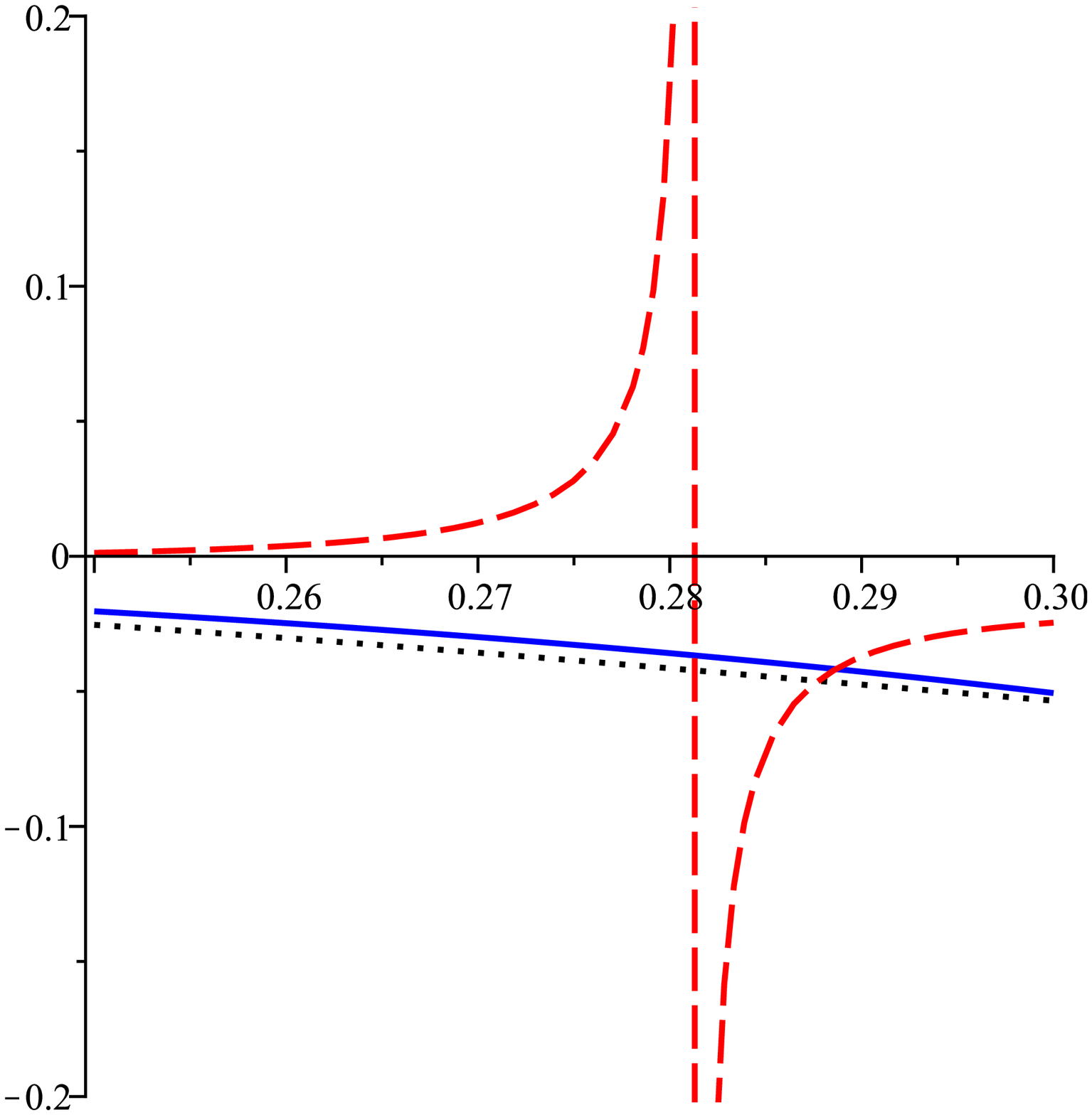}
	\epsfxsize=4.5cm \epsffile{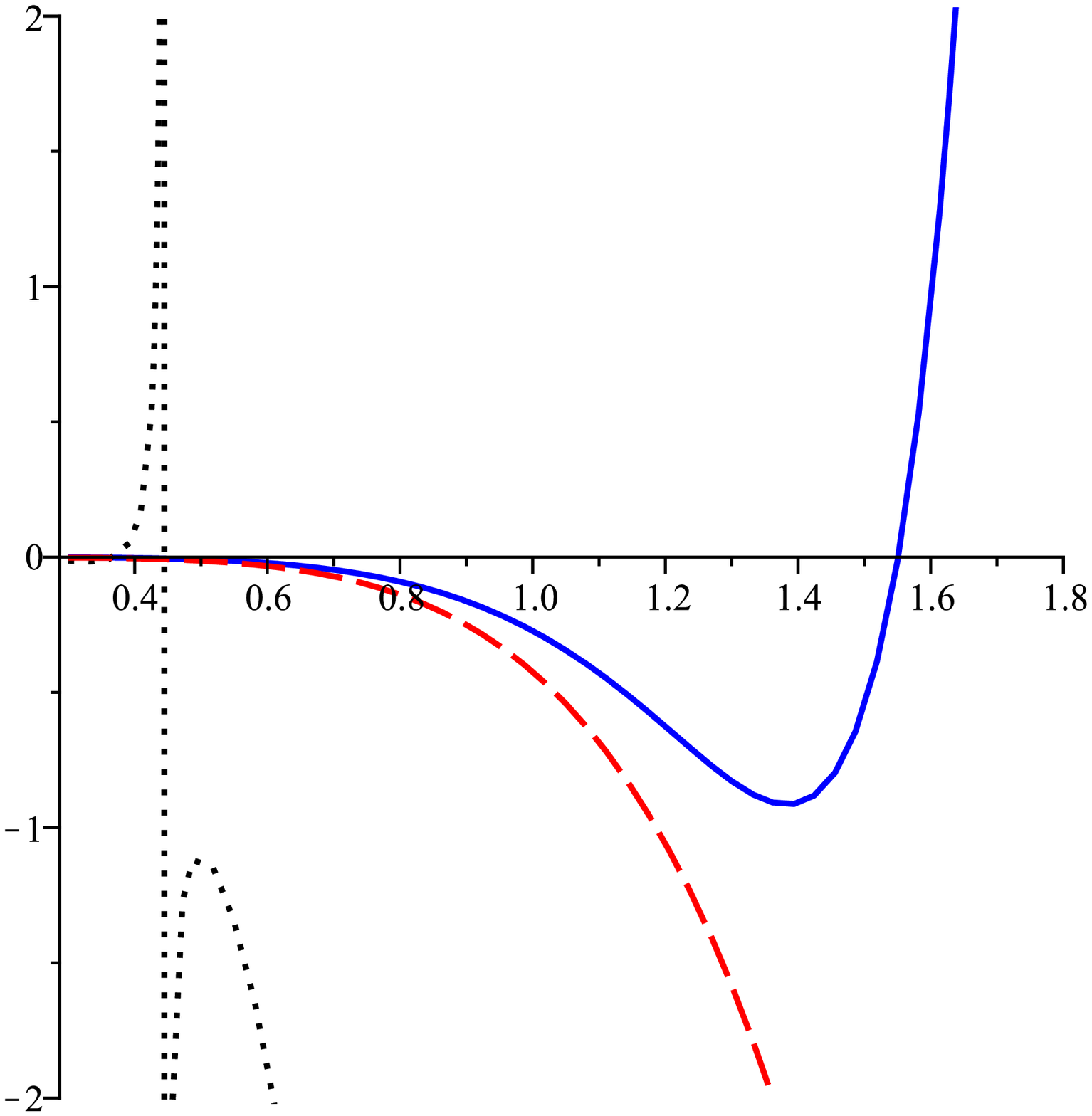}
	\epsfxsize=4.5cm \epsffile{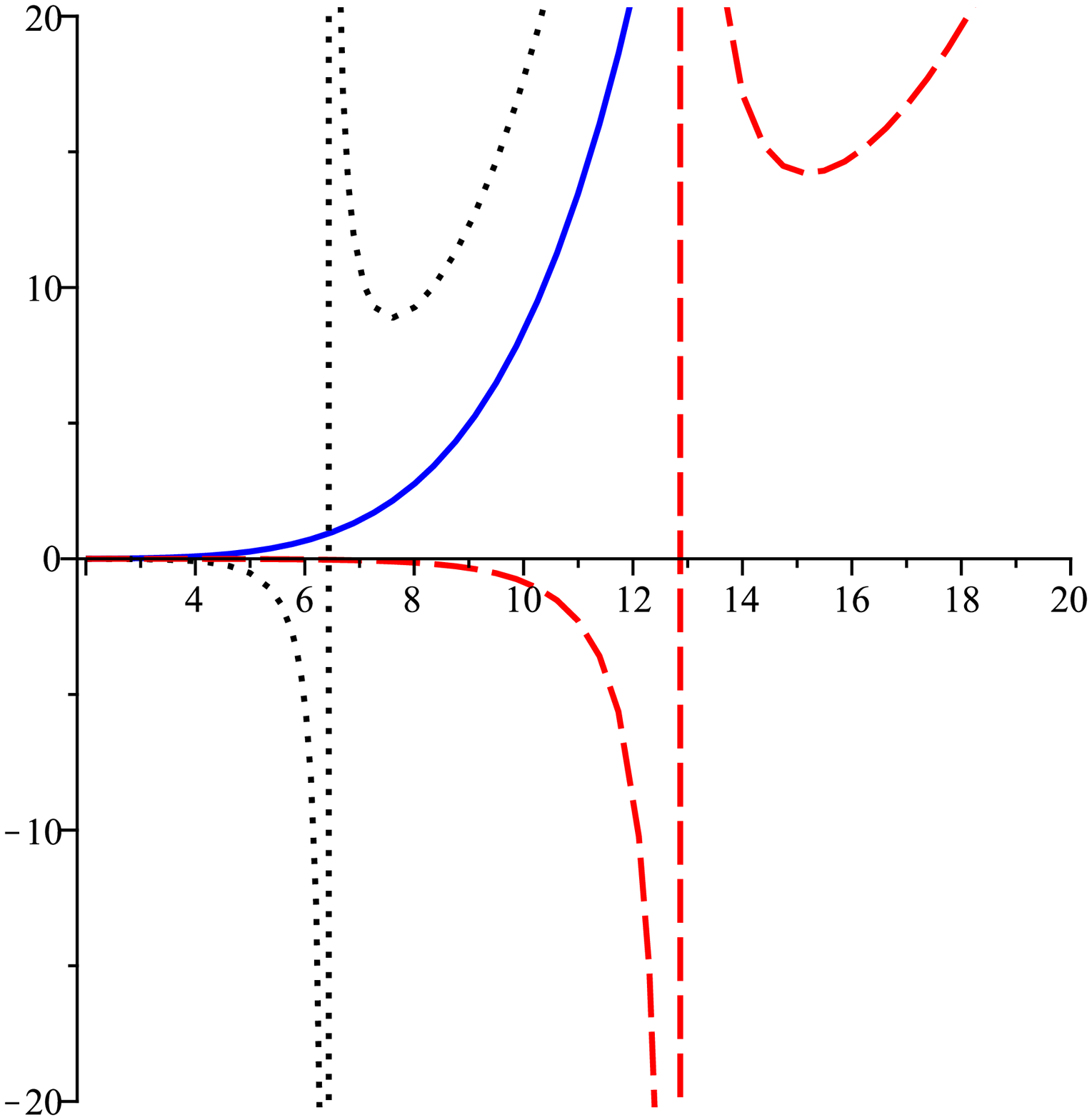}
	&  &  &
	\end{array}
	$%
	\caption{ $C_{Q}$ versus ${r_ + }$ for $k=0$, $d=7$, $q=1$, $\alpha = 1$, $\Lambda = - 1$, $m=0.1$, $c_{1}=c_{2}=c_{3}=c_{4}= 1$ and $c_{0}=1$ ({solid line}), $c_{0}=5$ ({dotted line}) and $c_{0}=10$ ({dashed line}).
		\textbf{Different scales:} \textit{first panel} ($0<r_{+}<0.25$), \textit{second panel} ($0.24<r_{+}<0.3$), \textit{third panel} ($0.3<r_{+}<1.8$) and  \textit{forth panel} ($1.8<r_{+}<20$).}
	\label{HC-k=0-massive}
\end{figure}

\textbf{Ricci flat black holes ($k=0$):}
In this case, Ricci flat black holes behaves typically like spherical black holes in Lovelock massive gravity. Again, the heat capacity always has only one root ($r_{b}$) with negative definite for regions $r_{+}<r_{b}$, and hence, Ricci flat black holes are thermally unstable (see Fig. \ref{HC-k=0-massive}). For region $r_{+}>r_{b}$, black holes are thermally stable if there is no divergence point in the heat capacity function. Depending on the values of parameters in the theory, the heat capacity may possess infinities (at one or two points) like those in the case of black holes with spherical horizons ($k=1$). In this regard, black holes are thermally stable in regions $r_{b}<r_{+}<r_{m}$ and $r_{+}>r_{u}$, and unstable for $r_{m}<r_{+}<r_{u}$.

In conclusion, the heat capacity of Ricci flat black holes qualitatively behaves like the case $k=1$ in a way reminiscent of spherical black holes in the Einstein gravity (see Figs. \ref{HC-k=0-massive} and \ref{HC-GR-k=1}). But they are completely different from Ricci flat black holes in Einstein gravity, in which the heat capacity cannot possess any divergent point.\\


\begin{figure}[!htbp]
	$%
	\begin{array}{ccc}
	\epsfxsize=5.5cm \epsffile{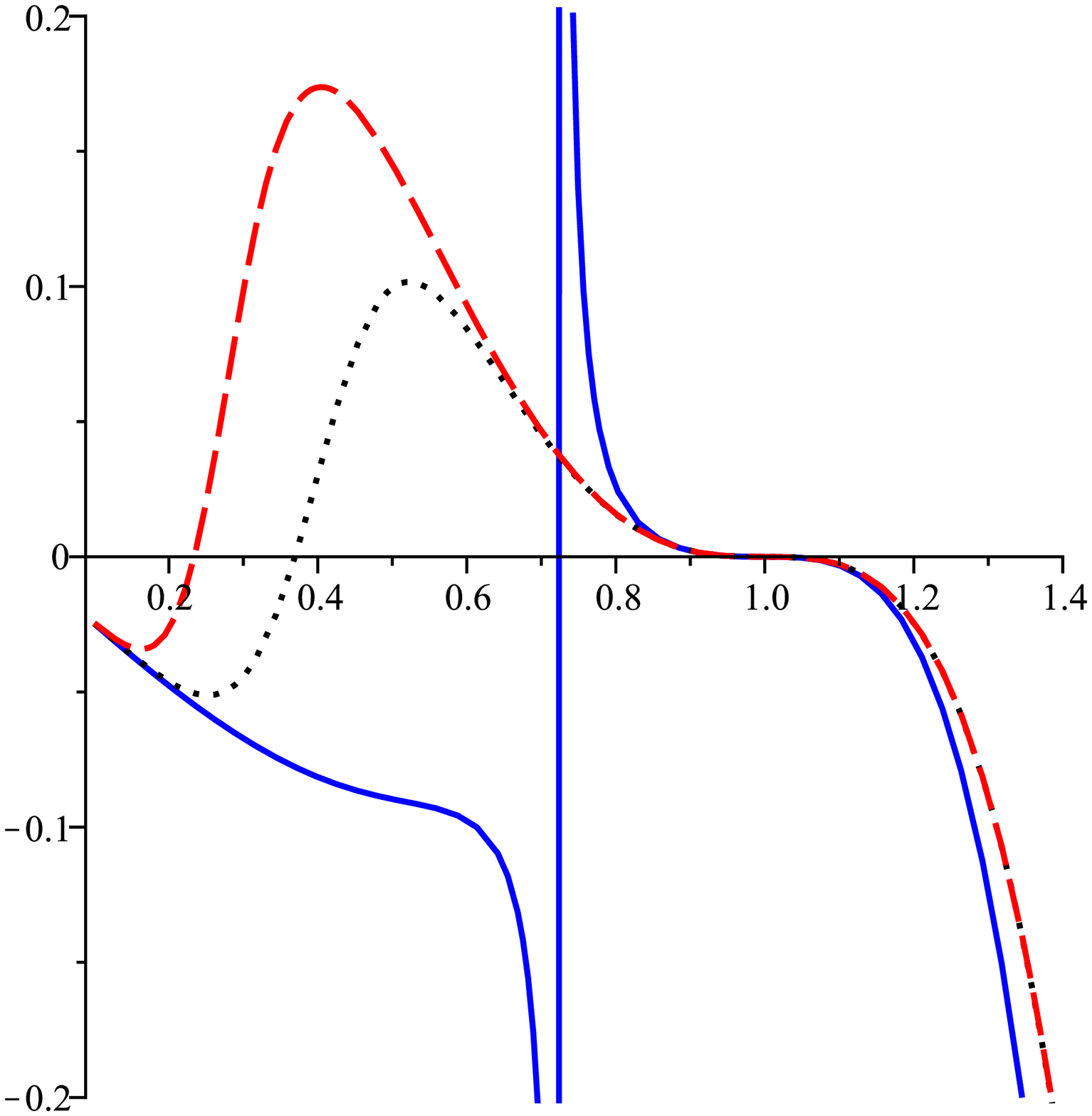}
	\epsfxsize=5.5cm \epsffile{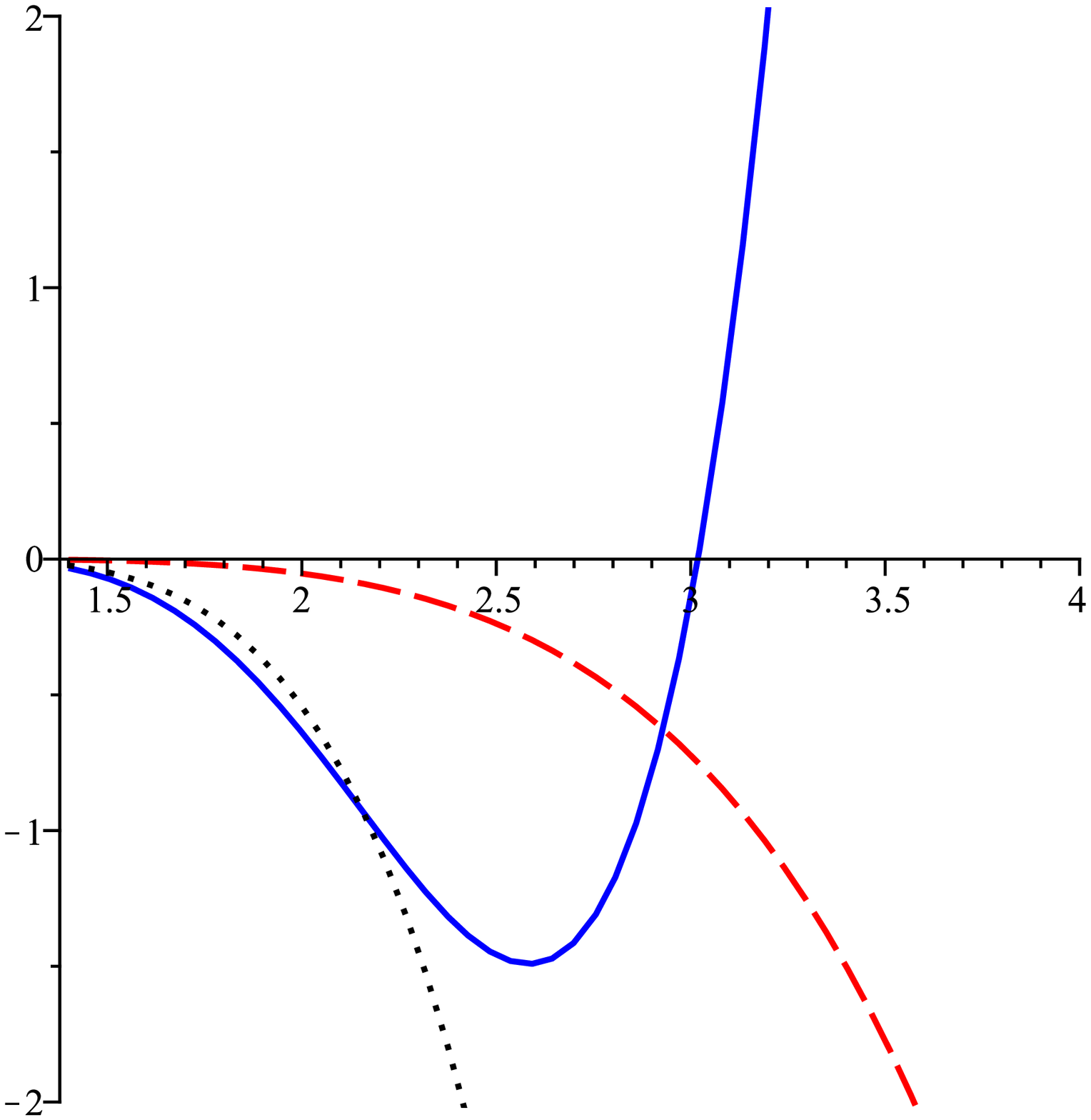}
	\epsfxsize=5.5cm \epsffile{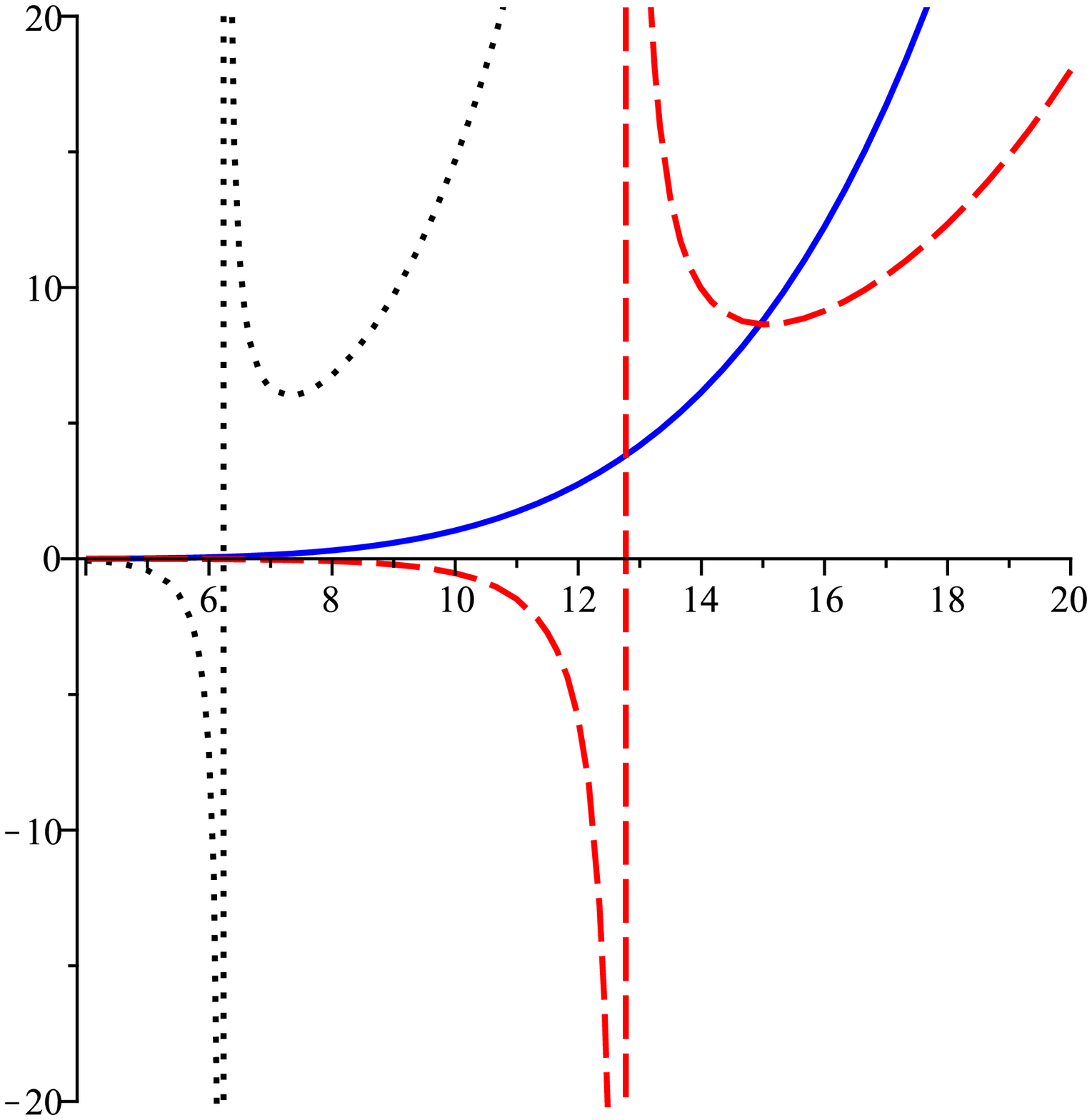}
	&  & 
	\end{array}
	$%
	\caption{ $C_{Q}$ versus ${r_ + }$ for $k=-1$, $d=7$, $q=1$, $\alpha = 1$, $\Lambda = - 1$, $m=0.1$, $c_{1}=c_{2}=c_{3}=c_{4}= 1$ and $c_{0}=1$ ({solid line}), $c_{0}=5$ ({dotted line}) and $c_{0}=10$ ({dashed line}).
		\textbf{Different scales:} \textit{left panel} ($0<r_{+}<1.4$), \textit{middle panel} ($1.4<r_{+}<4$) and \textit{right panel} ($4<r_{+}<20$).}
	\label{HC-k=-1-massive}
\end{figure}

\textbf{Hyperbolic black holes ($k=-1$):}
Interestingly, in the case of $k=-1$, there exists two positive roots and one divergent point ($r_{m}$) for the heat capacity (not seen in Einstein gravity). It is a general consequence of  higher curvature terms of TOL gravity and the results are radically different from hyperbolic black holes in Einstein gravity. As shown in Fig. \ref{HC-k=-1-massive}, the sequence of roots and divergent point is highly important to find stable regions. The smaller and larger roots are referred as $r_{b_{1}}$ and $r_{b_{2}}$ respectively. Investigations show two sequences are possible: i) the divergent point is smaller than the roots ($r_{m}<r_{b_{1}}<r_{b_{2}}$), and therefore, massive AdS black holes are thermally stable only in regions $r_{m}<r_{+}<r_{b_{1}}$ and $r_{+}>r_{b_{2}}$; ii) the divergent point is larger than the roots ($r_{b_{1}}<r_{b_{2}}<r_{m}$), and hence,  black hole solutions are thermally stable only in regions $r_{b_{1}}<r_{+}<r_{b_{2}}$ and $r_{+}>r_{m}$. In the other regions, hyperbolic black holes behave thermally unstable. The existence of two divergent points for the heat capacity of hyperbolic black holes will be clarified in a moment.\\

\begin{figure}[!htbp]
	$%
	\begin{array}{ccc}
	\epsfxsize=5.5cm \epsffile{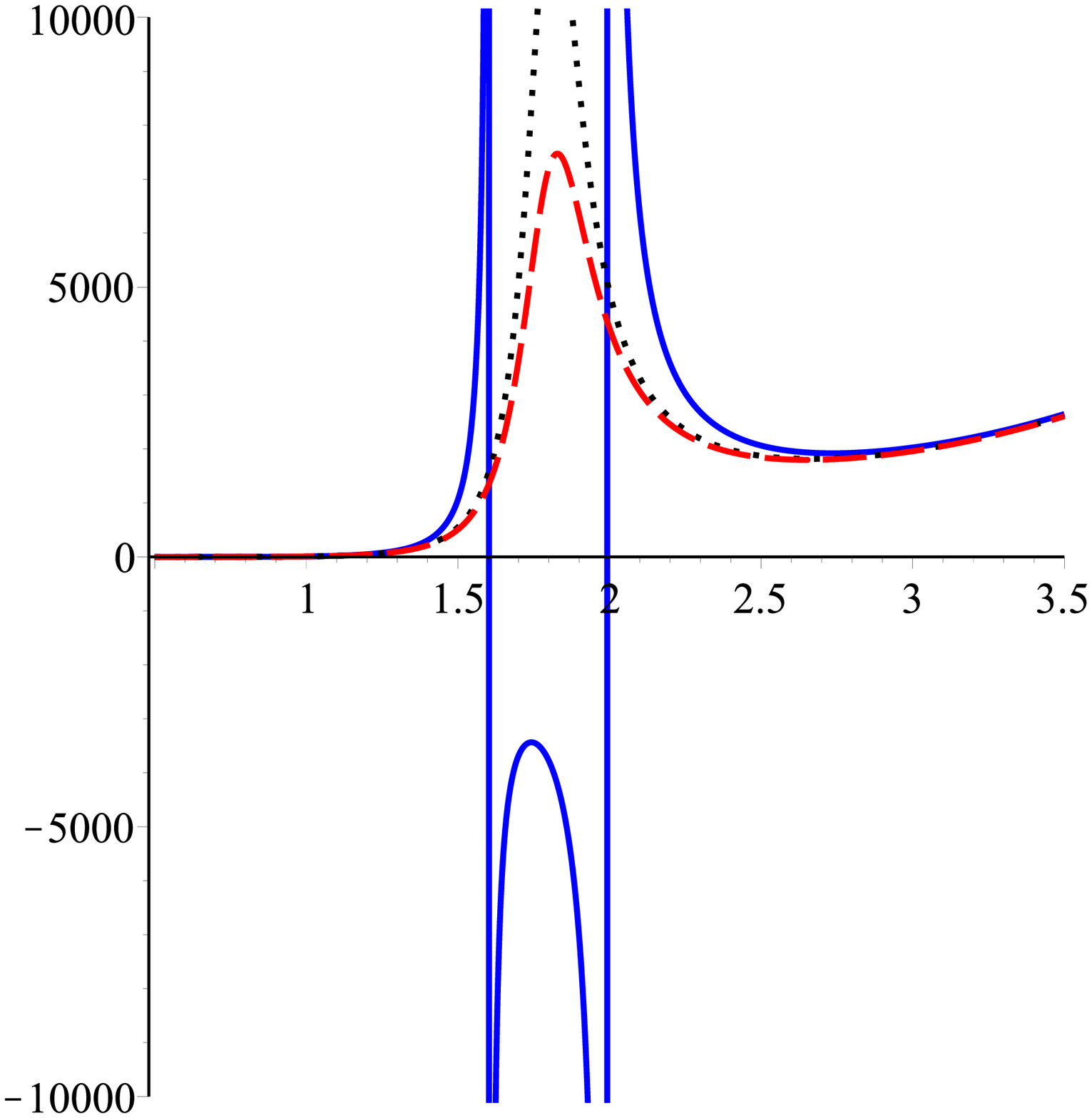}
	\epsfxsize=5.5cm \epsffile{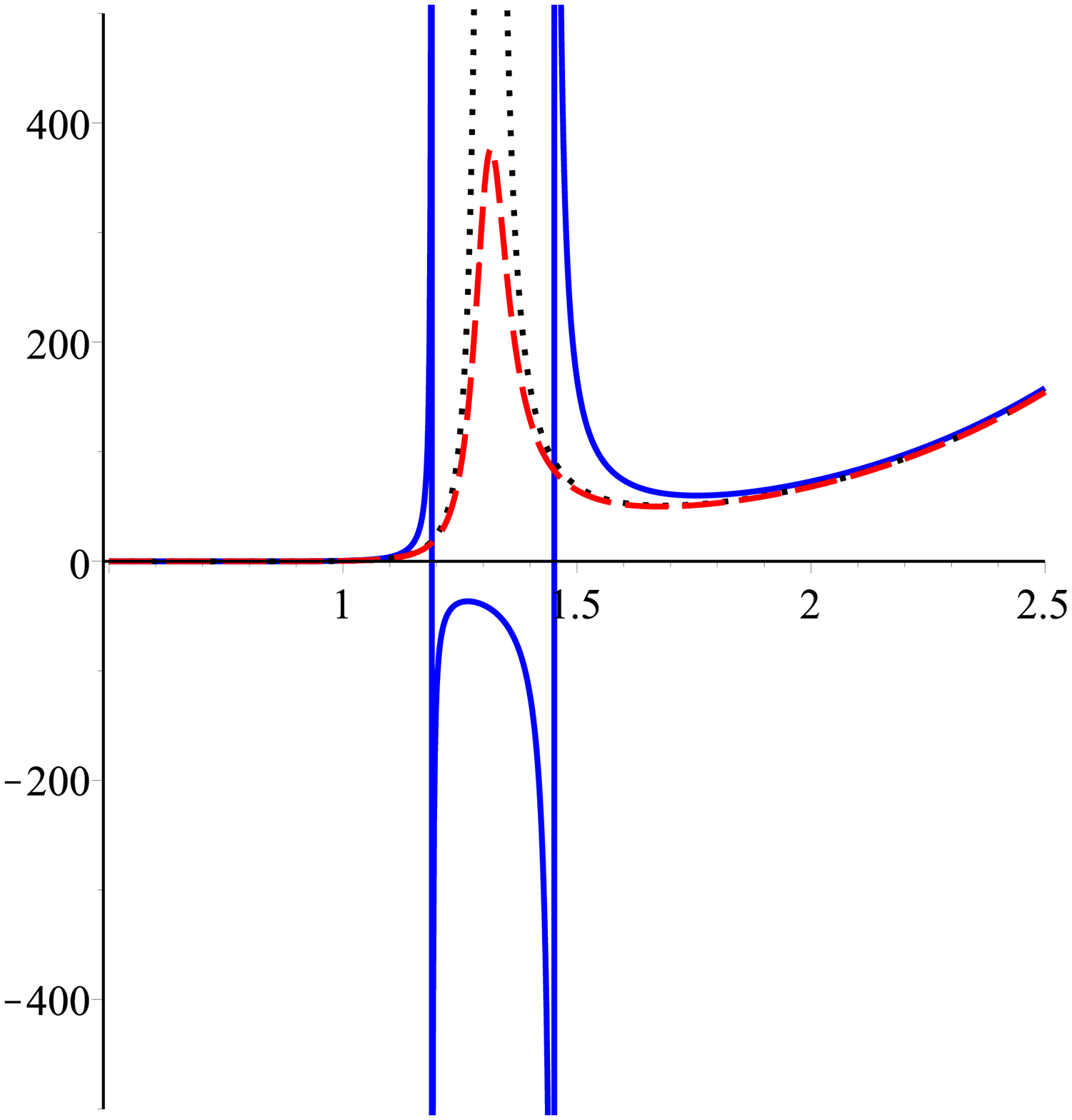}
	\epsfxsize=5.5cm \epsffile{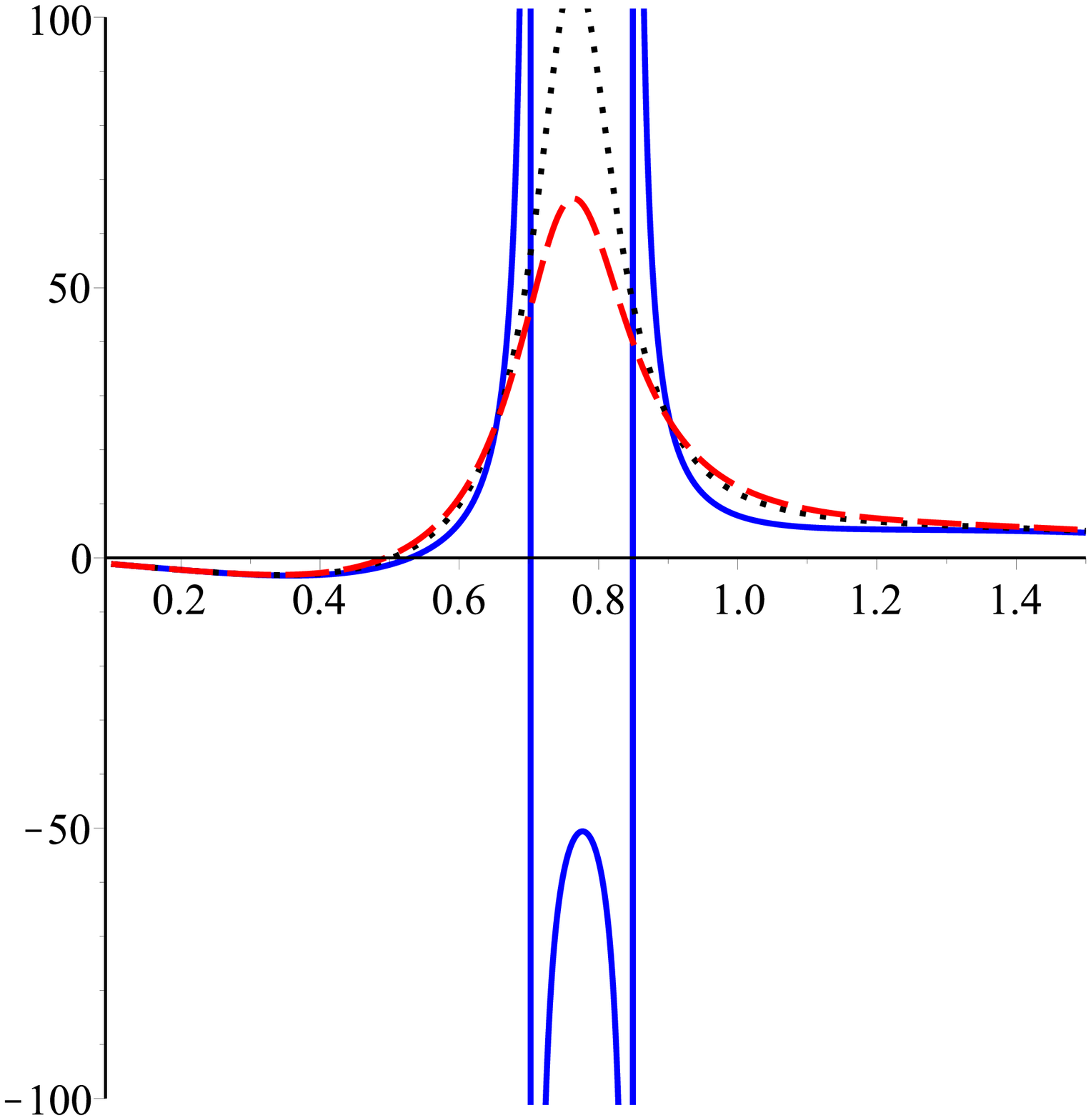}
	&  & 
	\end{array}
	$%
	\caption{ $C_{Q}$ versus ${r_ + }$ with one divergence point for topological black holes in $7$-dimensions. Solid lines display situation related to the two divergent points. By decreasing (for $k=+1,0$) and increasing ($k=-1$) the massive coupling $c_{4}$, the two divergent points (see solid lines) merge together to form one positive divergent point (see dotted and dashed lines). 
		\newline
		\textbf{Left panel:} for $k=+1$, $q=1$, $\alpha = 1$, $\Lambda = -1$, $m=0.1$, $c_{0}=1$, $c_{1}=-10$, $c_{2}=-10$, $c_{3}=-10$, $c_{4}= 7$ (solid line), $c_{4}= 6.4$ (dotted line) and $c_{4}= 6$ (dashed line).
		\newline 
		\textbf{Middle panel:} for $k=0$, $q=1$, $\Lambda = -1$, $m=1$, $c_{0}=1$, $c_{1}=20$, $c_{2}=10$, $c_{3}=-10$, $c_{4}= 2.4$ (solid line), $c_{4}= 2.09$ (dotted line) and $c_{4}= 2.04$ (dashed line).
		\newline
		\textbf{Right panel:} for $k=-1$, $q=1$, $\alpha = 6.7$, $\Lambda = -1$, $m=1$, $c_{0}=1$, $c_{1}=20$, $c_{2}=10$, $c_{3}=-10$, $c_{4}= 2.3$ (solid line), $c_{4}= 2.5$ (dotted line) and $c_{4}= 2.6$ (dashed line).}
	\label{HCODP}
\end{figure}

Now, we seek other qualitative behaviors of the topological black holes' heat capacity. As mentioned in section \ref{stabilty}, the heat capacity of spherically symmetric AdS black holes in Einstein gravity may have solely one divergence point ($r_{m}$) which is positive around such divergency. Interestingly, that behavior can occur for the obtained black hole solutions with various horizon topologies in LM gravity. In Fig. \ref{HCODP}, this possibility is depicted for all topological black holes in $7$-dimensions. Such single divergence point, with positive $C_{Q}$ around it, may signal critical behavior of the topological black holes. As seen, all plots have the same qualitative behavior as the spherically symmetric AdS ones in Einstein gravity. The only difference is the asymptotic behavior of the heat capacity of hyperbolic black hole in comparison with spherical and Ricci flat black holes. In fact, when $r_{+} \rightarrow \infty$, $C_{Q}$ approaches the large negative values for hyperbolic black hole solutions. Instead, the large positive values are observed for the asymptotic behavior of the heat capacity of AdS black holes with spherical and Ricci flat horizons. Consequently, there is a lower bound, referred as $r_{b}$ ($r_{b}<r_{m}$), where the heat capacity is positive for spherical and Ricci flat black holes in regions $r_{+}>r_{b}$. For hyperbolic black holes, there is an upper bound for the horizon radius ($r_{u}$) besides the lower bound ($r_{b}$), in which the heat capacity is positive in regions $r_{b}<r_{+}<r_{u}$ and negative for the other regions. \\

\begin{figure}[!htbp]
	$%
	\begin{array}{ccc}
	\epsfxsize=5.5cm \epsffile{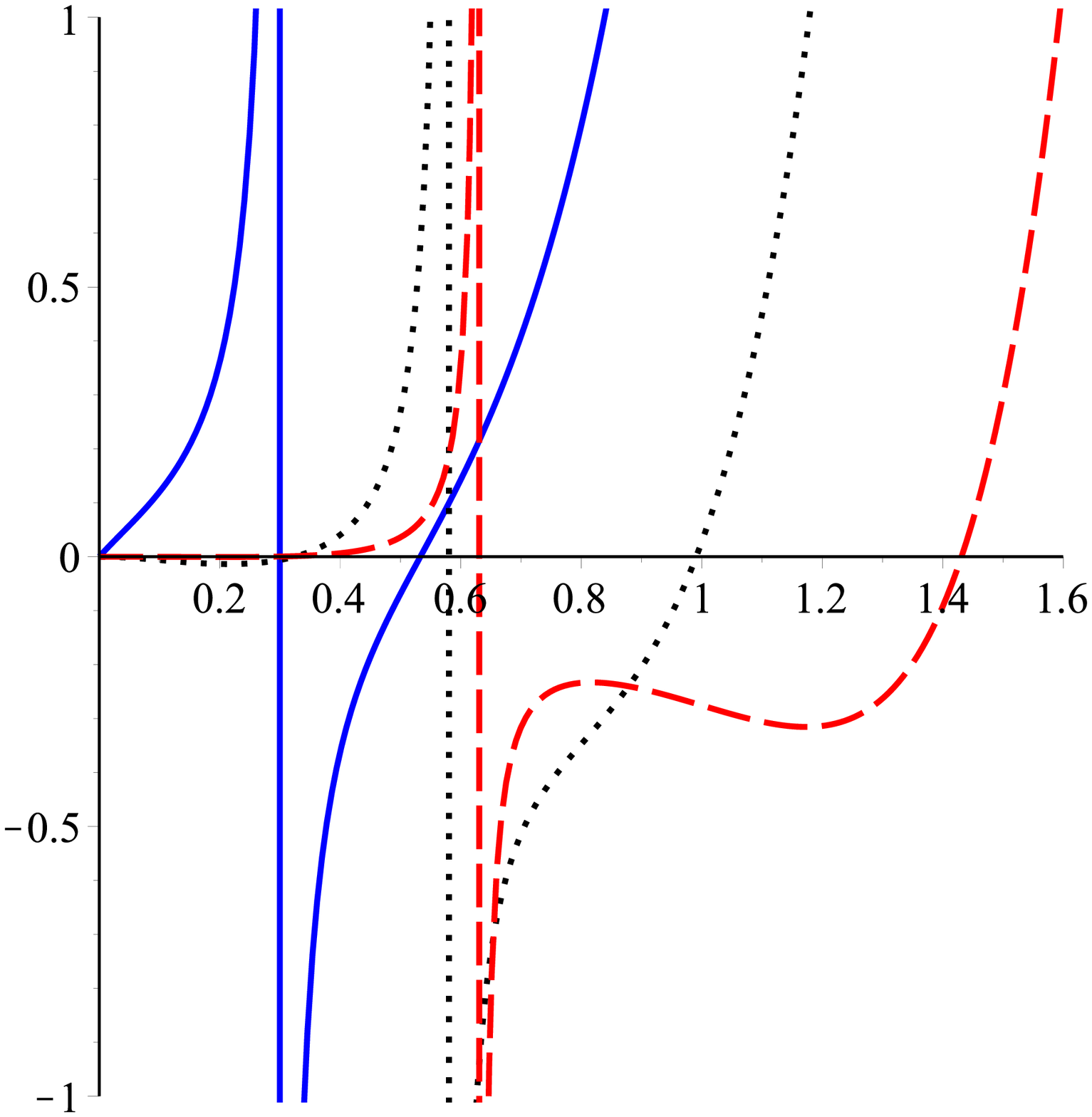}
	\epsfxsize=5.5cm \epsffile{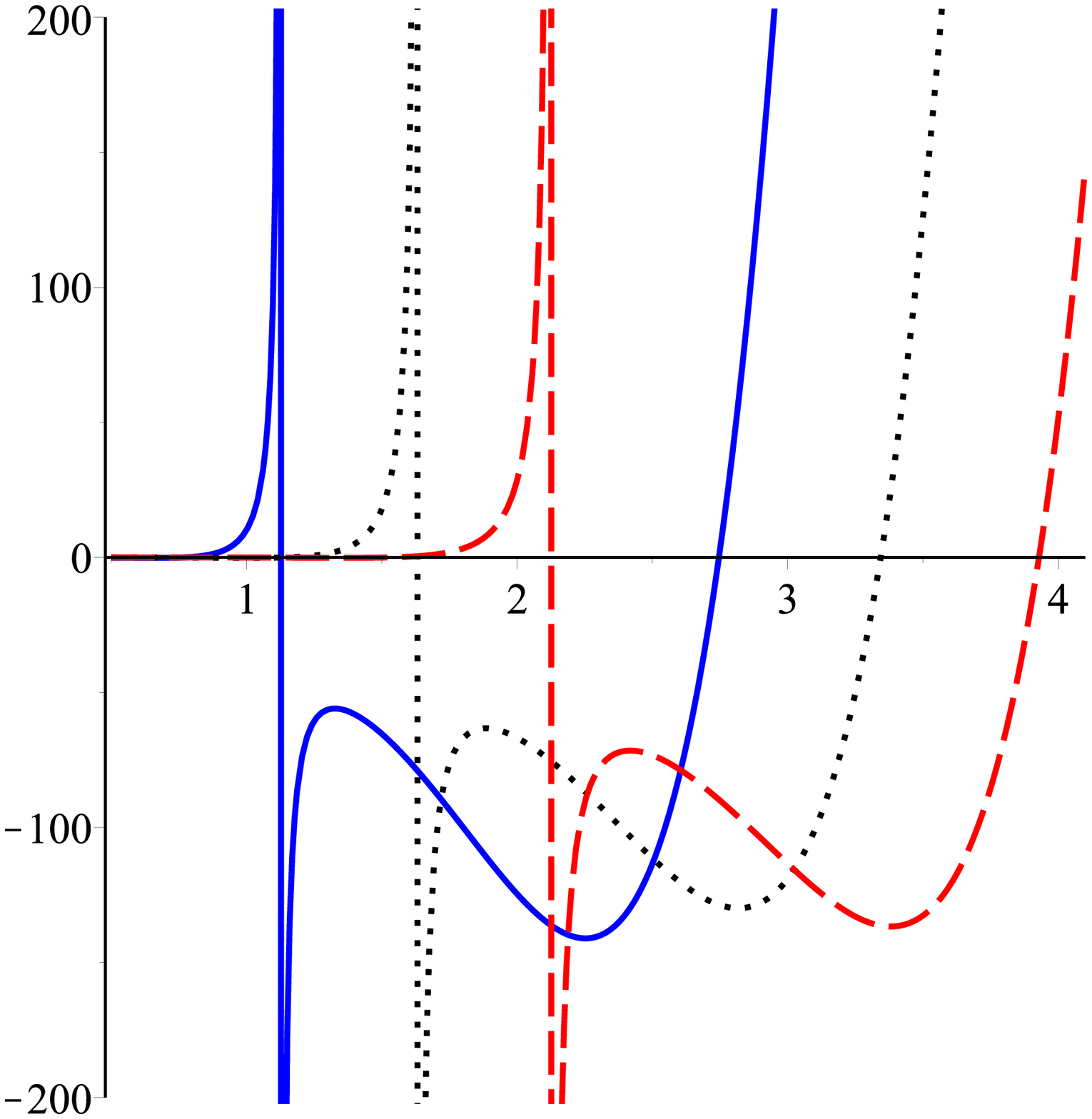}
	\epsfxsize=5.5cm \epsffile{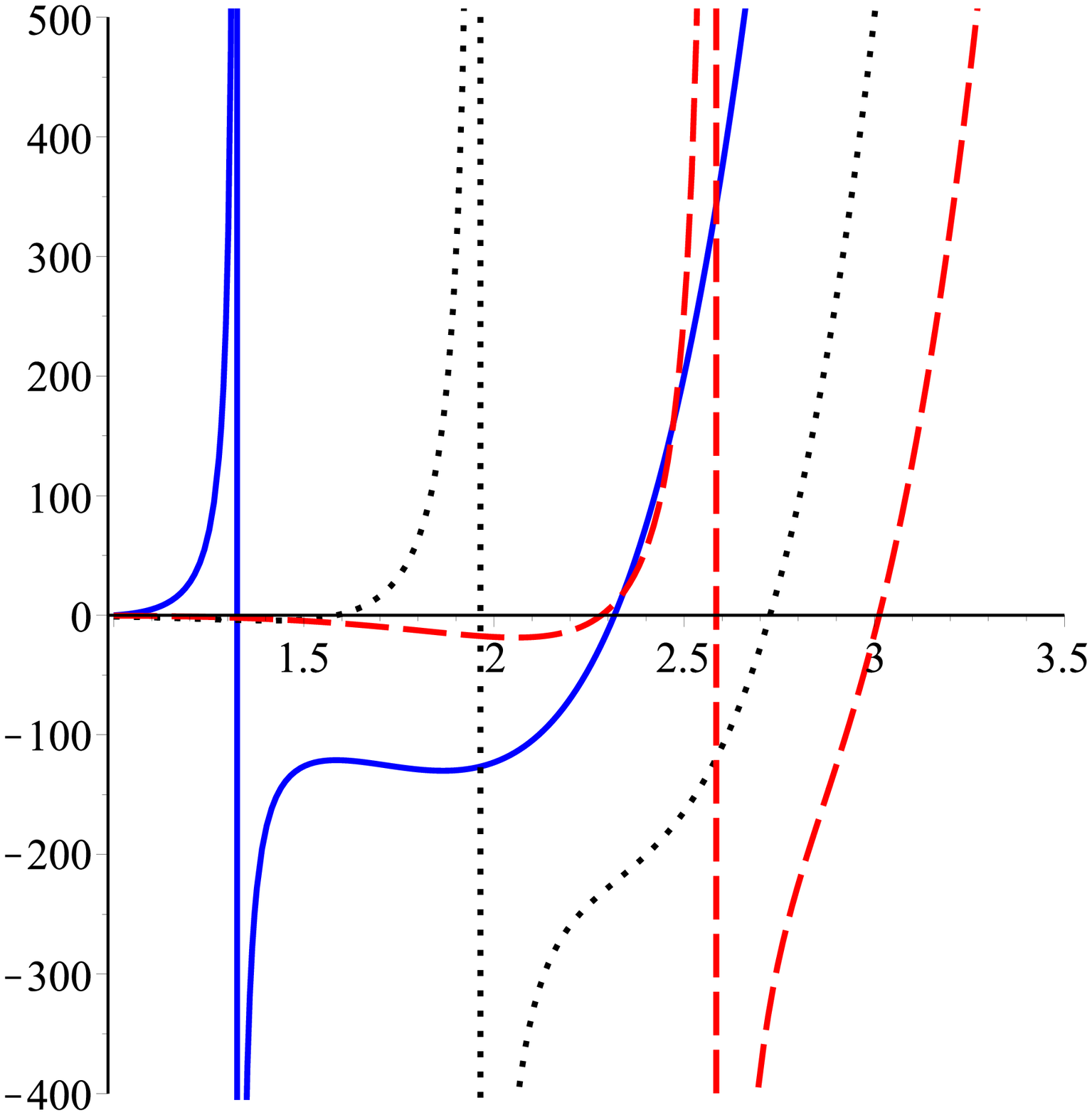}
	&  & 
	\end{array}
	$%
	\caption{ $C_{Q}$ versus ${r_ + }$ with one divergence point for topological black holes in $d=7$ ({solid lines}), $d=8$ ({dotted lines}) and $d=9$ ({dashed lines}).	Plots have been depicted with different scales.
	\newline
	\textbf{Left panel:} for $k=+1$, $q=0$, $\alpha = 1$, $\Lambda = -1$, $m=0.1$, $c_{0}=1$, $c_{1}=-10$, $c_{2}=-10$, $c_{3}=-10$ and $c_{4}= -10$.
	\newline 
	\textbf{Middle panel:} for $k=0$, $q=0$, $\Lambda = -1$, $m=0.5$, $c_{0}=1$, $c_{1}=2$, $c_{2}=-3$, $c_{3}=-2$ and $c_{4}= 1$.
	\newline
	\textbf{Right panel:} for $k=-1$, $q=0$, $\alpha = 0.1$, $\Lambda = -1$, $m=0.6$, $c_{0}=1$, $c_{1}=4$, $c_{2}=2$, $c_{3}=-4$ and $c_{4}= 1.9$.}

	\label{HCOneDivergency}
\end{figure}

In addition, there is another possibility for the case of one divergent point in the heat capacity, which indicates the Hawking-Page phase transition. According to Fig. \ref{HCOneDivergency}, around such divergency, the heat capacity is negative and positive corresponding to the area on the right and the area on the left, respectively. In all cases, $C_{Q}$ may have two positive roots (referred as $r_{b_{1}}$ and $r_{b_{2}}$) and a divergent point ($r_{m}$) between the roots. It should be noted the first or second roots could not exist depending on the parameters' values. Note that this situation is different from the previous case of hyperbolic black holes (see the items i and ii related to hyperbolic black holes). Assuming that there are two roots ($r_{b_{1}}<r_{b_{2}}$), an unstable region is observed for $r_{m}<r_{b_{2}}$ in the heat capacity's plots of Fig. \ref{HCOneDivergency}. Consequently, thermally stable regions correspond with $r_{b_{1}}<r_{+}<r_{m}$ and $r_{+}>r_{b_{2}}$.


\section{P-V criticality of LM charged-AdS black holes \label{criticality-massive Lovelock}}

\subsection{Extended phase space thermodynamics \label{extended thermodynamics-massive}}

As stated before, one may treat the negative cosmological constant as a thermodynamic pressure in an extended phase space and define its conjugate thermodynamic quantity as thermodynamic volume. Regarding this assumption, in this section, we reconsider the first law of black hole thermodynamics eq.(\ref{first law}) and makes it consistent with the Smarr formula. In order to extend the first law of thermodynamics for LM gravity, we also regard the massive couplings $c_{i}$ and Lovelock coefficient $\alpha$ as thermodynamic variables.

The thermodynamic quantities $M$, $T$, $S$, $Q$ and $\Phi$ have been already derived in section \ref{thermodynamics-massive Lovelock}. The conjugate quantity of the thermodynamic pressure, i.e., the thermodynamic volume, is defined as
\begin{equation} \label{volume}
V = {\left( {\frac{{\partial M}}{{\partial P}}} \right)_{S,Q,{\alpha _2},{\alpha _3},{c_i}}} = \frac{{{\omega _{{d_2}}}}}{{{d_1}}}r_ + ^{{d_1}}.
\end{equation}
Now, taking into account $c_{i}$'s and $\alpha$ as thermodynamic variables, the finite mass $M$ (enthalpy) will be a function of new variables, i.e. $M=M(Q,S,P,\alpha,c_{i})$. Consequently, the first law of thermodynamics in the extended phase is rewritten as
\begin{equation} \label{extended first law-massive}
dM = TdS + \Phi dQ + VdP + Ad\alpha  + \sum\limits_{i \ge 1} {{{\cal C}_i}d{c_i}},
\end{equation}
where
\begin{equation}
\Phi  = {\left( {\frac{{\partial M}}{{\partial Q}}} \right)_{S,P,{\alpha},{c_i}}},\,\,\,T = {\left( {\frac{{\partial M}}{{\partial S}}} \right)_{Q,P,{\alpha},{c_i}}},\,\,\,A = {\left( {\frac{{\partial M}}{{\partial \alpha }}} \right)_{S,Q,P,{c_i}}},
\end{equation}
and ${\cal C}_i$'s are conjugate quantities corresponding to the thermodynamic variables $c_{i}$ with the following explicit forms
\begin{equation}
{{\cal C}_i} = {\left( {\frac{{\partial M}}{{\partial {c_i}}}} \right)_{S,P,Q,{\alpha},{c_{j \ne i}}}} = \frac{{{\omega _{{d_2}}}}}{{16\pi }}{m^2}c_0^ir_ + ^{{d_{i + 1}}}\prod\limits_{j = 2}^i {{d_j}}.
\end{equation}
Moreover, after a long and tedious calculation, it can be derived that obtained thermodynamic quantities obey the Smarr relation as

\begin{equation} \label{Smarr for solutions}
(d - 3)M = (d - 2)TS + (d - 3)\Phi Q - 2PV + 2({A_1}\alpha  + {A_2}{\alpha ^2}) - {{\cal C}_1}{c_1} + {{\cal C}_3}{c_3} + 2{{\cal C}_4}{c_4}+...,
\end{equation}
in which
\begin{equation}
{A_1} = \frac{{{d_2}{k^2}r_ + ^{{d_5}}}}{{16\pi }} - \frac{{{d_2}kTr_ + ^{{d_4}}}}{{2{d_4}}},\,{A_2} = \frac{{{d_2}{k^3}r_ + ^{{d_7}}}}{{24\pi }} - \frac{{{d_2}{k^2}Tr_ + ^{{d_6}}}}{{2{d_6}}}\,.
\end{equation}
Since $c_{2}$ is a dimensionless parameter in the massive theory of gravity, it does not appear in the Smarr formula and has no thermodynamic contribution. Furthermore, one can get the Smarr relation by invoking the method of scaling argument, and performing the dimensional analysis for all variables in the theory as
\begin{equation}
\left[ M \right] = {L^{d - 3}},\,\,\left[ {{\alpha _k}} \right] = {L^{2(k - 1)}},\,\,\left[ {{c_i}} \right] = {L^{i - 2}},\,\,\left[ P \right] = {L^{ - 2}},\,\,\left[ S \right] = {L^{d - 2}},\,\,\left[ Q \right] = {L^{d - 3}}.
\end{equation}
As a result, we find that 
\begin{equation} \label{Smarr formula}
(d - 3)M = (d - 2)\left( {\frac{{\partial M}}{{\partial S}}} \right)S + (d - 3)\left( {\frac{{\partial M}}{{\partial Q}}} \right)Q - 2\left( {\frac{{\partial M}}{{\partial P}}} \right)P + \sum\limits_{i \ge 1} {(i - 2)\left( {\frac{{\partial M}}{{\partial {c_i}}}} \right){c_i}}  + \sum\limits_k {2(k - 1)} {\alpha _k}{\Psi _k},
\end{equation}
in which, evidently, $c_{2}$ has scaling weight equal to zero and the potentials ${\Psi _k}$ consist of three terms originating respectively from the dependence of the entropy, the mass and the bulk Hamiltonian on the Lovelock couplings $\alpha_{k}$'s (for more details see \cite{Kastor2010LovelockSmarr}). Regarding the condition (\ref{coefficient condition}), the general form of Smarr formula eq. (\ref{Smarr formula}) reduces to the our special case eq. (\ref{Smarr for solutions}) in which Lovelock coefficients are dependent to each other.

\subsection{Critical behavior and van der Waals phase transition \label{phase transition-massive Lovelock}}

In this section, we study the critical behavior of the LM charged AdS black holes in the canonical ensemble under certain conditions which van der Waals phase transition appears. Our starting point is assuming the negative cosmological constant as a thermodynamic pressure, i.e.,
\begin{equation}
P =  - \frac{\Lambda }{{8\pi }} = \frac{{{d_1}\,{d_2}}}{{16\pi {\ell ^2}}}.
\end{equation}
This means the thermodynamic volume can be defined as $V = {\left( {\partial H/\partial P} \right)_{{X_i}}}$ in which $H\equiv M$ in the extended phase space. Geometrical \textit{equation of state} for LM black hole solutions in the canonical ensemble (fixed charged $Q$) can be simply obtained from eq. (\ref{temperature - massive AdS BH}) as 
\begin{equation} \label{pressure - massive Lovelock}
P = \frac{{{d_2}{{\left( {k\alpha  + r_ + ^2} \right)}^2}T}}{{4r_ + ^5}} + \frac{{{q^2}}}{{8\pi r_ + ^{2{d_2}}}} - \frac{{{d_2}k\left( {{d_7}{\alpha ^2} + 3{d_5}k\alpha r_ + ^2 + 3{d_3}r_ + ^4} \right)}}{{48\pi r_ + ^6}} - \frac{{{d_2}{m^2}{{\cal B}_ + }}}{{16\pi }},
\end{equation}
where
\begin{equation}
{{\cal B}_ + } = {c_0}{c_1}r_ + ^{ - 1} + {d_3}c_0^2{c_2}r_ + ^{ - 2} + {d_3}{d_4}c_0^3{c_3}r_ + ^{ - 3} + {d_3}{d_4}{d_5}c_0^4{c_4}r_ + ^{ - 4} + O({c_5},r_ + ^{ - 5}),
\end{equation}
and $r_{+}$ is a function of thermodynamic volume, $V$. This is a worthwhile equation since one can easily recover the equation of states of charged AdS black holes in Einstein, Gauss-Bonnet and massive gravities (and also any possible combination of them). According to the Lovelock coefficient condition (\ref{coefficient condition}), since ${\alpha _2} \equiv {\alpha _{{\rm{GB}}}} \propto \alpha$ and ${\alpha _3} \equiv {\alpha _{{\rm{TOL}}}} \propto {\alpha ^2}$, by taking limits $\alpha  \to {d_3}{d_4}{\alpha _{{\rm{GB}}}}$ and ${\alpha ^2} \to 0$, our result recovers the Gauss-Bonnet-massive equation of state for charged-AdS black holes in Ref. \cite{Hendi2016JHEP}. In what follows, we restrict our study to ${\cal U}_i$ up to the fourth interaction term (${\cal U}_4$). Again, the physical equation of state can be obtained by translating the geometric version, eq. (\ref{pressure - massive Lovelock}). The same result is obtained for associated specific volume (by performing the steps as stated in \ref{PV for RN-AdS}), i.e. $v  = 4{r_ + }\ell _{\rm{P}}^{{d_2}}/{d_2}$. To do that, one has to define the shifted Hawking temperature \cite{PVmassive2015PRD,Reentrant-dRGTmassive-2017,Triple-BI-massive-2017} as 
\begin{equation}
\tilde T \equiv T - \frac{{{m^2}{c_0}{c_1}}}{{4\pi }},
\end{equation}
Rewriting the geometric equation of state \ref{pressure - massive Lovelock} as
\begin{equation} \label{pressure-massive Lovelock-physical}
P = \frac{{{d_2}\tilde T}}{{4{r_ + }}} - \frac{{{d_2}{d_3}(k + {m^2}c_0^2{c_2})}}{{16\pi r_ + ^2}} + \frac{{{d_2}(8\pi k\alpha T - {d_3}{d_4}{m^2}c_0^3{c_3})}}{{16\pi r_ + ^3}} - \frac{{{d_2}{d_5}({k^2}\alpha  + {d_3}{d_4}{m^2}c_0^4{c_4})}}{{16\pi r_ + ^4}} + \frac{{{d_2}{{(k\alpha )}^2}T}}{{4r_ + ^5}} - \frac{{{d_2}{d_7}k{\alpha ^2}}}{{48\pi r_ + ^6}} + \frac{{{q^2}}}{{8\pi r_ + ^{2{d_2}}}}
\end{equation}
and substituting the following quantities
\begin{equation}
P \Rightarrow {P_{phys}} = \frac{{\hbar c}}{{\ell _{\rm{P}}^{{d_2}}}}P,\,\,\,\tilde T \Rightarrow {\tilde T_{phys}} = {T_{phys}} - \frac{{\hbar c}}{{{k_{\rm{B}}}}}\frac{{{m^2}{c_0}{c_1}}}{{4\pi }},
\end{equation}
the physical equation of state is obtained as
\begin{equation}
{P_{phys}} = \frac{{{d_2}}}{{4{r_ + }\ell _{\rm{P}}^{{d_2}}}}{k_{\rm{B}}}{\tilde T_{phys}} + ...  .
\end{equation}
Hence, we can identify the specific volume $v$ as $v  = 4{r_ + }\ell _{\rm{P}}^{{d_2}}/{d_2}$. Since in the geometric units ${r_ + } ={d_2}v /4$, hereafter, we would rather use the event horizon radius ($r_ +$) instead of the specific volume ($v$) in order to analyze the criticality. Although we work with event horizon radius ($r_{+}$), but in what follows, we label the associated axes with thermodynamic volume ($V$) in $P-V$ and $T-V$ diagrams.\\

Let's compare the LM equation of state (\ref{pressure - massive Lovelock}) with the equation of state of RN-AdS black holes (\ref{pressure - RN-AdS}). In right hand side of RN-AdS equation of state, the first and the last terms are always positive and the second term can be positive, zero or negative. In fact, signs of the presented terms in the equation of state could ensure the critical behavior for a given black hole solution. We introduce the signature of equation of state of RN-AdS black holes as $P(+,\pm,+)$. We saw that there does not exist $P-V$ criticality for black holes with Ricci flat and hyperbolic horizons. In order to have $P-V$ criticality two positive terms and one negative term are needed, i.e., a equation of state with signature $P(+,-,+)$ which is the case with spherical horizon ($k=+1$). Therefore, at least, two positive terms and one negative term in equation of state can possibly ensure the critical behavior and phase transition for a given black hole. Regarding this, the equation of state of LM charged-AdS black holes (\ref{pressure-massive Lovelock-physical}) with signature $P(\pm,\pm,\pm,\pm,+,\pm,+)$ predicts critical behavior and phase transition for Ricci flat and hyperbolic black holes as well as spherically symmetric black holes depending on the massive coupling coefficients ($c_{i}$), Lovelock coefficient ($\alpha$) and topological factor ($k$). \\

We are looking for the inflection point of isothermal $P-V$ diagrams, the subcritical isobar of $T-V$ plots, and the characteristic swallow-tail form of $G-T$ diagrams for the obtained black hole solutions according to section \ref{PV for RN-AdS}. These pieces of evidence guarantee the existence of the phase transition and indicate the van der Waals like behavior for the LM AdS black holes. In our considerations, we suppose that all the massive coupling coefficients are simultaneously positive ($c_{1}=c_{2}=c_{3}=c_{4}=+ 1$) or negative ($c_{1}=c_{2}=c_{3}=c_{4}=- 1$) and keep track the effect of higher order curvature terms of the Lovelock Lagrangian on the outcomes of massive gravity. Later, we do not impose this assumption, and will summarize the results of arbitrary signs for the massive couplings ($c_{i}=\pm1$). \\

 First, we study the $P-V$ diagrams of LM AdS black holes with various topological factors. In an isothermal $P-V$ diagram, the critical point is an inflection point and can be obtained from eq. (\ref{critical point equation}). Investigation of the critical behavior is not possible, analytically, and so we apply the numerical analysis. Using eqs. (\ref{pressure - massive Lovelock}) and (\ref{critical point equation}), the critical point can be found numerically. At the critical point, we refer the value of horizon radius as critical horizon radius, $r_{c}$. The critical values for the topological black holes in $d=7$ dimensions are gathered in the table \ref{tab:topologicalBH-k} in which we have focused on the effect of horizon topology factor ($k$) and fixed other parameters.
  
  \begin{table} [!htbp]
  	\setlength\tabcolsep{8pt}
  	\caption{Topological black holes: $d=7$, $q=1$,  $m=1$, $c_{0}=c_{1}=c_{2}=c_{3}=c_{4}= 1$ and $\alpha=0.01$ (for $k=\pm1$).}
  	 \label{tab:topologicalBH-k} 
  	\begin{tabular}{ccccc}
  		\hline\hline\noalign{\smallskip}
  		$k$ & $P_{c}$ & $r_{c}$ & $T_{c}$ & $\frac{{{P_c}{r_c}}}{{{T_c}}}$ \\
  		\noalign{\smallskip}\hline\hline\noalign{\smallskip}
  		+1 & 22.8678 & 0.69318 & 19.7148 & 0.80404 \\
  		0 & 25.0668 & 0.68738 & 21.2994 & 0.80897 \\
  		-1 & 27.8071 & 0.68117 & 23.3249 & 0.81207 \\	
  		\noalign{\smallskip}\hline\hline
  	\end{tabular}
  \end{table}
 
\begin{figure}[!htbp]
	$%
	\begin{array}{ccc}
	\epsfxsize=5.5cm \epsffile{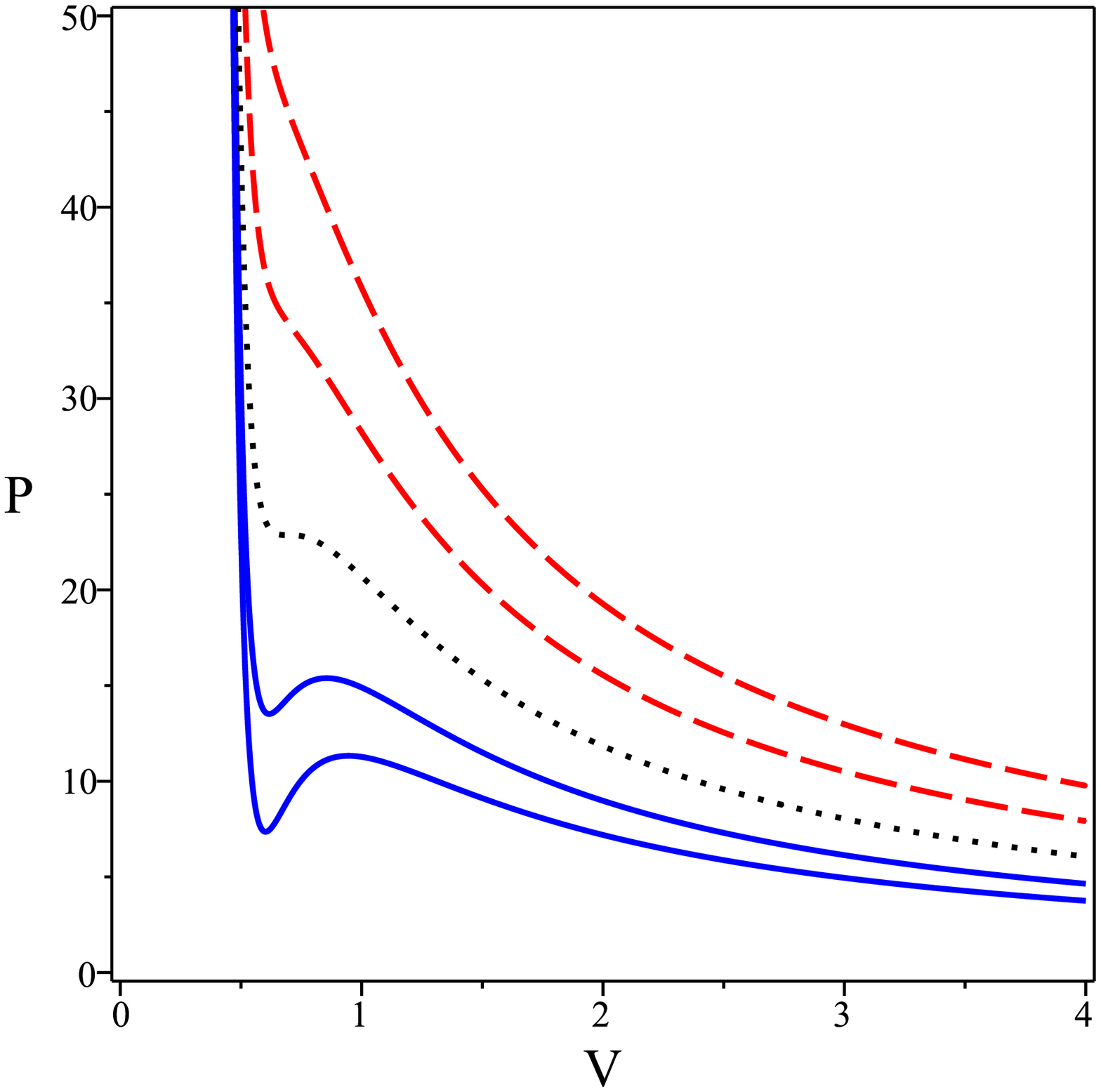}
	\epsfxsize=5.5cm \epsffile{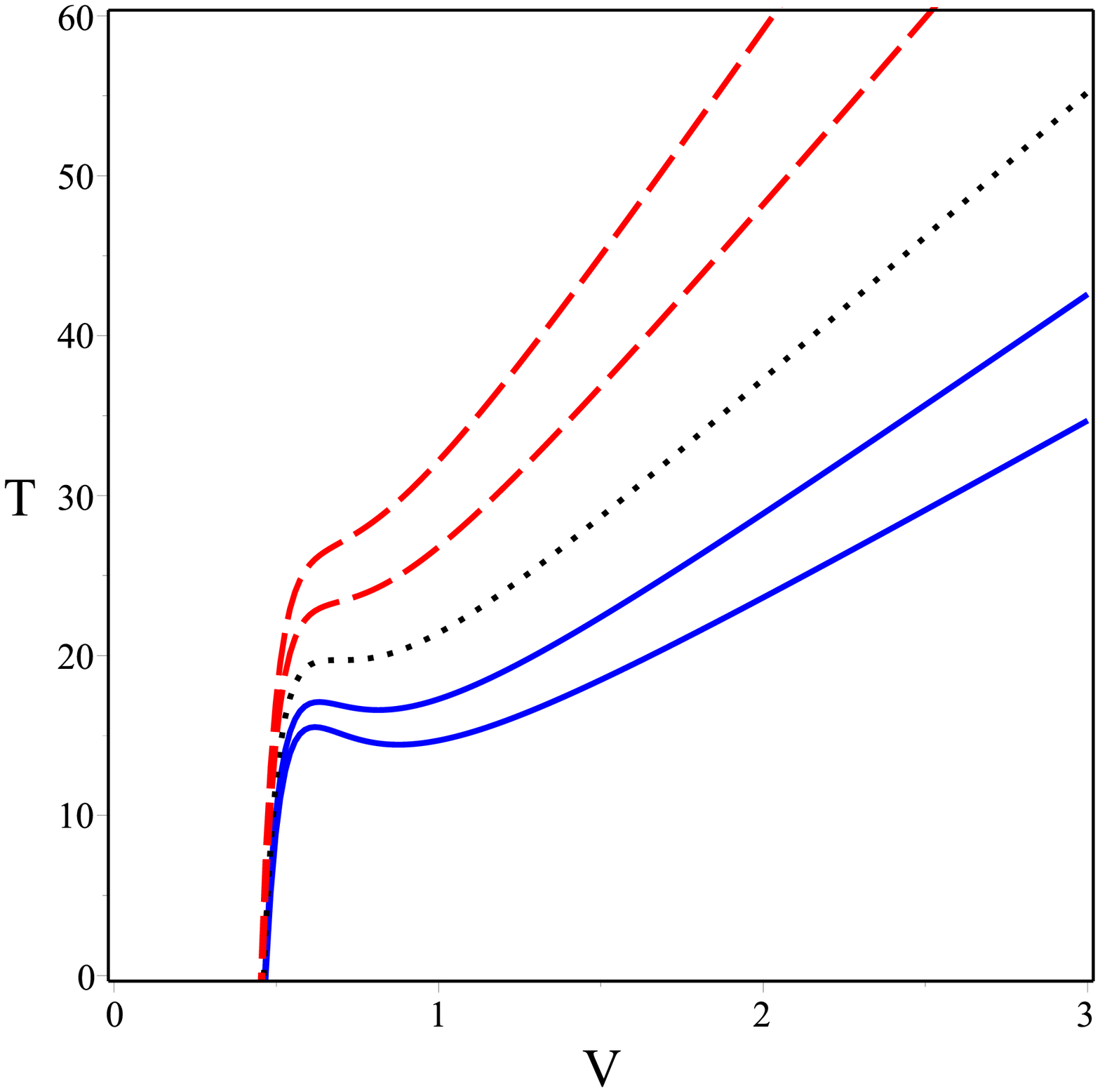}
	\epsfxsize=5.5cm \epsffile{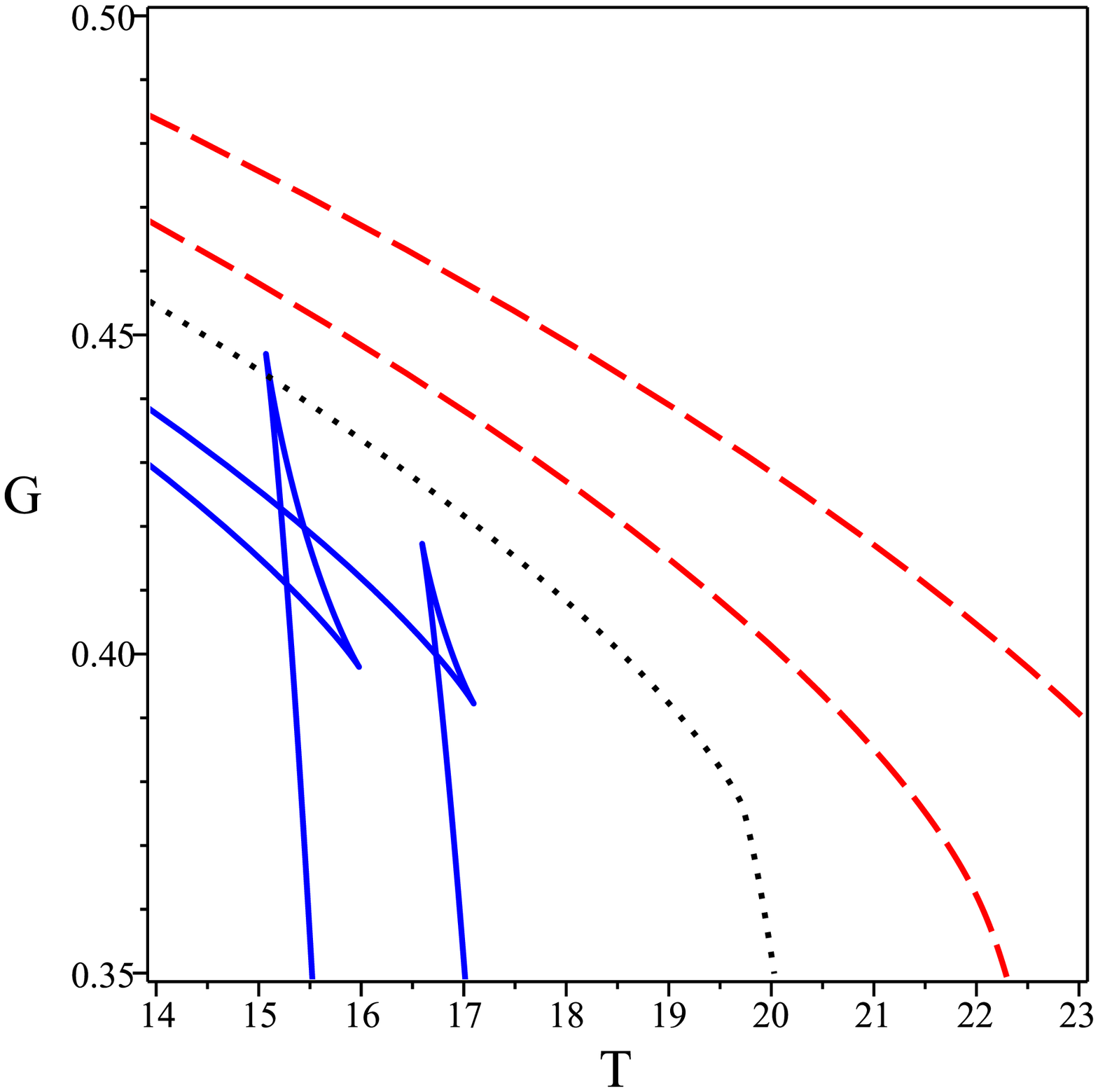}
	&  & 
	\end{array}
	$%
	\caption{\textbf{LM AdS black hole with spherical horizon:} $P-V$ (left), $T-V$
		(middle) and $G-T$ (right) diagrams; we have set $k=1$, $d=7$, $q=1$,  $m=1$, $c=c_{1}=c_{2}=c_{3}=c_{4}= 1$ and $\alpha=0.01$. \newline
		\textbf{Left panel:} $T<T_{c}$ (continuous lines), $T=T_{c}$ (dotted line) and $T>T_{c}$ (dashed lines).\newline
		\textbf{Middle and right panels:} $P<P_{c}$ (continuous lines), $P=P_{c}$ (dotted lines) and $P>P_{c}$ (dashed lines).}
	\label{PV-massiveBH-sphere}
\end{figure}

\begin{figure}[!htbp]
	$%
	\begin{array}{ccc}
	\epsfxsize=5.5cm \epsffile{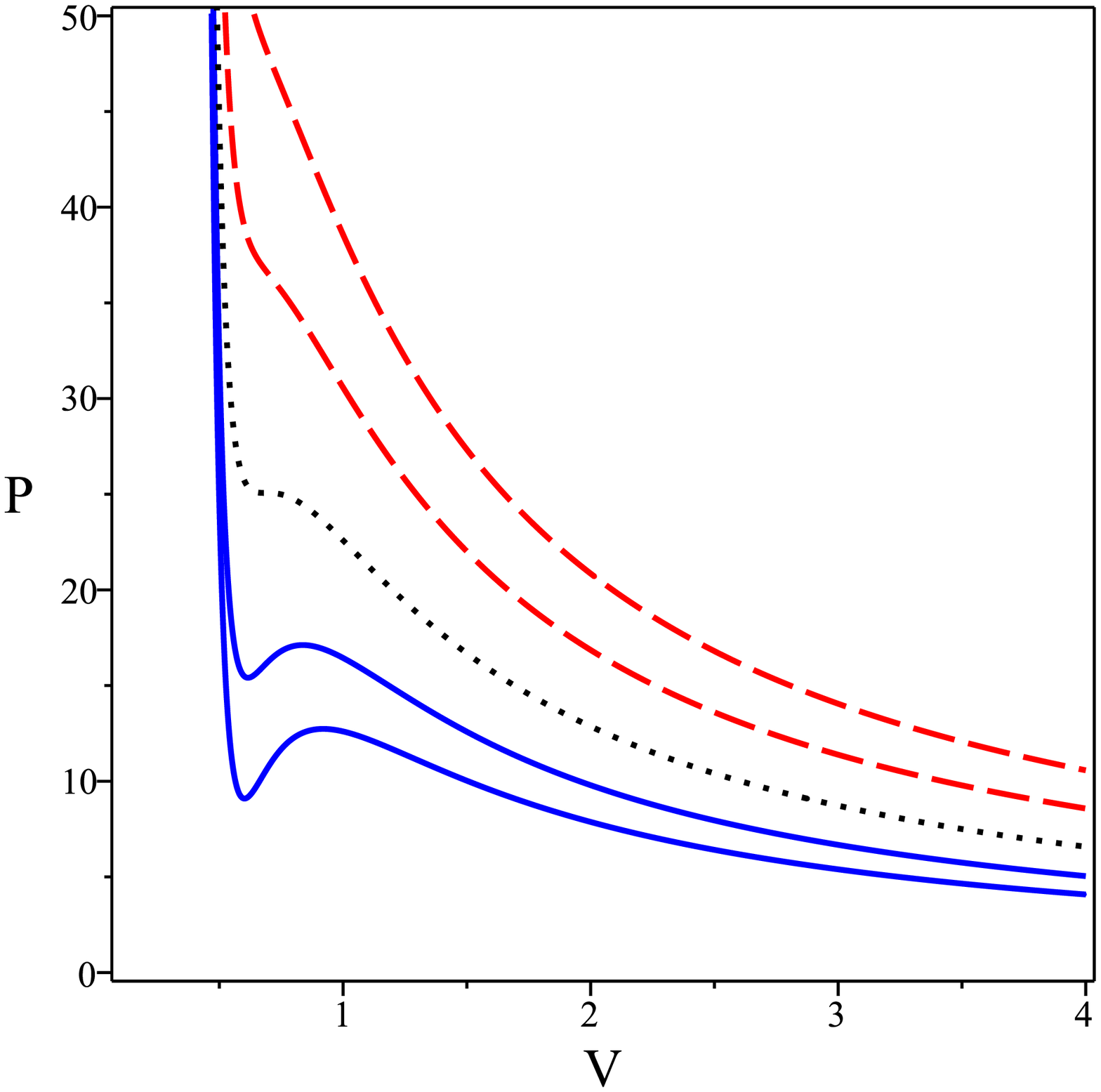}
	\epsfxsize=5.5cm \epsffile{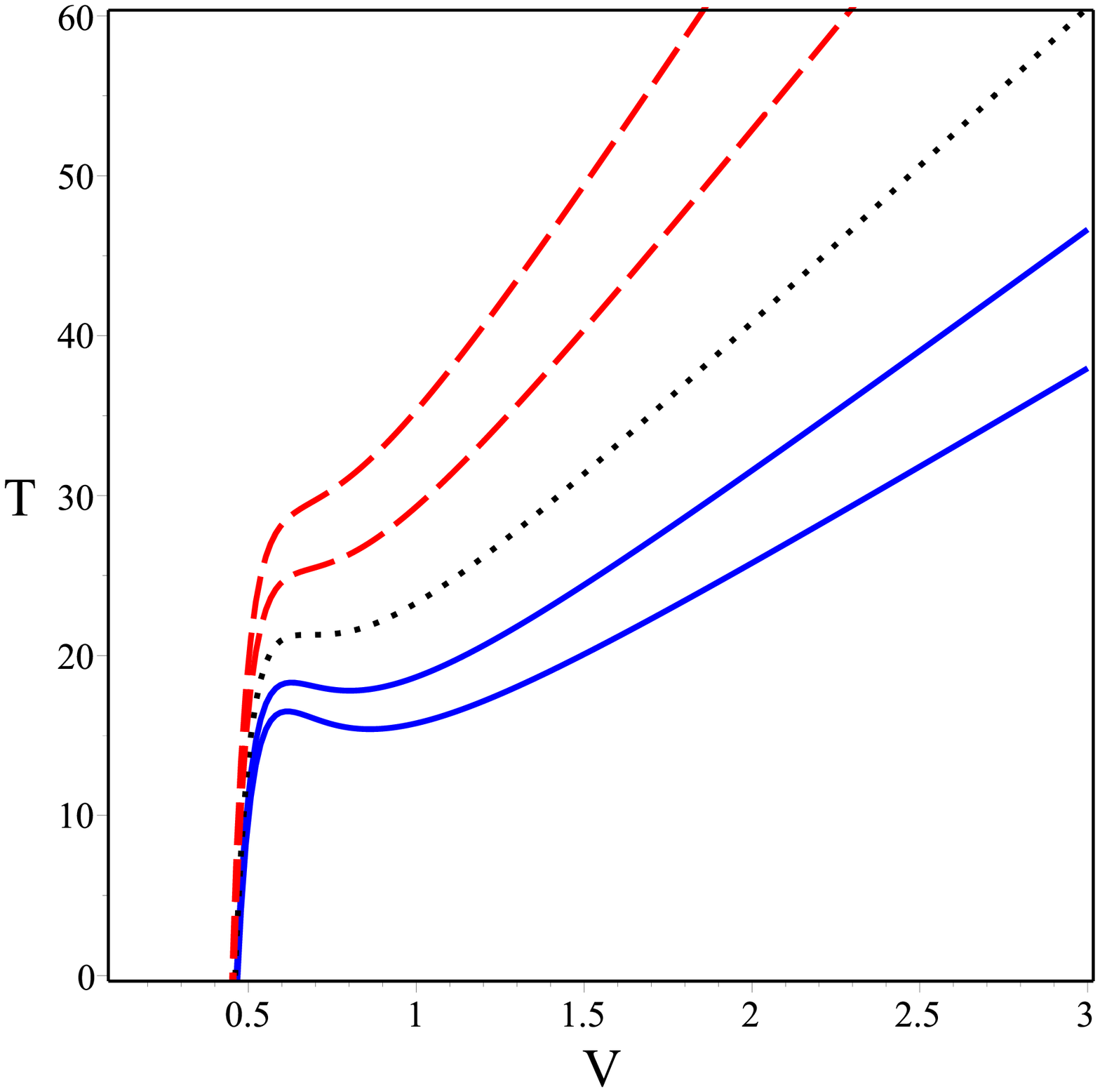}
	\epsfxsize=5.5cm \epsffile{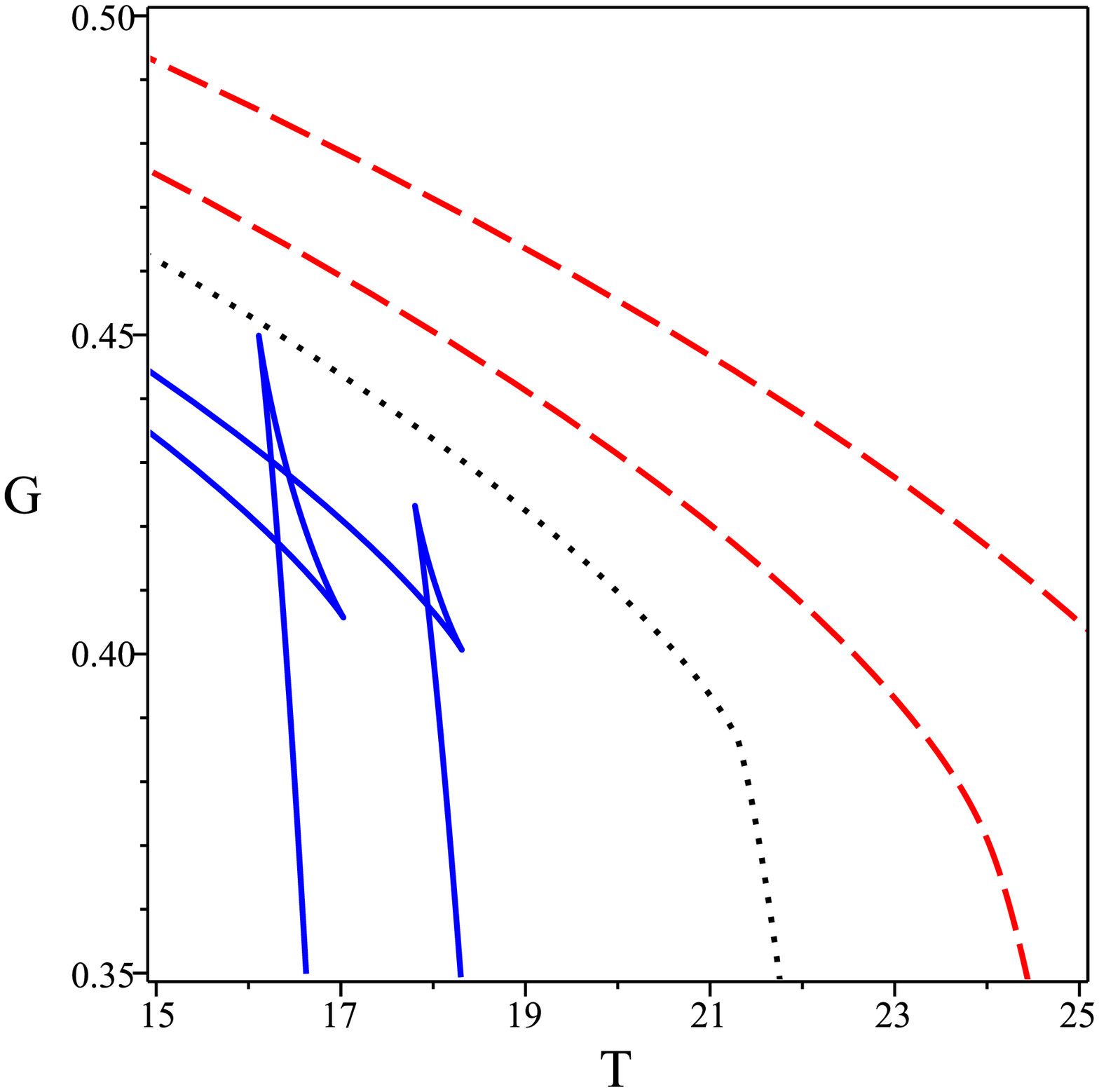}
	&  & 
	\end{array}
	$%
	\caption{\textbf{LM AdS black hole with Ricci flat horizon:} $P-V$ (left), $T-V$
		(middle) and $G-T$ (right) diagrams; we have set $k=0$, $d=7$, $q=1$,  $m=1$ and $c_{0}=c_{1}=c_{2}=c_{3}=c_{4}= 1$. \newline
		\textbf{Left panel:} $T<T_{c}$ (continuous lines), $T=T_{c}$ (dotted line) and $T>T_{c}$ (dashed lines).\newline
		\textbf{Middle and right panels:} $P<P_{c}$ (continuous lines), $P=P_{c}$ (dotted lines) and $P>P_{c}$ (dashed lines).}
	\label{PV-massiveBH-flat}
\end{figure}

\begin{figure}[!htbp]
	$%
	\begin{array}{ccc}
	\epsfxsize=5.5cm \epsffile{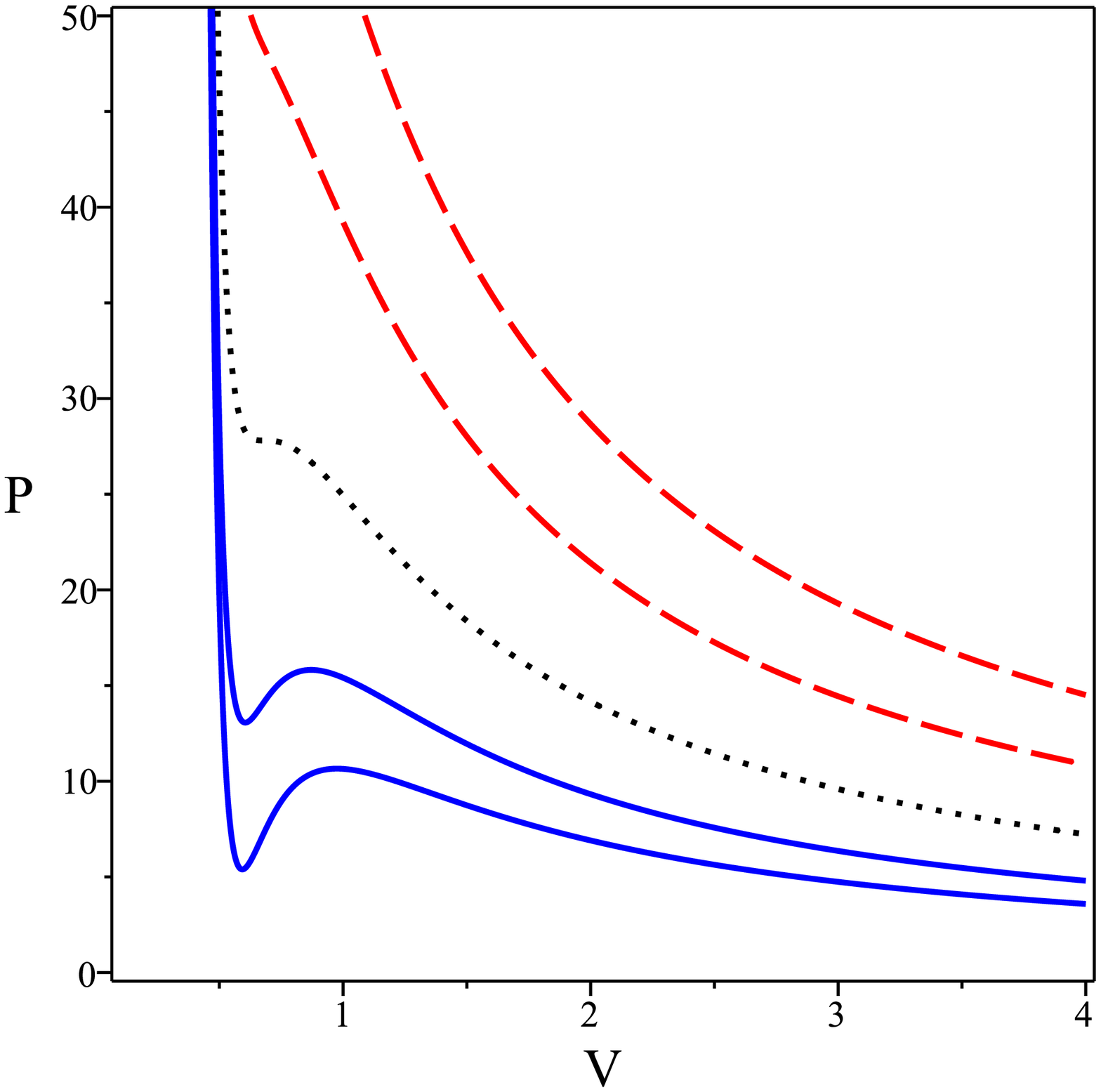}
	\epsfxsize=5.5cm \epsffile{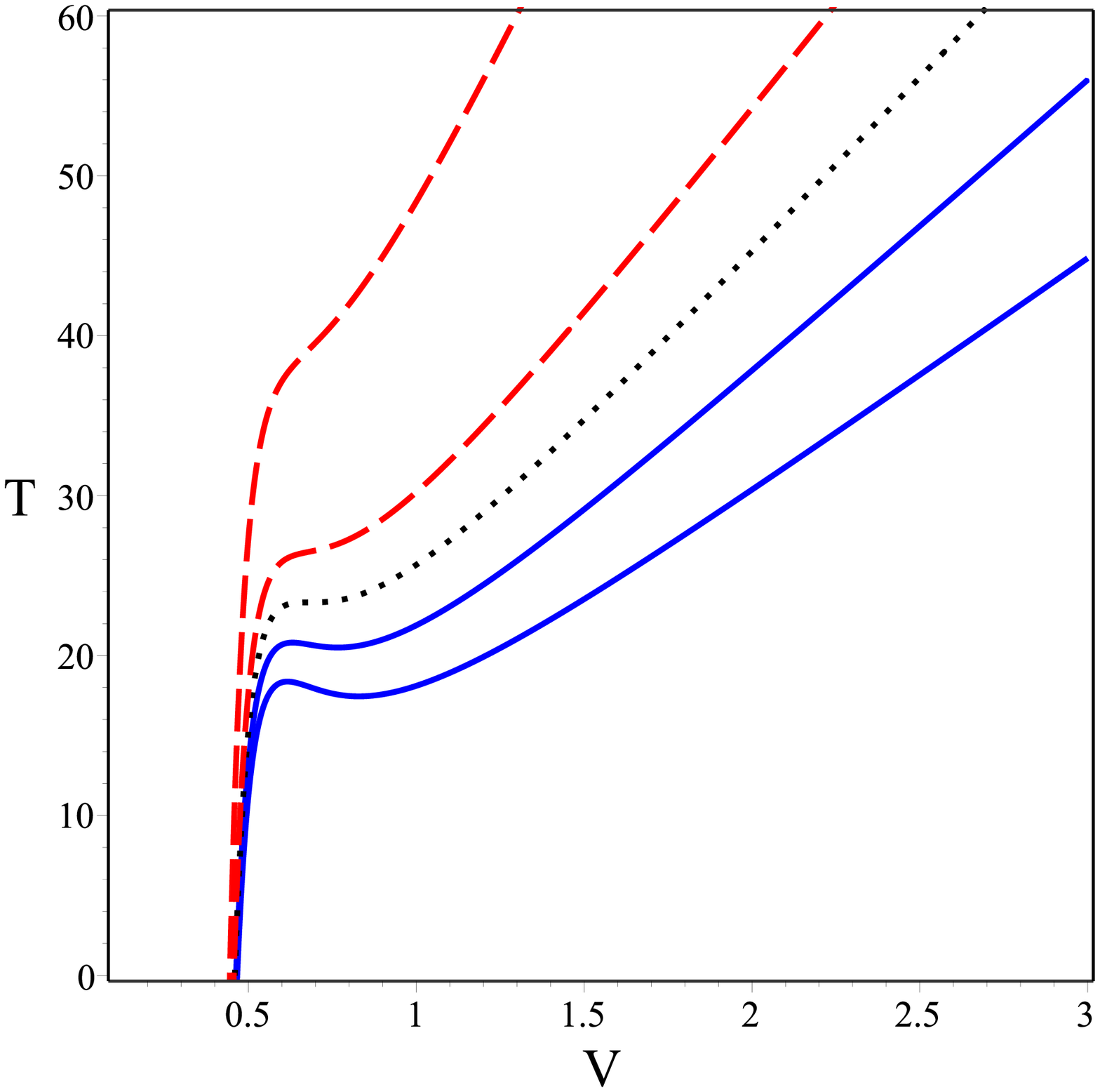}
	\epsfxsize=5.5cm \epsffile{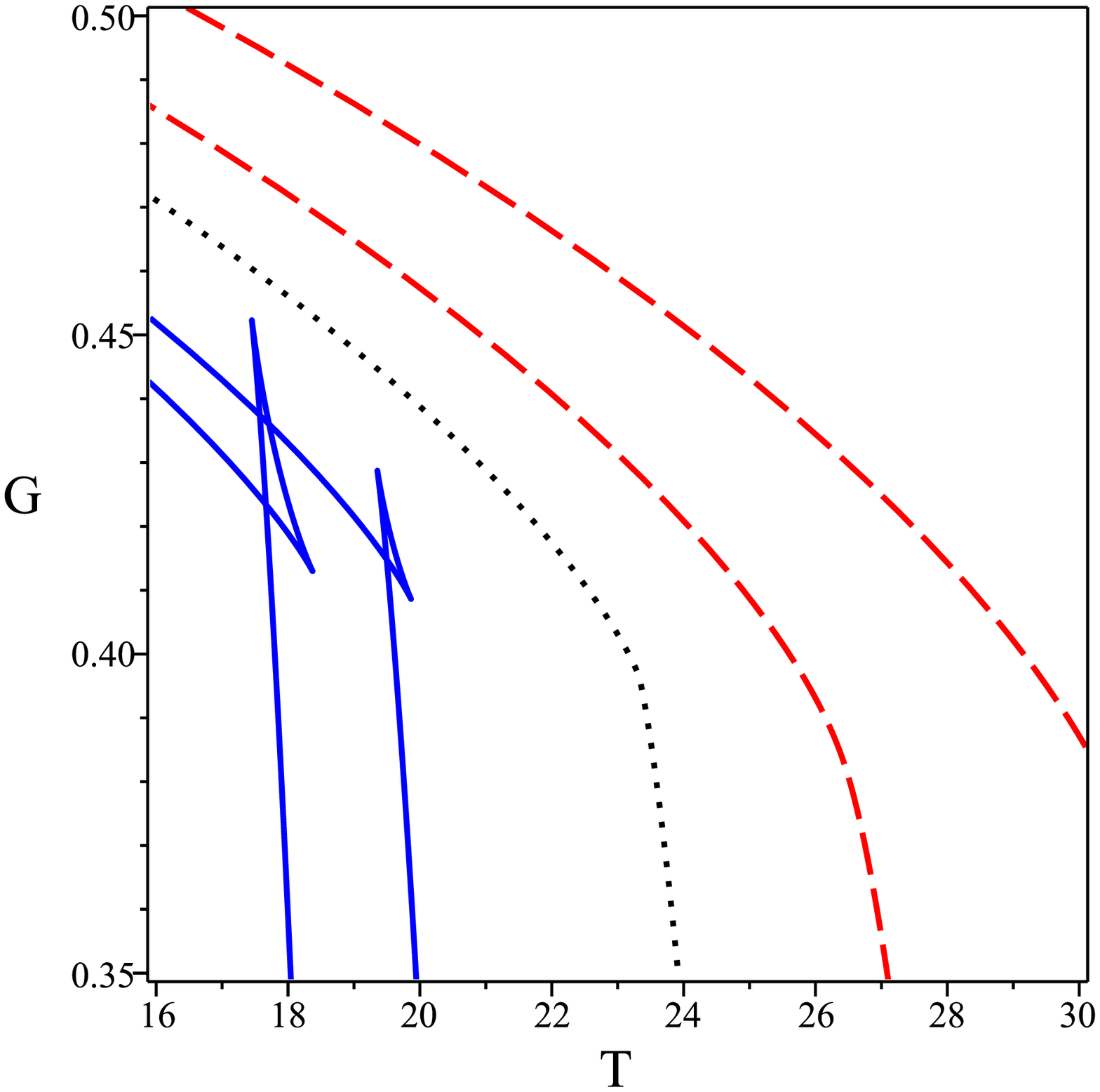}
	&  & 
	\end{array}
	$%
	\caption{\textbf{LM AdS black hole with hyperbolic horizon:} $P-V$ (left), $T-V$
		(middle) and $G-T$ (right) diagrams; we have set $k=-1$, $d=7$, $q=1$,  $m=1$, $c_{0}=c_{1}=c_{2}=c_{3}=c_{4}= 1$ and $\alpha=0.01$. \newline
		\textbf{Left panel:} $T<T_{c}$ (continuous lines), $T=T_{c}$ (dotted line) and $T>T_{c}$ (dashed lines).\newline
		\textbf{Middle and right panels:} $P<P_{c}$ (continuous lines), $P=P_{c}$ (dotted lines) and $P>P_{c}$ (dashed lines).}
	\label{PV-massiveBH-hyperbolic}
\end{figure}

 Considering the obtained critical values in the table \ref{tab:topologicalBH-k} for the pressure, horizon radius and temperature, one can plot their corresponding phase diagrams. In the left panels of Figs. \ref{PV-massiveBH-sphere}-\ref{PV-massiveBH-hyperbolic}, the characteristic behavior of pressure as function of event horizon radius ($r_{+}$) is depicted for the topological black holes. Compared to $P-V$ diagram of van der Waals fluid or RN-AdS black holes (see Fig. \ref{PV-RN-AdS}), it is seen that the associated $P-V$ diagrams for LM AdS black holes qualitatively behave like van der Waals fluid. Therefore, critical radius (inflection point) can be found for Ricci flat or hyperbolic black holes as well as spherically symmetric black holes. For all $P-V$ diagrams, the temperature of isotherms decreases from top to bottom. For $T>T_{c}$, the isotherms correspond to the \textit{ideal gas} with a single phase. For $T<T_{c}$, a two-phases behavior is seen, and in comparison with liquid/gas system, there exists (first order) small/large black hole phase transition for topological black holes. It is notable that, based on the Maxwell's equal-area law, the (unphysical) oscillatory part of each isotherm is replaced by a line of constant pressure. \\
  
Now, we are looking for the characteristic swallow-tail form of $G-T$ diagrams. The Gibbs free energy is given by computing on-shell action (${\cal I}_G$) or using the Legendre transformation, $G=M-TS$. Analytical calculation of the Gibbs free energy is too large and, therefore, we leave out the analytical result for reasons of economy. We have plotted the Gibbs free energy as a function of temperature for various pressures in right panels of Figs. \ref{PV-massiveBH-sphere}-\ref{PV-massiveBH-hyperbolic}. As seen, obviously, $G-T$ diagrams indicate the characteristic swallow-tail behavior for all types of topological black holes. This behavior demonstrates a first-order phase transition in the black hole systems. \\

For the sake of completeness, the qualitative behavior of temperature as a function of horizon radius (which corresponds to specific volume, $v$) are depicted in the middle panels of Figs. \ref{PV-massiveBH-sphere}-\ref{PV-massiveBH-hyperbolic}. Comparing Fig. \ref{PV-RN-AdS} with Figs. \ref{PV-massiveBH-sphere}-\ref{PV-massiveBH-hyperbolic}, $T-V$ diagrams shows a van der Waals like behavior. \\

To sum up, equation of state of LM AdS black holes (\ref{pressure - massive Lovelock}) can mimic the behavior of van der Waals fluid in physical regions (where topological black holes are thermally stable). In the table \ref{tab:topologicalBH-k}, the results of numerical calculation related to Figs. \ref{PV-massiveBH-sphere}-\ref{PV-massiveBH-hyperbolic} are presented. Evidently, Ricci flat and hyperbolic black holes can qualitatively imitate the critical behavior of spherical black holes in LM gravity (or Einstein gravity). It should be emphasized that the van der Waals like behavior persists in higher dimensions for all kinds of topological black holes. In addition, the universal ratio (i.e., $\frac{{{P_c}{r_c}}}{{{T_c}}}$) is a function of event horizon topology. \\

To be more specific, we analyze the equations of state and phase transitions for topological black holes case by case in details. We focus on the effects of Lovelock coefficient ($\alpha$), graviton mass parameter $m$, spacetime dimension ($d$) and topological factor ($k$) and present related tables (see tables \ref{tab:sphericalBH-d}, \ref{tab:sphericalBH-alpha}, \ref{tab:sphericalBH-m}, \ref{tab:RicciflatBH-d}, \ref{tab:RicciflatBH-m}, \ref{tab:hyperbolicBH-d (rc1)}, \ref{tab:hyperbolicBH-d (rc2)}, \ref{tab:hphericalBH-m}). In this regard, we reveal a peculiar phase transition and critical behavior for hyperbolic black holes in the LM gravity in higher dimensions of spacetime (i.e., $d\geqslant7$). \\

\textbf{Spherical horizon($k=+1$):}

The LM equation of state for spherical black holes with signature $P(\pm,\pm,\pm,\pm,+,-,+)$ reads
\begin{eqnarray}
P &=& \frac{{{d_2}(T - {m^2}{c_0}{c_1}/4\pi )}}{{4{r_ + }}} - \frac{{{d_2}{d_3}(1 + {m^2}c_0^2{c_2})}}{{16\pi r_ + ^2}} + \frac{{{d_2}(8\pi \alpha T - {d_3}{d_4}{m^2}c_0^3{c_3})}}{{16\pi r_ + ^3}} - \frac{{{d_2}{d_5}(\alpha  + {d_3}{d_4}{m^2}c_0^4{c_4})}}{{16\pi r_ + ^4}} + \frac{{{d_2}{\alpha ^2}T}}{{4r_ + ^5}} \nonumber \\
&& - \frac{{{d_2}{d_7}{\alpha ^2}}}{{48\pi r_ + ^6}} + \frac{{{q^2}}}{{8\pi r_ + ^{2{d_2}}}}.
\end{eqnarray}
Our investigations show this equation of state predicts the critical behavior in higher dimensions. In this case, if all the massive coupling coefficients ($c_{i}$) be positive (negative) definite, one (two) physical critical point(s) can be found at most. In table \ref{tab:sphericalBH-d}, the critical values for pressure, horizon radius and temperature have been computed for various dimensions. According to table \ref{tab:sphericalBH-d}, critical pressure and temperature are increasing functions of spacetime dimensions whereas critical horizon radius is a decreasing function of it. The universal ratio $\frac{{{P_c}{r_c}}}{{{T_c}}}$ is an increasing function of spacetime dimensions the same as RN-AdS black holes in Einstein gravity (see eq. \ref{universal ratio-RN-AdS}). 

 \begin{table}[!htbp]
 	\setlength\tabcolsep{8pt}
 	\caption{Spherical black holes: $k=1$, $q=1$,  $m=1$, $c_{0}=c_{1}=c_{2}=c_{3}=c_{4}= 1$ and $\alpha=1$.}
 	 \label{tab:sphericalBH-d} 
 	\begin{tabular}{ccccc}
 		\hline\hline\noalign{\smallskip}
 		$d$ & $P_{c}$ & $r_{c}$ & $T_{c}$ & $\frac{{{P_c}{r_c}}}{{{T_c}}}$ \\
 		\noalign{\smallskip}\hline\hline\noalign{\smallskip}
 		7 & 0.53489 & 1.17791 & 1.03718 & 0.60747 \\
 		8 & 1.96244 & 1.03318 & 2.19721 & 0.92278 \\
 		9 & 5.07146 & 0.97562 & 4.05437 & 1.22037 \\
 		10 & 10.7318 & 0.94705 & 6.76637 & 1.50208 \\
 		11 & 19.9878 & 0.93112 & 10.4942 & 1.77347 \\		
 		
 		\noalign{\smallskip}\hline\hline
 	\end{tabular}
 \end{table}
\begin{table}[!htbp]
	\setlength\tabcolsep{8pt}
	\caption{Spherical black holes: $k=1$, $d=7$, $q=1$,  $m=1$, and $c_{0}=c_{1}=c_{2}=c_{3}=c_{4}= 1$.}
	 \label{tab:sphericalBH-alpha} 
	\begin{tabular}{ccccc}
		\hline\hline\noalign{\smallskip}
		$\alpha$ & $P_{c}$ & $r_{c}$ & $T_{c}$ & $\frac{{{P_c}{r_c}}}{{{T_c}}}$ \\
		\noalign{\smallskip}\hline\hline\noalign{\smallskip}
		0.00000 & 25.9104 & 0.68616 & 22.2263 & 0.79989 \\
		0.01000 & 22.8678 & 0.69318 & 19.7148 & 0.80404 \\
		0.10000 & 10.4454 & 0.74202 & 9.39628 & 0.82488 \\
		1.00000 & 0.53489 & 1.17791 & 1.03718 & 0.60747\\
		10.0000 & 0.00341 & 9.05845 & 0.15170 & 0.20372 \\
		100.000 & 0.00004 & 78.3049 & 0.87750 & 0.03896 \\		
		\noalign{\smallskip}\hline\hline
	\end{tabular}
\end{table}
\begin{table}[!htbp]
	\setlength\tabcolsep{8pt}
	\caption{Spherical black holes: $k=1$, $d=7$, $q=1$, $c_{0}=c_{1}=c_{2}=c_{3}=c_{4}= 1$ and $\alpha=1$.}
	 \label{tab:sphericalBH-m} 
	\begin{tabular}{ccccc}
		\hline\hline\noalign{\smallskip}
		$m$ & $P_{c}$ & $r_{c}$ & $T_{c}$ & $\frac{{{P_c}{r_c}}}{{{T_c}}}$ \\
		\noalign{\smallskip}\hline\hline\noalign{\smallskip}
		0.000000 & 0.026994 & 2.247111 & 0.142259 & 0.426396 \\
		0.001000 & 0.026994 & 2.247103 & 0.142260 & 0.426395 \\
		0.010000 & 0.027016 & 2.246273 & 0.142327 & 0.426375 \\
		0.100000 & 0.029227 & 2.165797 & 0.149086 & 0.424592 \\
		1.000000 & 0.534893 & 1.177908 & 1.037184 & 0.607467 \\
		10.00000 & 60.18878 & 1.051550 & 93.72279 & 0.675305 \\		
		\noalign{\smallskip}\hline\hline
	\end{tabular}
\end{table}
According to numerical calculations (see table \ref{tab:sphericalBH-alpha}), there is an upper limit for the value of Lovelock parameter, $\alpha_{u}$, in which no phase transition could happen for $\alpha>\alpha_{u}$. This statement holds for LM AdS black holes with hyperbolic horizon. Indeed, inclusion of higher curvature terms (based on Lovelock Lagrangian) affects the criticality of AdS black holes in massive gravity; by tuning the Lovelock coefficient ($\alpha$) the first-order phase transition can be produced or ruined. In addition, critical pressure and temperature are decreasing functions of the Lovelock coefficient ($\alpha$) but critical horizon radius is an increasing function of it. The universal ratio $\frac{{{P_c}{r_c}}}{{{T_c}}}$ is an increasing and decreasing function of $\alpha$ in regions $0<\alpha<1$ and $1<\alpha<\alpha_{u}$, respectively.

The functional form of critical values with respect to the graviton mass parameter ($m$) are investigated in table \ref{tab:sphericalBH-m}. According to this table, critical pressure and temperature are increasing functions of $m$ and critical horizon radius is a decreasing function of it. \\

\textbf{Ricci flat horizon ($k=0$):}

The LM equation of state for Ricci flat black holes is given as
\begin{equation} \label{pressure-FlatBHs}
P = \frac{{{d_2}\tilde T}}{{4{r_ + }}} - \frac{{{d_2}{d_3}{m^2}c_0^2{c_2}}}{{16\pi r_ + ^2}} - \frac{{{d_2}{d_3}{d_4}{m^2}c_0^3{c_3}}}{{16\pi r_ + ^3}} - \frac{{{d_2}{d_3}{d_4}{d_5}{m^2}c_0^4{c_4}}}{{16\pi r_ + ^4}} + \frac{{{q^2}}}{{8\pi r_ + ^{2{d_2}}}},
\end{equation}
with the signature $P(\pm,\pm,\pm,\pm,+)$. As seen in eq. (\ref{pressure-FlatBHs}), the effect of higher order curvatures of TOL gravity, which encodes in the Lovelock coefficient $\alpha$, vanishes for Ricci flat black holes, and the only effect of them comes from the location of the event horizon according to $\psi(r_{+})=0$ (see eq. \ref{metric function}). In this case, one can show that there exists only one critical point depending on all massive coupling coefficients be positive or all of them be negative definite (see Sec. \ref{phase transition-massive}). Interestingly, in the case of Ricci flat black holes as shown in table \ref{tab:RicciflatBH-m}, there is a lower value for the graviton mass parameter, referred as $m_{b}$, in which no phase transition takes place in region $m<m_{b}$. Remarkably, Ricci flat black holes can experience critical behavior and small/large black hole phase transition by giving mass to gravitons. \\
\begin{table} 
	\setlength\tabcolsep{8pt}
	\caption{Ricci flat black holes: $k=0$, $q=1$, $m=1$ and $c_{0}=c_{1}=c_{2}=c_{3}=c_{4}= 1$.}
	\label{tab:RicciflatBH-d} 
	\begin{tabular}{ccccc}
		\hline\hline\noalign{\smallskip}
		$d$ & $P_{c}$ & $r_{c}$ & $T_{c}$ & $\frac{{{P_c}{r_c}}}{{{T_c}}}$ \\
		\noalign{\smallskip}\hline\hline\noalign{\smallskip}
		7  & 25.0668 & 0.68738 & 21.2994 & 0.80897 \\
		8  & 73.0874 & 0.69501 & 50.5821 & 1.00423 \\
		9  & 159.769 & 0.71105 & 95.0377 & 1.19535 \\
		10& 297.840 & 0.72793 & 156.551 & 1.38489 \\
		11& 501.676 & 0.74375 & 237.102 & 1.57368\\		
		
		\noalign{\smallskip}\hline\hline
	\end{tabular}
\end{table}
\begin{table} 
	\setlength\tabcolsep{8pt}
	\caption{Ricci flat black holes: $k=0$, $d=7$, $q=1$ and $c_{0}=c_{1}=c_{2}=c_{3}=c_{4}= 1$.}
	\label{tab:RicciflatBH-m} 
	\begin{tabular}{ccccc}
		\hline\hline\noalign{\smallskip}
		$m$ & $P_{c}$ & $r_{c}$ & $T_{c}$ & $\frac{{{P_c}{r_c}}}{{{T_c}}}$ \\
		\noalign{\smallskip}\hline\hline\noalign{\smallskip}
		0.00000& - & - & - & - \\
		0.01000& - & - & - & - \\
		0.05000& 0.00203 & 1.78699 & 0.00497 & 0.73219 \\
		0.10000& 0.017231 & 1.43748 & 0.03266 & 0.75834 \\
		0.50000& 2.72792 & 0.86004 & 2.93899 & 0.79827 \\	
		1.00000& 25.0668 & 0.68738 & 21.2994 & 0.80897 \\
		5.00000& 4599.43 & 0.40662 & 2267.57 & 0.82476 \\
		10.0000& 44299.1 & 0.32383 & 17304.4 & 0.82476 \\	
		
		\noalign{\smallskip}\hline\hline
	\end{tabular}
\end{table}

\textbf{Hyperbolic horizon ($k=-1$):}

The LM equation of state for hyperbolic black holes reads
\begin{eqnarray} \label{pressure-hyperbolicBHs}
P &=& \frac{{{d_2}(T - {m^2}{c_0}{c_1}/4\pi )}}{{4{r_ + }}} + \frac{{{d_2}{d_3}(1 - {m^2}c_0^2{c_2})}}{{16\pi r_ + ^2}} - \frac{{{d_2}(8\pi \alpha T + {d_3}{d_4}{m^2}c_0^3{c_3})}}{{16\pi r_ + ^3}} - \frac{{{d_2}{d_5}(\alpha  + {d_3}{d_4}{m^2}c_0^4{c_4})}}{{16\pi r_ + ^4}} + \frac{{{d_2}{\alpha ^2}T}}{{4r_ + ^5}} \nonumber \\
&& + \frac{{{d_2}{d_7}{\alpha ^2}}}{{48\pi r_ + ^6}} + \frac{{{q^2}}}{{8\pi r_ + ^{2{d_2}}}}, 
\end{eqnarray}
with the signature $P(\pm,\pm,\pm,\pm,+,+,+)$. Amazingly, when all the massive coupling coefficients are positive or all of them are negative, two (physical) critical points can be found at most. In fact, the equation of state (\ref{pressure-hyperbolicBHs}) could have three positive roots, but only two of them can be physical. We refer to the smaller and larger critical horizon radii as $r_{c_{1}}$ and $r_{c_{2}}$ respectively ($r_{c_{1}}<r_{c_{2}}$). It should be noted that the third critical point, which actually has the smallest value for critical horizon radius, is always unphysical since the associated black hole has a negative temperature. In the tables \ref{tab:hyperbolicBH-d (rc1)} and \ref{tab:hyperbolicBH-d (rc2)}, the smaller and larger critical horizon radii ($r_{c_{1}}$ and $r_{c_{2}}$) have been computed for the LM AdS black holes with hyperbolic horizons in higher dimensions. The hyperbolic black holes with the smaller critical horizons ($r_{c_{_{1}}}$) experience large values for the temperature and pressure, and therefore, we call them as the "hot black holes". In this sense, the hyperbolic black holes corresponding to the larger critical horizons (with small values for the temperature and pressure respect to the hot black holes) are referred as "Cold black holes".

\begin{figure}[!htbp]
	$%
	\begin{array}{ccc}
	\epsfxsize=5.5cm \epsffile{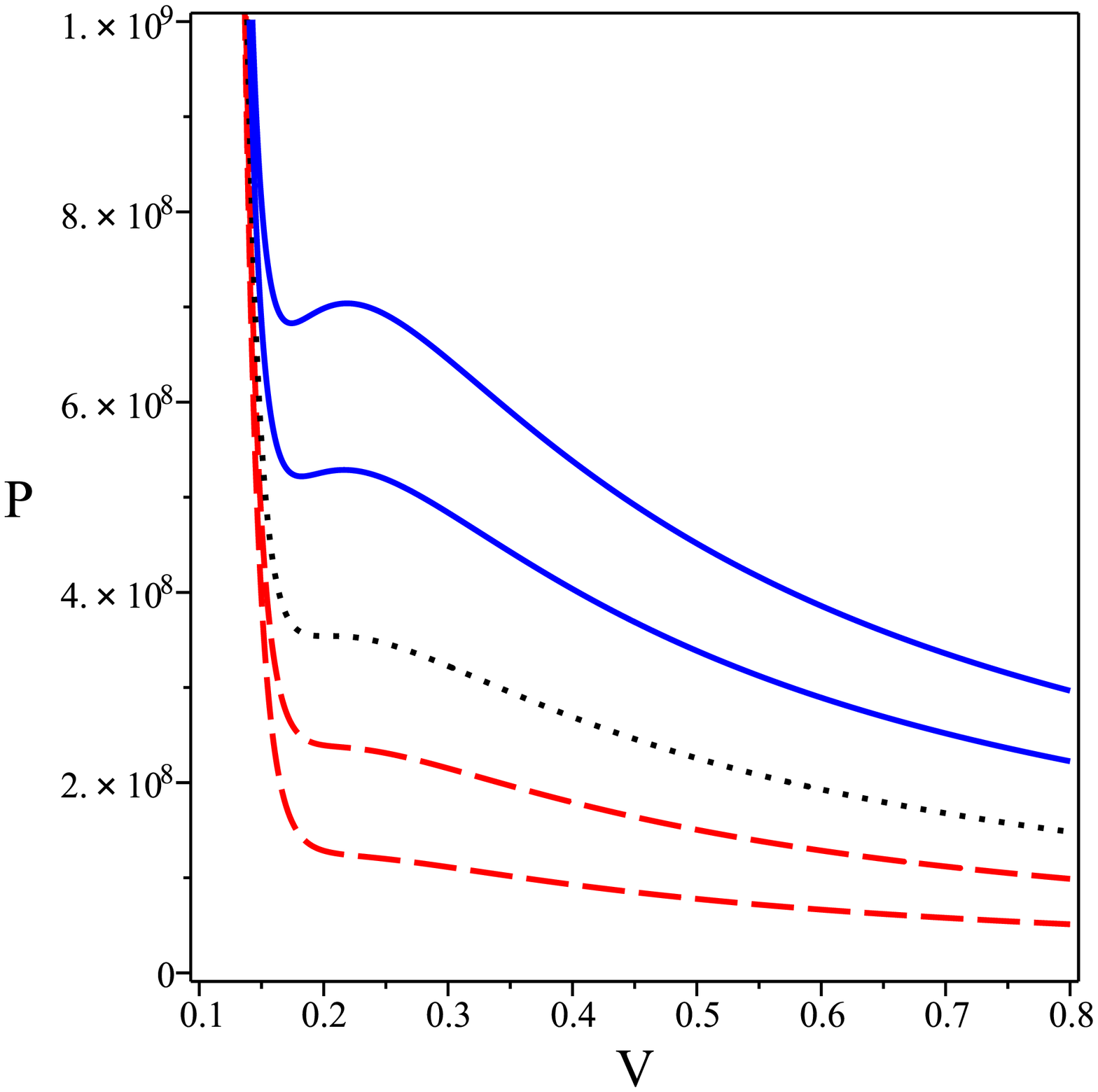}
	\epsfxsize=5.5cm \epsffile{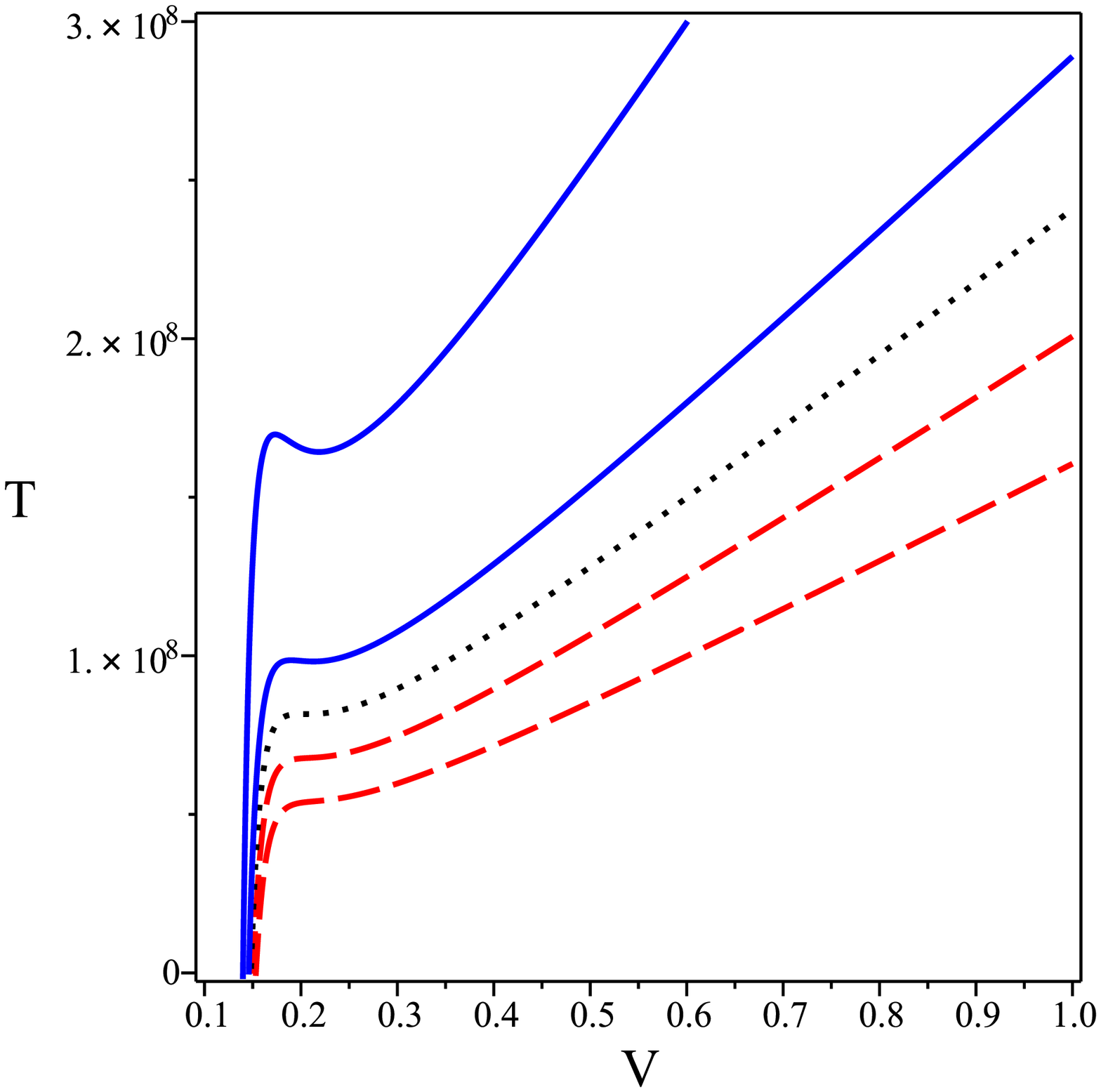}
	\epsfxsize=5.5cm \epsffile{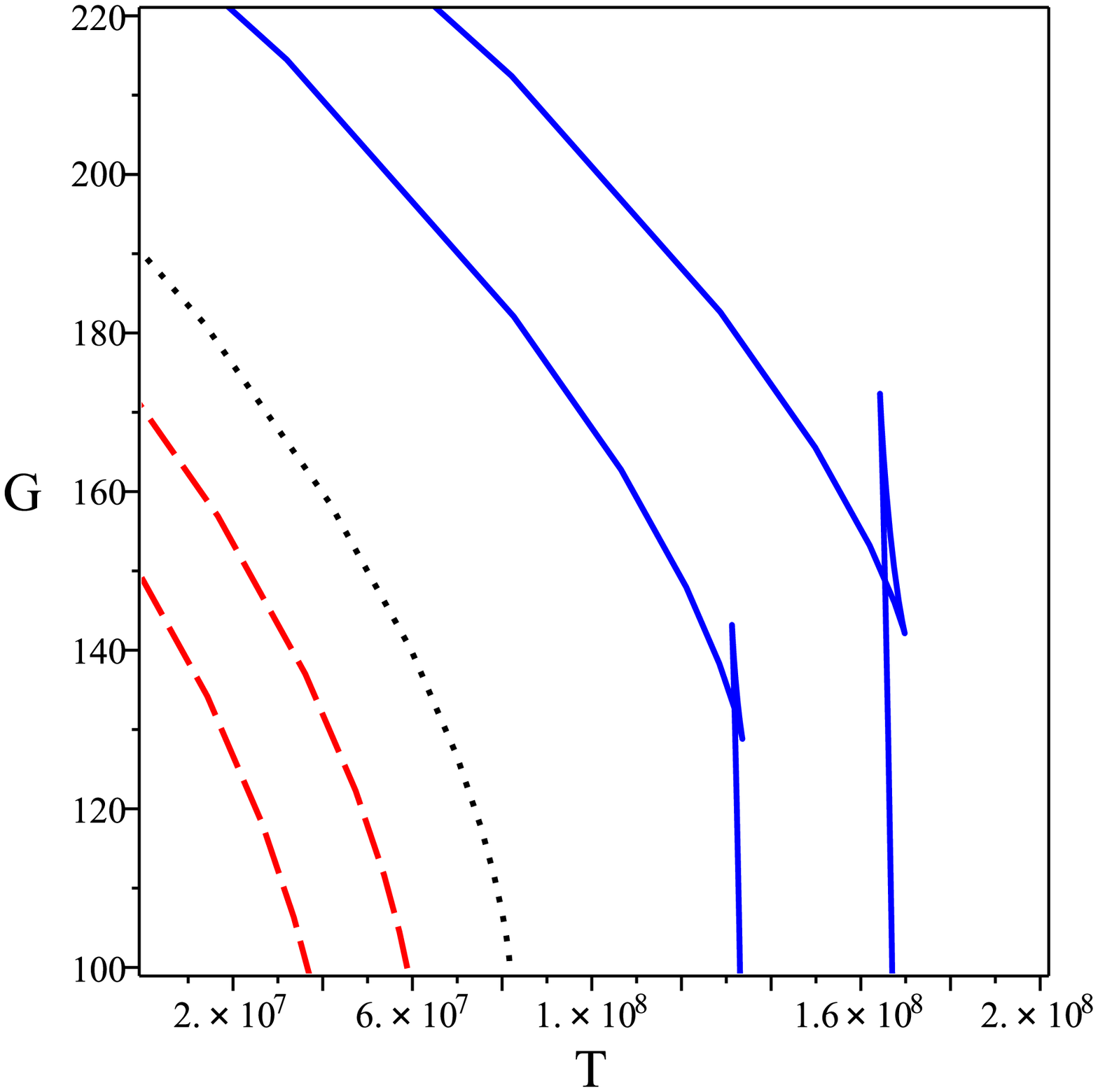}
	&  & 
	\end{array}
	$%
	\caption{\textbf{Hot hyperbolic black hole in LM gravity:} $P-V$ (left), $T-V$
		(middle) and $G-T$ (right) diagrams; we have set $k=-1$, $d=8$, $q=1$,  $m=1$, $c_{0}=c_{1}=c_{2}=c_{3}=c_{4}= 1$ and $\alpha=0.01$. \newline
		\textbf{Left panel:} $T<T_{c_{1}}$ (dashed lines), $T=T_{c_{1}}$ (dotted line) and $T>T_{c_{_{1}}}$ (continuous lines).\newline
		\textbf{Middle and right panels:} $P<P_{c_{1}}$ (dashed lines), $P=P_{c_{1}}$ (dotted lines) and $P>P_{c_{1}}$ (continuous lines).}
	\label{PV-tiny-hyperbolicBHs}
\end{figure}

\begin{figure}[!htbp]
	$%
	\begin{array}{ccc}
	\epsfxsize=5.5cm \epsffile{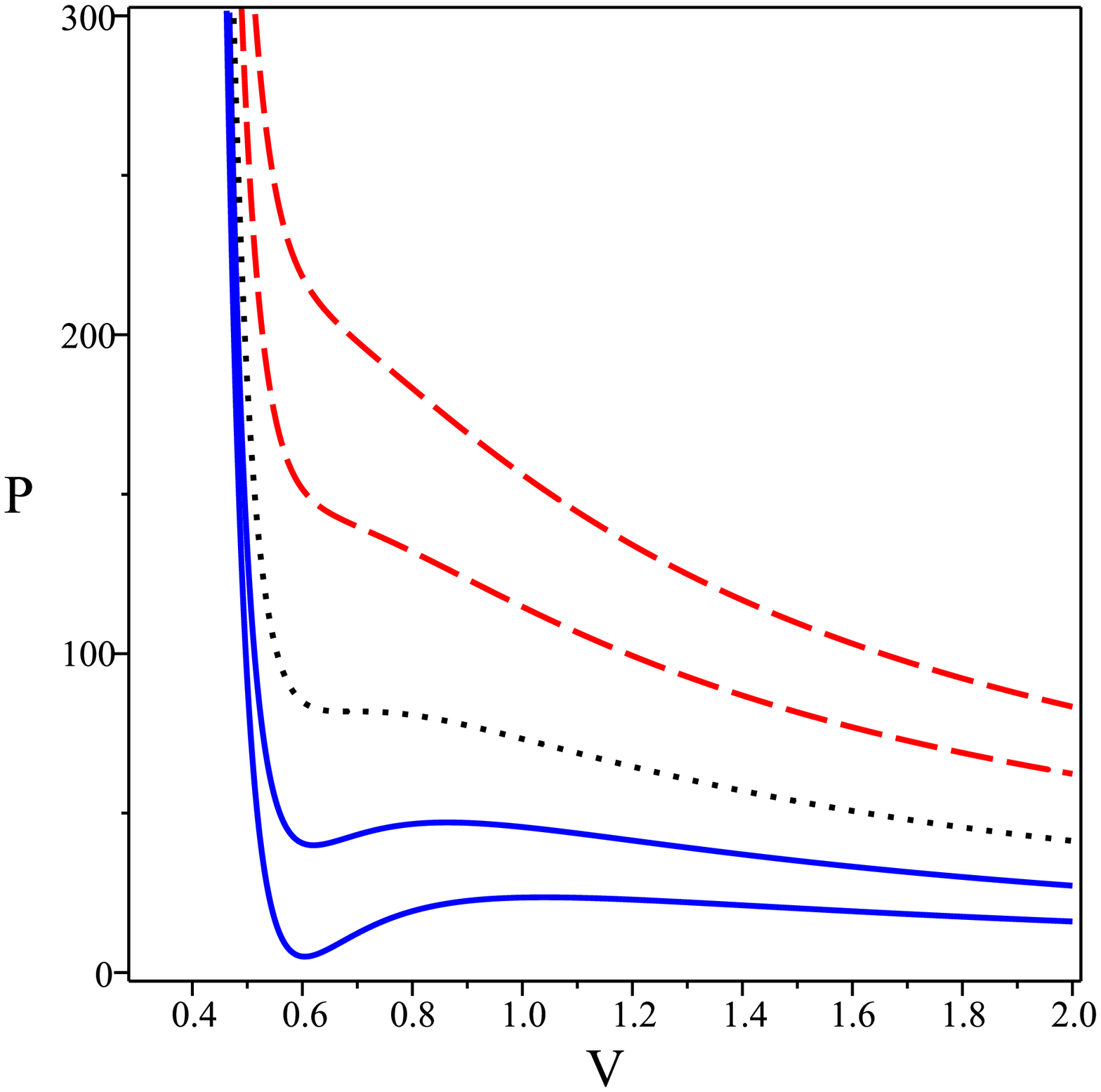}
	\epsfxsize=5.5cm \epsffile{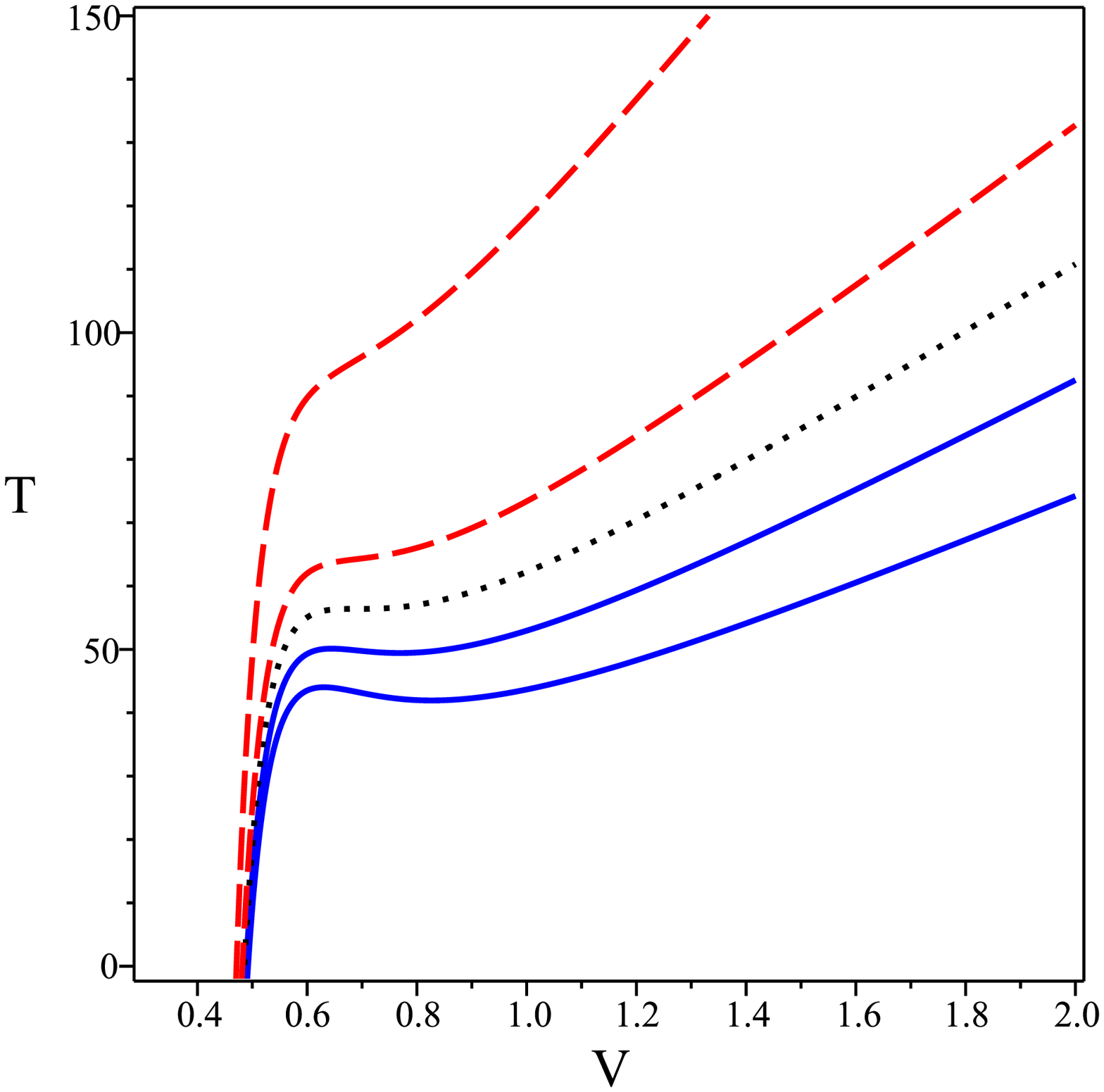}
	\epsfxsize=5.5cm \epsffile{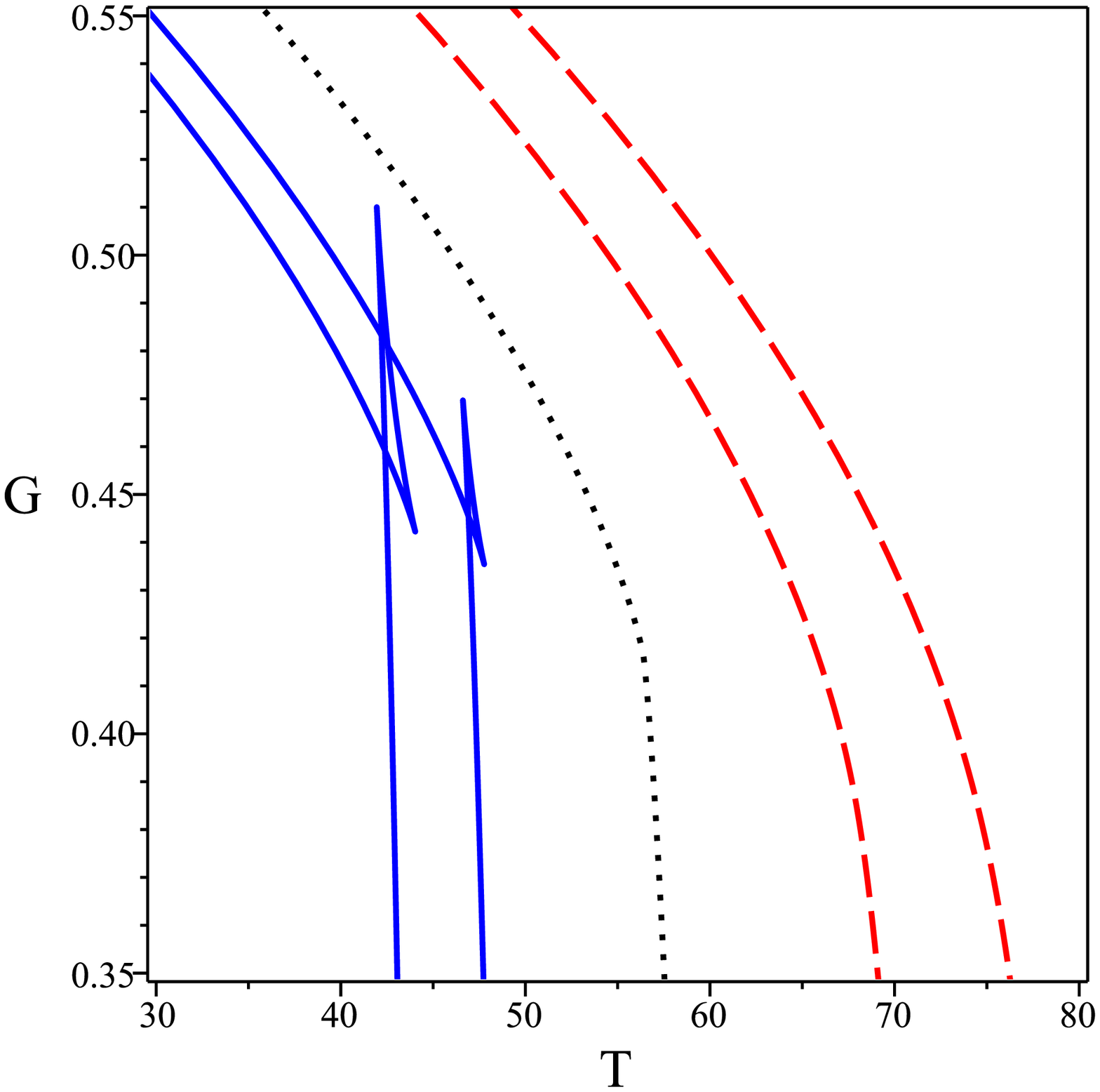}
	&  & 
	\end{array}
	$%
	\caption{\textbf{Cold hyperbolic black hole in LM gravity:} $P-V$ (left), $T-V$
		(middle) and $G-T$ (right) diagrams; we have set $k=-1$, $d=8$, $q=1$,  $m=1$, $c_{0}=c_{1}=c_{2}=c_{3}=c_{4}= 1$ and $\alpha=0.01$. \newline
		\textbf{Left panel:} $T<T_{c_{2}}$ (continuous lines), $T=T_{c_{2}}$ (dotted line) and $T>T_{c_{2}}$ (dashed lines).\newline
		\textbf{Middle and right panels:} $P<P_{c_{2}}$ (continuous lines), $P=P_{c_{2}}$ (dotted lines) and $P>P_{c_{2}}$ (dashed lines).}
	\label{PV-big-hyperbolicBHs}
\end{figure}

In Figs. \ref{PV-tiny-hyperbolicBHs} and \ref{PV-big-hyperbolicBHs}, we have depicted the typical behavior of $P-V$, $T-V$ and $G-T$ curves in the vicinity of the first and second (physical) critical points corresponds to $r_{c_{1}}$ and $r_{c_{2}}$ respectively. Also, the associated critical data for temperature, pressure and event horizon radius are presented in tables \ref{tab:hyperbolicBH-d (rc1)} and \ref{tab:hyperbolicBH-d (rc2)}. In Fig. \ref{PV-tiny-hyperbolicBHs}, For the first critical point (corresponds to the smaller horizon radius, denoted by  $P_{c_{1}}$, $r_{c_{_{1}}}$ and  $T_{c_{1}}$), we observe the swallow-tail behavior in $G-T$ diagrams which corresponds to the first order phase transition for $P>P_{c_{1}}$ in contrast to the van der Waals phase transition which only takes place  for $P<P_{c}$. In $P-V$ diagrams of hyperbolic black holes, interestingly, the (unphysical) oscillating part of isotherms take place for $T>T_{c_{1}}$ which means the existence of two phases behavior and according to Maxwell's equal area law, the oscillating part is replaced by an isobar. For region $T<T_{c_{1}}$ in $P-V$ diagrams, the one phase behavior corresponding to ideal gas is observed. Comparing with the van der Waals phase transition, this critical behavior is completely reverse. This evidence show hyperbolic black holes could potentially experience the reverse van der Waals like behavior for (first order) phase transition at high temperature and pressure which is a remarkable result. Further, we will uncover the theory dependency's origin of the ''reverse behavior'' in LM gravity. As far as we know, there is no reverse van der Waals phase transition in usual thermodynamic systems.

In Fig. \ref{PV-big-hyperbolicBHs}, the qualitative behavior of the hyperbolic (cold) black hole at the second critical point ($r_{c_{2}}$) is displayed. At this critical point, we observe the standard van der Waals phase transition which explained in details before. Interestingly, numerical calculations, which are presented in tables \ref{tab:hyperbolicBH-d (rc1)} and \ref{tab:hyperbolicBH-d (rc2)}, show that critical pressure, horizon radius, temperature and universal ratio ($\frac{P_{c_{i}} r_{c_{i}}}{T_{c_{i}}}$) are increasing functions of spacetime dimension ($d$) at the both critical points. In addition, no phase transition could happen for $\alpha>\alpha_{u}$ in which $\alpha_{u}$ is an upper limit for the Lovelock coefficient.

\begin{table} 
	\setlength\tabcolsep{8pt}
	\caption{(Hot) Hyperbolic black holes: $k=-1$, $q=1$,  $m=1$, $c_{0}=c_{1}=c_{2}=c_{3}=c_{4}= 1$ and $\alpha=0.01$.}
	\label{tab:hyperbolicBH-d (rc1)} 
	\begin{tabular}{cccccc}
		\hline\hline\noalign{\smallskip}
		$d$ & $P_{c_{1}}$ & $r_{c_{1}}$ & $T_{c_{1}}$ & $\frac{{{P_{c_{1}}}{r_{c_{1}}}}}{{{T_{c_{1}}}}}$ \\
		\noalign{\smallskip}\hline\hline\noalign{\smallskip}
		7& 1.201689804$\times 10^7$ & 0.199182022693580 & 3.308081979$\times 10^6$ & 0.7235461736\\
		8& 3.540682432$\times 10^8$ & 0.203488734706851 & 8.162818360$\times 10^7$ & 0.8826473362\\
		9&9.794234509$\times 10^9$& 0.206528200802543 & 1.941047787$\times 10^9$ & 1.042110166\\
		10& 2.589915504$\times 10^{11}$ & 0.208777110589713 & 4.499390023$\times 10^{10}$ & 1.201751955\\
		11& 6.619241728$\times 10^{12}$ & 0.210505697368947 & 1.023424155$\times 10^{12}$  & 1.361496198\\		
		
		\noalign{\smallskip}\hline\hline
	\end{tabular}
\end{table}

\begin{table} 
	\setlength\tabcolsep{8pt}
	\caption{(Cold) Hyperbolic black holes: $k=-1$, $q=1$,  $m=1$, $c_{0}=c_{1}=c_{2}=c_{3}=c_{4}= 1$ and $\alpha=0.01$.}
	 \label{tab:hyperbolicBH-d (rc2)}
	\begin{tabular}{cccccc}
		\hline\hline\noalign{\smallskip}
		$d$ & $P_{c_{2}}$ & $r_{c_{2}}$ & $T_{c_{2}}$ & $\frac{{{P_{c_{2}}}{r_{c_{2}}}}}{{{T_{c_{2}}}}}$\\
		\noalign{\smallskip}\hline\hline\noalign{\smallskip}
		7& 27.8071 & 0.68117 & 23.3249 & 0.81207\\
		8& 81.8748 & 0.68966 & 56.3992 & 1.00118\\
		9& 178.635 & 0.70655 & 106.200 & 1.18847\\
		10& 331.629 & 0.72410 & 174.622 & 1.37516\\
		11& 556.211 & 0.74044 & 263.733 & 1.56159\\		
		
		\noalign{\smallskip}\hline\hline
	\end{tabular}
\end{table}

\begin{table} 
	\setlength\tabcolsep{8pt}
	\caption{Hyperbolic black holes: $k=-1$, $d=7$, $q=1$, $c_{0}=c_{1}=c_{2}=c_{3}=c_{4}= 1$ and $\alpha=0.01$.}
	\label{tab:hphericalBH-m} 
	\begin{tabular}{ccccc}
		\hline\hline\noalign{\smallskip}
		$m$ & $P_{c}$ & $r_{c}$ & $T_{c}$ & $\frac{{{P_c}{r_c}}}{{{T_c}}}$ \\
		\noalign{\smallskip}\hline\hline\noalign{\smallskip}
		0.000000& 1.203736376$\times 10^{7}$ & 0.1991562990 & 3.313433416$\times 10^{6}$ & 0.7235144082 \\
		0.010000& 1.203737744$\times 10^{7}$ & 0.199156282030114 & 3.313437004$\times 10^{6}$ & 0.7235143849 \\
		0.100000& 1.203715916$\times 10^{7}$ & 0.199156555956048 & 3.313379914$\times 10^{6}$ & 0.7235147264 \\
		0.500000& 1.203224868$\times 10^{7}$ & 0.199162723328388 & 3.312095942$\times 10^{6}$ & 0.7235223426 \\
		
		\noalign{\smallskip}\hline\hline
	\end{tabular}
\end{table}

Until now we have considered the equations of state of AdS black holes in the LM gravity framework and disclosed a peculiar critical behavior and a strange phase transition for hyperbolic black holes at high temperatures. In order to have a better understanding of the theory dependency and nature of phase transitions, we will separately study the equations of state of AdS black holes in Lovelock and massive gravities. \\

\subsection{Lovelock gravity: the phase transition revisited\label{phase transition-Lovelock}}

Here, we explicitly show that the higher order curvatures are responsible for reverse van der Waals phase transition. In the context of TOL gravity (massless graviton case), by taking the limit $m \to 0$ of eq. (\ref{pressure-massive Lovelock-physical}), the equation of state for AdS black holes reads

\begin{equation} \label{pressure-Lovelock}
P = \frac{{{d_2}T}}{{4{r_ + }}} - \frac{{{d_2}{d_3}k}}{{16\pi r_ + ^2}} + \frac{{{d_2}k\alpha T}}{{2r_ + ^3}} - \frac{{{d_2}{d_5}{k^2}\alpha }}{{16\pi r_ + ^4}} + \frac{{{d_2}{{(k\alpha )}^2}T}}{{4r_ + ^5}} - \frac{{{d_2}{d_7}k{\alpha ^2}}}{{48\pi r_ + ^6}} + \frac{{{q^2}}}{{8\pi r_ + ^{2{d_2}}}}.
\end{equation}
The signature of the Lovelock equation of state (\ref{pressure-Lovelock}) implies no critical behavior and phase transition for Ricci flat black holes (i.e., $P(+,+)$). But, for Lovelock AdS black holes with spherical and hyperbolic horizons ($k=\pm1$), the equation of state (\ref{pressure-Lovelock}) with signature $P(+,\mp,\pm,-,+,\mp,+)$ predicts the possibility of finding critical point(s) for Lovelock AdS black holes (upper and lower signs in $P(+,\mp,\pm,-,+,\mp,+)$ are related to the spherical and hyperbolic geometries, respectively) . By using eq. (\ref{critical point equation}), the equation of critical point for Lovelock AdS-charged black holes is given as

\begin{equation} \label{critical point equation-Lovelock}
\begin{array}{l}
kr_ + ^{2d}\left\{ {12k\alpha r_ + ^6 + \left( {3(d + 5){k^2} - 5{d_7}} \right){\alpha ^2}r_ + ^4 + 2\left( {5{d_5}{k^2} - 9{d_7}} \right)k{\alpha ^3}r_ + ^2 - 5{d_7}{k^2}{\alpha ^4} - {d_3}r_ + ^8} \right\}\\
\,\,\,\,\, + 4{q^2}r_ + ^{10}\left\{ {6{d_{7/2}}k\alpha r_ + ^2 + 5{d_{9/2}}{k^2}{\alpha ^2} + {d_{5/2}}r_ + ^4} \right\} = 0 .
\end{array}
\end{equation}

\begin{figure}[!htbp]
	$%
	\begin{array}{ccc}
	\epsfxsize=5.5cm \epsffile{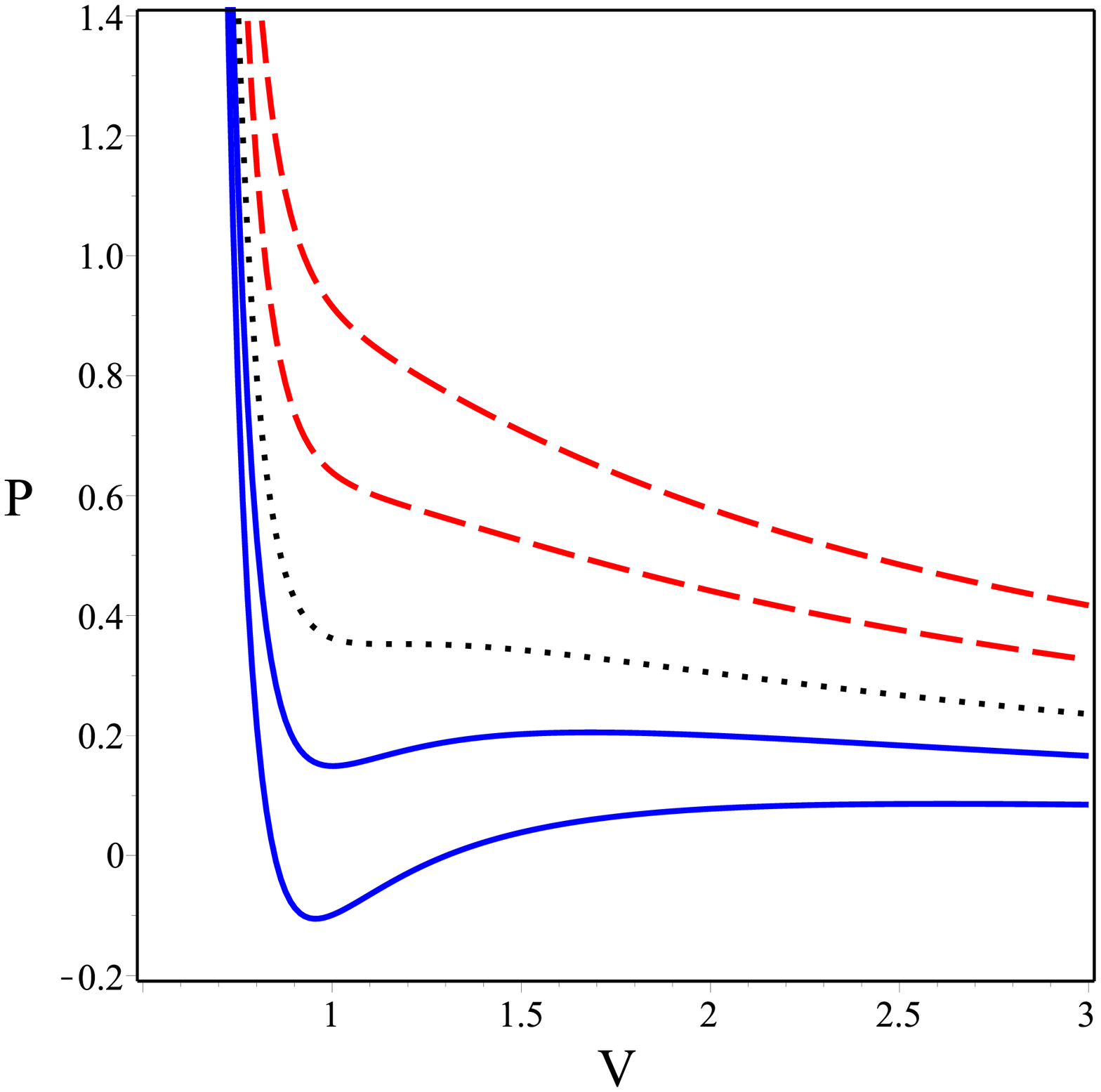}
	\epsfxsize=5.5cm \epsffile{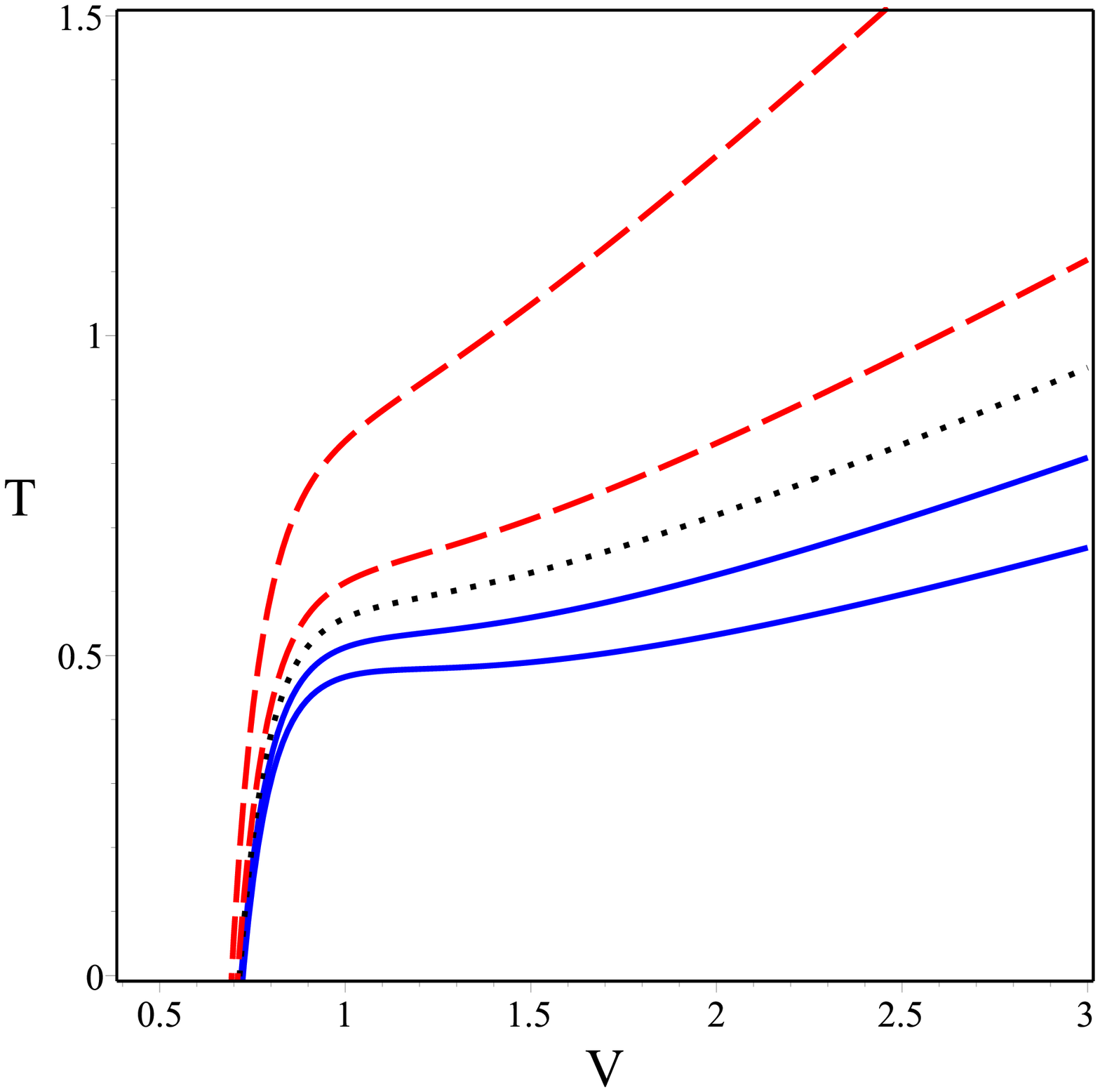}
	\epsfxsize=5.5cm \epsffile{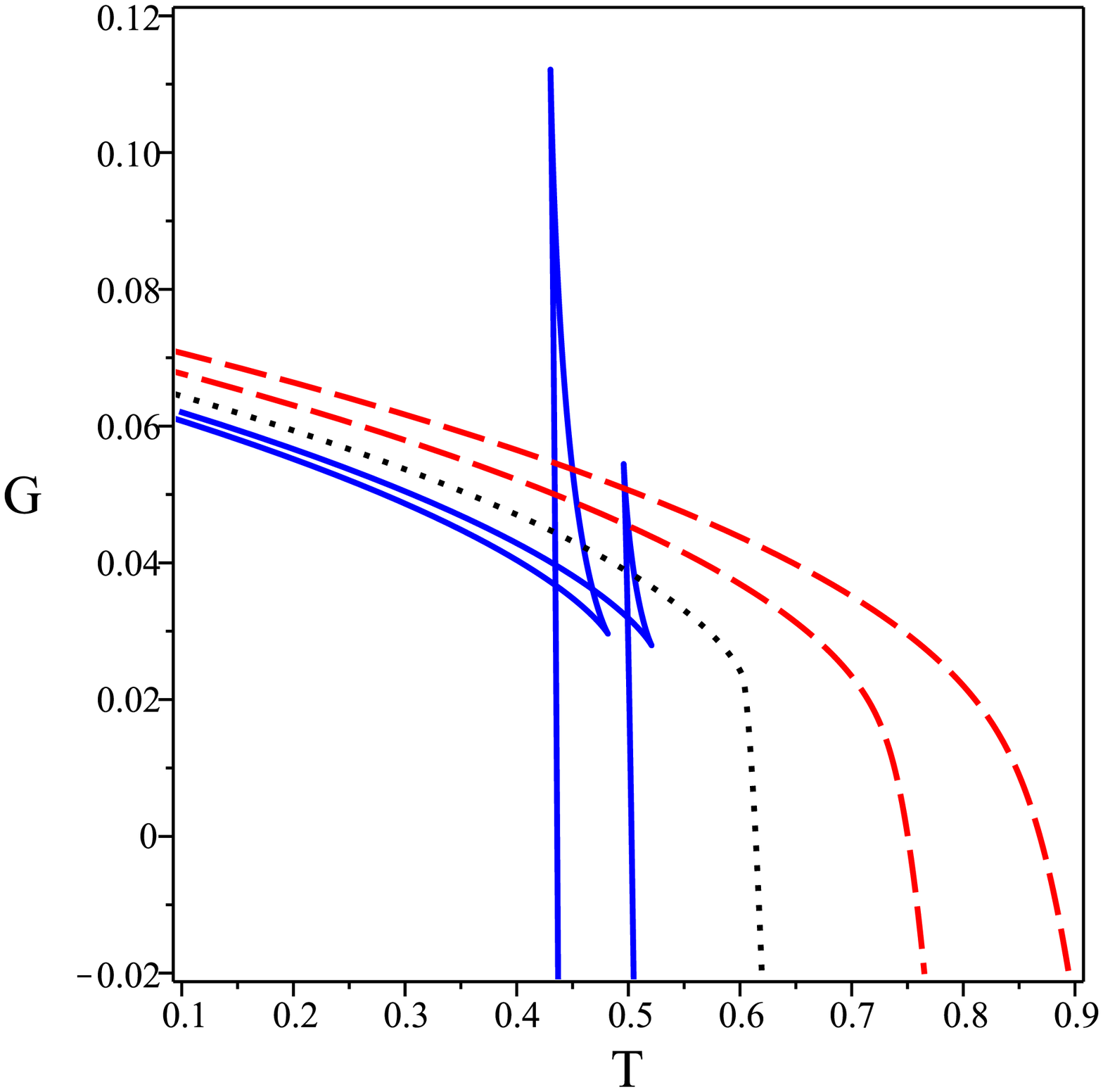}
	&  & 
	\end{array}
	$%
	\caption{\textbf{Spherical black hole in Lovelock gravity:} $P-V$ (left), $T-V$
		(middle) and $G-T$ (right) diagrams; we have set $k=+1$, $d=8$, $q=1$, and $\alpha=0.01$. \newline
		\textbf{Left panel:} $T<T_{c}$ (continuous lines), $T=T_{c}$ (dotted line) and $T>T_{c}$ (dashed lines).\newline
		\textbf{Middle and right panels:} $P<P_{c}$ (continuous lines), $P=P_{c}$ (dotted lines) and $P>P_{c}$ (dashed lines).}
	\label{PV-LovelockBHs-spherical}
\end{figure}

For simplicity, we first consider the neutral black holes ($q=0$), and then we discuss the charged case. Our investigations show the above equation (with $q=0$) admits the following solutions as critical radii

\begin{equation} \label{critical points-Lovelock}
{r_c} = \sqrt {\frac{{(d + 3)k \pm 2{k^2}\sqrt {{d_2}(12 - d)} }}{{{d_3}}}\alpha \,} \,,\,\,\,\,{r_c} = \sqrt { - k\alpha } ,
\end{equation}
in which only real-valued roots are permissible for the topological black holes. Obviously, there are at most one or two positive critical radii for Lovelock AdS (neutral) black holes. Here, we confront with an interesting situation: for spacetimes with $d>12$,  there does not exist any critical point for Lovelock AdS black holes with spherical horizon geometry, and so phase transition does not take place. In 7-dimensions, we observe only one critical point for spherical black holes, and in $8\leqslant d \leqslant11$ there are always two critical horizon radii, $r_{c_{1}}$ and $r_{c_{2}}$, which the corresponding critical points can be physical or unphysical (the critical data are presented in \cite{PV2014Lovelock}). As a result, in 7-dimensions, the van der Waals behavior, and in $8\leqslant d \leqslant11$, the reentrant behavior for phase transition are observed. In $d=12$ we do not observe $P-V$ criticality and instead there is a cusp-like behavior in the $G-T$ diagram. The effect of the $U(1)$ charge could drastically alter the number of critical point(s) and nature of phase transitions. In fact, in spacetime dimensions with the range $8\leqslant d \leqslant11$, there is a lower value for the electric charge ($Q_{b}$), where for $Q>Q_{b}$, one of the critical points disappears and consequently the reentrant behavior is replaced by the van der Waals like phase transition (some critical data associated with the Lovelock charged-AdS black holes in higher dimensions are presented in \cite{PV2015LovelockBI-Belhaj}). Interestingly, for $d\geqslant 12$, one critical point emerges when the $U(1)$ charge is considered and the van der Waals phase transition takes place. In addition, in $d=7$, the inclusion of  the $U(1)$ charge only changes the location of critical point.  In Fig. \ref{PV-LovelockBHs-spherical}, as an example, we have plotted the critical behavior of a spherical black hole. As seen, in this case $P-V$ criticality is qualitatively similar to the critical behavior of the van der Waals fluid, RN-AdS and LM AdS black holes. \\

\begin{figure}[!htbp]
	$%
	\begin{array}{ccc}
	\epsfxsize=5.5cm \epsffile{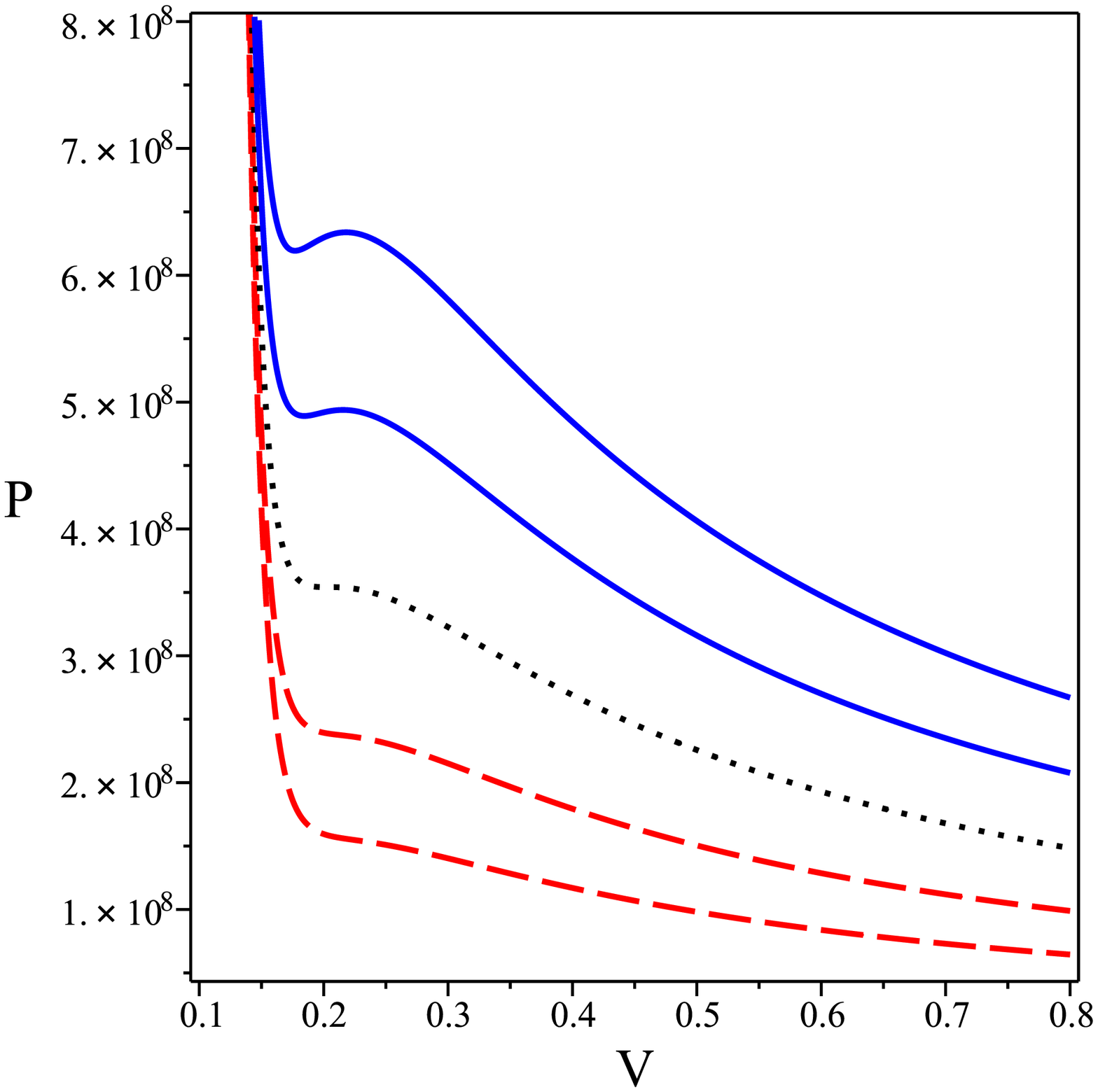}
	\epsfxsize=5.5cm \epsffile{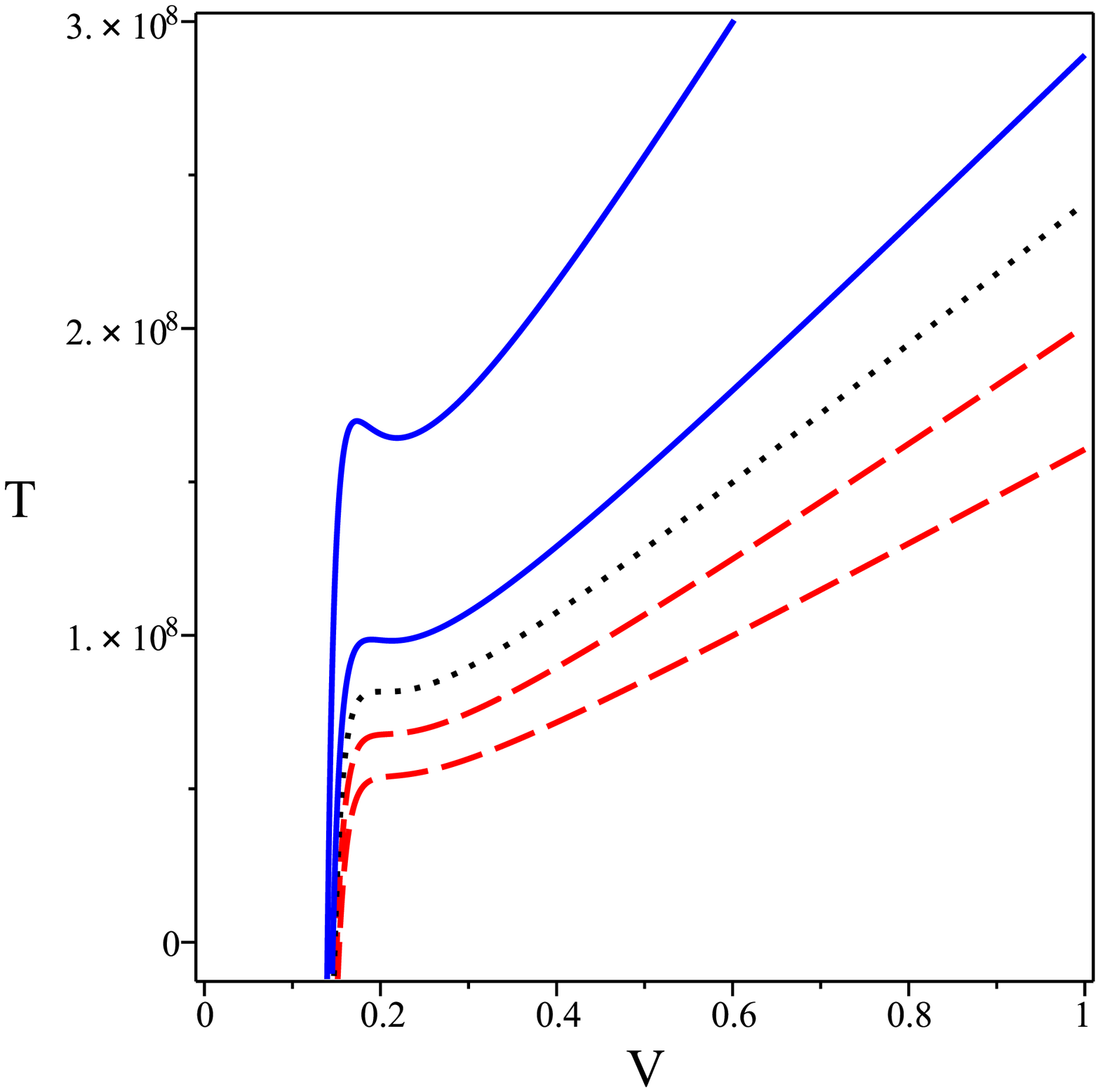}
	\epsfxsize=5.5cm \epsffile{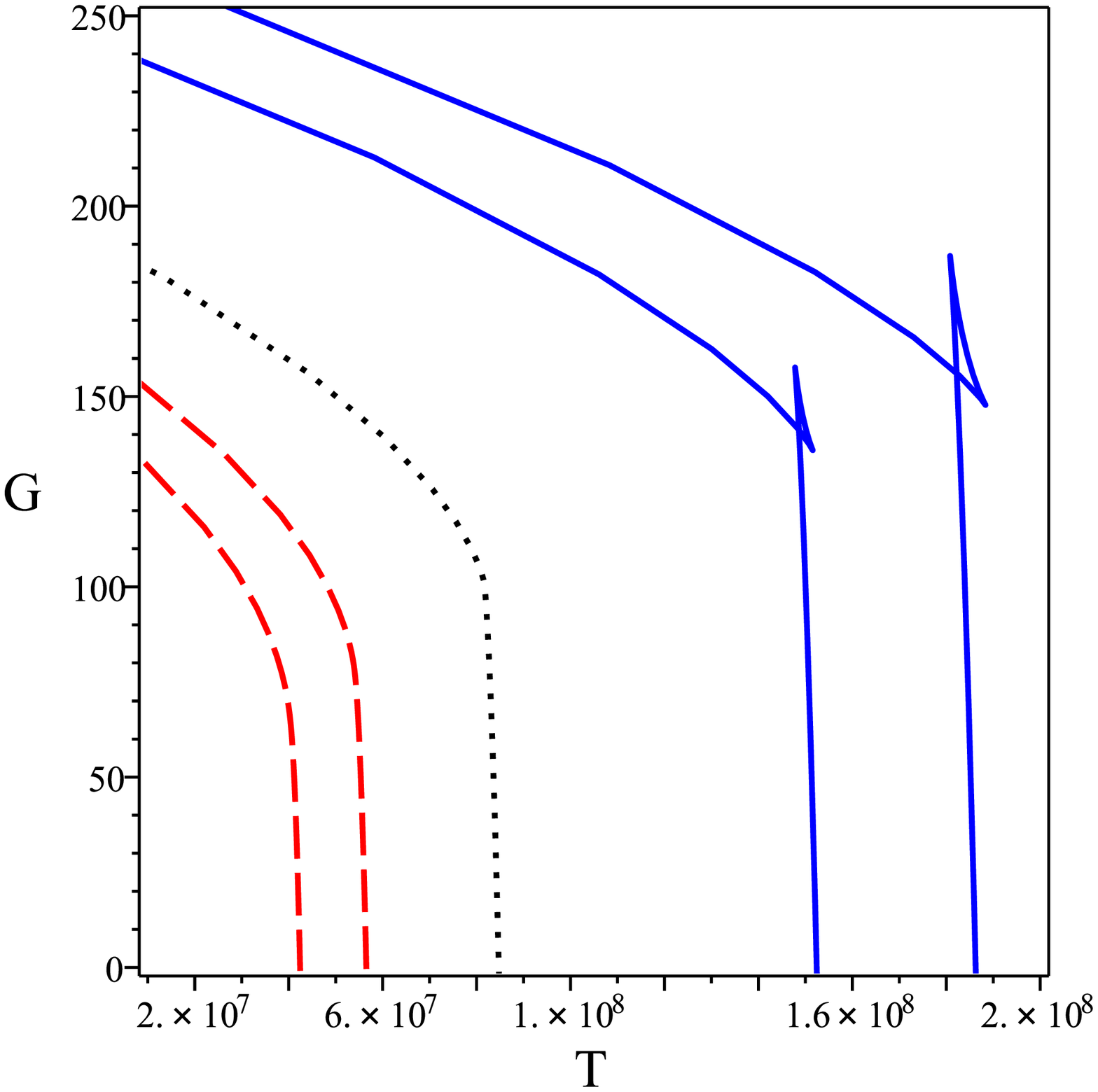}
	&  & 
	\end{array}
	$%
	\caption{\textbf{Hyperbolic black hole in Lovelock gravity:} $P-V$ (left), $T-V$
		(middle) and $G-T$ (right) diagrams; we have set $k=-1$, $d=8$, $q=1$, and $\alpha=0.01$. \newline
		\textbf{Left panel:} $T<T_{c}$ (dashed lines), $T=T_{c}$ (dotted line) and $T>T_{c}$ (continuous lines).\newline
		\textbf{Middle and right panels:} $P<P_{c}$ (dashed lines), $P=P_{c}$ (dotted lines) and $P>P_{c}$ (continuous lines).}
	\label{PV-LovelockBHs-hyperbolic}
\end{figure}

In the case of hyperbolic black holes, there always exists only one critical radius ($r_{c}=\sqrt{\alpha}$) in all spacetime dimensions ($d\geqslant 7$) according to eq. (\ref{critical points-Lovelock}). As stated in section \ref{stability}, the temperature of hyperbolic black holes blows up at the point ${r_{ + }}={r_ i} = \sqrt \alpha$ which is referred as the thermodynamic singularity \cite{Frassino2014} since all isotherms cross at $r_{c}=\sqrt{\alpha}={d_2}v_{c} /4$. Also, at this point the heat capacity of the hyperbolic black holes is zero. The critical point corresponding to this thermodynamic singularity is called the isolated critical point and is regarded as the first example of a critical point with critical exponents which are different from those of van der Waals fluid \cite{Frassino2014,DolanMann2014Lovelock}. When the $U(1)$ charge is considered, one additional critical point might emerge for the hyperbolic black holes. Regarding the charged case, if the value of $U(1)$ charge was more than a lower value ($Q>Q_{b}$), there would exist two critical radii which the smaller one is always unphysical (the corresponding black hole has a negative value for the temperature) and the larger one can be physical. Actually, for the small enough charges ($Q\simeq Q_{b}$), these critical points are created near the isolated critical point, and we observe that by increasing the value of the electric charge ($Q$) the distance between the thermodynamic coordinates of these points is increased in the extended phase space. As a matter of fact, one can grow up the thermodynamic quantities associated to the critical point by increasing the $U(1)$ charge and produce a first order phase transition at high temperature and pressure for hyperbolic black holes. Here, an interesting phenomenon for hyperbolic black holes in Lovelock gravity emerges and persists in higher dimensions ($d\geq7$). Our investigations show that the reverse van der Waals phase transition is a characteristic feature of Lovelock AdS black holes with hyperbolic horizon (this strange behavior is already pointed out in \cite{PV2014LovelockBI-Mo,PV2014Lovelock,Frassino2014}). As seen, the Fig. \ref{PV-LovelockBHs-hyperbolic} exposes the origin of the reverse van der Waals like behavior which has been found in section \ref{phase transition-massive Lovelock} for hyperbolic black holes in the LM gravity. In Fig. \ref{PV-LovelockBHs-hyperbolic}, we observe the existence of inflection point in the isothermal $P-V$ diagrams, the subcritical isobar of $T-V$ plots, and the characteristic swallow-tail form of $G-T$ diagrams, but, in contrast to the behavior of van der Waals fluid, in the opposite way. Therefore, we conclude higher order curvature terms based on the Lovelock Lagrangian are responsible for the reverse van der Waals phase transition. \\

It should be emphasized that the Lovelock equation of state (\ref{pressure-Lovelock}) has been obtained by the assumption of the Lovelock coefficient condition (\ref{coefficient condition}). In the more general case where Lovelock coefficients are independent, one may obtain up to three critical points for black holes which indicates the appearance of triple points \cite{Frassino2014}.

\subsection{Massive gravity: the phase transition revisited\label{phase transition-massive}}

In the context of massive gravity, the equation of state of charged-AdS black holes is given as 

\begin{equation} \label{pressure-massive gravity}
P = \frac{{{d_2}\tilde T}}{{4{r_ + }}} - \frac{{{d_2}{d_3}(k + {m^2}c_0^2{c_2})}}{{16\pi r_ + ^2}} - \frac{{{d_2}{d_3}{d_4}{m^2}c_0^3{c_3}}}{{16\pi r_ + ^3}} - \frac{{{d_2}{d_3}{d_4}{d_5}{m^2}c_0^4{c_4}}}{{16\pi r_ + ^4}} + \frac{{{q^2}}}{{8\pi r_ + ^{2{d_2}}}},
\end{equation}
in which we have used the zero higher curvature limit ($\alpha \to 0$) of the LM equation of state (\ref{pressure-massive Lovelock-physical}) and the shifted Hawking temperature is the same as before, i.e., $\tilde T = T - {m^2}{c_0}{c_1}/4\pi$. As seen, the topological factor ($k$) in the RN-AdS equation of state (\ref{pressure - RN-AdS}) is replaced by the combination $(k + {m^2}c_0^2{c_2})$ which suggests possible critical behavior for black holes with non-trivial horizon topologies. Clearly, the massive charged-AdS equation of state (\ref{pressure-massive gravity}) with signature $P(\pm,\pm,\pm,\pm,+)$ admits phase transition for black holes with spherical, Ricci flat and hyperbolic horizon geometries \cite{Hendi2017Mann-PRD}. In 4-dimensional spacetime, the critical point is obtained from the positive root of the equation ${r_{+}^2}(k + {m^2}c_0^2{c_2}) - 6{q^2} = 0$, in which we applied eqs. (\ref{critical point equation}) and (\ref{pressure-massive gravity}). Hence, there exists a critical horizon radius ($r_{c}$) when $(k + {m^2}c_0^2{c_2})\geqslant0$. It should be noted that, like RN-AdS black holes in Einstein gravity, the $U(1)$ charge is necessary to have critical behavior and phase transition in a 4-dimensional spacetime.  On the other hand, as indicated in \cite{PVmassive2015PRD}, in higher dimensional spacetimes ($d\geqslant5$), the $U(1)$ charge is unnecessary for the appearance of critical behavior and phase transition in the massive equation of state of AdS black holes since the third and forth terms (i.e., massive potential terms ${{m^2}c_0^3{c_3}}$ and ${{m^2}c_0^4{c_4}}$) in the right hand side of eq. (\ref{pressure-massive gravity}) can play the role of electric charge term (the last term). Regarding eqs. (\ref{critical point equation}) and (\ref{pressure-massive gravity}), the critical point(s) of the massive charged-AdS black holes can be obtained from the root(s) of the following equation

\begin{equation} \label{critical point equation-massive gravity}
{d_3}(k + {m^2}c_0^2{c_2})r_ + ^2 + 3{d_3}{d_4}{m^2}c_0^3{c_3}{r_ + } + 6{d_3}{d_4}{d_5}{m^2}c_0^4{c_4} - 4{d_{5/2}}{q^2}r_ + ^{ - 2{d_4}} = 0.
\end{equation}
This is a worthwhile and simple equation (which works in all spacetime dimensions) to investigate the critical point(s) of charged and uncharged-AdS black holes in massive gravity. Interestingly, the critical point(s) is independent of the first massive coupling coefficient ($c_{1}$). Considering the uncharged case ($q=0$) for the sake of simplicity, the critical point equation (\ref{critical point equation-massive gravity}) can be solved simply as
\begin{equation} \label{roots}
{r_c} = \frac{{ - 3{d_4}{m^2}c_0^3{c_3} \pm \sqrt {{{(3{d_4}{m^2}c_0^3{c_3})}^2} - 24(k + {m^2}c_0^2{c_2})({d_4}{d_5}{m^2}c_0^4{c_4})} }}{{2(k + {m^2}c_0^2{c_2})}}.
\end{equation}
It is obvious that there is (are) one or at most two positive critical radii for the equation of state of (neutral) AdS black holes in massive gravity. The conditions to have one positive critical radius are as
\begin{equation} \label{one critical radius-massive gravity(1)}
{r_c} = \frac{{ - 3{d_4}{m^2}c_0^3{c_3}}}{{2(k + {m^2}c_0^2{c_2})}} > 0\,,\,\,\,\,\frac{{\,3{d_4}{m^2}c_0^2c_3^2}}{{8{d_5}{c_4}(k + {m^2}c_0^2{c_2})}} = 1,
\end{equation}
or as
\begin{equation} \label{one critical radius-massive gravity(2)}
{r_ n}{r_c} = \frac{{6{d_4}{d_5}{m^2}c_0^4{c_4}}}{{k + {m^2}c_0^2{c_2}}} < 0\,,\,\,\,\,3{d_4}{m^2}c_0^2c_3^2 > 8{d_5}{c_4}(k + {m^2}c_0^2{c_2}),
\end{equation}
in which $r_ n$ is a negative definite root and $r_ c$ is the positive critical radius. In addition, in order to have two positive critical radii the following conditions must be satisfied
\begin{equation}  \label{two critical radius-massive gravity}
   3{d_4}{m^2}c_0^2c_3^2 > 8{d_5}{c_4}(k + {m^2}c_0^2{c_2})\,,\,\,\,\,{r_{{c_1}}} + {r_{{c_2}}} = \frac{{ - 3{d_4}{m^2}c_0^3{c_3}}}{{k + {m^2}c_0^2{c_2}}} > 0\,,\,\,\,\,{r_{{c_1}}}{r_{{c_2}}} = \frac{{6{d_4}{d_5}{m^2}c_0^4{c_4}}}{{k + {m^2}c_0^2{c_2}}} > 0,
\end{equation}
where $r_{c_1}$ and $r_{c_2}$ are the smaller and the larger critical radii. Consequently, by tuning the massive coupling coefficients ($c_{i}$) according to eqs. (\ref{roots}), (\ref{one critical radius-massive gravity(1)}), (\ref{one critical radius-massive gravity(2)}) and (\ref{two critical radius-massive gravity}), one can easily find one or two (physical) critical point(s) for all types of topological black holes depending on the values of $c_{i}$'s. \\

According to eqs. (\ref{roots}), (\ref{one critical radius-massive gravity(1)}), (\ref{one critical radius-massive gravity(2)}) and (\ref{two critical radius-massive gravity}), when all the massive coupling coefficients are simultaneously positive (negative) definite, there exists one critical radius and can be determined using eq. (\ref{one critical radius-massive gravity(2)}). In order to have two critical points, one should consider some specific signs for the massive couplings ($c_{i}$) based on eq. (\ref{two critical radius-massive gravity}. When the combination $(k + {m^2}c_0^2{c_2})$ is positive, two critical points can be found assuming that $c_{3}<0$ and  $c_{4}>0$, and when  $(k + {m^2}c_0^2{c_2})<0$, one has to assume  $c_{3}>0$ and  $c_{4}<0$. In the both cases, the massive coupling coefficients must satisfy the constraint $3{d_4}{m^2}c_0^2c_3^2 > 8{d_5}{c_4}(k + {m^2}c_0^2{c_2})$. \\

The existence of two critical points for (neutral) AdS black holes is possible when the spacetime dimensions are more than five ($d\geqslant6$) and an interesting phenomenon emerges which is called the reentrant of phase transition (for more details see \cite{Reentrant-dRGTmassive-2017}).  The inclusion of nonlinear electromagnetic fields can increased the number of critical point(s) \cite{Mann2012Gunasekaran}, and as a result, in the context of Born- Infeld-massive gravity \cite{Triple-BI-massive-2017}, the reentrant phase transition appears in 4-dimensions and the so-called triple point in spacetime dimensions more than five ($d\geqslant5$). It should be noted these considerations were done in the canonical ensemble.


\subsection{LM gravity: reentrant phase transitions and triple points \label{Lovelock massive revisited}}
This section is devoted to study the possibility of appearance the reentrant phase transition and triple point in the phase structure of the LM AdS black holes. In the previous section, we indicated that under certain conditions the equations of state of  topological black holes in pure massive gravity (without non trivial electromagnetic fields like BI electrodynamics) may have up to two critical points and thus exhibit van der Waals and reentrant phase transitions which the latter corresponds to three phase behavior. In Sec. \ref{phase transition-Lovelock}, we showed Lovelock (un)charged-AdS black holes with spherical horizon can exhibit van der Waals and reentrant phase transitions.  Also, for the Lovelock (un)charged-AdS black holes with hyperbolic horizon, the reverse van der Waals like behavior is observed which can be accompanied by a (normal) van der Waals phase transition. Consequently, since there are many thermodynamic variables in the extended phase space of the LM AdS black holes, one expects these black holes may enjoy a vast range of thermodynamic behaviors which found in the other gravitational theories similar to those in usual thermodynamics. Here we intend to examine these possibilities.

 First, we consider Ricci flat black holes in the LM gravity. As stated, the effect of higher curvature terms which is encoded in the Lovelock coefficient ($\alpha$) vanishes for Ricci flat black holes. As a result, using eq. (\ref{two critical radius-massive gravity}), one can find the reentrant phase transition in spacetime dimensions $d\geqslant6$ for the neutral black holes \cite{Reentrant-dRGTmassive-2017}, and by inserting nonlinear electromagnetic fields the associated triple point can be found \cite{Triple-BI-massive-2017}.

\begin{figure}[!htbp]
	$%
	\begin{array}{ccc}
	\epsfxsize=5.5cm \epsffile{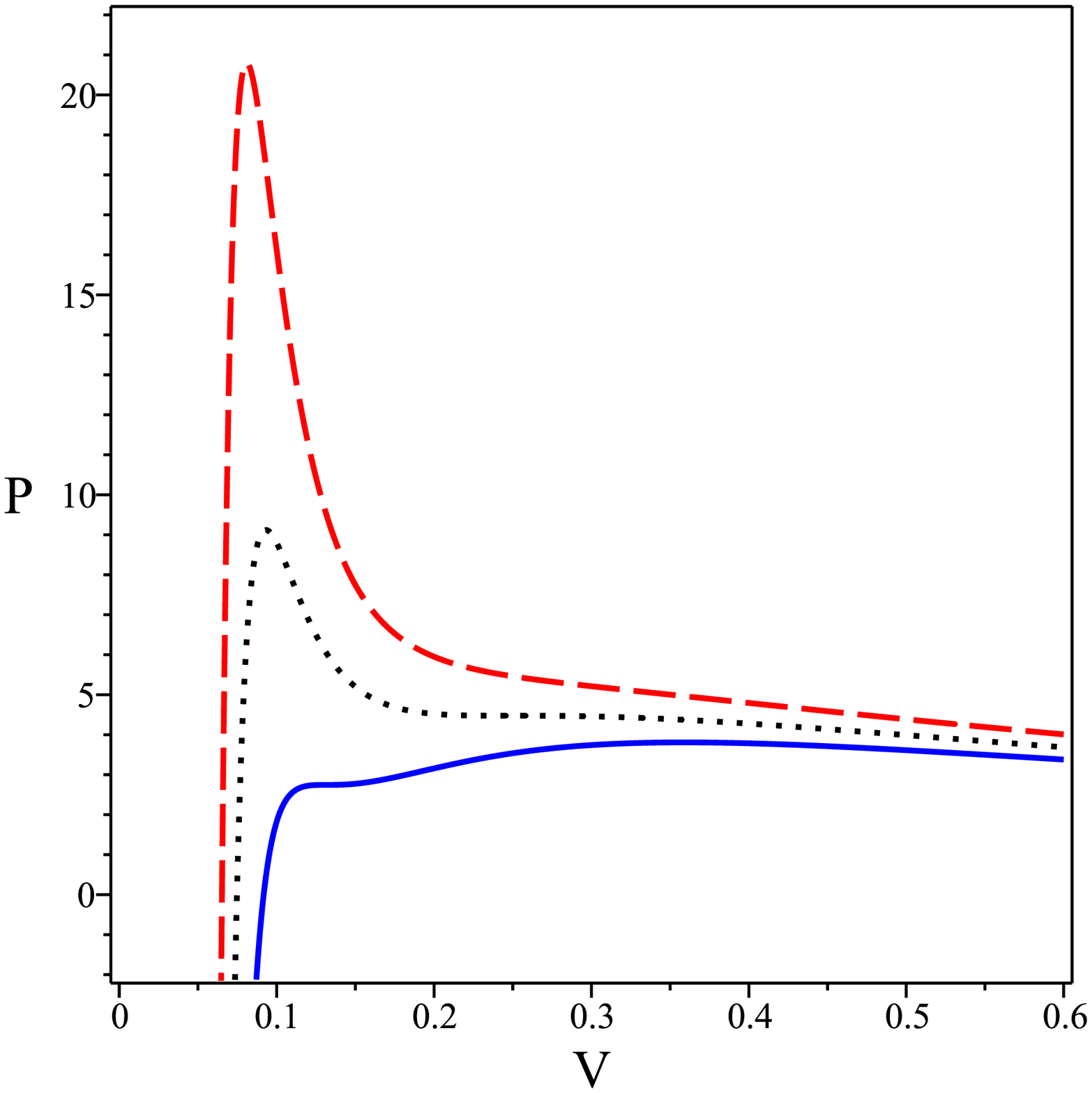}
	\epsfxsize=5.5cm \epsffile{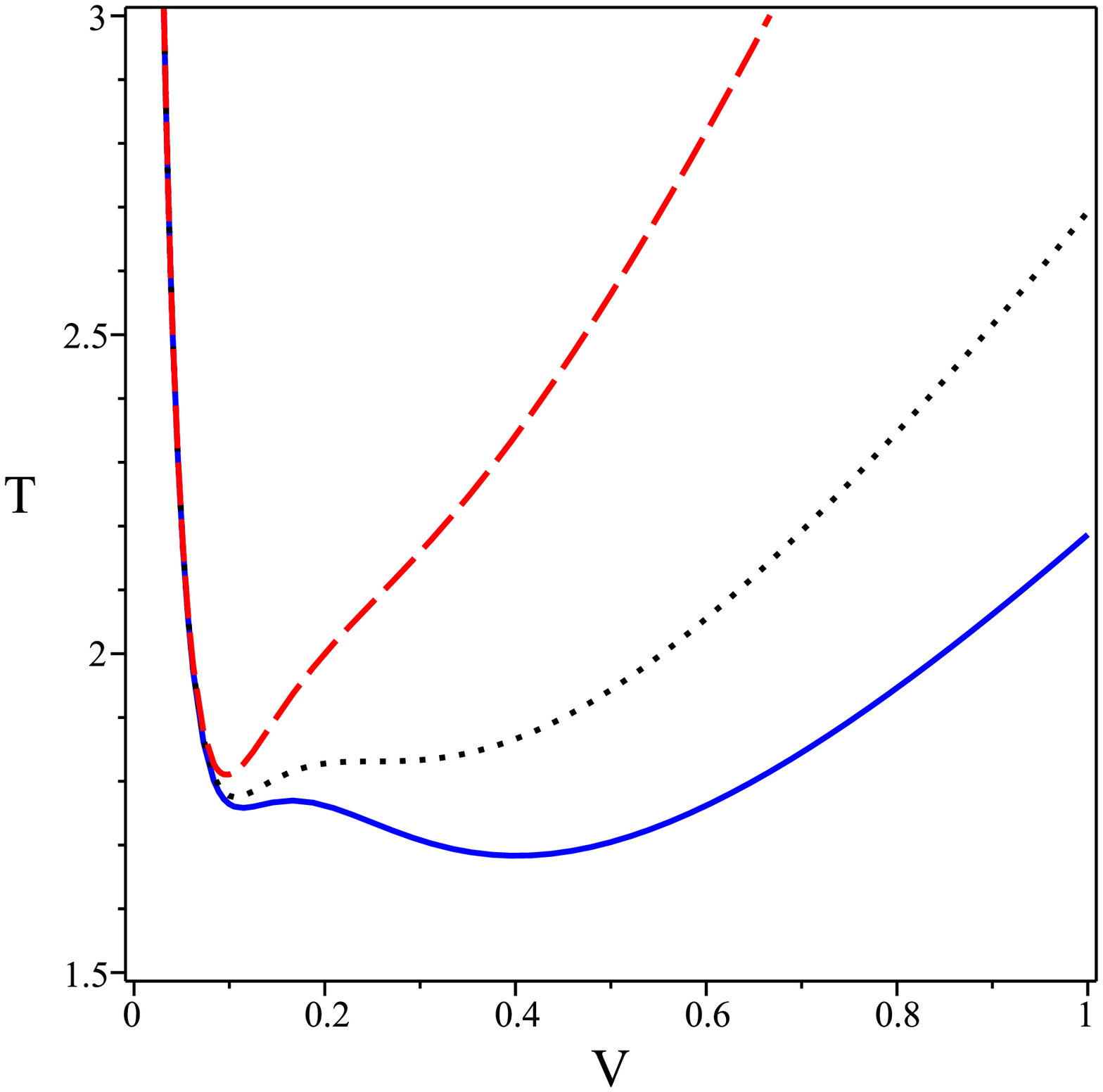}
	\epsfxsize=5.5cm \epsffile{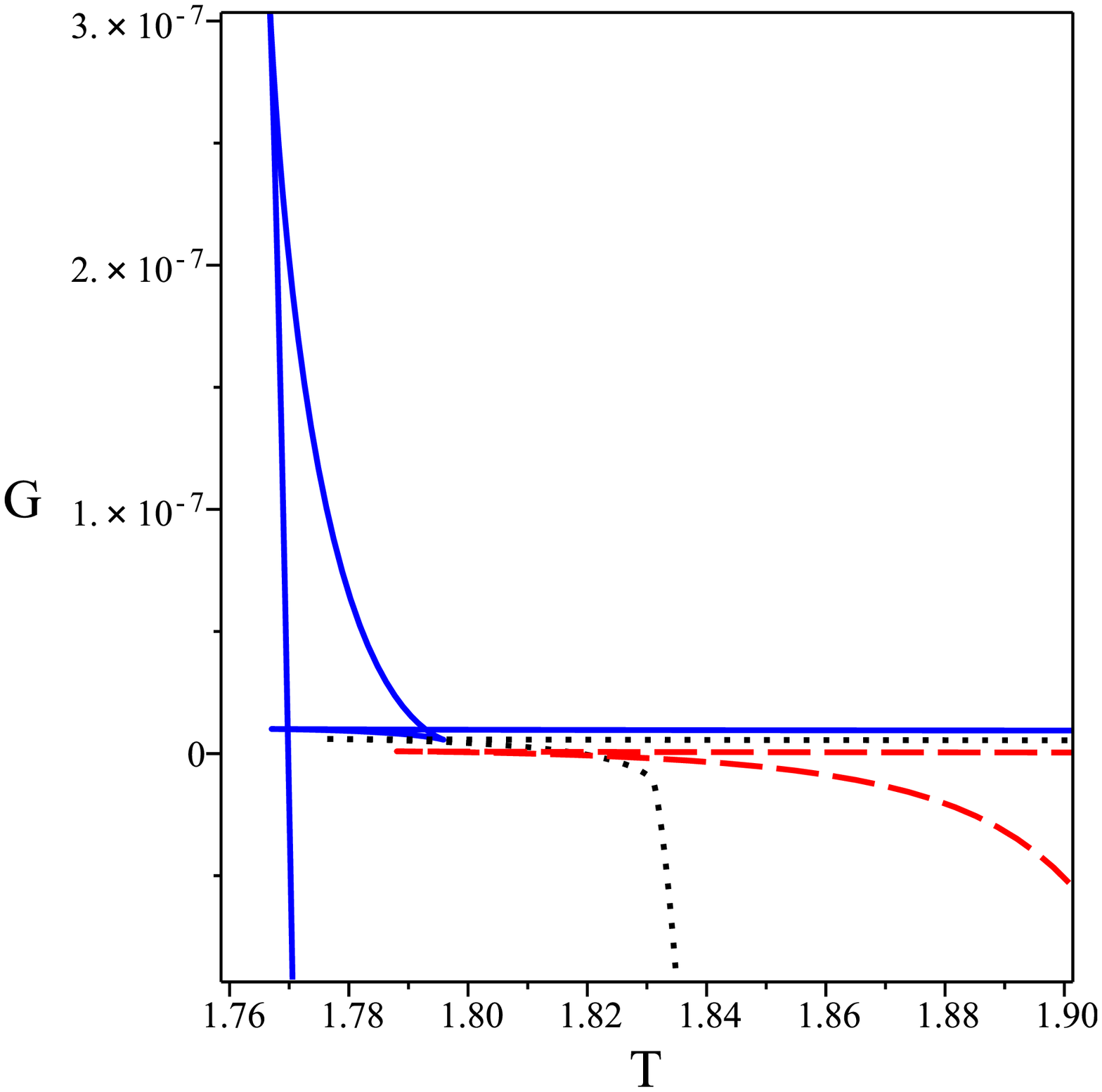}
	&  & 
	\end{array}
	$%
	\caption{\textbf{Reentrant phase transition of spherical black holes:} $P-V$ (left), $T-V$
		(middle) and $G-T$ (right) diagrams; we have set $k=1$, $d=10$, $q=0$, $m=0.1$, $c_{0}=1$, $c_{1}=-9$, $c_{2}=-2$, $c_{3}=-1$, $c_{4}= 0$ and $\alpha=0.01$. \newline
		\textbf{Left panel:} $T<T_{c}$ (continuous line), $T=T_{c}$ (dotted line) and $T>T_{c}$ (dashed line).\newline
		\textbf{Middle and right panels:} $P<P_{c}$ (continuous lines), $P=P_{c}$ (dotted lines) and $P>P_{c}$ (dashed lines).  \newline
	\textbf{Critical data:} $(P_{c_{1}}=2.74376, r_{c_{1}}=0.131663, T_{c_{1}}=1.74384)$ and $(P_{c_{2}}=4.47518, r_{c_{2}}=0.249414, T_{c_{2}}=1.83107)$. \newline
     \underline{Note}: The values of Gibbs free energies in the $G-T$ diagram (continuous and dashed lines) are slightly shifted up since their isobaric curves overlapped.}
	\label{PV-Reentrant-spherical}
\end{figure}

\begin{figure}[!htbp]
	$%
	\begin{array}{ccc}
	\epsfxsize=5.5cm \epsffile{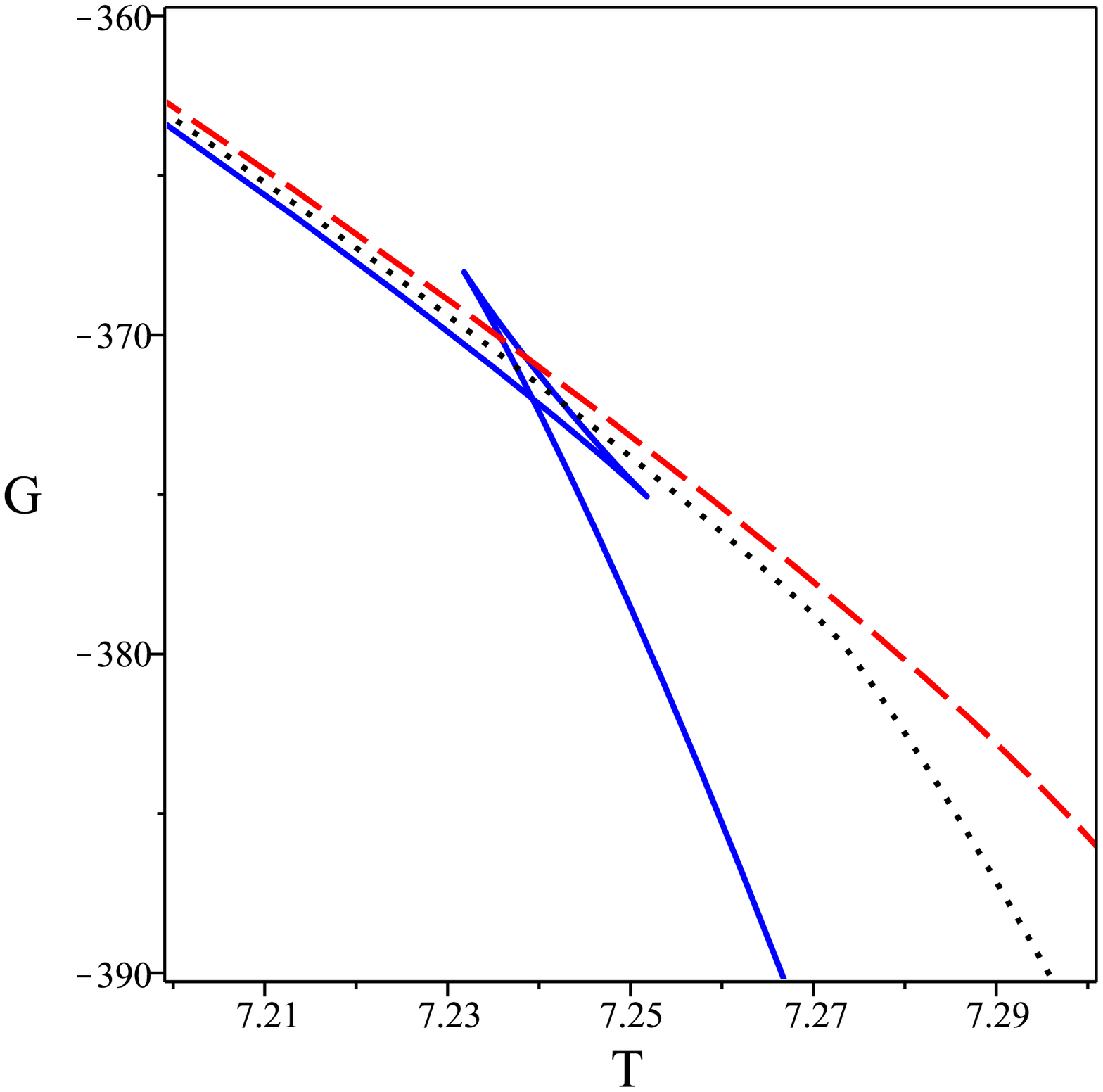}
	\epsfxsize=5.5cm \epsffile{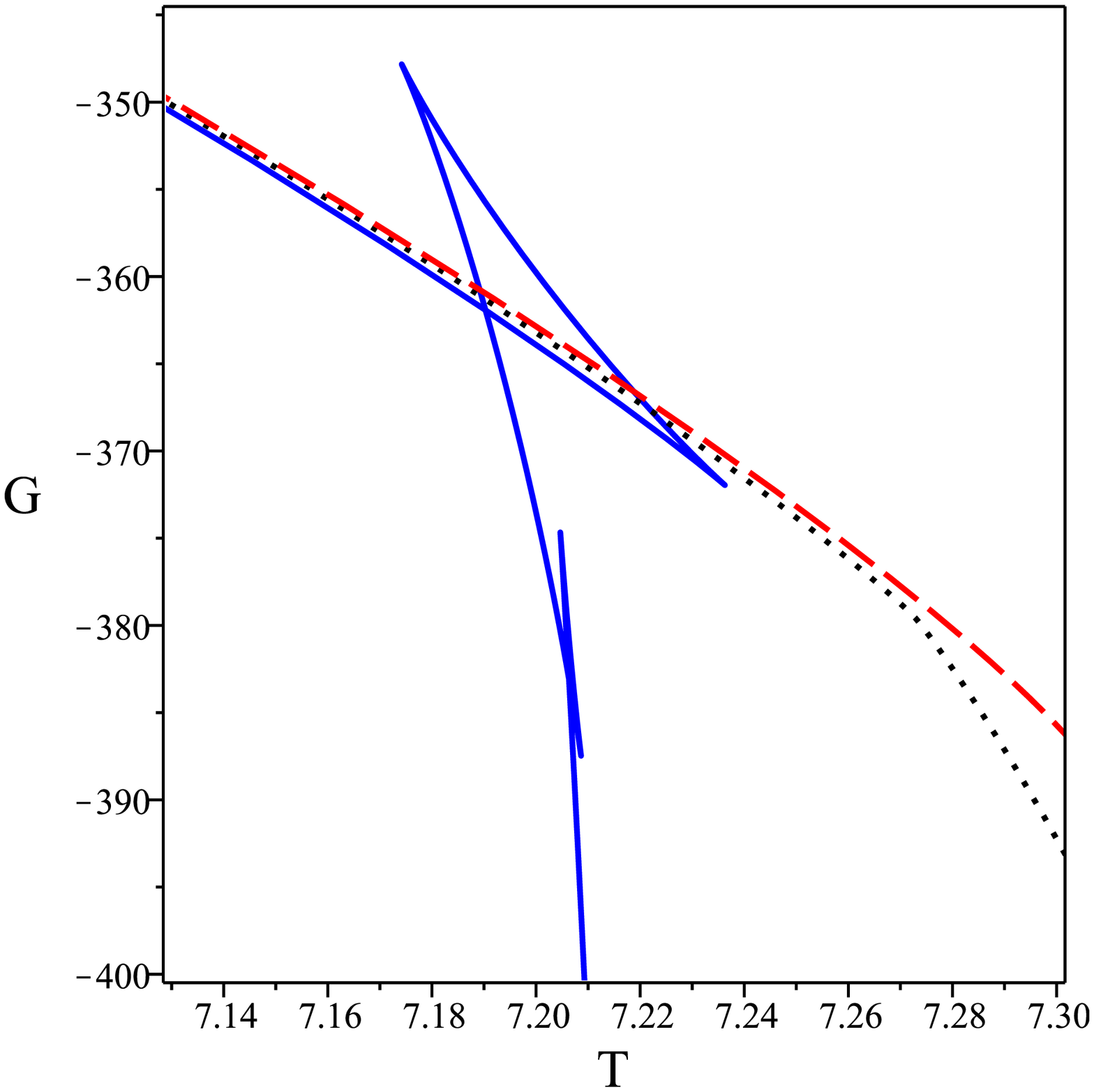}
	\epsfxsize=5.5cm \epsffile{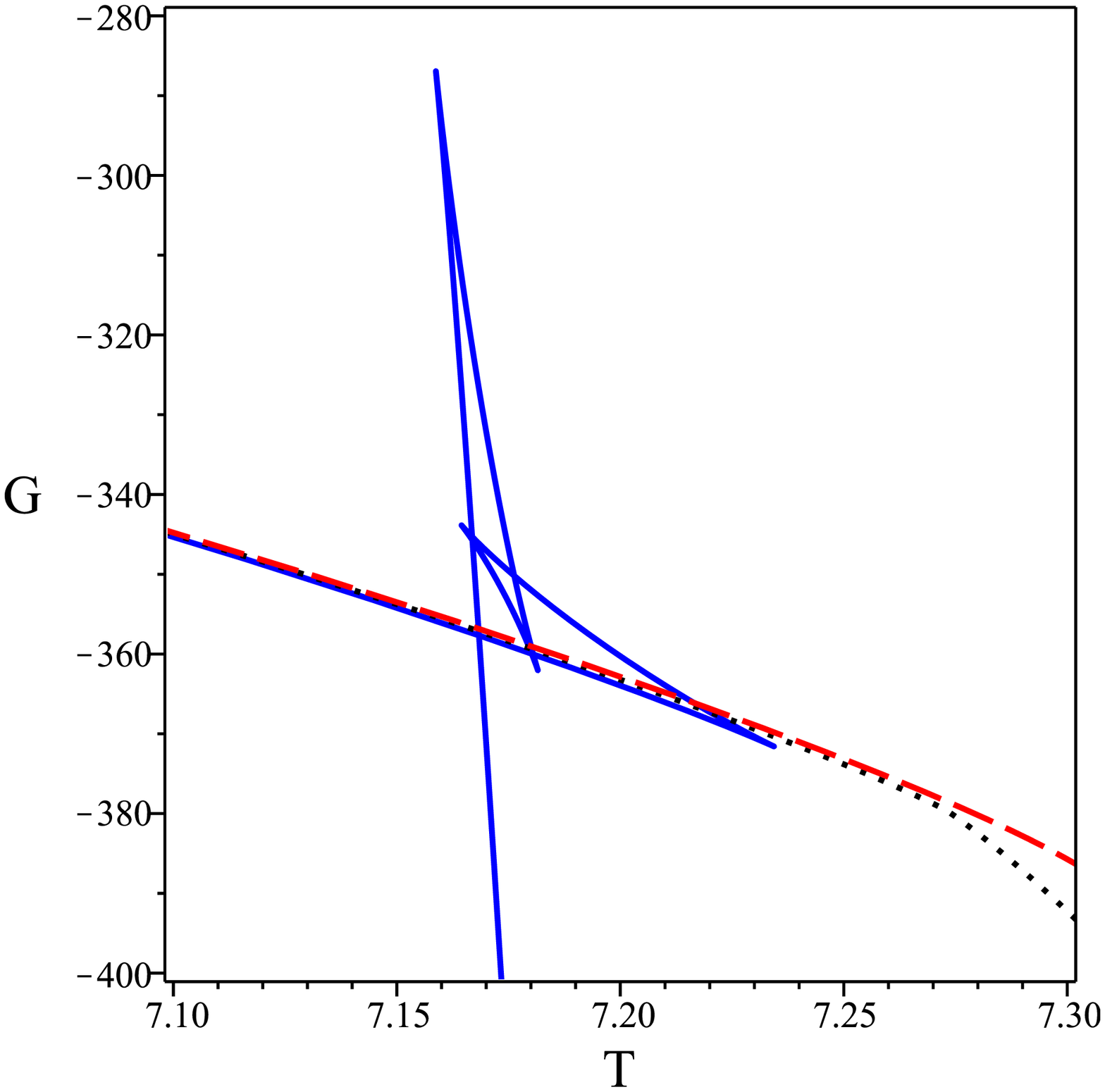}
	&  & 
	\end{array}
	$%
	\caption{\textbf{Analogue of triple point in spherical black holes:} $G-T$ diagrams; we have set $k=1$, $d=7$, $q=0$, $m\simeq5.98$, $c_{0}=1$, $c_{1}\simeq -3.98$, $c_{2}\simeq 12.8$, $c_{3}\simeq12.02$, $c_{4}= 0$ and $\alpha \simeq 9.08$. \newline
		\textbf{Left panel:} $P_{c_{2}}<P<P_{c_{1}}$ (continuous line), $P=P_{c_{1}}$ (dotted line) and $P>P_{c_{1}}$ (dashed line).\newline
		\textbf{Middle panel:} $P_{tr}<P<P_{c_{2}}$ (continuous line), $P=P_{c_{1}}$ (dotted line) and $P>P_{c_{1}}$ (dashed line).\newline
		\textbf{Right panel:} $P\approx P_{tr}<P_{c_{2}}$ (continuous line), $P=P_{c_{1}}$ (dotted line) and $P>P_{c_{1}}$ (dashed line). \newline
		\textbf{Critical data:} $(P_{c_{1}}=1.35184, r_{c_{1}}=2.23576, T_{c_{1}}=7.27288)$,  $(P_{c_{2}}=1.126321, r_{c_{2}}=5.54664, T_{c_{2}}=7.22339)$ and $(P_{tr}=1.08035, r_{tr}=3.73653, T_{tr}=7.14256)$.}
	\label{GT-Triple-spherical}
\end{figure}

Investigation shows the LM AdS black holes with spherical horizon may have up to three physical critical points for charged and uncharged cases. In order to observe the reentrant behavior of phase transition for spherical black holes, we have tuned the massive coupling coefficients to produce two physical critical points and plotted $P-V$, $T-V$ and $G-T$ diagrams in Fig. \ref{PV-Reentrant-spherical}. In this case, there are two (physical) critical points, referred as $r_{c_{1}}$ and  $r_{c_{2}}$. As seen, by monotonic decreasing the temperature, the black hole system undergoes a reentrant phase transition for the certain range of pressure.
By another tuning, we arrive at one triple point ($P_{tr}$) and two critical points ($r_{c_{1}}$ and $r_{c_{2}}$). This situation is depicted in Fig. \ref{GT-Triple-spherical} in which the Gibbs free energy is displayed near the critical points for various pressures. It should be mentioned that there is a lower value for the $U(1)$ charge ($Q_{b}$), where for $Q>Q_{b}$, one of the critical points disappears. Hence, in the case of charged-AdS black holes, the analogue of triple point and solid/liquid/gas phase transition can be found only for small enough values of the electric charge, $Q$.

\begin{figure}[!htbp]
	$%
	\begin{array}{ccc}
	\epsfxsize=5.5cm \epsffile{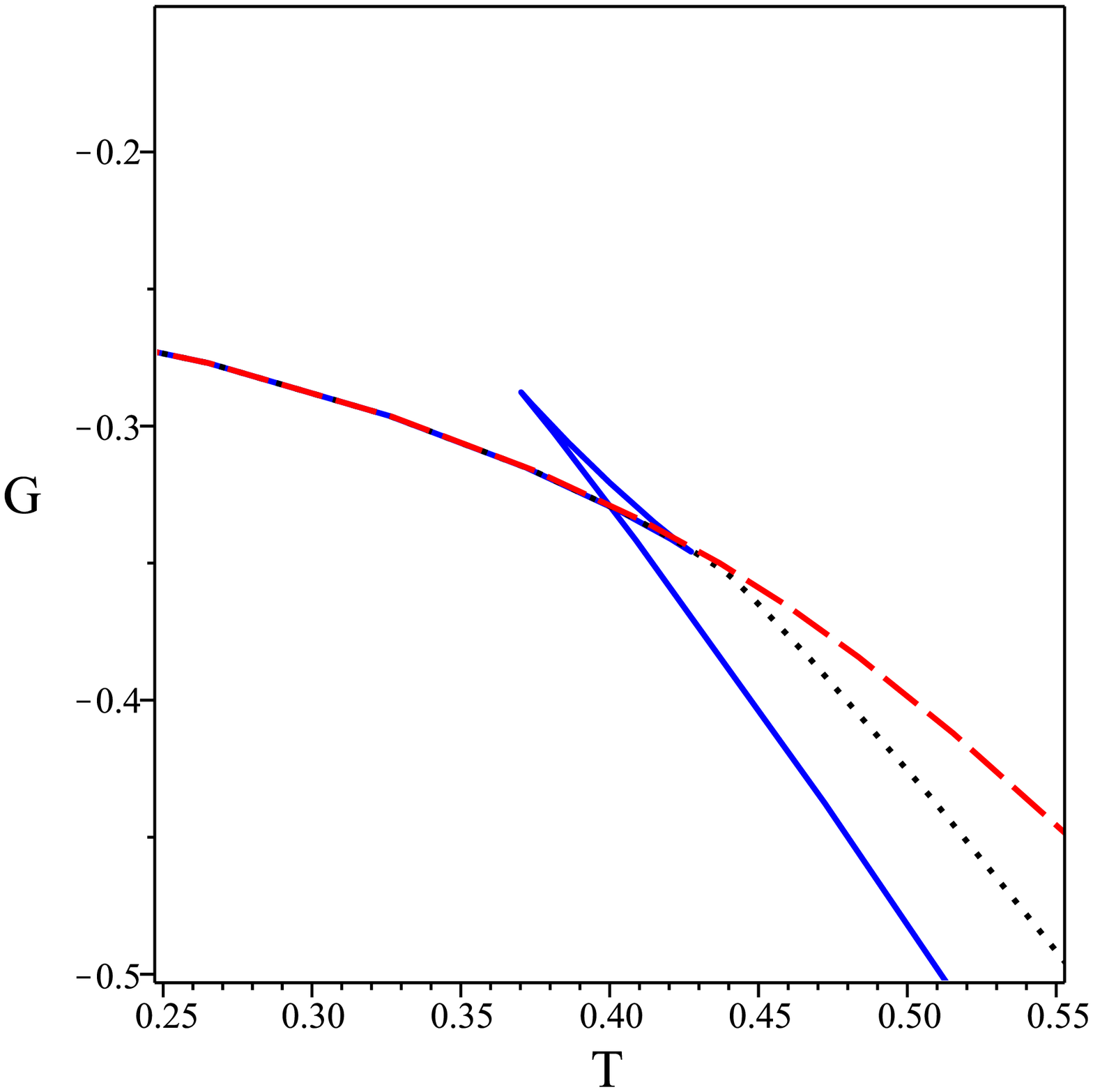}
	\epsfxsize=5.5cm \epsffile{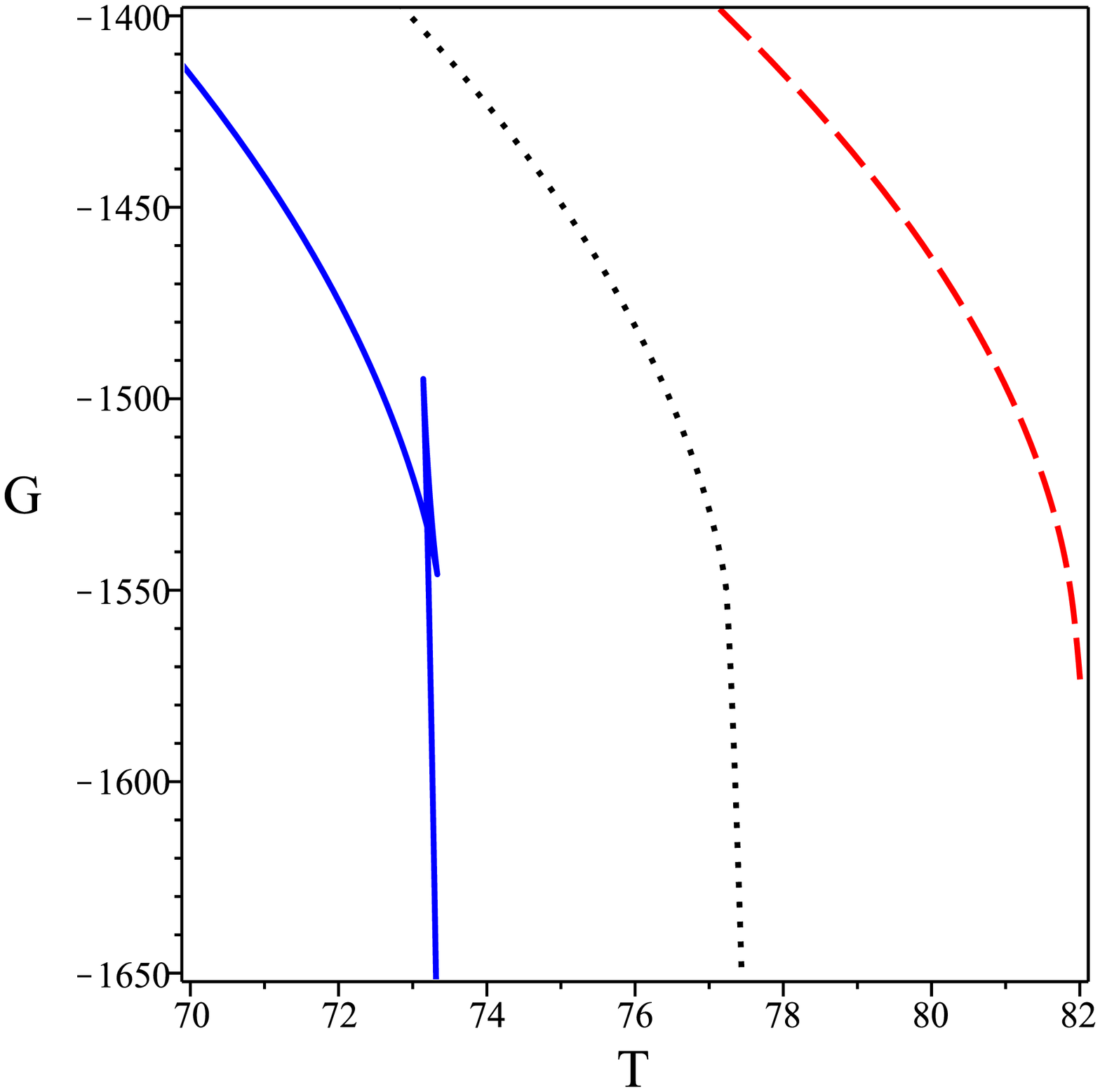}
	\epsfxsize=5.5cm \epsffile{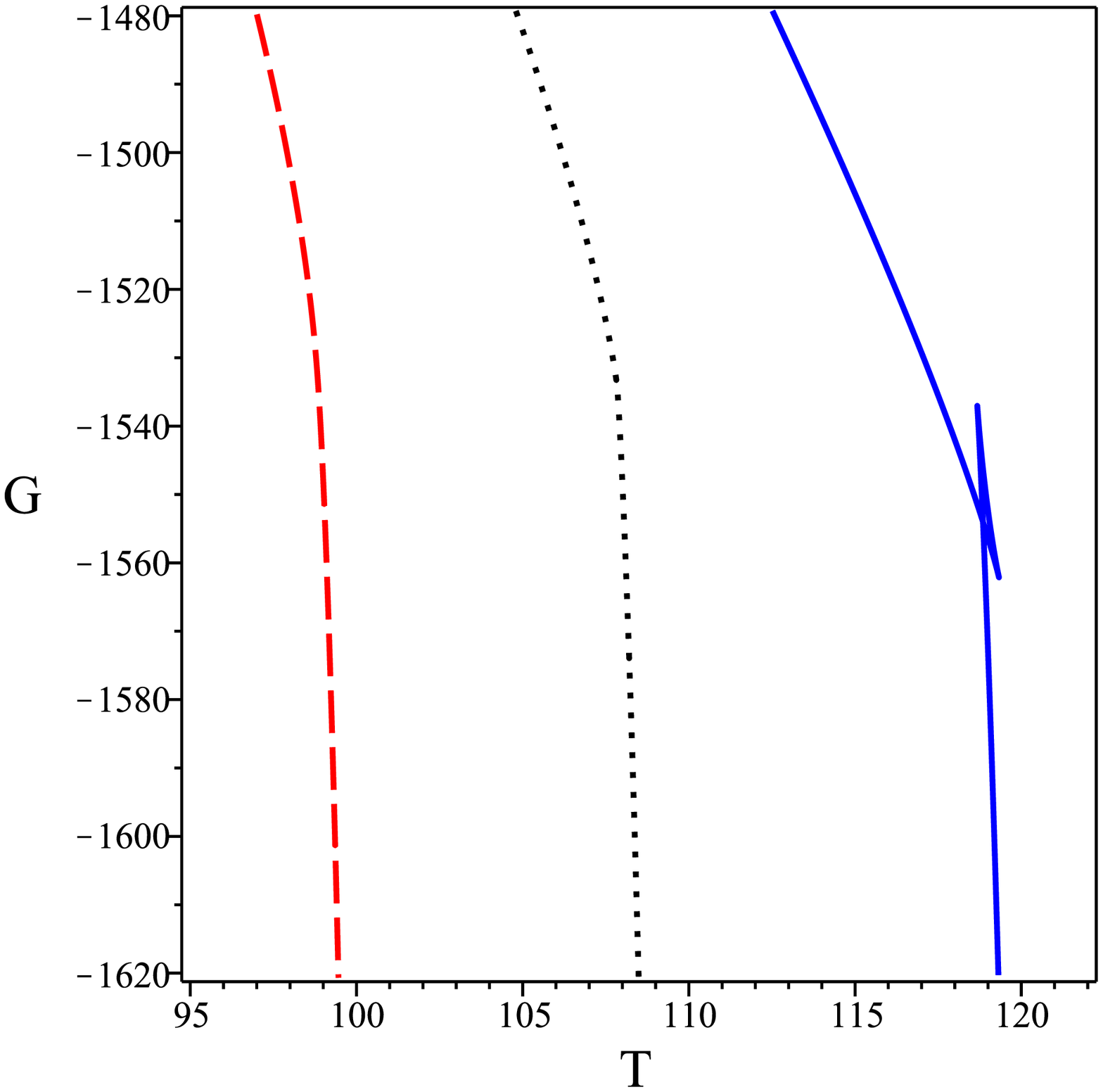}
	&  & 
	\end{array}
	$%
	\caption{\textbf{Hyperbolic black holes with three critical points:} $G-T$ diagrams; we have set $k=1$, $d=7$, $q=0$, $m\simeq5.52$, $c_{0}=1$, $c_{1}\simeq 12.8$, $c_{2}\simeq 16.9$, $c_{3}\simeq -19.9$, $c_{4}\simeq 1.42$ and $\alpha \simeq 2.71$. \newline
		\textbf{Left panel:} $P<P_{c_{1}}$ (continuous line), $P=P_{c_{1}}$ (dotted line) and $P>P_{c_{1}}$ (dashed line).\newline
		\textbf{Middle panel:} $P<P_{c_{2}}$ (continuous line), $P=P_{c_{2}}$ (dotted line) and $P>P_{c_{2}}$ (dashed line).\newline
		\textbf{Right panel:} $P<P_{c_{3}}$ (continuous line), $P=P_{c_{3}}$ (dotted line) and $P>P_{c_{3}}$ (dashed line). \newline
		\textbf{Critical data:} $(P_{c_{1}}=68565.2, r_{c_{1}}=0.09863, T_{c_{1}}=0.44008)$,  $(P_{c_{2}}=5.03994, r_{c_{2}}=4.18428, T_{c_{2}}=77.2272)$ and $(P_{c_{3}}=11.7248, r_{c_{3}}=3.33423, T_{c_{3}}=107.841)$.}
	\label{GT-3points-hyperbolic}
\end{figure}

The phase structure of hyperbolic black holes is really rich and drastically different from those with spherical and Ricci flat horizons. In both charged and uncharged cases, three (physical) critical points can be found for hyperbolic black holes. Interestingly, the analogue of triple point does not exist in the phase structure of these black holes. In fact, besides the existence of the two critical points correspondence to two distinct first order transition, we arrive at an additional reverse van der Waals phase transition associated to the third critical point in the phase space. That situation is illustrated in Fig. \ref{GT-3points-hyperbolic} which is a generic feature of this model and persists in all dimensions. This is the first example of such phase structure and not possible for spherical and Ricci flat black holes.

We could not find any evidence related to the existence of four critical points in this model. The existence of four critical points may be potentially possible when the phase space of the spherically symmetric AdS black holes in LM gravity is enriched by adding nonlinear $U(1)$ gauge fields in the theory. 

\section{Concluding remarks} \label{conclusion}

The effects of massive and Lovelock gravities are encoded in the deformation parameters $m$ and $\alpha$ respectively. In the Lovelock massive (LM) gravity, one can simply recovered the outcomes of Einstein (by $\alpha, m\longrightarrow0$), Lovelock (by $m\longrightarrow0$) and massive (by $\alpha\longrightarrow0$) theories of gravity. Considering LM gravity, in this paper, we have introduced topological black hole solutions and then analyzed thermodynamic properties and critical behavior of AdS black holes in the extended phase space. The asymptotic behavior of the black hole solutions may be (A)dS or flat, and by computing the thermodynamic quantities, we have shown they satisfied the first law of thermodynamics. \\

Next, by treating the cosmological constant as a thermodynamic variable (pressure), we extended the thermodynamic phase space, and proved the massive coupling and Lovelock coefficients as well as cosmological constant are required for consistency of the extended first law of thermodynamics with the Smarr formula. In addition, we examined thermal stability of the Lovelock massive AdS black holes in the canonical ensemble and showed the qualitative behavior of heat capacity for AdS black holes with different horizon topologies. In this regard, we mainly focused on the topology of event horizons and showed in what regions the topological black holes are thermally stable. \\

In LM gravity, critical behavior and phase transition occur for all types of AdS black holes (with spherical, Ricci flat and hyperbolic topologies for event  horizon) in contrast to Einstein gravity which only admits phase transition for spherically symmetric ones. In addition, phase transitions occur in both canonical and grand canonical ensembles in contrast to Reissner-Nordstr\"{o}m-AdS black hole where criticality cannot happen in grand canonical ensemble \cite{KubiznakMann2012}. For Ricci flat black holes, phase transition originated only from the interacting terms of massive gravitons and the effect of higher curvature terms vanishes. Interestingly, we found that there is a lower value for the graviton mass parameter, referred as $m_{b}$, in which no phase transition takes place in region $m<m_{b}$. This is one of the remarkable characteristics of massive gravity. For hyperbolic black holes, two radically different first order transitions are observed: i) a (normal) van der Waals like behavior, and ii) reverse van der Waals like behavior. The reverse behavior of van der Waals phase transition completely comes from the higher curvature terms of Lovelock Lagrangian which is not seen in Gauss-Bonnet gravity (as indicated in \cite{Frassino2014,Cai2013GaussBonnet}, Gauss-Bonnet black holes with hyperbolic horizon do not admit physical phase transition). The reverse behavior predicts that hyperbolic black holes could experience first order phase transition at high temperature and pressure, which is a novel effect. It was shown the inclusion of higher curvature terms (based on Lovelock Lagrangian) affects the criticality. In fact, for LM AdS black holes with diverse horizon topologies, depending on the chosen parameters, there is always an upper limit for the value of Lovelock coefficient ($\alpha_{u}$) in which no phase transition could happen for $\alpha>\alpha_{u}$. Considering tables, we found that the universal ratio, i.e. $\frac{{{P_c}{v_c}}}{{{T_c}}}$, is function of spacetime dimensions ($d$), topological factor ($k$), graviton mass parameter ($m$) and strength of higher curvature terms (captured with Lovelock coefficient, $\alpha$). \\

In addition, the van der Waals, reentrant and analogue of solid/liquid/gas phase transitions were found in the extended phase space of (un)charged-AdS black holes with spherical horizon. But, in the case of hyperbolic black holes, reentrant and small/intermediate/large phase transitions were not found. Indeed, the reverse van der Waals phase transition in the phase space of hyperbolic black holes is accompanied with one or two distinct (standard) van der Waals phase transitions. To our knowledge, this is the first example of such phase structure. These pieces of evidence shows the generic features of different theories of gravitation can be summed into a unique model to produce more complex structures for thermodynamic phase space of black holes.


\begin{acknowledgements}
We wish to thank Shiraz University Research Council. We acknowledge M. Momennia for reading the manuscript and useful comments. AD would like to thank S. Zarepour for useful discussions and providing Mathematica programming codes. S.H.H would like to thank the hospitality of the Institute of Physics, University of Oldenburg during his short visit. This work has been supported financially by the Research Institute for Astronomy and Astrophysics of  Maragha, Iran.
\end{acknowledgements}


\end{document}